\documentclass[12pt, a4paper]{article}
\usepackage{hyperref}
\usepackage{graphicx}
\usepackage{mathrsfs}
\usepackage{amssymb}
\usepackage{color}
\usepackage{amsmath}
\usepackage{ascmac}
\usepackage{amsthm}
\usepackage{booktabs}
\usepackage{lscape}
\usepackage{dcolumn}
\usepackage{longtable}
\usepackage{dcolumn}
\usepackage{arydshln}
\usepackage{bm}
\usepackage{caption}
\usepackage{subfig}
\usepackage{setspace}
\usepackage{cases}
\usepackage{comment}
\usepackage{bigfoot}
\usepackage{multirow}
\usepackage{booktabs}
\usepackage[square, sort,comma,numbers]{natbib}
\bibpunct[, ]{(}{)}{;}{a}{}{,}
\usepackage{url}
\usepackage{authblk}

\makeatletter

\begin{document}

\title{
\LARGE{Risk-coping Behaviors in Metropolis:\\ Evidence from Working-class Households\\ in Prewar Tokyo}}

\author{Kota Ogasawara\thanks{
Department of Industrial Engineering and Economics, School of Engineering, Institute of Science Tokyo, 2-12-1, Ookayama, Meguro-ku, Tokyo 152-8552, Japan (E-mail: ogasawara.k.ab@m.titech.ac.jp).\newline
I wish to thank Stephen Broadberry, Andrew Foster, Oscar Gelderbloom, Bishnupriya Gupta, Janet Hunter, Kris Inwood, Tom Learmouth, Ralf Meisenzahl, Safya Morshed, Naofumi Nakamura, Yasuyuki Sawada, Tavneet Suri, Masayuki Tanimoto, and two anonymous referees.
I would also like to thank the seminar participants at the University of Tokyo, Meiji, Toronto, Strathclyde, and Hohenheim for their helpful comments.
I am thankful to the Ohara Institute for Social Research for providing access to the archives of the Tsukishima Survey and Gonda Yasunosuke.
I am grateful to the CIRJE at the University of Tokyo for granting me the opportunity to be a short-term visitor in 2024.
There are no conflicts of interest to declare.
All remaining errors are my own.\newline
}
}

\affil{Institute of Science Tokyo}

\date{\today}

\maketitle

\begin{abstract}
\begin{spacing}{0.85}
I analyze the risk-coping strategies of factory-worker households in early twentieth-century Tokyo. I digitized a unique daily longitudinal household budget survey conducted in Tsukishima, a representative manufacturing area, to examine how consumption was affected by idiosyncratic shocks. I find that although the households were vulnerable and their consumption levels were impacted by these shocks, the estimated income elasticity of indispensable consumption was relatively low in the short run. The results of the mechanism analysis suggest that credit purchases from local retailers helped smooth short-run consumption, highlighting the role of informal credit institutions in mitigating vulnerability among urban worker households.
\end{spacing}
\bigskip

\noindent\textbf{Keywords:}
Consumption smoothing;
credit institution;
urban local community;
measurement error;
risk-coping strategy;
risk-sharing
\bigskip


\end{abstract}

\newpage
\section{Introduction}

During the early stages of industrialization, workers who migrated to large cities in search of employment faced various uncertainties.
Lacking the communal ties found in rural areas, they were forced to manage on their own.
The vulnerability to idiosyncratic income shocks makes it challenging to smooth consumption and increases the risk of falling into poverty.
Therefore, understanding how these workers managed their livelihoods helps clarify the mechanisms behind improvements in living standards in industrializing cities.

This study examines risk-coping behaviors among working-class households in Tsukishima, a representative manufacturing area in early twentieth-century Tokyo City.
Against the backdrop of industrialization during World War I (WWI), many workers migrated to the city, forming a notable portion of the population.
Despite the absence of public insurance systems, evidence shows that these workers were able to form and maintain families during this period \citep{Nakagawa1985}.
This historical context offers a rare empirical opportunity to understand the risk-coping strategies of urban workers during industrialization.

To provide micro-level evidence on risk-coping behaviors, I digitize a unique daily longitudinal budget survey of factory-worker households.
Using a systematic consumption-smoothing regression, I find that the full insurance hypothesis is strongly rejected for overall consumption expenditure.
However, I also find evidence that expenses on indispensable consumption categories are partially smoothed in the short run: the estimated income elasticity for the food category is less than half of that for overall consumption.
I then use several temporary income categories within the same regression framework to assess the risk-coping behaviors.
Considering both favorable and adverse income shocks, I find that savings, credit purchases, and gifts responded to these shocks.
To focus specifically on responses to adverse income shocks, I use the household head's absence due to the influenza pandemic as an instrument.
Results from the instrumental variable approach show that credit purchases from local retailers played a significant role in mitigating adverse idiosyncratic shocks, whereas other risk-mitigation strategies were less practical.

Given these results, I analyze the institutional aspects of credit purchases descriptively.
First, I find that credit purchases were common among workers and that volatility in household heads' earnings did not correlate with average credit purchases.
Detailed cases suggest that risk premiums were not charged as explicit interest but were instead embedded in retail prices owing to imperfect customer information.
Second, I examine conditions among retailers using several official survey reports.
These reports provide suggestive evidence that retailers competed intensely within the same sales areas.
Because credit purchases did not require explicit contracts, they often led to losses from unpaid sales.
However, to retain regular customers and build loyalty, local retailers had strong incentives to offer credit transactions.

The primary contribution of this study is the revelation of the role of informal credit institutions in a historical metropolis.
By considering both formal and informal financial institutions simultaneously, this study documents how informal credit transactions with local retailers helped mitigate idiosyncratic income losses.
This finding adds to the literature on consumption-smoothing behavior across various economies,\footnote{The early representative literature includes \citet{Townsend1994-kq}, \citet{Townsend1995-om}, and \citet{Ravallion1997-kz}. \citet{Morduch1999-qq}, \citet{Townsend1995-om}, and \citet{Dercon2004} review earlier studies. \citet{Attanasio2016-fi} and \citet{Meyer2023} provide recent discussions.}
as well as the studies investigating the mechanisms behind human capital formation through consumption smoothing.\footnote{Evidence shows that markets in contemporary developing countries are often imperfect. Consequently, households remain vulnerable to risk, and idiosyncratic shocks affect their consumption levels, influencing human capital formation \citep{Foster1995-co, Rose1999-ro, Gertler2002-tu}.}
Specifically, I present a case study of an urban setting characterized by weak communal ties, which has often been overlooked in the literature.
For example, development economics research shows that informal credit institutions based on relatives and neighbors help stabilize consumption in rural economies, especially where formal financial institutions are scarce \citep{Udry1994-vp, Udry1990-fs, Fafchamps2003-rt}.\footnote{Related studies examine risk-sharing among households with similar cultural and institutional backgrounds \citep{Grimard1997-uz, Munshi2016-ac}. Recent field experiments also focus on village economies \citep{Comola2023-sx, Macours2022-iv}. \citet{Townsend1995-om}, \citet{Skoufias2005-nb}, \citet{Aguila2017-jv}, and \citet{Ogasawara2024} provide case studies on middle-sized contemporary and historical economies, but evidence on agglomeration economies such as Tokyo City remains scarce.}
Another important strand of the literature highlights the role of social insurance provided by religious communities in rural contexts \citep{AgerUnknown-qj, Ferrara2023-jm, Chen2010-yb}.
My findings suggest that, in the case of a metropolis, workers primarily relied on informal credit institutions arising from local commercial practices, as social institutions based on communal ties and religious organizations were largely absent.
This result may also offer empirical lessons for rapidly urbanizing cities in developing countries.\footnote{The real GDP per capita in 2011 dollars was $3,217$ in 1919 Japan \citep{Bolt2025-fv}. This figure is close to the $3,537$ reported for South and Southeast Asian countries in 2010, where rapid urbanization has been observed \citep{Henderson2020-su}.}

The second contribution is the incorporation of analytical perspectives on consumption-smoothing behavior into economic history studies.
I examine a set of homogeneous households in a representative manufacturing area, which provides a valuable empirical setting for testing consumption smoothing.\footnote{Focusing on a sample that shares similar occupational and regional characteristics systematically reduces bias when estimating income elasticity \citep{Schulhofer-Wohl2011-fo, Mazzocco2012-jl}.}
This study advances economic history research that documents compensatory responses to income loss among working-class households but lacks direct evidence on consumption smoothing \citep{Scott2012-pc, Horrell2000-nb, Kiesling1996-ku, Boyer1997-th}.
I implement a systematic panel data analysis to provide comparable estimates of income elasticity.
This comparability improves understanding of the availability of risk-coping institutions at different stages of economic development and strengthens the potential for cross-period and cross-economy comparisons.
It also adds an analytical perspective to other strands of literature on the supply-side evidence of financial institutions \citep{OConnell2005-wy} and historical patterns of consumption and production \citep{Francks2009, Gordon2013}.

Lastly, examining how, and to what extent, local society could secure household consumption offers a new perspective within the urban history literature.
A representative historical study on urban households shows that workers in Tokyo were able to establish and sustain their families around WWI \citep{Nakagawa1985}.
The findings of this study suggest that the expansion of urbanization and credit transactions supports this traditional view.
Another implication relates to the findings of \citet{Munshi2016-ac}, who link spatial wage inequality and men's mobility in India to the availability of risk-mitigation networks among workers within the same caste.
From this perspective, the existence of local credit institutions with lower entry barriers may have encouraged rural workers to migrate to urban areas, helping to explain Tokyo's rapid growth around WWI.

\section{Historical Background} \label{sec:sec2}

\subsection{Japanese Economy \textit{circa} the First World War} \label{sec:sec21}

\begin{figure}[hbtp]
\centering
\captionsetup{justification=centering,margin=1.5cm}
\includegraphics[width=0.45\textwidth]{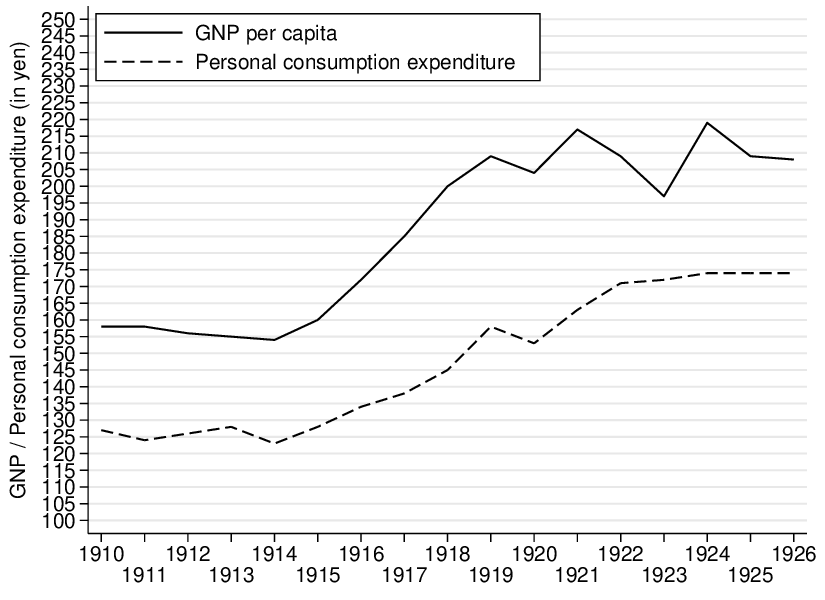}
\caption{Gross National Product per Capita and Personal Consumption Expenditure (in yen)}
\label{fig:gnp_pce}
\scriptsize{\begin{minipage}{350pt}
\setstretch{0.85}
Note:
This figure shows Japan's GNP per capita (solid line) and personal consumption expenditure (dashed line) from 1910 to 1926.
Prices are deflated using the 1934--1935 consumer price index (CPI).
Source: Created by the author based on Ohkawa et al. (1974, p.~237).
\end{minipage}}
\end{figure}

Between 1885 and 1915, Japan's real gross national product (GNP) grew at an annual rate of 2.6\%, while the population increased by 1.1\%.\footnote{Faster economic growth relative to population growth indicates that modern economic growth had begun during this period, as defined by Kuznets \citep[][pp.~60; 64]{Miyamoto2008}. In 1915, the employed population was distributed across the primary, secondary, and tertiary sectors at 63\%, 20\%, and 18\%, respectively. Based on domestic genuine production, these sectors accounted for just under 30\%, over 30\%, and approximately 40\%, respectively. Descriptions in this subsection are based on \citet[pp.~128; 142; 145; 147]{Nakamura1971}, \citet[pp.~10; 21--23]{Nakamura1989}, and \citet[][pp.~250; 253--254]{Sawai2016}.}
Soon after the outbreak of WWI, European exports declined.
Japan increased its exports to Asian markets in place of European suppliers and exported munitions and foodstuffs to European allies, resulting in a current account surplus.
Import substitution further expanded the domestic market, driving growth in the machinery, metal, and chemical sectors.
A global shortage of ships also spurred rapid development in shipbuilding, which expanded significantly during this period.
The growth of the shipbuilding industry, in turn, stimulated demand in the machinery and steel industries.
Figure~\ref{fig:gnp_pce} shows that Japan's nominal GNP grew by $11.5$\% from 1915 to 1919 ($4.4$\% in real terms), marking the highest growth rate observed throughout the 1910s and 1920s.

The rise of business accelerated urbanization as the rural population moved to cities in search of employment.
The share of large cities (with a population of $50,000$ or more) in the total population was less than $10$\% in 1889 but rose to nearly $16$\% in 1920 (Online Appendix Figure~\ref{fig:citizens}).
Growth was particularly rapid during WWI, indicating that structural changes during the war accelerated urban population growth.
Urbanization also increased public investment in large cities, especially in social infrastructure such as roads and water supply systems.

Exports in the 1920s were lower owing to the appreciation of the yen in real terms, caused by the accumulation of domestic and foreign currencies during WWI.
The current account deficit was settled through the disbursement of foreign currencies held by the government and the Bank of Japan, so the domestic money supply remained unchanged despite the sizeable deficit.
As a result, domestic prices stayed high.
The first half of the 1920s saw a temporary decline in industrial investment as the WWI-era capital investment boom ended under falling demand.

Overall, while a short-term business cycle emerged--marked by the WWI-induced boom and subsequent recession--Japan achieved a high economic growth rate relative to other countries in the early twentieth century.
The average growth rates of personal consumption expenditure reached 112\% in 1915--1919 and 118\% in 1920--1924 (Figure~\ref{fig:gnp_pce}), reflecting a high level by international standards.\footnote{Total consumption as a percentage of the GNP remained above 80\%. Although this level is not internationally high-standard \citep{Kuznets1962-qi}, the growth rate of real personal consumption per capita was relatively high; from 1875 to 1939, the average growth rate of personal consumption was 1.36\%, and it was never negative except during the Russo--Japanese War period \citep[pp.~2--5; 27; 30--31]{Nakamura1971}.}

\subsection{Tsukishima: A Representative Manufacturing Area} \label{sec:sec22}

\begin{figure}[hbtp]
\centering
\captionsetup{justification=centering,margin=1.5cm}
\includegraphics[angle =270, width=0.70\textwidth]{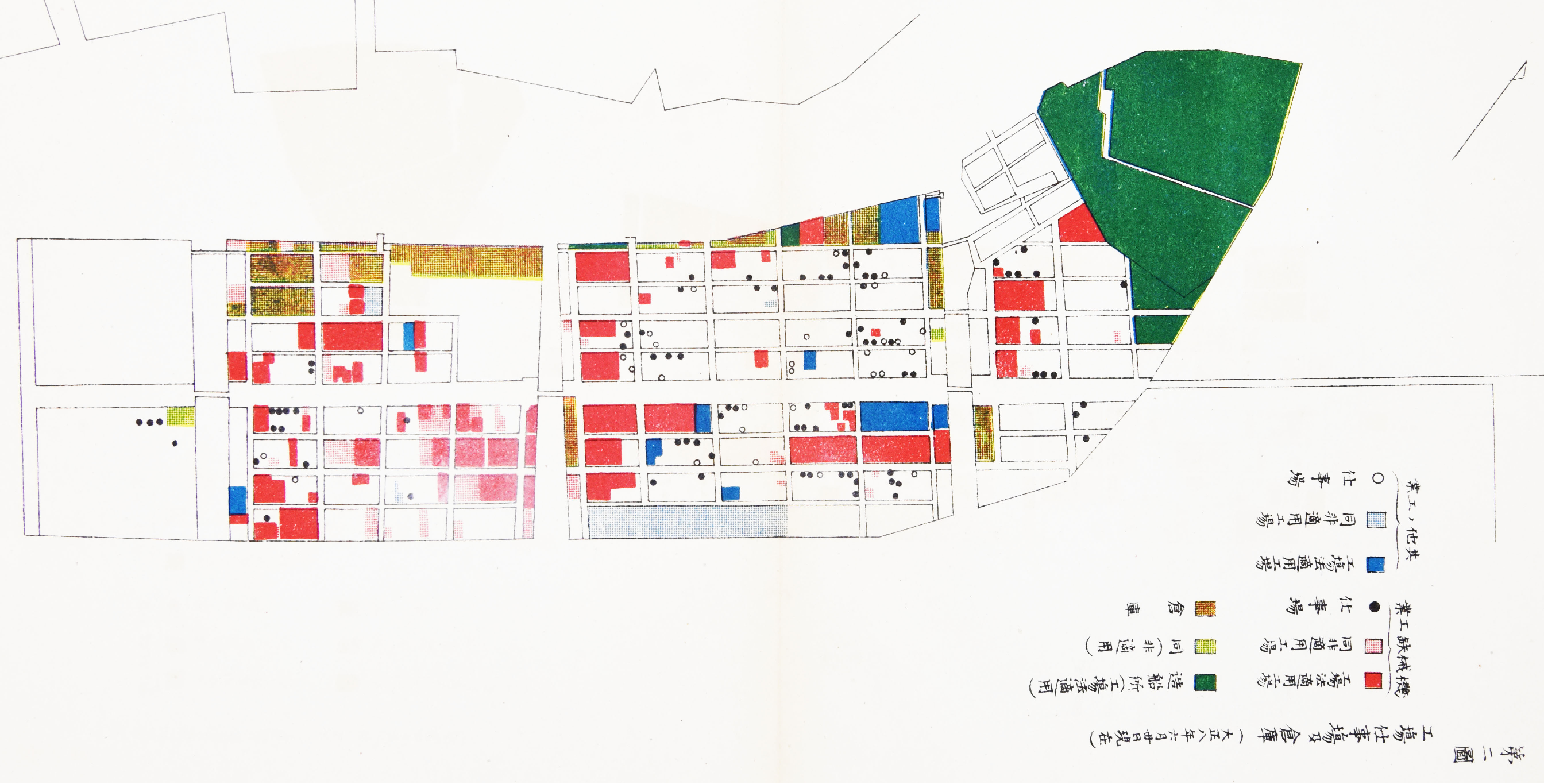}
\caption{Factories in Tsukishima}
\label{fig:tsukishima_factory}
\scriptsize{\begin{minipage}{350pt}
\setstretch{0.85}
Notes: 
Blocks in green and light green represent shipyards.
Blocks in red and pink represent factories in the machinery sector.
Blocks in blue and light blue represent factories in other manufacturing sectors.
Blocks in brown indicate warehouses.
Small circles colored black and white show workplaces in the machinery and other manufacturing sectors, respectively.
The upper right corner indicates north.
Source: 
Department of Health, Ministry of the Interior 1923c, second map.
The tone was adjusted by the author using Adobe Photoshop 24.7.0.
\end{minipage}}
\end{figure}

The population of Tokyo City had been growing since the end of the 19th century, and industrialization during WWI further spurred its growth.\footnote{Tokyo City's population share increased from $2.5$\% in 1889 to approximately $4$\% by 1920. Its population exceeded Osaka's, the second-largest city, by about 1 million. In 1920, Japan had more than $10,000$ municipalities, underscoring Tokyo's relative size. Online Appendix Figure~\ref{fig:citizens} summarizes the population trends of the top five cities between 1889 and 1920.}
Throughout the 1910s, the share of industrial and commercial workers in Tokyo City increased.
According to the 1920 census, nearly half of the male workers in Tokyo were employed in the industrial sector (Column 1 of Panel A in Table 1).

The subject of this study is Tsukishima, an island located southeast of Kyobashi Ward.\footnote{Online Appendix Figure~\ref{fig:map_tokyo} shows the location of Tsukishima. The development history of Tsukishima is summarized in Online Appendix~\ref{sec:seca_tsukishima}.}
Kyobashi has an industrial composition similar to that of Tokyo City (Column 2), and Tsukishima represents a typical industrial area within the ward, with approximately $68$\% of male workers employed in the manufacturing sector (Column 3).
In 1920, Tsukishima had $214$ factories, of which approximately $80$\% belonged to the machinery and related sectors.\footnote{Department of Health, Ministry of the Interior (1923a, pp.~385--389). According to the 1920 census, the metal and machinery equipment manufacturing industry ($27.1$\%) employed the largest share of workers in Tokyo. A similar pattern was observed in Kyobashi Ward ($35.5$\%) (Statistics Bureau of the Cabinet 1929c, pp.~85--86; 108--109).}
Figure~\ref{fig:tsukishima_factory} shows the spatial distribution of factories in Tsukishima as of June 30, 1918, demonstrating the concentration of machine factories and shipyards.\footnote{A representative shipbuilding plant, Ishikawajima Shipyard, was located on Ishikawajima, an island north of Tsukishima. Import substitution during this period strongly accelerated domestic ship production (Section~\ref{sec:sec21}). The production of a hydraulic 80-ton hardened crane by Ishikawajima Shipyard in 1917 exemplifies domestic output of a product that had previously relied on imports (Nakamura and Odaka 1989, pp.~10; 32).}

\begin{table}[h!]
\def\arraystretch{1.0}
\centering
\begin{center}
\caption{Industrial and Family Structures in Tsukishima}
\label{tab:tab1}
\scriptsize
\scalebox{0.94}[1]{
\begin{tabular}{lrrrr}
\toprule[1pt]\midrule[0.3pt]
\multicolumn{5}{l}{\textbf{Panel A: Industrial structure}}\\
&&&&\\
\cmidrule(rrr){3-5}
\multicolumn{2}{l}{Name of survey}	&(1) 1920 Population Census	&(2) 1920 Population Census	&(3) 1920 Population Census\\
\multicolumn{2}{l}{Survey area}		&Tokyo City						&Kyobashi Ward						&Tsukishima\\

\multicolumn{2}{l}{Survey subject}			&Complete survey&Complete survey&Complete survey\\
\multicolumn{2}{l}{Survey month and year}	&October 1920	&October 1920	&October 1920\\\hline

\multicolumn{2}{l}{Agriculture}						&1.0	&0.6	&0.4	\\ 
\multicolumn{2}{l}{Fisheries}						&0.1	&0.3	&0.8	\\
\multicolumn{2}{l}{Mining}							&0.4	&0.4	&0.8	\\
\multicolumn{2}{l}{Manufacturing}					&44.5	&45.2	&67.7	\\
\multicolumn{2}{l}{Commerce}						&32.4	&34.0	&16.9	\\
\multicolumn{2}{l}{Transport}						&7.6	&10.2	&6.1	\\
\multicolumn{2}{l}{Public service and professions}	&11.3	&10.2	&4.6	\\ 
\multicolumn{2}{l}{Housework}						&0.3	&0.1	&0.2	\\
\multicolumn{2}{l}{Other industry}					&2.4	&2.0	&2.5	\\\hline
&&&&\\
\multicolumn{5}{l}{\textbf{Panel B: Demographic structure}}\\
&&&&\\
\cmidrule(rrr){3-5}
\multicolumn{2}{l}{Name of survey}	&(1) 1920 Population Census	&(2) 1920 Population Census	&(3) 1920 Population Census\\
\multicolumn{2}{l}{Survey area}		&Tokyo City				&Kyobashi Ward						&Tsukishima\\

\multicolumn{2}{l}{Survey subject}			&Complete survey&Complete survey&Complete survey\\
\multicolumn{2}{l}{Survey month and year}	&October 1920	&October 1920	&October 1920\\\hline

\multicolumn{2}{l}{Average household size (in person)}	&4.6	&4.7	&4.3	\\ 
\multicolumn{2}{l}{Share of married males (\%)}			&38	&36	&40	\\ 
\multicolumn{2}{l}{Sex ratio (males/females)}			&1.2	&1.2	&1.3	\\ 
\multicolumn{2}{l}{\hspace{5pt} 0--13 years}			&1.1	&1.1	&1.1	\\ 
\multicolumn{2}{l}{\hspace{5pt} 14--19 years}			&1.5	&1.7	&2.2	\\ 
\multicolumn{2}{l}{\hspace{5pt} 20--29 years}			&1.3	&1.4	&1.6	\\ 
\multicolumn{2}{l}{\hspace{5pt} 30--39 years}			&1.2	&1.2	&1.3	\\ 
\multicolumn{2}{l}{\hspace{5pt} 40--49 years}			&1.2	&1.2	&1.3	\\ 
\multicolumn{2}{l}{\hspace{5pt} 50--59 years}			&1.1	&1.1	&1.2	\\ 
\multicolumn{2}{l}{\hspace{5pt} 60+}					&0.8	&0.8	&0.9	\\
\multicolumn{2}{l}{Average age of males}				&25.3	&25.3	&25.1	\\\midrule[0.3pt]\bottomrule[1pt]
\end{tabular}
}
{\scriptsize
\begin{minipage}{448pt}
\setstretch{0.85}
Notes:
\textbf{Panel A}: This panel summarizes the industrial structure based on the occupations of male workers recorded in the 1920 Population Census.
Each share is calculated as the number of males employed as regular workers in a given sector divided by the total number of male workers (in percentage).
All figures exclude unemployed (\textit{mugy\=o sha}) males.
\textbf{Panel B}: The average household size is calculated as the total number of people divided by the number of households.
The share of married males is the number of married males, living with/without spouses, divided by the total number of males (in percentage).
The average age of males is calculated using the population tables by age group reported in the census.
For open-ended age categories, the class for those over 60 years old is rounded to $64.5$, following the range of the previous category (50--59).
All figures in this panel exclude a small number of quasi-households (\textit{jyun setai}), which include individuals such as students in dormitories and patients in hospitals.
The figures for Tsukishima listed in Column 3 of both panels are based on the total sum of statistics across all blocks (\textit{ch\=ome}) in Tsukishima (i.e., from \textit{Tsukudajima} to \textit{Tsukishima d\=ori jy\=uni ch\=ome}).
Sources:
Columns 1 and 2 of Panel A: Statistics Bureau of the Cabinet (1929c, pp.~18--19).
Columns 1 and 2 of Panel B: Statistics Bureau of the Cabinet (1929c, pp.~38--43); Tokyo City Office (1922b, pp.~2--3; 42--46).
Column 3 of Panel A: Tokyo City Office (1922c, pp.~42--47).
Column 3 of Panel B: Tokyo City Office (1922a, pp.~26--31); Tokyo City Office (1922b, pp.~2--3; 42--46).
\end{minipage}
}
\end{center}
\end{table}

Panel B in Table 1 summarizes population and occupational statistics from the census to examine the characteristics of households in Tsukishima.
The average household size ($4.3$) is similar to, but slightly smaller than, the values in Tokyo City and Kyobashi Ward.
The percentage of married males ($40$\%) is comparable to the figures in Columns (1) and (2), while the average age of men ($25.1$) is slightly lower.
Sex ratios follow a similar trend to those of Tokyo and Kyobashi, although the ratios for the $14$--$19$ and $20$--$29$ age bins are higher.
Overall, these statistics suggest that the typical household consists of a couple with two children.
The slightly smaller household size in Tsukishima can be attributed to the relatively large number of young factory-worker households, and the percentage of married men indicates that a portion of households in their early 20s may have been couples before starting a family.

Finally, I provide a brief overview of life in factory workers' households.
Factories and workers' residences were scattered throughout Tsukishima.
Factory workers lived in similar tenements and made use of a wide variety of commercial stores in the three lots: Tsukudajima and Shin-tsukudajima in the north, Lot 1 in the middle, and Lot 2 in the south (Figure~\ref{fig:tsukishima_factory}).\footnote{The northernmost Ishikawajima island contains only the Ishikawajima Shipyard, and the southernmost rectangular Lot 3 is uninhabited. Online Appendix Figures~\ref{fig:map_tsukishima} and~\ref{fig:map_retailers} provide detailed maps of the lots and retail shops, respectively. Online Appendix~\ref{sec:seca_housing} summarizes housing in Tsukishima.}
The main routes to the mainland were the wooden Aioi Bridge (\textit{Aioi bashi}) and three ferries (\textit{watashi bune}).
The average daily number of ferry users in 1918 was $53,318$, more than twice the population of Tsukishima.\footnote{Department of Health, Ministry of the Interior (1923a, p.~30). The number of people and households in Tsukishima at the end of December 1918 was $24,399$ and $5,562$, respectively (National Police Agency 1920a, pp.~50--51). Online Appendix Figure~\ref{fig:map_tsukishima} shows the ferry routes.}
These figures illustrate Tsukishima as a busy machine manufacturing area during industrialization.

\subsection{Risk-coping Institutions} \label{sec:sec23}

Generally, social insurance such as health, accident, and unemployment insurance can mitigate idiosyncratic income shocks \citep{Gertler2002-tu, Kantor1996-qp, Gruber1997-fa}.
However, comprehensive social insurance systems did not exist in Japan in the early 1920s.
Health insurance was not enacted until 1927, and unemployment insurance was absent throughout the prewar period.
The Factory Act enacted in 1916 did not specify compensation for workers' sickness.
As a result, factory workers had no public support for illness, injuries, or unemployment.

Instead, several financial institutions were available for factory workers.\footnote{Finer details of these institutions and the statistics used in this subsection are summarized in Online Appendix~\ref{sec:seca_city_fi}.}
A representative savings institution was postal savings (\textit{y\=ubin chokin}).
In Tokyo City, approximately $1.4$ million people held postal savings accounts, covering approximately $67$\% of the citizens in 1920.
The average savings per capita in Kyobashi Ward was $44.5$ yen, less than one month's earnings for skilled factory workers (Section~\ref{sec:sec32}).
Savings banks (\textit{chochiku gink\=o}) were another well-known savings institution.
About $1.4$ million people held accounts, representing approximately $67$\% of Tokyo's population.
However, $84.8$\% of all depositors were commercial and miscellaneous workers, while only $11.5$\% belonged to the manufacturing sector in Kyobashi Ward.
This reflects that savings banks were designed for lower-income workers such as those in miscellaneous industries and day laborers, rather than factory workers (Tokyo Institute for Municipal Research 1925b, p.~81).
This is also consistent with the fact that Tsukishima had no savings bank branches but did have a few post offices.
Another savings institution, the mutual loan association (\textit{mujin}), was mainly used by small business owners and merchants to run their businesses.\footnote{Ordinary banks in prewar Japan provided large loans to businesses \citep{Teranishi2011}. Tsukishima had no bank, since workers did not borrow money from banks to make ends meet (Department of Health, Ministry of the Interior 1923a, p.~48).}

Credit purchases (\textit{kakegai}) with retailers were the most common informal credit institution.
People often used credit purchases at various retailers for rice, fish, vegetables, firewood, and charcoal in Tokyo City.
This institution is described in detail in Section~\ref{sec:sec5}.
Pawnshops (\textit{shichiya}) were widely distributed lending institutions in the city.\footnote{The number of pawnshops in Tokyo was $1,334$ and $1,261$ in 1918 and 1919, with total transactions of $8,226,883$ and $7,573,406$ each year, respectively--more than three times the population of Tokyo City.}
However, evidence shows that the most frequent users were day laborers and workers classified as ``miscellaneous,'' accounting for $63.4$\% of all users.
By contrast, the shares for agriculture, commerce, and manufacturing sectors were only $1.7$, $19.5$, and $15.6$\%, respectively (Tokyo Institute for Municipal Research 1926, p.~25).
In practice, pawnshop lenders did not need to screen borrowers' credit, and borrowers did not face the risk of heavy debt, as the main items pawned were inexpensive clothes \citep{Shibuya1982}.
Interest rates were regulated by the Pawnbroker Regulation Act of 1895, and the average redemption rate was high under these lower rates.\footnote{The average redemption rate was approximately $94$\% in 1920 among $88$ pawnshops in Kyobashi ($401,222$/$427,265$ cases) (Tokyo City Office 1922b, pp.~888--889).}
In short, the primary users of pawnshops were low-income workers who needed credit for daily necessities, rather than factory workers (Tokyo City Social Affairs Bureau 1921, p.~9).
Other lending institutions included moneylenders (\textit{kinsen kashitsuke gy\=o}) and credit unions (\textit{shiny\=o kumiai}).
Moneylenders were generally used by business owners and charged very high interest rates (Shibuya 2000, pp.~184; 248).
Credit unions had very limited coverage, likely serving only a small share of male workers in the manufacturing sector.

Informal gifts may have served as another type of risk-mitigation institution.
While systematic statistics are unavailable, a household survey of $185$ factory-worker households in Tokyo found that the average monthly gifts accounted for only $3$\% of total income in November 1922 (Social Affairs Division 1925, pp.~58--59).
Therefore, whether these small informal transfers functioned as effective mutual aid remains unclear.

Life insurance was not designed to compensate for temporary income losses.
Life insurance companies generally provided death insurance, which did not cover short-term income reductions, and they did not target working-class customers \citep[pp.~87--88]{Usami1984}.
Although postal life insurance (\textit{kani seimei hoken}) operated by the government began in 1916, it offered only life and endowment insurance and did not cover illness or unemployment \citep[pp.~102--107]{Usami1984}.\footnote{Several employment agencies existed in Tokyo to introduce jobs to low-income job seekers. However, many were unscrupulous and not commonly used at the time \citep{Machida2016}.}

Finally, cooperative societies were available to help worker households purchase daily commodities.
In 1924, Tokyo prefecture had $26$ cooperative societies, of which 23 were organized by citizens and workers (Central Federation of Industrial Associations 1925, p.~57--59).
While the two largest societies had approximately $4,000$ members each, the others were generally small, with fewer than 100 members.
Overall, this accounted for only a small share of workers in the city.

To summarize, although several risk-coping institutions existed, few were accessible to factory workers.
Postal savings were the main official financial institution available, while credit purchases were the most plausible informal mechanism.
Factory workers in Tsukishima likely combined these devices to manage idiosyncratic income shocks.

\section{Data}\label{sec:sec3}

\subsection{Tsukishima Household Budget Survey}\label{sec:sec31}

The Tsukishima Survey was the first urban social survey conducted in Japan by the Ministry of Home Affairs \textit{circa} 1919.
Its purpose was to examine the living conditions of urban worker households amid rapid industrialization during WWI.
The survey was designed as a field study with multiple components, including a budget survey (hereinafter THBS).
The Ministry of Home Affairs published several official reports in 1921, which contained aggregated information on the target households.\footnote{The THBS was the first household survey to use the budget book method and is considered the prototype for subsequent household surveys in Japan \citep[][p.~43]{Sekiya1970}. The official reports published in 1921 by the Ministry of Home Affairs are included in the archives of Gonda Yasunosuke and are available to the public (The OISR, Archives of Gonda Yasunosuke (7-2; 7-3; 7-4)).}
However, household-level budget data are necessary to analyze consumption-smoothing strategies.

Fortunately, the household budget books (\textit{kinsendeiri hikaech\=o}) collected in the THBS are preserved in the archives of the Ohara Institute of Social Research (OISR).
These budget books were transferred to the OISR by Professor Iwasaburo Takano of Tokyo Imperial University.
He served on the research committee of the Ministry of Home Affairs responsible for the Tsukishima Survey and later became the first director of the OISR.\footnote{Online Appendix~\ref{sec:secb1} provides further details on how this survey was conducted as part of Iwasaburo Takano's ambitious social survey project.}

I collected and digitized all forty THBS budget books for the factory-worker households.\footnote{The author was permitted access to the unreleased materials related to the Tsukishima Survey by the OISR: Archives of the Tsukishima Survey (THBS, unreleased). The characteristics of the budget books are summarized in detail in Online Appendix~\ref{sec:secb2}.}
During the study period, each household recorded its daily income and expenses in a budget book.
Notably, a survey office was established in Tsukishima, where a few surveyors lived and maintained close contact with the surveyed households.
Weekly administrative meetings were held to monitor the survey's progress \citep[][p.~38]{Miyoshi1980}.

\subsection{Sample Characteristics}\label{sec:sec32}

\subsubsection*{Analytical Sample}

The estimation strategy requires panel structures for aggregate income and expenditure (Section~\ref{sec:sec4}).
I use the semi-monthly series for the baseline analysis, as the household head's income was paid twice a month.\footnote{Section~\ref{sec:sec33} provides details on the definitions of aggregate income and expenditure.}
Among the $40$ THBS households, five cross-sectional units appear in my semi-monthly data.
I exclude these, leaving $35$ households.
I then remove a household with incomplete income information and another for which no information on payments for credit purchases are recorded.
Consequently, the THBS sample for the semi-monthly series includes $33$ households with panel structures from January 12 to November 11, 1919.\footnote{I also use the adjusted monthly panel dataset for the sub-analysis in Section~\ref{sec:sec4}. Following the same trimming steps, the adjusted monthly panel dataset includes $26$ households for the same period. Note that several units have only three semi-month cells, for example, less than two adjusted months. These units have a panel structure in the semi-monthly dataset but not in the adjusted monthly panel. Online Appendix~\ref{sec:secb_trim} summarizes the original data structure and trimming in detail.}

Before assessing the representativeness of the THBS sample, I evaluate potential selection bias due to unit attrition.
Panels with shorter (or longer) time-series observations might have different consumption preferences than those with longer (or shorter) observations.
In other words, households with shorter panels should have family size characteristics similar to those with longer panels.
To test this, I regress an indicator variable for the shorter panel units on family size variables.
Online Appendix Table~\ref{tab:balancing} summarizes the results.
All estimated coefficients on the covariates are close to zero and statistically insignificant, and the Wald statistics support the null hypothesis.
This result is robust to using different thresholds such as the first quantile, median, and third quantile.
These findings suggest that household preferences are unlikely to be correlated with panel length.

\subsubsection*{Representativeness}

\begin{table}[htbp]
\def\arraystretch{0.95}
\centering
\begin{center}
\caption{Assessing Representativeness: Tsukishima Household Budget Survey}
\label{tab:tab2}
\scriptsize
\scalebox{0.96}[1]{
\begin{tabular}{lrrrr}
\toprule[1pt]\midrule[0.3pt]
\multicolumn{5}{l}{\textbf{Panel A: Manufacturing sector}}\\
\cmidrule(rrrrr){2-5}
Name of survey			&(1) 1919 Statistics of 	&\multicolumn{2}{r}{(2) 1920 Tsukishima Factory}	&(3) The THBS\\
						&National Police Agency	&\multicolumn{2}{r}{Survey}						&\\
Survey area				&Tsukishima			&\multicolumn{2}{r}{Tsukishima}					&Tsukishima\\
Survey month and year	&December 1919		&\multicolumn{2}{r}{November 1920}				&1919\\
Survey subject			&All factories with		&\multicolumn{2}{r}{All factories with \# of}			&33 households\\
\cmidrule(rr){3-4}
						&\# of workers $\geq 15$	&workers $\geq 15$	&workers $< 15$	&\\\hline

Textile					&0.2	&2.9	&0.0	&0		\\
Machinery				&87.4	&85.3	&91.8	&93.9	\\
Chemical				&1.1	&5.9	&5.5	&0		\\
Food					&1.2	&1.5	&0.0	&3.0	\\ 
Miscellaneous			&10.2	&4.4	&2.7	&3.0	\\
Observations			&$7,647$ workers	&68 factories&146 factories	&33 heads\\
&&&&\\
\multicolumn{5}{l}{\textbf{Panel B: Family structure}}\\
\cmidrule(rrrr){2-4}
Name of survey			&(1) 1920 Population 		&(2) 1919 Statistics of 	&(3) The THBS	&\\
						&Census					&National Police Agency	&				&\\
Survey area				&Tsukishima				&Tsukishima			&Tsukishima	&\\
Survey month and year	&October 1920				&December 1919		&1919			&\\
Survey subject			&Complete survey			&Complete survey		&33 households	&\\\hline
Average household size			&4.3	&4.3	&4.3	&\\
Sex ratio						&1.3	&1.2	&1.2	&\\
Household size (\% share)		&&&&\\
\hspace{10pt}1					&6.0			&--		&0&\\
\hspace{10pt}2					&16.7 (18.7)	&--		&15.2&\\
\hspace{10pt}3--5				&50.8 (57.1)	&--		&57.6&\\
\hspace{10pt}6--8				&21.5 (24.2)	&--		&27.3&\\
\hspace{10pt}9+				&4.9			&--		&0&\\
Pearson $\chi^{2}$ statistic $p$-value		&--		&--		&0.948	&\\
&&&&\\
\multicolumn{5}{l}{\textbf{Panel C: Monthly earnings}}\\
\cmidrule(rrrrr){2-5}
Name of survey				&\multicolumn{2}{r}{(1) The THBS}			&\multicolumn{2}{r}{(2) Manufacturing Census} 	\\
Survey area					&\multicolumn{2}{r}{Tsukishima}			&\multicolumn{2}{r}{Tokyo City} 				\\
Survey/equivalent year		&\multicolumn{2}{r}{1919}					&\multicolumn{2}{r}{1919}							\\
Survey subject				&\multicolumn{2}{r}{THBS heads}				&\multicolumn{2}{r}{Male factory workers} 		\\
							&\multicolumn{2}{r}{(Ave. age: 33.2)}			&\multicolumn{2}{r}{in machinery factories} 		\\
							&\multicolumn{2}{r}{}						&\multicolumn{2}{r}{(Age range: 30--40)} 		\\\hline

Average monthly earnings	&\multicolumn{2}{r}{59.1 (median = 56.2)}	&\multicolumn{2}{r}{56.0}				\\
							&\multicolumn{2}{r}{95\%CI [53.7, 64.5]}	&\multicolumn{2}{r}{}					\\
\midrule[0.3pt]\bottomrule[1pt]
\end{tabular}
}
{\scriptsize
\begin{minipage}{423pt}
\setstretch{0.85}
Notes:
\textbf{Panel A}: 
Column 1 shows the share of factory workers in each sector relative to the total number of factory workers (percentage).
This includes both female and male workers in the 68 factories with 15 or more employees.
Column 2 indicates the share of factories in each sector by factory size (percentage).
Column 3 summarizes the manufacturing sectors of the THBS household heads.
Columns 2 and 3 follow the classification used in the Statistics of the National Police Agency (Column 1).
According to this classification, several ``woodworking'' factories recorded in the Tsukishima Factory Survey are included in the machinery sector in Column 2, as it covers shipbuilding and wooden pattern factories.
The report also confirms that wooden patterns are used in the casting process and should therefore be included in the machinery sector (Department of Health, Ministry of the Interior 1923a, pp.~388--389).
\textbf{Panel B}: 
The first and second rows in Columns 1 and 2 show the average household size and sex ratio from the 1920 Population Census and the 1919 Statistics of the National Police Agency, respectively.
Rows four to eight in Columns 1 and 3 present household size distributions in the census and the THBS sample, respectively.
Figures in parentheses in Column 1 indicate the percentage share for family size bins from 2 to 8 people.
Row nine in Column 3 reports the $p$-value from the Pearson $\chi^{2}$ test for equality of household size distributions between 2 and 8 people.
The census shares in each household size bin are used as the theoretical probabilities in calculating the test statistic (Online Appendix~\ref{sec:secb_size}).
\textbf{Panel C}: 
Column 1 lists the average monthly earnings of THBS household heads, calculated from the adjusted monthly panel dataset.
Of the 33 THBS households, data from 26 households--where adjusted monthly income can be traced--are used in this panel to match units in the manufacturing census.
Comparison using average semi-monthly income from all 33 households produces similar results.
Gaussian-based 95\% confidence intervals (CI) using bootstrap standard errors are reported in brackets.
Column 2 shows the average monthly wage of male factory workers aged 30--40 in machinery factories in Tokyo City.
This figure is calculated using average monthly wages and ancillary wages from the manufacturing censuses.
All wage statistics are deflated using the CPI for cities from Ohkawa et al. (1967, p.~255).
Online Appendix~\ref{sec:secb_est_wage} provides further details on the wage estimation process.
Sources:
Panel A:
Column 1: National Police Agency (1920b, p.~236);
Column 2: Department of Health, Ministry of the Interior (1923a, p.~385--386); 
Column 3: Department of Health, Ministry of the Interior (1923a, p.~154).
Panel B:
Column 1: Tokyo City Office (1922a, pp.~26--31; 262--283); Tokyo City Office (1922b, pp.~2--3; 42--46);
Column 2: National Police Agency (1920b, p.~53);
Column 3: Calculated by the author from the THBS dataset.
Panel C:
Column 1: THBS dataset;
Column 2: Tokyo City Statistics Division (1926a, pp.~16; 244--245); Tokyo City Office (1921, pp.~726--757).

\end{minipage}
}
\end{center}
\end{table}

Panel A of Table~\ref{tab:tab2} summarizes the share of male workers in the manufacturing industry by sector.
Statistics from the National Police Agency (SNPA) for factories with 15 or more workers show that approximately $90$\% of male workers were employed in the machinery sector in 1919 (Column 1).
Column 2 presents statistics from the Tsukishima Factory Survey.
The proportion of machinery factories is $85$\% among larger factories and $92$\% among smaller factories, indicating that most male workers were in the machinery sector.\footnote{Paternal occupational statistics for primary school students in Tsukishima yield materially similar results (Online Appendix~\ref{sec:seca_factory}).}
Column 3 confirms that THBS household heads show a similar pattern, with $94$\% working in the machinery sector.
The report states: 
``It is not too much to say that most of the industry in Tsukishima is machinery manufacturing. Therefore, to describe the labor situation in Tsukishima, it is sufficient to describe its machinery industry'' (Department of Health, Ministry of the Interior 1923a, p.~389).
The rest of the THBS heads work in food or miscellaneous industries, with proportions similar to those in Column 2.

Panel B of Table~\ref{tab:tab2} summarizes family structure.
Average household sizes are identical across the Population Census (Column 1), SNPA (Column 2), and THBS (Column 3).
Sex ratios are also similar across the surveys ($1.2$--$1.3$).
THBS households range from 2 to 8 members, excluding single-person and very large households.
In other words, the THBS mainly captures consumption behavior in couples and households with a few children.
For this range, the Pearson $\chi^{2}$ test does not reject the null hypothesis of equality with the census household size distribution ($p$-value = $0.948$), indicating that the THBS sample reasonably approximates the distribution of family households in Tsukishima.
Online Appendix~\ref{sec:secb_size} provides a more detailed view of household size distributions.
This supports the report's statement that THBS households can be ``regarded as representative of the family form in Tsukishima'' (Department of Health, Ministry of the Interior 1923a, p.~145).

The official Tsukishima Survey report suggests that approximately nine out of ten factory workers in the skilled-age range are classified as skilled workers.
Approximately six of these skilled workers are employed in medium- to large-scale factories, and three in small-scale factories.\footnote{Online Appendix~\ref{sec:seca_factory} provides detailed evidence using the complete survey of primary school students in Tsukishima.}
Accordingly, the THBS is designed to examine household budgets of skilled workers.
The average age of THBS household heads is $33$, which falls within the representative age range of skilled workers in machinery factories (Kitazawa 1924).\footnote{Specific occupations are not available for most heads, as they are generally listed as ``factory workers.'' However, sub-categories available for some heads indicate typical skilled machinery factory roles such as lathe operators and finishers. Online Appendix~\ref{sec:seca_work} summarizes work in the machinery factories in Tsukishima.
The background and rationale for focusing on skilled factory workers in this survey are detailed in Online Appendix~\ref{sec:secb1}.}
Column 1 of Panel C in Table~\ref{tab:tab2} shows that the average monthly earnings of THBS heads is $59$ yen (median = $56$ yen).
Column 2 reports the average monthly earnings of male machinery factory workers in the skilled-age range, calculated using the manufacturing censuses, which is similar at $56$ yen.\footnote{This figure is calculated using the average daily wage of male factory workers aged 30--40 and the average annual working days measured in the manufacturing censuses. Online Appendix~\ref{sec:secb_est_wage} provides detailed calculation steps. Average wages for male factory workers follow an inverted-U-shaped distribution with respect to age, peaking in the 30s (Online Appendix Figure~\ref{fig:hist_daily_wage}).}
These results indicate that the monthly earnings of THBS household heads closely match those of skilled workers' households in Tsukishima.
Overall, this confirms that the THBS sample is suitable for analyzing the mean tendencies of skilled factory workers' households in Tsukishima.

\subsection{Aggregation}\label{sec:sec33}

\subsubsection*{Measurement Error}

\begin{figure}[htbp]
\captionsetup{justification=centering,margin=0.25cm}
\centering
\subfloat[Daily panel dataset:\\ income and expenditure]{\label{fig:tsd_minc_mexp}\includegraphics[width=0.4\textwidth]{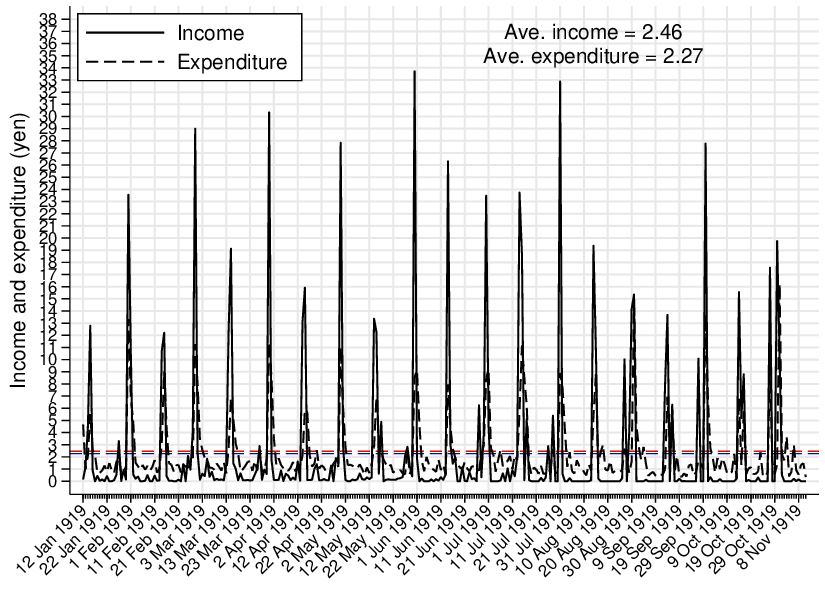}}
\subfloat[Daily panel dataset:\\ net income]{\label{fig:tsd_mexpdef}\includegraphics[width=0.4\textwidth]{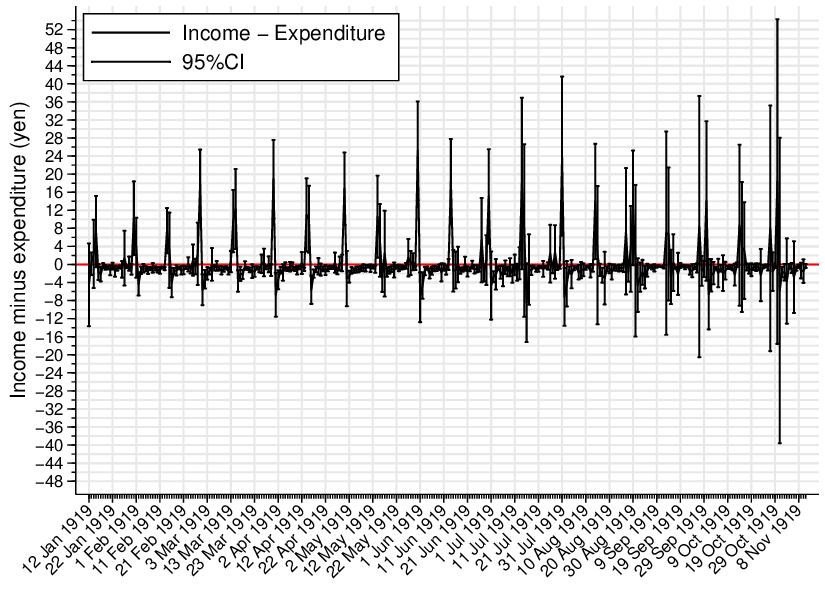}}\\
\subfloat[Semi-monthly panel dataset:\\ income and expenditure]{\label{fig:tssm_minc_mexp}\includegraphics[width=0.4\textwidth]{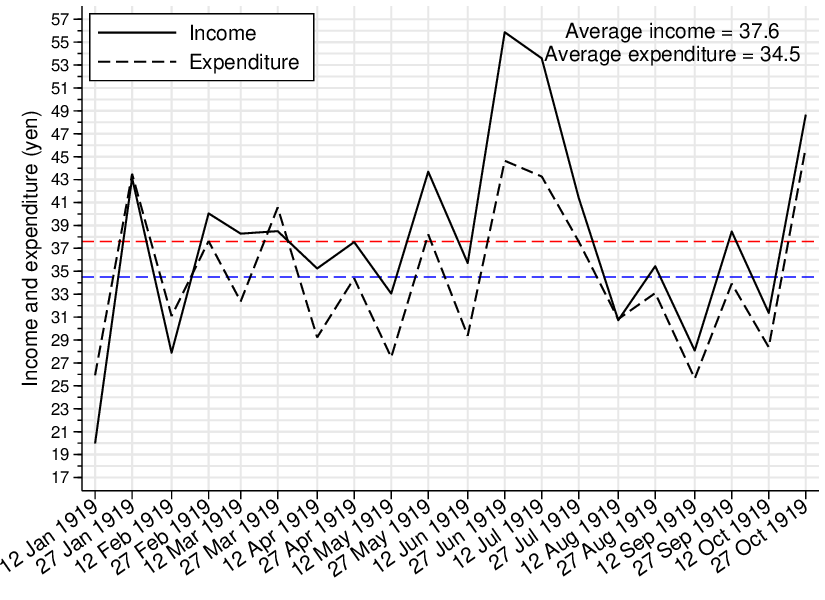}}
\subfloat[Semi-monthly panel dataset:\\ net income]{\label{fig:tssm_mexpdef}\includegraphics[width=0.4\textwidth]{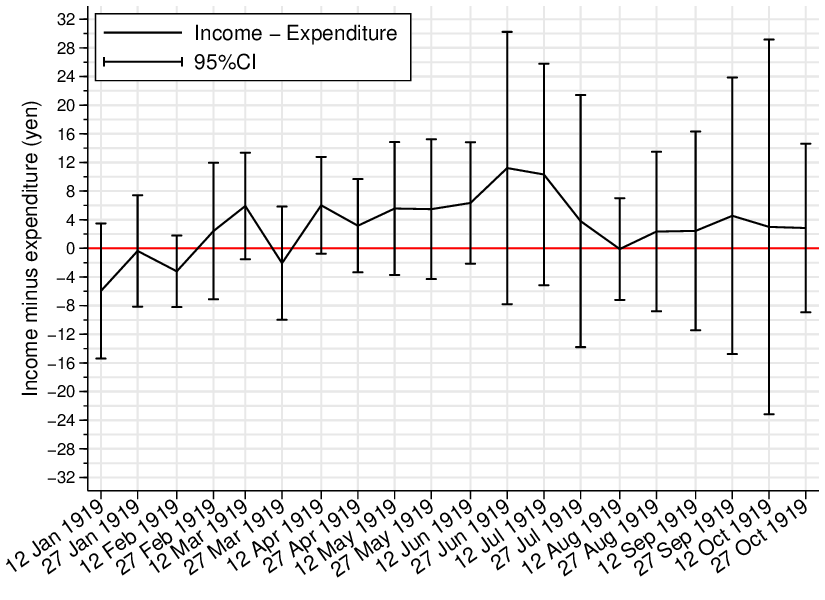}}\\
\subfloat[Unadj. semi-monthly panel dataset: income and expenditure]{\label{fig:tssm_minc_mexp_raw}\includegraphics[width=0.4\textwidth]{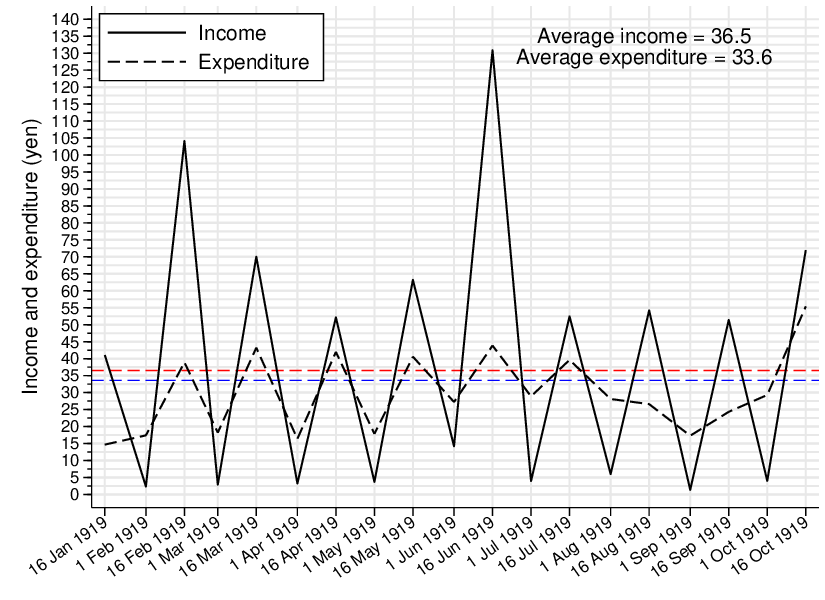}}
\subfloat[Unadj. semi-monthly panel dataset: net income]{\label{fig:tssm_mexpdef_raw}\includegraphics[width=0.4\textwidth]{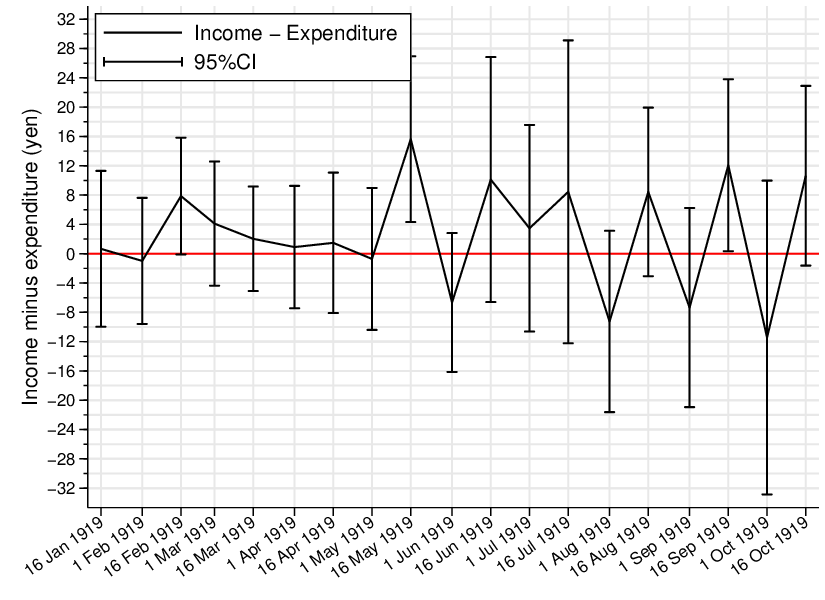}}

\caption{Relationships between Average Income and Expenditure\\ by Different Time-series Frequencies}
\label{fig:ts_overview}
\scriptsize{\begin{minipage}{400pt}
\setstretch{0.85}
Notes:
Figures~\ref{fig:tsd_minc_mexp},~\ref{fig:tssm_minc_mexp}, and~\ref{fig:tssm_minc_mexp_raw} present time-series plots of average daily, semi-monthly, and unadjusted semi-monthly income and expenditure, respectively.
Figures~\ref{fig:tsd_mexpdef},~\ref{fig:tssm_mexpdef}, and~\ref{fig:tssm_mexpdef_raw} show the corresponding time-series plots of average daily, semi-monthly, and unadjusted semi-monthly net income (income minus expenditure).
Figures~\ref{fig:tsd_minc_mexp} and~\ref{fig:tsd_mexpdef} display daily data from January 12 to November 11, 1919.
Figures~\ref{fig:tssm_minc_mexp} and~\ref{fig:tssm_mexpdef} show the semi-monthly series calculated from the daily data over the same period.
In the adjusted semi-monthly series, the first half covers the 12--26th and the second half the 27--11th.
Figures~\ref{fig:tssm_minc_mexp_raw} and~\ref{fig:tssm_mexpdef_raw} present the unadjusted semi-monthly series calculated from daily data between January 15 and October 31, 1919, using the 15th of each month as the threshold.
Source: Created by the author using the THBS sample.
\end{minipage}}
\end{figure}

Summarizing daily budget information over a longer time span is necessary because households need time to smooth their consumption given their realized income.
Figure~\ref{fig:tsd_minc_mexp} shows that the head's income is paid twice a month, typically in the middle (14th--16th) and at the end of the month (30th or 31st).\footnote{The head's income is the primary source of earnings, as most households are typical breadwinner households (Panel B in Table~\ref{tab:sum}).}
Importantly, Figure~\ref{fig:tsd_mexpdef} indicates that households spend most of their income immediately on payday and gradually use the remainder until the next payday.
In other words, expenditure does not increase sharply just before paydays.
This suggests that households plan their consumption schedules based on the income they received.
Calendar-based (semi-)monthly aggregation cannot capture this behavior because it does not align the payday with the start of the aggregation period.\footnote{For example, a calendar semi-month cell in January (1st--14th) starts on the 1st, but the payday falls on the 14th. Aggregating in this way assumes that the consumption within the cell is based on income at the end of the cell, which is unrealistic.}
Another source of measurement error arises when the payday shifts: occasionally, the payday for the latter half of a month falls on the first day of the next month.
While infrequent, this can distort calendar-month aggregation, which may then include three paydays in some months and only one in others.
Such mismatches between income and expenditure timing can induce Type II errors when testing for consumption smoothing \citep{Nelson1994-js}.

\subsubsection*{Adjustment}

To address the timing mismatch, I aggregate daily observations from the 12th to the 26th for the first half of each month and from the 27th to the 11th of the following month for the second half.\footnote{This means that the number of days in a semi-month ranges from 13 (February) to 16 (January). This heterogeneity is constant across households, and differences are fully captured by semi-month fixed effects in the regression.}
This range is based on the observed paydays in the THBS dataset, which fall between the 27th and 31st for the second half of each month.\footnote{I do not use a semi-month definition starting on the 2nd (2nd--16th) because it places the payday at the end of the semi-month cell. Online Appendix~\ref{sec:secb_me} shows that although this alternative improves the income series, expenditure cannot correctly follow the income, as expenses measured under this definition are based on income from the previous semi-month.}
Figure~\ref{fig:tssm_minc_mexp} presents the time-series plots of income and expenditure using this adjusted semi-monthly panel, whereas Figure~\ref{fig:tssm_minc_mexp_raw} shows the unadjusted (calendar-based) semi-monthly series, using the 15th of each month as the threshold.
The unadjusted series clearly provides a rough measure of income and expenditure owing to measurement errors from misaligned paydays.
The main improvement in Figure~\ref{fig:tssm_minc_mexp} is that expenditure now aligns with income in each month, capturing households' consumption strategies based on the timing of actual income.
As a result, the net income series shows a much smoother trend in Figure~\ref{fig:tssm_mexpdef}, which is not evident in the unadjusted series (Figure~\ref{fig:tssm_mexpdef_raw}).
Online Appendix~\ref{sec:secb_ab} demonstrates that measurement errors in the unadjusted dataset strongly attenuate the estimates.

\subsection{Data Description}\label{sec:sec34}

\subsubsection*{Variables}

The THBS budget book has two main sections: expenditure and income.
Panel A of Table~\ref{tab:sum} lists the $11$ expenditure subcategories: food, housing, utilities, furniture, clothes, education, medical, entertainment, transportation, gifts, and miscellaneous.
These subcategories are used to test the smoothness of the consumptions.
To examine the role of risk-coping strategies, I consider five net income categories: savings, insurance, borrowing, credit purchases, and gifts.
Net variables are defined as income minus expenditure in each category; for example, net savings refer to withdrawals minus deposits to savings.
I also include the sales of miscellaneous goods and family members' (excluding the household head's) labor earnings to assess the contribution of small asset sales and intrahousehold labor supply adjustments.
Panel B of Table~\ref{tab:sum} summarizes these variables.\footnote{Online Appendix~\ref{sec:secb_st_sc} presents the results of panel unit root tests. The null hypothesis of unit roots is rejected at conventional levels for all these variables.}

Changes in family size may affect household preferences \citep{Jappelli2017-ta}.\footnote{In village economies, landholding and assets such as livestock and grain may also correlate with households' risk-coping strategies \citep[e.g.,][]{Udry1994-vp, Udry1995-ri}. As this study focuses on urban factory-worker households, the roles of these assets are not relevant.}
Fortunately, the archives include sheets documenting changes in the composition of THBS households (\textit{Kazoku id\=o hy\=o}).\footnote{OISR, Archives of the Tsukishima Survey (THBS, unreleased).}
I find that household size remains stable over time.
During the sample period, only one household experienced a change in family size due to the birth of a son.
This indicates that preference shifts rarely occurred among THBS households.
Although family size is included in all regressions, it has minimal effect on the results.\footnote{The family size variable is omitted in some regressions owing to multicollinearity in fixed-effects models, given its small within-household variation. As an alternative time-varying measure of preference shocks, I use the share of children below primary school age. Including this child-share variable does not affect the results; estimates are materially similar to the baseline specification (Online Appendix Table~\ref{tab:cs_baseline_child}).}
Panel C of Table~\ref{tab:sum} presents the summary statistics for family size.

\begin{table}[htbp]
\def\arraystretch{0.88}
\centering
\captionsetup{justification=centering}
\begin{center}
\caption{Summary Statistics}
\label{tab:sum}
\footnotesize
\scalebox{0.89}[1]{
{\setlength\doublerulesep{2pt}
\begin{tabular}{lrrrrrrrr}
\toprule[1pt]\midrule[0.3pt]
\multicolumn{8}{l}{\textbf{Panel A: Variables for testing consumption smoothing (in yen)}}\\
&\multicolumn{4}{c}{Semi-monthly panel dataset}&\multicolumn{4}{c}{Monthly panel dataset}\\
\cmidrule(rrrr){2-5}\cmidrule(rrrr){6-9}
&\multicolumn{2}{c}{Level}&\multicolumn{2}{c}{Log}
&\multicolumn{2}{c}{Level}&\multicolumn{2}{c}{Log}\\
\cmidrule(rr){2-3}\cmidrule(rr){4-5}\cmidrule(rr){6-7}\cmidrule(rr){8-9}
&Mean&Obs.&Mean&Obs.&Mean&Obs.&Mean&Obs.\\\hline
Total expenditure				&34.5	&289&3.4		&289		&69.3	&124&4.2	&124		\\
\hspace{10pt}Food				&10.9	&289&2.3		&270		&21.5	&124&3.0	&115		\\
\hspace{10pt}Housing			&2.7		&289&1.6		&134		&5.8		&124&1.8	&105		\\
\hspace{10pt}Utilities				&1.2		&289&-0.1	&232		&2.3		&124&0.6	&111		\\
\hspace{10pt}Furniture			&0.4		&289&-1.6	&177		&0.8		&124&-1.1&99		\\
\hspace{10pt}Clothes			&2.8		&289&0.3		&248		&5.8		&124&1.2	&113		\\
\hspace{10pt}Education			&1.3		&289&-0.0	&245		&2.6		&124&0.6&111	\\
\hspace{10pt}Medical			&1.7		&289&0.2		&269		&3.3		&124&0.9	&115		\\
\hspace{10pt}Entertainment		&1.3		&289&-0.2	&255		&2.6		&124&0.5	&114		\\
\hspace{10pt}Transportation		&0.4		&289&-1.3	&195		&0.9		&124&-0.7&98		\\
\hspace{10pt}Gifts				&3.1		&289&0.5		&243		&6.3		&124&1.3	&114		\\
\hspace{10pt}Miscellaneous		&0.5		&289&-1.3	&222		&1.0		&124&-0.7&111		\\
Disposable income				&37.6	&289&3.5		&278		&76.8	&124&4.2	&123		\\
&&&&&&&&\\
\multicolumn{8}{l}{\textbf{Panel B: Variables for testing risk-coping mechanisms (in yen)}}\\
&\multicolumn{3}{c}{Semi-monthly panel dataset}&\multicolumn{3}{c}{Monthly panel dataset}&&\\
\cmidrule(rrr){2-4}\cmidrule(rrr){5-7}
&Mean&Std. Dev.&Obs.&Mean&Std. Dev.&Obs.&&\\
\cmidrule(rrrrrrr){1-7}
\multicolumn{8}{l}{Panel B-1: Net savings, insurance, borrowing, and gifts (income $-$ expenditure)}\\
\hspace{10pt}Net savings			&-0.01	&3.97	&289		&0.08	&5.78	&124		&&\\
\hspace{10pt}Net insurance		&-0.03	&6.67	&289		&0.26	&10.15	&124		&&\\
\hspace{10pt}Net borrowing		&0.31	&5.03	&289		&1.16	&7.63	&124		&&\\
\hspace{10pt}Net credit purchase	&-2.47	&11.33	&289		&-5.05	&20.92	&124		&&\\
\hspace{10pt}Net gifts			&-2.04	&6.11	&289		&-3.83	&9.82	&124		&&\\
\multicolumn{8}{l}{Panel B-2: Labor supply adjustments \& sales of miscellaneous assets}\\
\hspace{10pt}Other family members' earnings	&3.74	&6.11	&289		&8.21	&16.44	&124	&&\\
\hspace{10pt}Sales of miscellaneous assets				&0.02	&0.12	&289		&0.06	&0.17	&124	&&\\
\multicolumn{8}{l}{Panel B-3: Income variable}\\
\hspace{10pt}Head's earnings							&29.43	&18.11	&289		&59.2	&29.9	&124	&&\\
&&&&&&&&\\
\multicolumn{8}{l}{\textbf{Panel C: Family size (people)}}\\
&\multicolumn{5}{c}{Semi-monthly panel dataset}&&&\\
\cmidrule(rrrrr){2-6}
&Mean&Std. dev.&Min&Max&Obs.&&&\\
\cmidrule(rrrrrr){1-6}
Number of family members			&4.3	&1.6	&2.0	&8.0	&289	&&&\\
&&&&&&&&\\
\multicolumn{8}{l}{\textbf{Panel D: Health shocks}}\\
&\multicolumn{5}{c}{Semi-monthly panel dataset}&&&\\
\cmidrule(rrrrr){2-6}
&Mean&Std. dev.&Min&Max&Obs.&&&\\
\cmidrule(rrrrrr){1-6}
Number of days with illness (head)		&0.1	&0.8	&0	&11	&254	&&&\\
Number of days with illness				&0.2	&1.7	&0	&16	&254	&&&\\
 (other family members)				&&&&&&&&\\
\midrule[0.3pt]\bottomrule[1pt]
\end{tabular}
}
}
{\scriptsize
\begin{minipage}{435pt}
\setstretch{0.85}
Notes:
\textbf{Panel A}: This panel presents summary statistics for the variables used to test consumption smoothing for the 33- and 26-panel units in the adjusted semi-monthly and adjusted monthly panel datasets, respectively.
Disposable income is defined as total income minus tax payments.
\textbf{Panel B}: This panel presents summary statistics for the variables used to examine risk-coping strategies for the 33- and 26-panel units in the adjusted semi-monthly and adjusted monthly panel datasets, respectively.
Each net income variable is defined as income minus expenditure.
``Net savings'': withdrawals minus deposits to savings.
``Net insurance'': insurance payments received minus expenses paid for insurance.
``Net borrowing'': the amount borrowed minus debt repayments.
This category includes net income from both pawnshops and other lending institutions.
``Net credit purchase'': credit purchases minus credit repayments.
``Net gifts'': total gifts received, including both pecuniary and non-pecuniary, minus any payments for gifts.
``Other family members' earnings'': total income earned by all family members except the household head.
``Sales of miscellaneous assets'': include the sale of goods such as old newspapers and empty bottles.
\textbf{Panel C}: 
This panel shows summary statistics for the family size variable for the 33-panel units in the adjusted semi-monthly panel dataset.
Statistics for the adjusted monthly panel dataset are not reported because they are materially similar.
\textbf{Panel D}: 
This panel shows summary statistics for the number of days family members were ill among the 33-panel units in the adjusted semi-monthly panel dataset.
Since these are lagged variables, the initial period is omitted for all units.
Accordingly, households with only two semi-month observations are excluded.
Source:
Created by the author from the THBS sample.
\end{minipage}
}
\end{center}
\end{table}

\subsubsection*{Trend}

Figure~\ref{fig:tssm_minc_mexp} shows a slight upward trend in average income, consistent with the overall macroeconomic increase in nominal income during this period.
However, this growth in earnings is offset by steep inflation following postwar expansions (Section~\ref{sec:sec21}).
In Figure~\ref{fig:tssm_mexpdef}, the deviation of expenditure from income in January reflects spending on goods for New Year events.
The gaps from June to the first half of July likely correspond to bonus payments.
Peaks in March may indicate preparation for the new fiscal and academic year starting in April, suggesting a higher share of temporary income from withdrawals and loans.
These patterns demonstrate that the THBS household budget data capture key aspects of macroeconomic trends.

\subsubsection*{Idiosyncratic Shocks}

Online Appendix Figure~\ref{fig:hist_is} illustrates the residuals from the regression of the first difference in semi-monthly income on the first difference in aggregate semi-monthly income.\footnote{Let $n_{t}$ be the number of households in semi-month $t$ in 1919. For household $i$ in semi-month $t$, the residual is defined as $\Delta \hat{\upsilon}_{i,t} = \Delta \textit{Income}_{i,t} - \hat{\alpha} \Delta \overline{\textit{Income}}_{.,t}$, where $\overline{\textit{Income}}_{.,t} = \sum_{i=1}^{n_{t}}\frac{\textit{Income}_{i,t}}{n_{t}}$ and $\hat{\alpha}$ is the estimated coefficient.}
Although the residuals satisfy the zero-average property, they occasionally take large positive or negative values, indicating that workers experienced idiosyncratic income shocks in some cells.
These deviations from the mean help examine household responses to idiosyncratic shocks.
Fortunately, the THBS archives (\textit{Kazoku id\=o hy\=o}) document the exact timing and duration of family members' illnesses.
To understand the mechanisms behind consumption smoothing, I treat illness as the least predictable adverse shock \citep{Gertler2002-tu}.
While unemployment could also act as a shock, THBS household heads rarely changed occupations during the survey period, as they were skilled workers unlikely to move to other jobs.\footnote{Household heads who changed jobs may have been omitted from the survey. However, households with longer and shorter time-series observations exhibit similar family characteristics (Online Appendix~\ref{sec:secb_trim}), so this omission does not create selection bias regarding vulnerability to idiosyncratic risk. Moreover, dismissals at the Tsukishima factories were rare. For example, the average monthly dismissal rate from June 1919 to May 1920 at the one factory with available data was $1.5$\% (Department of Health, Ministry of the Interior 1923a, p.~415).}
Panel D of Table~\ref{tab:sum} reports summary statistics for the health shock variables.

\subsubsection*{Risk Preferences}

Different attitudes toward risk across households with varying occupations and in diverse regions can bias standard risk-sharing regressions \citep{Schulhofer-Wohl2011-fo, Mazzocco2012-jl}.
However, this study focuses on consumption-smoothing strategies among homogeneous households---skilled factory workers in the machinery manufacturing sector in Tsukishima.
To assess the homogeneity of risk preferences among THBS households, I test whether the earnings of household heads in specific factories or sectors responded differently to aggregate shocks, following \citet{Schulhofer-Wohl2011-fo}.
The results show that the sensitivities to aggregate shocks were statistically similar.\footnote{The heads' earnings are regressed on aggregate consumption, its interaction with an indicator for large-scale or governmental factories (or for smithing and non-machinery factories), the family size variable, and household fixed effects. The results remain unchanged when a few semi-months with smaller cross-sectional samples are trimmed.}
THBS households shared similar risk preferences, conditional on household fixed effects.
Online Appendix~\ref{sec:secb_preference} provides additional details of the analysis.

\section{Empirical Analysis}\label{sec:sec4}

\subsection{Consumption Smoothing}\label{sec:sec41}

\subsubsection*{Estimation Strategy}\label{sec:sec411}\

I use a linear fixed-effects model to test consumption smoothing.
Online Appendix~\ref{sec:secc0} presents the conceptual framework and the derivation of my empirical specification.
For household $i$ at time $t$, the reduced form equation is
\begin{equation}\label{eqn:eq_cs}
\log c_{i,t} = \gamma \log y_{i,t} +  \delta x_{i,t} + \mu_{i} + \phi_{t}+ u_{i,t},
\end{equation}
where $c_{i,t}$ is consumption, $y_{i,t}$ is disposable income, $x_{i,t}$ is the family size control, $\mu_{i}$ is the household fixed effect, $\phi_{t}$ is the time fixed effect, and $u_{i,t}$ is a random error term.
The household fixed effect captures time-invariant unobservables such as permanent income and consumption preferences.
The time fixed effect accounts for unobservable macroeconomic shocks and trends described in Section~\ref{sec:sec34}.
The estimate of $\gamma$ represents income elasticity, ranging from zero under perfect insurance to one in the absence of insurance.
Since identification relies on both adverse and favorable idiosyncratic income shocks, the estimated elasticity reflects the overall response to these shocks.
A cluster-robust variance-covariance matrix estimator is used to address heteroskedasticity and serial correlation \citep{Arellano1987-sc}.

Another specification commonly used in the literature is the first-difference model, which uses aggregate consumption to capture macroeconomic shocks \citep{Mace1991-lo}.
However, the aggregate measure may not accurately reflect macroeconomic shocks, particularly when the number of households is relatively small.
To be conservative, I therefore employ the two-way fixed-effects model, which systematically controls for macroeconomic shocks through the time fixed effect \citep{Cochrane1991-th, Ravallion1997-kz}.

\subsubsection*{Results}\label{sec:sec412}

\def\arraystretch{1.0}
\begin{table}[htbp]
\begin{center}
\caption{Results of Estimating Income Elasticities}
\label{tab:cs_baseline}
\footnotesize
\scalebox{1.0}[1]{
\begin{tabular}{lrlcrlcrlc}
\toprule[1pt]\midrule[0.3pt]
&\multicolumn{3}{c}{(1) Semi-monthly panels}&\multicolumn{3}{c}{(2) Monthly panels}\\
\cmidrule(rll){2-4}\cmidrule(rll){5-7}
&\multicolumn{2}{c}{Disposable income}&\multirow{2}{*}{Obs.}&\multicolumn{2}{c}{Disposable income}&\multirow{2}{*}{Obs.}\\
\cmidrule(rl){2-3}\cmidrule(rl){5-6}
&Coef.&Std. error&&Coef.&Std. error&\\\hline
Total consumption 						&0.358&[0.033]***	&278	&0.433&[0.059]***	&123	\\
\hspace{10pt}Food						&0.136&[0.030]***	&259	&0.293&[0.066]***	&114	\\
\hspace{10pt}Housing					&0.099&[0.415]		&134	&0.013&[0.084]		&105	\\
\hspace{10pt}Utilities						&-0.003&[0.138]	&223	&0.343&[0.236]		&110	\\
\hspace{10pt}Furniture					&0.426&[0.223]*	&169	&-0.114&[0.706]	&98	\\
\hspace{10pt}Clothes					&0.378&[0.190]*	&238	&0.347&[0.424]		&112	\\
\hspace{10pt}Education					&0.101&[0.051]*	&236	&0.162&[0.128]		&110	\\
\hspace{10pt}Medical expenses			&0.039&[0.066]		&258	&0.194&[0.120]		&114	\\
\hspace{10pt}Entertainment expenses		&0.307&[0.084]***	&250	&0.716&[0.181]***	&113	\\
\hspace{10pt}Transportation				&0.278&[0.110]**	&187	&0.605&[0.264]**	&97	\\
\hspace{10pt}Gifts						&0.523&[0.106]***	&234	&0.469&[0.197]**	&113	\\
\hspace{10pt}Miscellaneous				&-0.076&[0.147]	&214	&0.522&[0.252]**	&110	\\\midrule[0.3pt]\bottomrule[1pt]
\end{tabular}
}
{\scriptsize
\begin{minipage}{375pt}
\setstretch{0.85}
***, **, and * denote statistical significance at the 1\%, 5\%, and 10\% levels, respectively.
Standard errors in brackets are clustered at the household level.\\
Notes: 
This table presents the results of equation~\ref{eqn:eq_cs}.
It reports regressions of 11 measures of log-transformed consumption on log-transformed disposable income, the family size control, the household fixed effect, and the time fixed effect.
Columns 1 and 2 show the results for the adjusted semi-monthly and adjusted monthly panel datasets, respectively.
The estimated coefficients on log-transformed disposable income are shown in the columns labeled ``Coef.''
The estimated elasticities reflect the overall responses of households to both adverse and favorable idiosyncratic income shocks.
\end{minipage}
}
\end{center}
\end{table}

Column 1 of Table~\ref{tab:cs_baseline} presents the results from the semi-monthly panel dataset.
The estimated coefficient for total consumption indicates that a one percentage point decrease in disposable income reduces total expenditure by $0.358$ percentage points, suggesting that factory-worker households could not fully smooth their consumption.
An important finding is that the estimate for food categories is smaller ($0.136$) but highly statistically significant.
This suggests that while households reduced food expenses in response to adverse income shocks, they still had some means to smooth food consumption.
The estimates for entertainment and transportation expenses ($0.307$ and $0.278$) are also statistically significant.
This is understandable because these expenses include family travel and theater visits.\footnote{On this small island, workers could walk to the factories, so commuting costs were typically not recorded in the budget books.}
Similarly, the larger estimate for gifts ($0.523$) reflects that these expenses were discretionary.
The results for other subcategories are statistically insignificant or ambiguous.
Estimates for furniture and clothing are weakly statistically significant, which may reflect that both are luxury categories.
The estimate for the education category is small but also weakly statistically significant ($0.101$), as it includes an allowance for children that can be easily reduced.\footnote{Toyotaro Miyoshi, who served as a research assistant at the Tsukishima Survey, provided important retrospective accounts (Online Appendix~\ref{sec:secb1}). According to his book, children in these factory-worker households regularly received a small allowance to play outside while their parents worked, since no nurseries were available \citep[pp.~383--384; 386; 387]{Miyoshi1989}. \label{miyoshi}}

Column 2 reports the results for the monthly panel dataset examining whether income elasticities change over a longer period.
As expected, the monthly series yields larger estimates than the semi-monthly series.
The estimate for total consumption is $0.433$, approximately $1.2$ times higher than that from the semi-monthly series.
One possible explanation is that the sources of temporary income are limited and insufficient to offset adverse income shocks lasting at least one month.
The most notable example is the food category, which shows more than twice the estimate ($0.293$) compared with the semi-monthly series.
This implies that households' food expenditures could be insured for a few weeks but not over a full month.
Similarly, expenses for entertainment and transportation are much more sensitive ($0.716$ and $0.605$, respectively) to shocks in the monthly series.
The estimate for gifts is slightly lower but remains in a similar range.
Expenses on these discretionary categories are particularly responsive to relatively persistent shocks.

By contrast, some subcategories are not sensitive to idiosyncratic shocks across different time-series frequencies.
For example, estimates for housing, utilities, and medical expenses are not statistically significant in either the semi-monthly or monthly series.
This may reflect the fact that rent and utility payments are relatively fixed.\footnote{Rent is usually fixed and paid once per month. Online Appendix~\ref{sec:secc1} provides quantitative evidence that the housing category should be analyzed using the monthly series rather than the semi-monthly series.}
Medical expenses include fees for bathing (\textit{yuya}).
People bathe every few days, a habit unlikely to change, although the frequency with which women opt for hairdressing services(\textit{kamiyui}) in the same category can reduce.
The estimates for furniture, clothing, and education are statistically insignificant in the monthly series.
While this may partly reflect efficiency issues, these expenses are generally insensitive to unexpected demand shocks such as unanticipated losses of household goods, clothing, or stationery, including writing paper.\footnote{An alternative explanation for the clothing category relates to seasonality. Workers bought new work clothes for the season (OISR, Archives of the Tsukishima Survey (THBS \#40, March 31, 1919)). As these clothes were necessary, seasonal purchases attenuate the estimates.}
This systematically reduces the estimates for specific categories in the monthly series.\footnote{Panels with short-run time bins are more likely to be influenced by unexpected consumption shocks, which attenuate income elasticity \citep{Nelson1994-js}. When this mechanism operates, subcategories sensitive (insensitive) to unexpected consumption shocks are more (less) attenuated. In my results, however, estimates for the monthly panels are higher than those for the semi-monthly panels for many subcategories, suggesting that households faced greater difficulties smoothing consumption over the longer term.}

\subsection{Risk-coping Mechanisms}\label{sec:sec42}

The preceding results suggest that workers were able to cope with short-term shocks to some extent.
In this section, I examine the mechanisms underlying their consumption-smoothing behavior using information on risk-coping strategies.

\subsubsection*{Estimation Strategy}\label{sec:sec421}

To test the roles of risk-coping strategies, I employ the empirical specification proposed by \citet{Fafchamps2003-rt}.
For household $i$ at time $t$, the specification is
\begin{equation}\label{eqn:eq_risk}
r_{i,t} = \kappa \tilde{y}_{i,t} + \eta x_{i, t} + \nu_{i} + \zeta_{t} + \epsilon_{i,t},
\end{equation}
where $\tilde{y}_{i,t}$ is the head's earnings, $x_{i, t}$ is the family size control, $\nu_{i}$ is the household fixed effect, $\zeta_{t}$ is the time fixed effect, and $\epsilon_{i,t}$ is a random error term.
Online Appendix~\ref{sec:secc0} summarizes the derivation of this specification.
I use the head's income instead of total disposable income in Equation~\ref{eqn:eq_risk} to avoid potential endogeneity.\footnote{For example, total household income is the outcome of intra-household responses to shocks, such as labor supply adjustments \citep{Asdrubali2020-ac}. The results are largely unchanged if disposable income is used instead of the head's income (not reported). This is consistent with the absence of labor supply adjustments among THBS households (Table~\ref{tab:risk}). Nevertheless, I use the conservative approach to reduce potential biases and maintain comparability with the structural equation in Section~\ref{sec:sec43}.}
Regarding $r_{i,t}$, I consider several variables to capture different risk-coping strategies, including net income from savings, insurance, borrowing, credit purchases, and gifts.
Each net variable is defined as income minus expenses.
The estimates of $\kappa$ are expected to be negative, since net income is typically negative in good times and positive in unfavorable times.
I also consider earnings of other family members (excluding the head) and income from sales of miscellaneous goods \citep{Udry1995-ri, Blundell2016-ea}.\footnote{\citet{Rosenzweig1989-zz} show that marriage combined with migration in South Indian villages reduced variability in household food consumption. Since the THBS tracked households living in Tsukishima throughout the sample period, ex-ante risk mitigation via migration is not relevant here.}
Estimated coefficients for sales and labor supply adjustment variables may be negative, reflecting additional earnings when the head's income declines.

\subsubsection*{Results}\label{sec:sec422}

\def\arraystretch{1.0}
\begin{table}[htbp]
\begin{center}
\caption{Results of Testing the Risk-coping Mechanisms}
\label{tab:risk}

\footnotesize
\scalebox{1.0}[1]{
\begin{tabular}{lrlcrlc}
\toprule[1pt]\midrule[0.3pt]
&\multicolumn{3}{c}{(1) Semi-monthly panels}&\multicolumn{3}{c}{(2) Monthly panels}\\
\cmidrule(rll){2-4}\cmidrule(rll){5-7}
&\multicolumn{2}{c}{Head's earnings}&\multirow{2}{*}{Obs.}&\multicolumn{2}{c}{Head's earnings}&\multirow{2}{*}{Obs.}\\
\cmidrule(rl){2-3}\cmidrule(rl){5-6}
&Coef.&Std. error&&Coef.&Std. error&\\ \hline
\multicolumn{7}{l}{\textbf{Panel A: Net savings, insurance, borrowing, and gifts (income $-$ expenditure, yen)}}\\
Net savings					&-0.043	&[0.018]**		&289  	&-0.035	&[0.028]	&124	\\
Net insurance					&0.001	&[0.013]		&289  	&0.002	&[0.033]	&124	\\
Net borrowing					&-0.021	&[0.027]		&289  	&-0.027	&[0.017]	&124	\\
Net credit purchase				&-0.128	&[0.039]***	&289 	 &-0.011	&[0.042]	&124	\\
Net gifts						&-0.052	&[0.019]**		&289 	 &-0.006	&[0.029]	&124	\\
&&&&&&\\
\multicolumn{7}{l}{\textbf{Panel B: Labor supply adjustments \& sales of miscellaneous assets (yen)}}\\
Other members' earnings			&0.001	&[0.010]		&289		&0.036	&[0.033]	&124	\\
Sales of miscellaneous assets		&-0.000	&[0.000]		&289		&0.000	&[0.001]	&124	\\
\midrule[0.3pt]\bottomrule[1pt]
\end{tabular}
}
{\scriptsize
\begin{minipage}{425pt}
\setstretch{0.85}
***, **, and * denote statistical significance at the 1\%, 5\%, and 10\% levels, respectively.
Standard errors in brackets are clustered at the household level.\\
Notes: 
This table reports the results of equation~\ref{eqn:eq_risk}.
All regressions include the family size control, household fixed effects, and time fixed effects.
Columns 1 and 2 present estimates for the adjusted semi-monthly and adjusted monthly panel datasets, respectively.
The estimated coefficients on the head's earnings are shown in the ``Coef.'' columns.
\textbf{Panel A}: 
Each net income variable is defined as income minus expenditure.
``Net savings'': withdrawals minus deposits to savings.
``Net insurance'': insurance payouts minus insurance expenses.
``Net borrowing'': borrowing minus debt repayments.
This category includes net income from both pawnshops and other lending institutions.
``Net credit purchase'': credit purchases minus credit repayments.
``Net gifts'': the total received gifts, both pecuniary and non-pecuniary, minus payments for gifts.
\textbf{Panel B}: ``Other members' earnings'': total income earned by all household members except the head.
``Sales of miscellaneous assets'': include the sale of daily items such as newspapers and empty bottles.
All dependent variables are measured in yen.
\end{minipage}
}
\end{center}
\end{table}

Table~\ref{tab:risk} presents the results.
Columns 1 and 2 summarize the findings for the semi-monthly and monthly panel datasets, respectively.
Panel A reports the results for the net income variables, while Panel B reports those for labor supply adjustments and sales of small assets.

The estimated coefficient for net savings is negative and statistically significant in Column 1.
This suggests that a one yen decrease (increase) in the head's earnings increased (decreased) net savings by $0.043$ yen.
The estimate from the monthly series is similar but statistically insignificant (Column 2).
While this reflects that many workers had savings accounts in postal savings and savings banks (Section~\ref{sec:sec23}), the magnitude was small, indicating that withdrawals from savings did not fully compensate for earnings losses.

Estimates for the insurance category are near zero and statistically insignificant in both columns.
The budget books indicate that the primary source was mutual loan associations (\textit{mujin}).
However, these associations' rules were unsuitable for offsetting idiosyncratic shocks.
For example, a THBS household won a lottery and received $100$ yen in a monthly cell, which was independent of the worker's earnings.\footnote{OISR, Archives of the Tsukishima Survey (THBS \#5, May 17, 1919). The bidding rule could support a specific member if all members have a common belief in aid, but most income from \textit{mujin} recorded in the budget books was less than one yen, which is not representative on average.}

For net borrowing, estimated coefficients are close to zero and statistically insignificant in both columns.
Although THBS households occasionally used pawnshops, these instances were infrequent.\footnote{OISR, Archives of the Tsukishima Survey (THBS \#2; 19; 22).}
This is understandable because skilled factory workers were not classified as low-income, who were the primary users of pawnshops (Section~\ref{sec:sec23}).

The estimated coefficient for net income from credit purchases is negative and strongly statistically significant in Column 1.
This suggests that a one yen decrease (increase) in the head's earnings increased (decreased) net income from credit purchases by $0.13$ yen.
This magnitude is the largest among all categories.
According to the budget books, credit purchases were primarily used to buy rice, fish, vegetables, and fuel (firewood and charcoal).
As rice---a staple food---was usually expensive, credit purchases played an important role for these households.
Credit purchases were also occasionally used for medical expenses such as doctor visits and pharmacy purchases, consistent with the baseline results for medical expenses (Column 1 in Table~\ref{tab:cs_baseline}).
While these results indicate that credit purchases were the primary risk-mitigation strategy in the short term, Column 2 shows that credit ceases to be effective over the longer term.
This is reasonable, as borrowers typically made monthly or semi-monthly payments and did not borrow for periods longer than a month (Section~\ref{sec:sec5}).

Estimates for the gifts category are negative and statistically significant in Column 1.
The budget books show that the gifts include seasonal gifts, celebration money, and return gifts.
Most of these gifts were not aid from kinship networks, including rural relatives, but rather monetary and material exchanges with acquaintances based on a culture of gift-giving.
According to an official report from Tokyo City, average monthly gifts accounted for approximately $3$\% of total income (Social Affairs Division 1925, pp.~58-59).
Similarly, this share is approximately $4$\% in my monthly panel dataset.
Even if mutual aid from personal networks is included in this small figure, it likely provided insufficient support for households facing economic hardship.
Accordingly, the estimate ceases to be statistically significant in the monthly series (Column 2) and becomes negligible when focusing on the impact of adverse income shocks (Section~\ref{sec:sec43}).

Finally, Panel B tests the roles of intra-household labor supply adjustments and sales of miscellaneous assets.
The estimated coefficients in both columns of the first row are close to zero and statistically insignificant.
\citet[pp.~383--386]{Miyoshi1989} documented that wives regularly engaged in side jobs such as sewing, rather than taking temporary work to compensate for income losses.
However, their earnings were too small to replace the lost income from their breadwinning husbands.\footnote{OISR, Archives of the Tsukishima Survey (THBS\#1, January 13, 1919).}
In the second row, sales of miscellaneous assets such as newspapers, bottles, and clocks are shown to be ineffective as a risk-mitigation strategy.

\subsubsection*{Heterogeneity}\label{sec:sec423}

The residential area of Tsukishima spanned less than $1.5$ kilometers, with a width of $500$ meters (Ch\=u\=o Ward 1994).
Within this narrow area, workers had easy access to shops in different lots (Department of Health, Ministry of the Interior 1923a, pp.~40--46).
Nevertheless, I conduct several additional tests to explore heterogeneous responses using maps of retailers and pawnshops documented in the report.

First, I calculate the density of retail shops for daily necessities in each lot and include it in the regression for credit purchases.
The estimate from this expanded regression is largely consistent with my baseline estimate (Column 1 in Panel A of Table~\ref{tab:risk}).
As an alternative test, I add an interaction term between the head's earnings and the density of retailers into Equation~\ref{eqn:eq_risk}.
The interaction effect is approximately zero and remains unchanged when using an alternative definition of density.
These results indicate that shop accessibility was homogeneous for the workers.
Online Appendix~\ref{sec:secc_heterogeneity} provides detailed summaries of these analyses.

Second, I examine the role of local pawnshops separately from the net borrowing regression in Table~\ref{tab:risk}.
In European countries and the U.S., religious communities played an important role in providing social insurance \citep{Ferrara2023-jm, AgerUnknown-qj}.
Although some philanthropists assisted the poor before government-managed social welfare became available, few social facilities run by religious communities existed in Japan \citep{Yoshida2004}.
Unlike in European countries \citep{Murhem2016-nl}, pawnshops in Japan were private, profit-oriented enterprises until public pawnshops with social functions were established in 1927 under the Public Pawnbroking Act.
Despite this, local pawnshops may have had informal ties with workers, as they served the same residents of Tsukishima.

When I use temporary income from pawnshops as the dependent variable in Equation~\ref{eqn:eq_risk}, the estimated coefficient on the head's earnings is close to zero and statistically insignificant.\footnote{The cooperative society is another local risk-coping institution, though it is rarely used (Section~\ref{sec:sec23}). Using total purchases from the cooperatives as a dependent variable in Equation~\ref{eqn:eq_risk} yields a coefficient on the head's earnings that is nearly zero and statistically insignificant (Online Appendix~\ref{sec:secc_coop}).}
This result remains unchanged after including the density of pawnshops in each lot in the same regression.
I then estimate an expanded regression of Equation~\ref{eqn:eq_risk} for temporary income from pawnshops, which includes an interaction term between the head's earnings and the density of pawnshops.
Although the interaction effect is negative, it is small and statistically insignificant.
A similar result is obtained when using an alternative definition of the density variable.
These findings suggest that the use of pawnshops was independent of idiosyncratic shocks.
Online Appendix~\ref{sec:secc_coop} summarizes the results.

\subsection{Response to Adverse Health Shocks}\label{sec:sec43}

The identification in the analyses above relies on both adverse and favorable idiosyncratic shocks.
To focus on the mechanisms behind consumption smoothing in response to adverse shocks, I use short-term income losses due to the temporary illnesses of household heads for the identification in this section.

\subsubsection*{Estimation Strategy}\label{sec:sec431}

\begin{figure}[htb]
\centering
\captionsetup{justification=centering,margin=1.5cm}
\includegraphics[width=0.50\textwidth]{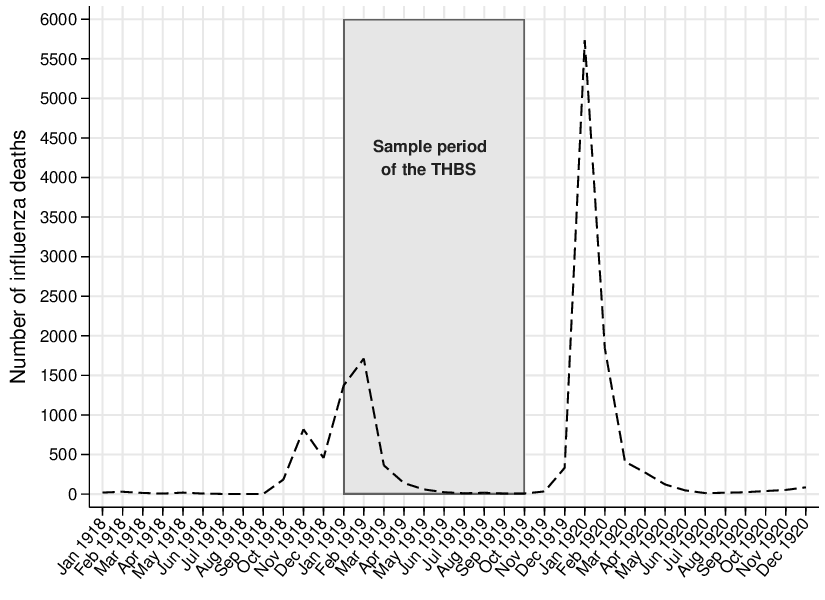}
\caption{Number of Influenza Deaths in Tokyo between 1918 and 1920}
\label{fig:ts_flu}
\scriptsize{\begin{minipage}{400pt}
\setstretch{0.85}
Notes: 
This figure shows time-series plots of the monthly death counts from influenza (\textit{ry\=uk\=osei kanb\=o}) in Tokyo prefecture from January 1918 to December 1920.
The death statistics are collected through the comprehensive registration system.
Source: 
Created by the author using the 1918, 1919, and 1920 editions of the Statistics of Causes of Deaths of the Empire of Japan (Statistics Bureau of the Cabinet 1921; 1922; 1923).
\end{minipage}}
\end{figure}

I begin my analysis by testing whether the head's illness affected earnings.
Figure~\ref{fig:ts_flu} shows the monthly death counts from influenza (\textit{ry\=uk\=osei kanb\=o}) in Tokyo between 1918 and 1920.
Early 1919 saw a moderate spike, partially captured in the THBS period.
Indeed, a few THBS budget books indicate that family members contracted influenza during this time.
According to the \textit{Kazoku id\=o hy\=o} (Section~\ref{sec:sec34}), five households experienced illness during the sample period, with illness affecting the heads concentrated between February and May 1919, when influenza exposure was most prevalent.
The average duration of illness for the heads was $5.8$ days (Std.~Dev.~$= 3.6$), consistent with Type A influenza, which causes fever lasting less than six days \citep{Uyeki2022-jt}.
Therefore, influenza likely caused the illnesses during this period.

Earnings in a semi-month are based on work performed during the previous semi-month.
Table~\ref{tab:r_hs} presents the regression results of the head's earnings on the number of illness days in the previous semi-monthly cell.
While instantaneous health shocks have little impact on earnings (Column 1), lagged health shocks have a statistically significant effect (Column 2).
This result remains unchanged when both variables are included in the same regression (Column 3), supporting the evidence that health shocks in the previous cell reduce the head's earnings.
The estimate in Column 2 suggests that a one day increase in illness reduces the head's semi-monthly income by approximately $2.4$ yen.
Notably, this is close to the average daily wage of skilled factory workers, reflecting salary reductions due to absence (Online Appendix~\ref{sec:secb_est_wage}).
This aligns with anecdotal evidence that the head's illness directly caused income loss due to absence \citep{Miyoshi1989}.
Finally, Column 4 adds the health shocks of other family members to the specification in Column 3.
The estimated coefficients on these placebo variables are statistically insignificant, supporting the robustness of the results in Column 3.

\def\arraystretch{0.95}
\begin{table}[htbp]
\begin{center}
\captionsetup{justification=centering,margin=1.5cm}
\caption{Results of Testing the Impacts of Adverse Health Shocks\\ on the Head's Earnings}
\label{tab:r_hs}
\footnotesize
\scalebox{1.0}[1]{
\begin{tabular}{lD{.}{.}{-2}D{.}{.}{-2}D{.}{.}{-2}D{.}{.}{-2}}
\toprule[1pt]\midrule[0.3pt]
&\multicolumn{4}{c}{Dependent variable: semi-monthly earnings of the head}\\
\cmidrule(rrrr){2-5}
&\multicolumn{1}{c}{(1)}			&\multicolumn{1}{c}{(2)}		&\multicolumn{1}{c}{(3)}		&\multicolumn{1}{c}{(4)}		\\\hline
Health shocks on:&&&&\\
\hspace{5pt}The head [level]	&-0.061	&				&-0.972		&-1.319*	\\
								&[0.679]	&				&[0.699]		&[0.666]	\\
\hspace{5pt}The head [lag]		&			&-2.371$***$	&-2.868$***$	&-2.994$***$	\\
								&			&[0.644]		&[0.734]		&[0.765]	\\
\hspace{5pt}Others [level]	&			&				&				&0.889		\\
								&			&				&				&[1.275]	\\
\hspace{5pt}Others [lag]	&			&				&				&3.675		\\
								&			&				&				&[3.312]	\\\hline
Observations 					&\multicolumn{1}{c}{287}		&\multicolumn{1}{c}{254}		&\multicolumn{1}{c}{254}		&\multicolumn{1}{c}{254}		\\
\midrule[0.3pt]\bottomrule[1pt]
\end{tabular}
}
{\scriptsize
\begin{minipage}{370pt}
\setstretch{0.85}
***, **, and * denote statistical significance at the 1\%, 5\%, and 10\% levels, respectively.
Standard errors in brackets are clustered at the household level.\\
Notes: 
This table presents results from regressions of the head's earnings on health shock variables using the adjusted semi-monthly panel dataset.
Households with only two semi-month observations are excluded from all regressions because lagged health shock variables are included.
All regressions include the family size control, household fixed effects, and time fixed effects.
``Health shocks on the head'': the number of days the head was ill.
``Health shocks on others'': the number of days other family members (excluding the head) were ill.
``[level]'': the variable measured in the same semi-month cell.
``[lag]'': the variable from the previous semi-month cell.
Although the correlation between the level and lag of health shocks for other family members is high (Pearson's coefficient $=0.99$), including either variable does not change the finding that health shocks for other family members have negligible effects.
\end{minipage}
}
\end{center}
\end{table}

I then characterize equation~\ref{eqn:eq_cs} as a structural equation by using the first-stage reduced form equation as follows:
\begin{equation}\label{eqn:eq_cs_fs}
\log y_{i,t} = \beta h_{i,t-1} +  \psi x_{i,t} + \varpi_{i} + \tau_{t}+ e_{i,t},
\end{equation}
where $h_{i,t-1}$ is the number of days the head was ill in the previous semi-month cell, $\varpi_{i}$ is the household fixed effect, $\tau_{t}$ is the semi-month fixed effect, and $e_{i,t}$ is a random error term.
For the analysis using equation~\ref{eqn:eq_risk}, the log-transformed disposable income in equation~\ref{eqn:eq_cs_fs} ($y_{i,t}$) is replaced with the head's earnings ($\tilde{y}_{i,t}$).
The health shock variable is plausibly exogenous because the illnesses resulted from influenza epidemics.\footnote{As noted in Section~\ref{sec:sec23}, factories did not provide sick leave compensation because the Factory Act did not specify benefits for workers' illness (Study Group of the Factory Act 1916). Even at the largest factory in Tsukishima, the Ishikawajima Shipyard, the company did not record any compensation for influenza epidemics (Arai 1930).}
In addition, the health shock variable is statistically correlated with the head's earnings (Table~\ref{tab:r_hs}).
Together, this evidence supports the exogeneity of the instrument and satisfies the relevance condition in this just-identified system.

\subsubsection*{Results: To What Extent do Illnesses Affect Consumption?}\label{sec:sec432}

\def\arraystretch{1.0}
\begin{table}[htb]
\begin{center}
\captionsetup{justification=centering,margin=1.5cm}
\caption{Results of Estimating Income Elasticities:\\ Smoothing Consumption against Adverse Health Shocks}
\label{tab:cs_sm_health}
\footnotesize
\scalebox{1.0}[1]{
\begin{tabular}{lrlcrlcc}
\toprule[1pt]\midrule[0.3pt]
&\multicolumn{7}{c}{Semi-monthly panel dataset}\\
\cmidrule(rllrlll){2-8}
&\multicolumn{3}{c}{(1) Reduced-form approach}&\multicolumn{4}{c}{(2) Instrumental variable approach}\\
\cmidrule(rll){2-4}\cmidrule(rlll){5-8}
&\multicolumn{2}{c}{Head's illness}&\multirow{2}{*}{Obs.}&\multicolumn{2}{c}{Disposable income}&\multirow{2}{*}{Obs.}&First-stage\\
\cmidrule(rl){2-3}\cmidrule(rl){5-6}
&Coef.&Std. error&&Coef.&Std. error&&$F$-statistics\\ \hline
Total consumption 									&-0.118	&[0.015]***	&254  &0.543	&[0.076]***	&245&17.6\\
\hspace{10pt}Food									&-0.075	&[0.017]***	&236  &0.355	&[0.075]***	&227&18.4\\
\hspace{10pt}Housing								&-0.017	&[0.070]		&126  &0.118	&[0.477]		&126&7.6\\
\hspace{10pt}Utilities									&-0.051	&[0.066]		&199  &0.201	&[0.331]		&192&12.4\\
\hspace{10pt}Furniture								&-0.157	&[0.106]		&156  &0.761	&[0.801]		&150&6.2\\
\hspace{10pt}Clothes								&-0.339	&[0.050]***	&219  &1.233	&[0.222]***	&211&65.4\\
\hspace{10pt}Education								&-0.097	&[0.024]***	&216  &0.462	&[0.101]***	&209&18.9\\
\hspace{10pt}Medical expenses						&-0.002	&[0.028]		&236  &-0.001	&[0.142]		&227&18.4\\
\hspace{10pt}Entertainment expenses					&-0.094	&[0.055]*		&227  &0.427	&[0.501]**		&218&20.2\\
\hspace{10pt}Transportation							&-0.100	&[0.094]		&170  &0.436	&[0.510]		&162&20.5\\
\hspace{10pt}Gifts									&-0.325	&[0.093]***	&212  &1.567	&[0.797]*		&204&17.8\\
\hspace{10pt}Miscellaneous							&-0.075	&[0.027]***	&195  &0.343	&[0.101]***	&189&17.3\\\midrule[0.3pt]\bottomrule[1pt]
\end{tabular}
}
{\scriptsize
\begin{minipage}{440pt}
\setstretch{0.85}
***, **, and * denote statistical significance at the 1\%, 5\%, and 10\% levels, respectively.
Standard errors in brackets are clustered at the household level.\\
Notes: 
\textbf{Column 1} reports results of equation~\ref{eqn:eq_cs}, replacing log-transformed disposable income with the head's illness (number of days ill in the previous semi-month).
Dependent variables are the 11 measures of log-transformed consumption.
All regressions include the family size control, household fixed effects, and time fixed effects.
The coefficient on the health shock variable is shown in the ``Coef.'' sub-column.
\textbf{Column 2} treats log-transformed disposable income as endogenous, using the head's illness as an instrument in the first-stage equation (\ref{eqn:eq_cs_fs}).
The estimated coefficient on disposable income appears in the ``Coef.'' sub-column, and first-stage $F$-statistics for the instrument are listed in the final column.
\end{minipage}
}
\end{center}
\end{table}

I first consider a reduced-form specification using the head's illness as the exposure variable in equation~\ref{eqn:eq_cs}, following \citet{Cochrane1991-th}.
Column 1 in Table~\ref{tab:cs_sm_health} presents the results, providing evidence that adverse health shocks to the heads reduced consumption levels.
Next, I explicitly model the structure underlying the relationship between health shocks and earnings using the reduced-form equation in (\ref{eqn:eq_cs_fs}).
If the negative impacts of health shocks on consumption derive from income losses, the results from the instrumental variable approach should correspond to those in Column 1.
The estimates are listed in Column 2 of Table~\ref{tab:cs_sm_health}.
As shown, these results align with the reduced-form findings in Column 1.
Importantly, the estimates are systematically larger than those reported in Table~\ref{tab:cs_baseline}, with total and food consumption estimates of $0.543$ and $0.355$, respectively.
While this partly reflects bias reduction under instrumental variable estimation, it also reflects the system's design to reduce bias by specifying the channel from adverse (rather than favorable) income shocks.
For example, allowances for children can be cut during unfavorable periods but may not increase during favorable ones.
Similarly, furniture expenses are more likely to be reduced during shocks, but households purchase furniture infrequently, providing few observations for rare health shocks.\footnote{In other words, furniture may only have been purchased when households had extra income.}
This pattern may partly explain the smaller first-stage $F$-statistics for the furniture subcategory.\footnote{Note that the logarithmic transformation retains cells with furniture expenses. Thus, the reduced-form equation captures the relationship between health shocks and earnings in cells with any furniture expenses. While the first-stage $F$-statistic is $35.7$ in the linear-linear model, the second-stage result is robust to the functional form assumption. This may reflect the asymptotic advantage of panel data models under nearly weak instruments \citep{Staiger1997-tu, Cai2012-ea, Baltagi2012-hm}. The same argument applies to the housing category.}
A similar interpretation applies to the transportation category, as family travel is infrequent and unlikely to coincide with illness events.
By contrast, entertainment expenses are still likely to be cut in case of illness, reflecting reductions in tobacco use, theater visits, and allowances for heads.
This may explain the differences from the baseline results in Table~\ref{tab:cs_baseline}.

Although the estimates for clothes and gifts now exceed one, the null hypothesis of equality to the population mean of one is not statistically rejected for either category.
While the large estimates may partly result from the instrument based on rare health shock events, the findings are broadly consistent with the baseline results---both categories remain sensitive to shocks, similar to other dispensable categories.

\subsubsection*{Results: How did the Households Compensate for Losses from Illness?}\label{sec:sec433}

\def\arraystretch{1.0}
\begin{table}[htb]
\begin{center}
\captionsetup{justification=centering,margin=1.5cm}
\caption{Results of Testing the Risk-coping Mechanisms:\\ Adverse Health as the Idiosyncratic Shock}
\label{tab:risk_hs}
\footnotesize
\scalebox{1.0}[1]{
\begin{tabular}{lrlcrlc}
\toprule[1pt]\midrule[0.3pt]
&\multicolumn{6}{c}{Semi-monthly panels}\\
\cmidrule(rllrll){2-7}
&\multicolumn{3}{c}{(1) Reduced-form approach}&\multicolumn{3}{c}{(2) Instrumental variable approach}\\
\cmidrule(rll){2-4}\cmidrule(rll){5-7}
&\multicolumn{2}{c}{Head's illness}&\multirow{2}{*}{Obs.}&\multicolumn{2}{c}{Head's earnings}&\multirow{2}{*}{Obs.}\\
\cmidrule(rl){2-3}\cmidrule(rl){5-6}
&Coef.&Std. error&&Coef.&Std. error&\\ \hline
\multicolumn{7}{l}{\textbf{Panel A: Net savings, insurance, borrowing, and gifts (income $-$ expenditure, yen)}}\\
Net savings					&0.397		&[0.318]	&254	&-0.167	&[0.117]		&254	\\
Net insurance					&-0.168	&[0.204]	&254	&0.071		&[0.100]		&254	\\
Net borrowing					&0.116		&[0.261]	&254	&-0.049	&[0.103]		&254	\\
Net credit purchase				&1.294		&[0.674]*	&254	&-0.546	&[0.203]***	&254	\\
Net gifts						&0.181		&[0.177]	&254	&-0.077	&[0.090]		&254	\\
&&&&&&\\
\multicolumn{7}{l}{\textbf{Panel B: Labor supply adjustments \& sales of miscellaneous assets (yen)}}\\
Other members' earnings	&0.001	&[0.081]	&254	&-0.001	&[0.034]		&254 	\\
Sales of miscellaneous assets		&0.002		&[0.002]	&254	&-0.001	&[0.001]		&254	\\
\midrule[0.3pt]\bottomrule[1pt]
\end{tabular}
}
{\scriptsize
\begin{minipage}{430pt}
\setstretch{0.85}
***, **, and * denote statistical significance at the 1\%, 5\%, and 10\% levels, respectively.
Standard errors in brackets are clustered at the household level.\\
Notes:
\textbf{Column 1} reports the results of equation~\ref{eqn:eq_risk}, where the head's earnings are replaced by the head's illness (i.e., the number of days ill in the previous semi-month cell).
Family size, household fixed effects, and time fixed effects are included in all regressions.
The estimated coefficients on the head's illness are listed in the sub-column labeled ``Coef.''
\textbf{Column 2} reports the results of equation~\ref{eqn:eq_risk}, treating the head's earnings as the endogenous variable.
The head's illness serves as an instrumental variable in the first-stage reduced-form equation.
The first-stage results match those in Column 2of Table~\ref{tab:r_hs}, with a first-stage $F$-statistic of $13.6$.
The estimated coefficients on the head's earnings are listed in the sub-column labeled ``Coef.''
\textbf{Panel A}: 
Each net income variable is defined as income minus expenditure.
``Net savings'': withdrawals minus deposits.
``Net insurance'': insurance receipts minus premiums paid.
``Net borrowing'': borrowing minus debt repayments.
This category considers net income from both pawnshops and other lending institutions.
``Net credit purchase'': total credit purchases minus repaid credit.
``Net gifts'': total received gifts, including pecuniary and non-pecuniary, minus any gift payments.
\textbf{Panel B}: ``Other members' earnings'': the total income earned by all household members except the head.
``Sales of miscellaneous assets'': include sales of daily items such as newspapers and empty bottles.
All dependent variables are in yen.
\end{minipage}
}
\end{center}
\end{table}

Table~\ref{tab:risk_hs} presents the analysis of the risk-coping strategies for mitigating adverse health shocks. 
Column 1 presents the reduced-form results.
The estimated coefficients on the net income variables are positive in most cases, indicating households' responses to the shocks.
The estimate for the credit purchase category is weakly statistically significant and has the largest magnitude.
This suggests that a one standard deviation increase in the duration of the head's flu exposure increases net credit purchases by $4.6$ yen ($1.29 \times 3.6~\text{days}$), which represents approximately $16$\% of their average semi-monthly earnings of the head (Table~\ref{tab:sum}).
Column 2 reports the estimates from the instrumental variable approach, treating the head's earnings as endogenous.
The estimate for the credit purchase category is approximately four times larger than that reported in Column 1 of Table~\ref{tab:risk}.
It implies that a one yen decrease in the head's earnings increases net income from credit purchases by $0.55$ yen.
This magnitude aligns with the direct effect of health shocks on net credit purchases estimated in Column 1 of Table~\ref{tab:risk_hs}.\footnote{Given their average daily wage of $2.56$ (Online Appendix~\ref{sec:secb_est_wage}), $2.56 \times 0.55 = 1.4$ yen, which is close to $1.3$ yen in Column 1.}
The estimates for the savings and gifts categories remain negative but are now statistically insignificant.

Overall, these results suggest that the primary response to adverse idiosyncratic income shocks among THBS households was to use credit purchases.
Compensation via credit purchases accounted for roughly half of their temporary income losses, helping to smooth consumption in the short term.

\section{Credit System in the Local Economy}\label{sec:sec5}

The results highlight the importance of the relationships between workers and local retailers.
To the best of my knowledge, however, little historical research exists on credit purchases in prewar Japan, and the details of this credit institution remain unclear.
As with state-contingent loans in the rural credit market \citep{Udry1994-vp}, information on credit purchases is personal and poorly documented.
Accordingly, I analyze the institution using the THBS dataset along with a few available official documents.

\subsection{Availability and Premiums: Evidence from the THBS Dataset}\label{sec:sec51}

\begin{figure}[htb]
\captionsetup{justification=centering,margin=0.25cm}
\centering
\subfloat[Daily panel dataset:\\ purchase and payment]{\label{fig:tsd_mcp}\includegraphics[width=0.4\textwidth]{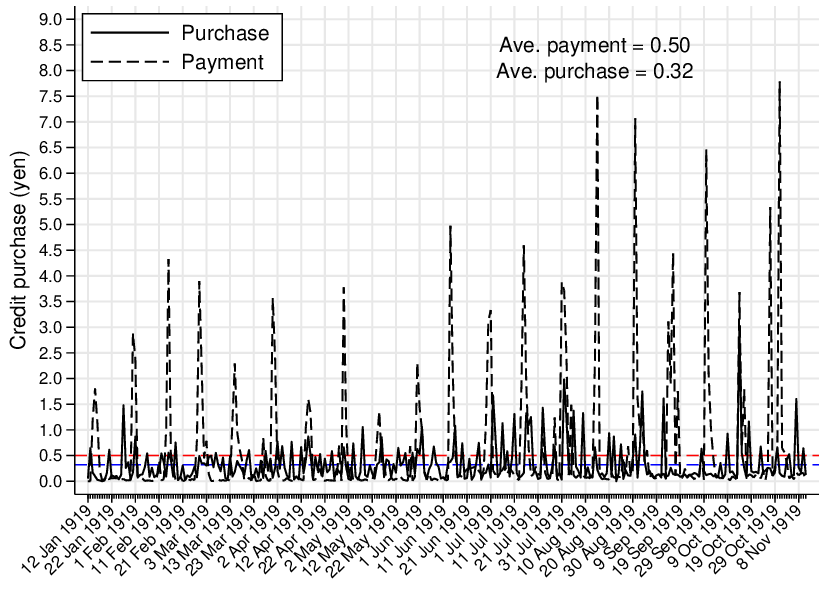}}
\subfloat[Daily panel dataset:\\ net purchase]{\label{fig:tsd_mcpdef}\includegraphics[width=0.4\textwidth]{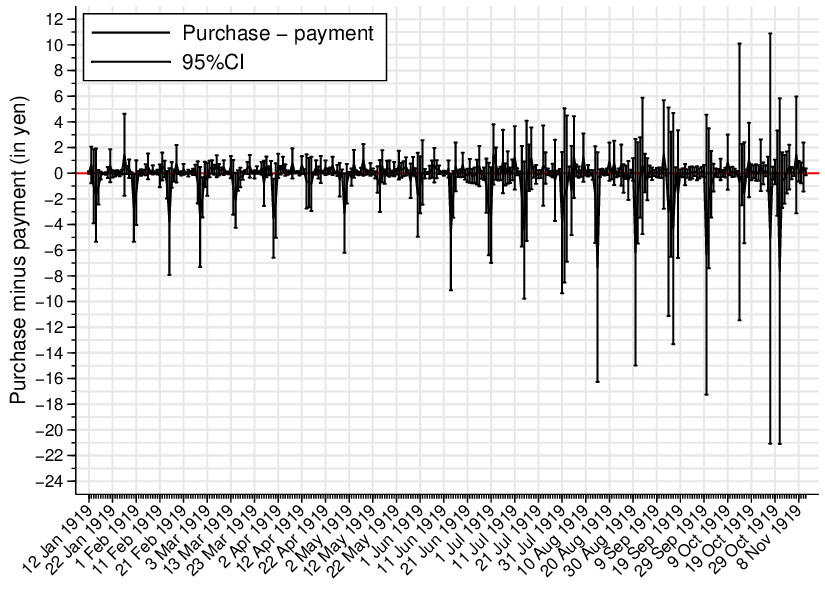}}\\
\subfloat[Semi-monthly panel dataset:\\ purchase and payment]{\label{fig:tssm_mcp}\includegraphics[width=0.4\textwidth]{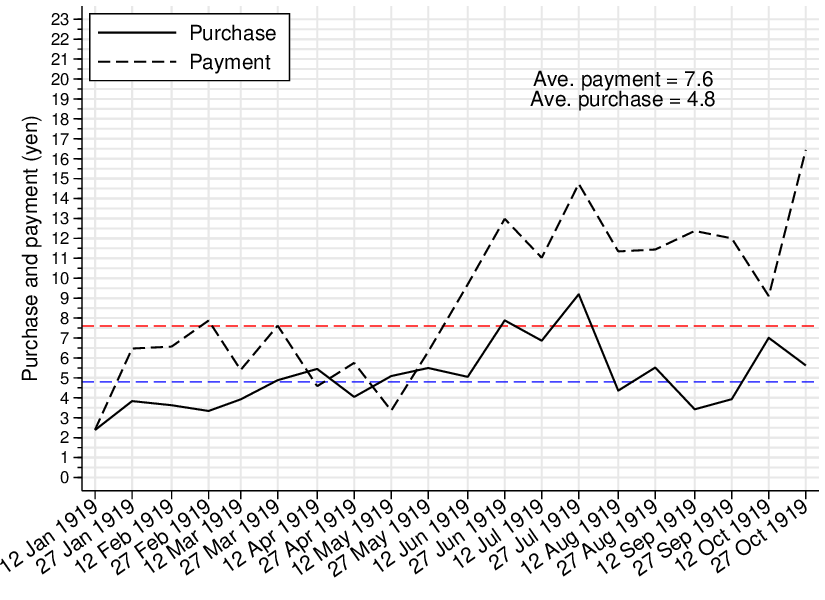}}
\subfloat[Semi-monthly panel dataset:\\ net purchase]{\label{fig:tssm_mcpdef}\includegraphics[width=0.4\textwidth]{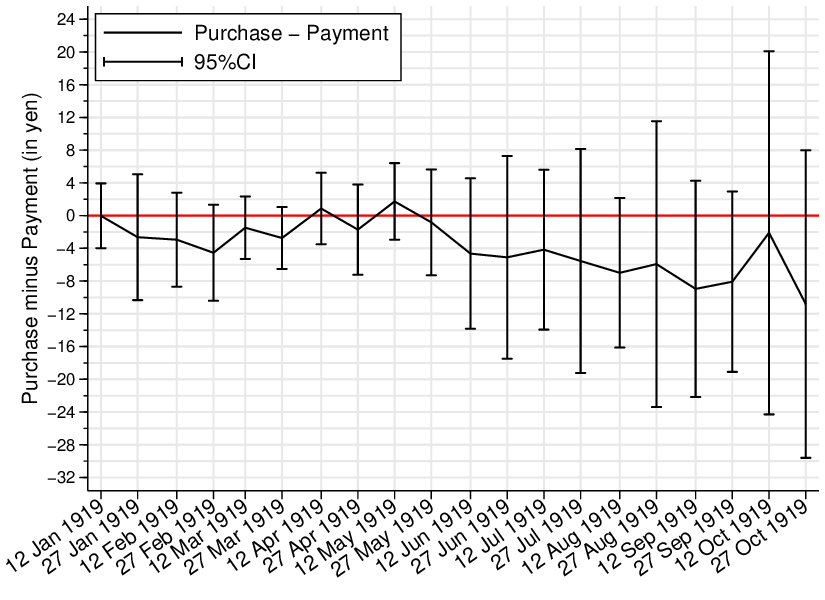}}

\caption{Relationships between Average Credit Purchases and Payments\\ by Different Time-series Frequencies}
\label{fig:ts_cp}
\scriptsize{\begin{minipage}{400pt}
\setstretch{0.85}
Notes:
Figures~\ref{fig:tsd_mcp} and~\ref{fig:tssm_mcp} plot average daily and semi-monthly credit purchases and payments, respectively. The blue and red dashed lines indicate the average credit purchases and payments, respectively. Figures~\ref{fig:tsd_mcpdef} and~\ref{fig:tssm_mcpdef} plot credit purchases minus payments. The red solid line marks zero.
Daily data cover January 12--November 11, 1919.
Figures~\ref{fig:tssm_mcp} and~\ref{fig:tssm_mcpdef} show the semi-monthly series derived from the same daily data.
All figures use the 29 THBS households that made credit purchases.
Source: Created by the author using the THBS sample.
\end{minipage}}
\end{figure}

I begin this descriptive analysis by examining credit purchases in the THBS dataset.
Most THBS households ($29/33$) made credit purchases during the sample period, confirming that this practice was common among factory workers.
Figure~\ref{fig:tsd_mcp} illustrates the average daily credit purchases and payments.
Payments generally peak semi-monthly, reflecting the practice of settling credit purchases on or shortly after paydays.
The purchase series shows a similar pattern but with smaller peaks and greater variance.
This indicates that households first make purchases on or soon after paydays and then consume perishable goods such as fish, fruit, and vegetables.
Figure~\ref{fig:tsd_mcpdef}, which illustrates net purchases, confirms this behavior: households used credit daily and repaid it near paydays.
Figure~\ref{fig:tssm_mcp} shows that payments slightly exceed purchases, implying near-full loan repayment for most households.
Although the difference between payment and purchase becomes more noticeable after August 1919, it is not statistically significant in all semi-months (Figure~\ref{fig:tssm_mcpdef}).\footnote{Strictly, payments can include accrued liabilities from prior periods. Therefore, average payments reasonably exceed average purchases in most semi-month cells. Note that the wider variances in the later sample period are owing to the smaller number of available households (Online Appendix Figure~\ref{fig:panel_structure}).}

Next, I analyze whether earnings volatility restricts credit purchases.
Because household credit is unobservable, I use earnings volatility as a proxy for credit risk in the following empirical specification:
\begin{equation}\label{eqn:eq_cc}
I(\overline{CP}_{i} = 0) = \varkappa_{0} + \varkappa_{1} CV_{i} + \bar{\mathbf{x}}_{i}' \boldsymbol{\chi} + v_{i},
\end{equation}
where $I(\cdot)$ is an indicator function, $\overline{CP}$ is the (unit-)average credit purchases, $CV$ is the coefficient of variation for the head's earnings, $\bar{\mathbf{x}}$ is a vector of (unit-)average family composition variables, and $v$ is a random error term.
All variables are calculated using the semi-monthly series.\footnote{For example, the coefficient of variation for household $i$ is defined as $CV_{i} = \{\sum_{t}^{T_{i}}(\tilde{y}_{i,t}-\bar{\tilde{y}}_{i,.})^{2}/(T_{i}-1)\}^{\frac{1}{2}}/ \bar{\tilde{y}}_{i,.}$, where $T_{i}$ indicates the number of time-series observations, and $\bar{\tilde{y}}_{i,.}$ is the time-series average of the head's income.}
In this cross-sectional model, I include family composition variables to control for unobserved preferences, since household fixed effects cannot be included as an error component.
If retailers allow households to use credit regardless of earnings variability, the estimate of $\varkappa_{1}$ should be close to zero.
By contrast, if households with high earnings volatility are restricted from trading on credit, the estimate should be positive and sizable.
Online Appendix~\ref{sec:secc_volatility} summarizes the results.
In short, the estimated coefficient ($\hat{\varkappa}_{1}$) is close to zero and statistically insignificant.
This finding remains unchanged when alternative thresholds for the dependent indicator variable are applied, including households below the first quartile of average credit purchases.
This suggests that retailers were generous in considering earnings volatility when extending credit to factory workers, consistent with their incentive to retain local customers (Section~\ref{sec:sec52}).

This, in turn, suggests that retailers may factor risk into retail prices or by charging interest on loans.
Unfortunately, no systematic statistics exist on interest charged for credit purchases.
One might attempt to reverse-calculate risk premiums by differencing payments and purchases.
However, this approach requires information on both the price and type of retailer.
Budget books usually record prices by item, but not by \textit{retailer}, which is problematic because the same items are often sold by different types of retailers--for example, eggs are commonly sold in greengrocers (\textit{yaoya}) but occasionally also in general stores (\textit{aramonoya}).
Fortunately, a specific month in a budget book provides price-by-retailer information.\footnote{The OISR, Archives of the Tsukishima Survey (THBS \#4, February 1919).}
According to this record, reimbursements were made semi-monthly, as shown in Figure~\ref{fig:tsd_mcpdef}.
Between February 1 and February 14, 1919, the household purchased various goods from greengrocers, fish stores, and general stores, and paid exactly the same amounts on the 15th.\footnote{The repayment period for credit purchases among retailers and wholesalers (footnote~\ref{cpr}) was often set at approximately one month (Tokyo City Office 1932, pp.~157-159). Therefore, customers likely made repayments at semi-monthly or monthly intervals.}
In the latter half of February, the household paid the same amounts to greengrocers and general stores on the 28th, but less to fish stores.
Because the March budget book lacks information on retailer type, determining exactly how the reimbursement shortfall was handled is difficult.
Nevertheless, this case suggests that risk premiums were incorporated into retail prices rather than charged as explicit interest.

\subsection{Local Environment and Retailers' Incentives: Evidence from the Official Surveys}\label{sec:sec52}

Next, I examine the circumstances among retailers using a survey report published by the Tokyo City Office in 1930 (hereafter, the 1930 Survey Report).\footnote{This survey was conducted in February and March 1930 by Tokyo City and covered $3,892$ small and medium enterprises (Tokyo City Office 1932a, p.~1), including retailers, wholesalers, and other small businesses. The number of respondents varies by question. The statistical table on payment methods includes $2,654$ respondents, of whom $1,197$ were retailers (Tokyo City Office 1932b, p.~273). To the best of my knowledge, this is the most comprehensive survey of retailers in Tokyo circa 1930.}
Although a similar survey around 1920 is unavailable, the industrial and household structures in 1920 and 1930 were generally comparable (Online Appendix Table~\ref{tab:tabc_1930}).
Thus, the 1930 report can serve as a reference to illustrate how credit purchases functioned in the city.
According to the report, credit purchases were a popular payment method across consumer transactions: $99$\%($155/157$) of grain retailers, $81$\%($35/43$) of fish retailers, $78$\%($64/82$) of greengrocers, and $94$\%($45/48$) of fuel retailers used credit purchases as part of their selling methods (Tokyo City Office 1932b, pp.~260; 263).
This widespread use indicates that credit purchases were neither region- nor religion-specific customs.

\def\arraystretch{1.0}
\begin{table}[htb]
\begin{center}
\captionsetup{justification=centering,margin=1.5cm}
\caption{Customers, Sales Methods, and Competitiveness\\ among Retailers in Tokyo City}
\label{tab:cp_tokyo}
\scriptsize
\scalebox{0.89}[1]{
\begin{tabular}{llrrrrrrrr}
\toprule[1pt]\midrule[0.3pt]

Selling item				&Number of 	&Share of sales		&\multicolumn{3}{l}{Social class of}			&\multicolumn{3}{l}{Sales method (\%)}	&Number of \\
\cmidrule(rrr){7-9}
						&stores			&for regular		 	&\multicolumn{3}{l}{ regular customers (\%)}	&Cash		&Credit		&Monthly	&competitors\\
\cmidrule(rrr){4-6}
								&&customers (\%)		&Upper		&Middle		&Other				&			&purchase	&installment			\\\hline
White Rice				&101			&73.1				&16		&45.7		&38.2				&26.9		&71.5		&1.7			&8.5		\\
Fish						&59			&64.6				&6.8		&57.6		&35.6				&43.9		&56.1		&0				&5.0		\\
Fruits and vegetables	&66			&45.4				&9.4		&40.4		&50.2				&65.8		&34.2		&0				&6.2		\\
Firewood and charcoal	&100			&51.1				&15.7		&50.9		&33.4				&20.4		&78.4		&1.3			&8.6		\\\midrule[0.3pt]\bottomrule[1pt]
\end{tabular}
}
{\scriptsize
\begin{minipage}{445pt}
\setstretch{0.85}
Notes: 
This table summarizes sales statistics for $326$ retailers in Tokyo surveyed in 1935, covering white rice, fish, fruits and vegetables, and firewood and charcoal. The number of stores reporting the share of regular customers (Column 2) is $90$, $56$, $61$, and $93$ for the white rice, fish, fruits and vegetables, and fuel categories, respectively. The ``share of sales for regular customers'' refers to the percentage of annual sales made to regular customers. Social classes (``Upper,'' ``Middle,'' and ``Other'') are not defined in the report. The share of sales by method is based on the percentage of annual sales conducted using each method. The number of competitors indicates peers within a three-\textit{cho} radius (approximately $327$ meters). Department stores and retail markets are excluded from this survey.
Source: Tokyo Chamber of Commerce and Industry 1937, pp.~legend; 20--21; questionnaire.
\end{minipage}
}
\end{center}
\end{table}

A survey of retailers in 1935, published by the Tokyo Chamber of Commerce and Industry, provides detailed information on sales practices.
Table~\ref{tab:cp_tokyo} summarizes these statistics.\footnote{The survey covered $939$ retailers, and the report was finalized on December 31, 1935. It includes detailed questions regarding sales and purchases (Tokyo Chamber of Commerce and Industry 1937, pp.~legend, questionnaire). Online Appendix Table~\ref{tab:cp_tokyo_full} decomposes the statistics in Table~\ref{tab:cp_tokyo} by business scale, confirming that the main findings are materially similar.}
First, a large share of annual sales come from regular customers, most of whom belonged to non-upper classes.
$73$\% of annual sales by rice retailers come from regular customers, and more than $80$\% of these customers were from non-upper classes, including factory workers.
Second, credit purchase was a popular sales method, particularly among rice and fuel retailers: more than $70$\% of sales in both categories were conducted on credit.\footnote{These figures are systematically lower than those in the 1930 Survey Report. Both surveys, however, show the same general tendency toward credit purchases; differences may reflect variations in the samples, retailer category definitions, or the definition of credit purchase itself.}
Cash was also accepted by greengrocers, but $34$\% of their sales were on credit.
This confirms that retailers engaged in credit transactions, often maintaining long-term relationships with regular customers.\footnote{A 1924 textbook for retailers explains the psychology behind these long-term relationships: credit purchases are convenient for customers from a budget perspective and confer ``vanity'' or status as regular customers (Shimizu 1924, pp.~115--116). Retailers also commonly used credit purchases to buy source goods from wholesalers (Tokyo City Office 1932a, pp.~154--157), making credit a familiar payment method for both retailers and consumers. \label{cpr}}
The final column of Table~\ref{tab:cp_tokyo} provides suggestive evidence on competition.
The survey asks whether the sales area is competitive and, if so, how many competitors are located within a three-\textit{cho} (approximately $327$ meters) radius.
Rice and fuel stores have two to three more competitors than retailers selling fish or fruits and vegetables.
Given the relatively high use of credit among rice and fuel retailers, this suggests that they may have responded to competition by extending credit to their customers.
Online Appendix~\ref{sec:secc_1935survey} presents regression-based evidence of a positive relationship between credit use and the number of local competitors.

Finally, I assess the losses suffered by retailers from unpaid credit purchases.
According to the 1930 Survey Report, $99$\% of rice retailers and $97$\% of fuel retailers experienced losses on credit sales.\footnote{Online Appendix~\ref{sec:secc_1930survey} summarizes detailed statistics on losses from credit purchases in the 1930 Survey Report.}
This indicates that credit purchases were inherently loss-making for retailers.
However, these losses were generally small, typically capped at 5\% of total sales.
For example, available statistics show that $87$\% of rice retailers and $65$\% of fuel retailers suffered losses amounting to $5$\% of sales.
Although retailers could not avoid losses from credit sales, they may have set prices to anticipate these expected losses.
This appears to be a plausible strategy for managing adverse selection under imperfect consumer information (Section~\ref{sec:sec51}).

In summary, the available documents show that retailers competed with others in the same sales area and provided regular customers the opportunity to trade on credit without interest for at least one month.
The 1930 Survey Report noted that ``intense competition among competitors has led to a high frequency of credit purchases'' (Tokyo City Office 1932b, p.~415).
This informal credit system could help workers manage consumption following unexpected individual setbacks in the short term.\footnote{Another potential explanation for the popularity of credit purchases might be theft. However, the number of thefts per $100$ people in 1919 was $0.5$ in Tsukishima, close to or slightly lower than the city average of $1.3$ (National Police Agency 1920b, pp~53--54; 244--245). The Tsukishima Police Department does not report that the area had an unusually high rate of theft (Tsukishima Police Department 1976, pp.~244--245). Therefore, the crime channel does not explain the use of credit purchases.}

\section{Conclusion}\label{sec:sec6}

This study investigates the consumption-smoothing strategies of skilled factory-worker households by digitizing the longitudinal budget survey conducted in Tsukishima in 1919.
Despite this narrow geographical focus, the analytical sample represents average skilled factory-worker households in this representative manufacturing area of Tokyo.

The estimated elasticity for total consumption indicates that households experienced measurable consumption losses when facing idiosyncratic income shocks.
The estimated elasticity for total expenditure ($0.43$) is similar to estimates found in developing countries.
The elasticity for rural households in Indian villages between 1975 and 1984 is reported to be approximately $0.5$.
In Bangkok, which had a GDP per capita similar to that of Japan in 1920, the estimated elasticity for urban households is $0.4$ between 1975 and 1990.\footnote{The estimates for India, Bangkok, and Osaka are from \citet{Ravallion1997-kz}, \citet{Townsend1995-om}, and \citet{Ogasawara2024}, respectively.}

A recent study reports an income elasticity of $0.39$ for factory-worker households in Osaka around 1920.
My estimate from the monthly panels is slightly higher.
Taken literally, this suggests that Tsukishima factory workers were slightly more vulnerable than those in Osaka.
This indicates that higher levels of urbanization do not necessarily correspond to greater availability of risk-coping institutions.
A limitation of this study is that it does not conduct inter-city and cross-national comparisons of income elasticities, nor does it analyze the institutional characteristics affecting consumption; these are left for future research.

Despite this vulnerability, results from the mechanism analyses show that credit purchases were commonly used by workers and helped mitigate adverse idiosyncratic shocks in several consumption subcategories, particularly food.
These findings offer a new historical perspective on the interdependence between workers and local retailers.
As researchers develop a better understanding of local credit institutions, we may gain insight into regional consumption behavior and the factors that improved living standards in prewar Japan.

\bibliographystyle{plainnat}
\bibliography{reference.bib}
\renewcommand{\refname}{Documents and Database}

\renewcommand{\refname}{Archival Sources}

\newpage
\clearpage
\thispagestyle{empty}

\begin{center}
\qquad

\qquad

\qquad

\qquad

\qquad

\qquad

{\LARGE \textbf{
Online Appendix to ``Risk-coping Behaviors in Metropolis: Evidence from Working-class Households in Prewar Tokyo'' by Kota Ogasawara
}}
\end{center}
\clearpage
\appendix
\def\thesection{Appendix~\Alph{section}}
\def\thesubsection{\Alph{section}.\arabic{subsection}}
\setcounter{page}{1}

\section{Background Appendix}\label{sec:seca}
\setcounter{figure}{0} \renewcommand{\thefigure}{A.\arabic{figure}}
\setcounter{table}{0} \renewcommand{\thetable}{A.\arabic{table}}

\subsection{Development of the Cities}\label{sec:seca_cities}

\begin{figure}[htbp]
\centering
\captionsetup{justification=centering,margin=0.2cm}
\subfloat[Number of people in the large cities]{\label{fig:citizens}\includegraphics[width=0.40\textwidth]{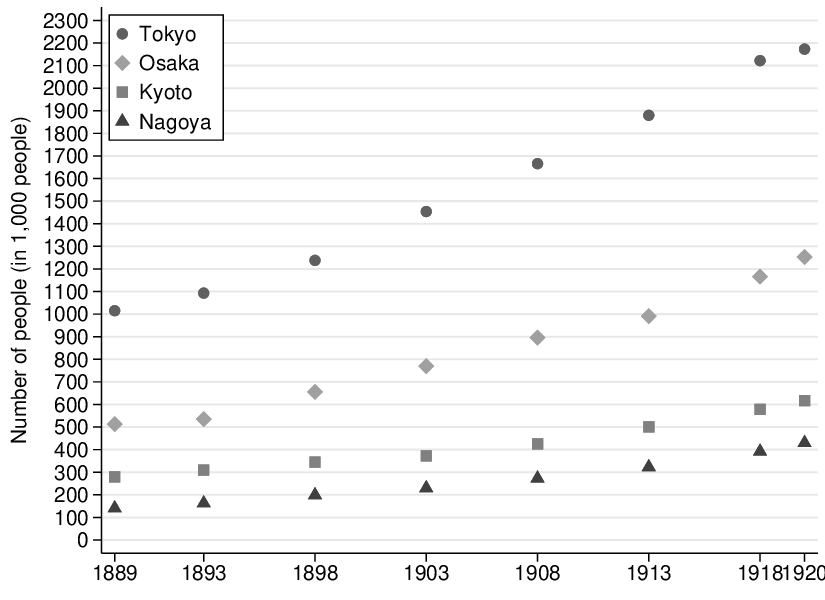}}
\subfloat[Share of people in the large cities (\%)]{\label{fig:citizens_share}\includegraphics[width=0.40\textwidth]{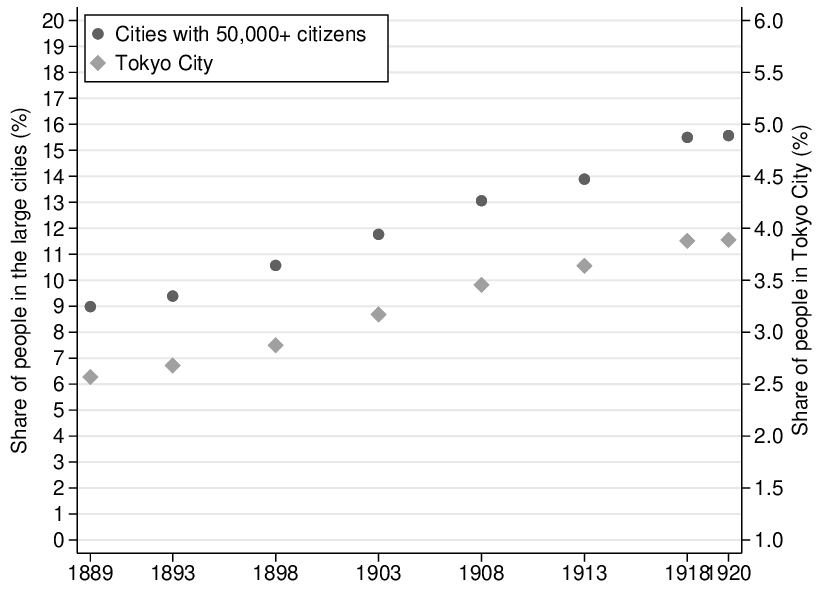}}
\caption{Development of the Large Cities in Japan}
\label{fig:citizens}
\scriptsize{\begin{minipage}{450pt}
\setstretch{0.85}
Notes: This figure illustrates the development of large cities in Japan between 1889 and 1920.
Figure~\ref{fig:citizens} shows the number of people in Tokyo, Osaka, Kyoto, and Nagoya cities.
Figure~\ref{fig:citizens_share} shows the share of people in Tokyo City (diamond) and the cities with more than $50,000$ people (circle) in total population in Japan.
Sources: Umemura et al. (1983, pp.~303--305); Umemura et al. (1988, pp.~166; 168).
\end{minipage}}
\end{figure}

Figure~\ref{fig:citizens} shows the developments of the large cities in Japan between 1889 and 1920.

\subsection{Financial Institutions in the Cities}\label{sec:seca_city_fi}

\begin{table}[htb]
\def\arraystretch{0.94}
\centering
\captionsetup{justification=centering}
\begin{center}
\caption{Financial Institutions in Tokyo, Osaka, \\Kyoto, and Nagoya Cities in 1920}
\label{tab:comparison_fi}
\footnotesize
\scalebox{0.95}[1]{
{\setlength\doublerulesep{2pt}
\begin{tabular}{lrrrr}
\toprule[1pt]\midrule[0.3pt]
&\multicolumn{4}{c}{Name of cities}\\
\cmidrule(rrrr){2-5}
													&Tokyo			&Osaka			&Kyoto		&Nagoya\\\hline
\multicolumn{4}{l}{\textbf{Panel A: Savings Institutions}}\\
Postal saving										&&&&\\
\hspace{5pt} Number of depositors	&1,462,604	&725,642		&305,927		&172,183		\\
\hspace{15pt} Population percentage (\%)			&67.3			&57.9			&51.74			&40.04			\\
\hspace{5pt} Total amounts (yen)					&94,330,789	&34,742,000	&12,036,439	&9,540,725	\\
\hspace{15pt} Amount per people (yen)				&64.50			&47.88			&39.34			&55.41			\\
&&&&\\
Savings bank&&&&\\
\hspace{5pt} Number of depositors					&1,461,389	&1,648,090	&663,831		&482,960		\\
\hspace{15pt} Population percentage (\%)			&67.25			&131.53		&112.26		&112.32		\\
\hspace{5pt} Total amounts (yen)					&48,682,260	&50,674,000	&28,495,407	&24,014,781	\\
\hspace{15pt} Amount per people (yen)				&33.31			&30.75			&42.93			&49.72			\\
&&&&\\
\multicolumn{4}{l}{\textbf{Panel B: Pawnshop}}\\
\hspace{5pt} Number of retailers						&1,253			&867			&441			&412\dag		\\
\hspace{5pt} Number of units lent					&6,042,501	&2,027,690	&1,251,283	&319,609		\\
\hspace{5pt} Amount of loans (yen)					&38,667,993	&3,785,186	&4,424,357	&2,983,250	\\
\hspace{5pt} Amount per loan (yen)					&6.40			&1.87			&3.54			&9.33			\\
\midrule[0.3pt]\bottomrule[1pt]
\end{tabular}
}
}
{\scriptsize
\begin{minipage}{355pt}
\setstretch{0.85}
Notes:
\textbf{Panel A}:
The statistics on postal savings were measured at the end of March 1921.
The statistics on savings banks were measured at the end of 1920.
\textbf{Panel B}: 
Pawnshops in Tokyo, Osaka, and Kyoto were measured in 1920.
Those for Nagoya (suggested in $\dag$) were measured in 1923 because the statistics in the former years are unavailable.\\
Sources:
Panel A: 
Tokyo City Office (1923, pp.~969; 977);
Osaka City Office (1922, pp.~6(61); 6(67));
Kyoto Prefecture (1922, p.~257); Kyoto City Office (1926, pp.~188);
Nagoya City Office (1922, pp. 286; 290).
Panel B: 
Tokyo City Office (1923, pp.~888--889);
Osaka City Office (1922, pp.~6(68)--(69));
Kyoto City Office (1922, p.~191);
Tokyo City Statistics Division (1926b,~p.~12).
\end{minipage}
}
\end{center}
\end{table}

A representative savings institution was postal savings (\textit{y\=ubin chokin}).
Panel A of Table~\ref{tab:comparison_fi} shows that approximately $1.46$ million people held postal savings accounts in Tokyo in 1920,
covering about $67$\% of the city's population.
However, the average savings per capita was insufficient.
In Kyobashi Ward, the average balance of postal savings accounts was only $44.5$ yen in 1919 (Tokyo City Office 1921, pp.~892--893).
This amount is lower than the average monthly earnings of skilled factory workers (Section~\ref{sec:sec32}), suggesting that postal savings may not have provided adequate temporary income.\footnote{Although statistics on postal savings by depositors' occupation are unavailable, some nationwide data exist. According to the Postal Savings Bureau (1924, p.~72), depositors in the manufacturing sector accounted for $4.8$\% of the total, and their share based on deposit amounts was $5.5$\%).}

Savings banks (\textit{chochiku gink\=o}) were another common savings institution (Ito and Saito 2019, pp~79--81).
Panel A of Table~\ref{tab:comparison_fi} shows that the number of people with savings accounts was approximately $1.46$ million, representing approximately $67$\% of Tokyo's total population.
Although this figure is similar to that for postal savings, savings banks were designed as financial institutions for low-income workers (Tokyo Institute for Municipal Research 1925b, p.~81).\footnote{There were 123 savings banks in Tokyo City in 1922, and 10 of them were in Kyobashi Ward (Tokyo Institute for Municipal Research 1925, p.~76). However, Tsukishima---a representative manufacturing area---did not have savings banks at the time.}
Table~\ref{tab:sb_occ} summarizes the number of depositors and the amounts of deposits in savings banks by occupation in Tokyo City (Panel A) and Kyobashi Ward (Panel B).
In Kyobashi, the commerce and miscellaneous sectors together accounted for about $84.8$\% of all depositors, while the manufacturing sector accounted for only $11.5$\%.
These shares remain unchanged when calculated based on deposit amounts.

Mutual loan associations (\textit{mujin}) represent another type of savings institution.
Members contribute a fixed amount of money over several periods, and, under specific rules, each member withdraws the total contributions once during the cycle.
A lottery or bidding system determines withdrawal order, which means that workers could not rely on these associations to manage sudden adverse income shocks.\footnote{This is the most representative mutual loan system called \textit{ts\=ujy\=o} (i.e., common) \textit{mujin}. There was a different type called \textit{sueoki chokin} (or \textit{tsumitate kai}) from which people could receive their stakes with interests after a fixed amount had been accumulated for a certain period (Tokyo Legislative Research Association 1915, pp.~3--4). However, these were also not flexible for compensating for income shocks.}
Instead, they were mainly used by small business owners and merchants to generate revenue (Tokyo Chamber of Commerce~1918, p.~121).
Workers in the manufacturing sector accounted for only $10$\% of all mutual loan associations in Tokyo City (Tokyo City Office 1935, p.~25).

Cooperative societies were also available for workers.
In Tokyo, these societies used credit as their payment method; households could purchase goods in advance and settle their payments at the end of the month or in two installments, mid-month and month-end.
Commodity prices in these cooperatives were similar to, or slightly lower than, those in retail shops (Central Federation of Industrial Associations 1925, pp.~18--19).
In 1924, Tokyo Prefecture had $26$ cooperative societies, of which 23 were organized by citizens and workers (Central Federation of Industrial Associations 1925, p.~57--59).
Among these $23$ cooperatives, two were large, with approximately $4,000$ members each, while the rest were small, with fewer than $100$ members.
Given that the number of male workers exceeded $768$ thousand (Statistics Bureau of the Cabinet 1929a, pp.~20--21; 26--29), these cooperatives accounted for only a small share of all workers in the city.

Another borrowing option was purchasing on credit from retailers.
Details on credit purchases are poorly documented, and systematic statistics are unavailable because transaction records are personal.
Nevertheless, a few existing sources indicate that credit purchases were widely used across various retailers (Section~\ref{sec:sec5}).
The THBS households frequently used credit to buy goods, including rice, fish, vegetables, alcohol, firewood, charcoal, and medicine, from different retailers.

\begin{table}[htb]
\def\arraystretch{0.95}
\centering
\captionsetup{justification=centering}
\begin{center}
\caption{Deposits in the Savings Banks in Tokyo City by Occupations}
\label{tab:sb_occ}
\footnotesize
\scalebox{0.92}[1]{
{\setlength\doublerulesep{2pt}
\begin{tabular}{lrrrrr}
\toprule[1pt]\midrule[0.3pt]
&\multicolumn{5}{c}{Occupations}\\
\cmidrule(rrrrr){2-6}
&Agriculture&Commerce&Manufacturing&Miscellaneous&Total\\\hline
\textbf{Panel A: Entire Tokyo City}		&&&&&\\
\hspace{10pt}Number of depositors		&76,210&550,784&166,498&619,963&1,413,455\\
\hspace{10pt}Amount of deposits 		&2,002,124&19,112,713&6,663,091&19,809,591&47,587,519\\
\hspace{10pt}Number of depositors (\%)	&5.4&39.0&11.8&43.9&100.0\\
\hspace{10pt}Amount of deposits (\%) 	&4.2&40.2&14.0&41.6&100.0\\
&&&&&\\
\textbf{Panel B: Kyobashi Ward}		&&&&&\\
\hspace{10pt}Number of depositors		&3,809&34,665&11,664&51,481&101,619\\
\hspace{10pt}Amount of deposits 		&130,781&1218,081&450,805&1,696,086&3,495,753\\
\hspace{10pt}Number of depositors (\%)	&3.7&34.1&11.5&50.7&100.0\\
\hspace{10pt}Amount of deposits (\%) 	&3.7&34.8&12.9&48.5&100.0\\

\midrule[0.3pt]\bottomrule[1pt]
\end{tabular}
}
}
{\scriptsize
\begin{minipage}{435pt}
\setstretch{0.85}
Notes: 
This table summarizes the number of depositors and the total deposits in the savings banks by occupation.
In the document, the miscellaneous category is not clearly defined but named as ``Miscellaneous (\textit{zatsu gy\=o}).''
These statistics were measured on December 31, 1919.
Panel A and B list the statistics for Tokyo City and Kyobashi Word, respectively.
Source:
Tokyo City Office (1922a, pp.~864--865).
\end{minipage}
}
\end{center}
\end{table}

\begin{figure}[]
\centering
\captionsetup{justification=centering,margin=1.5cm}
\includegraphics[angle =270, width=0.65\textwidth]{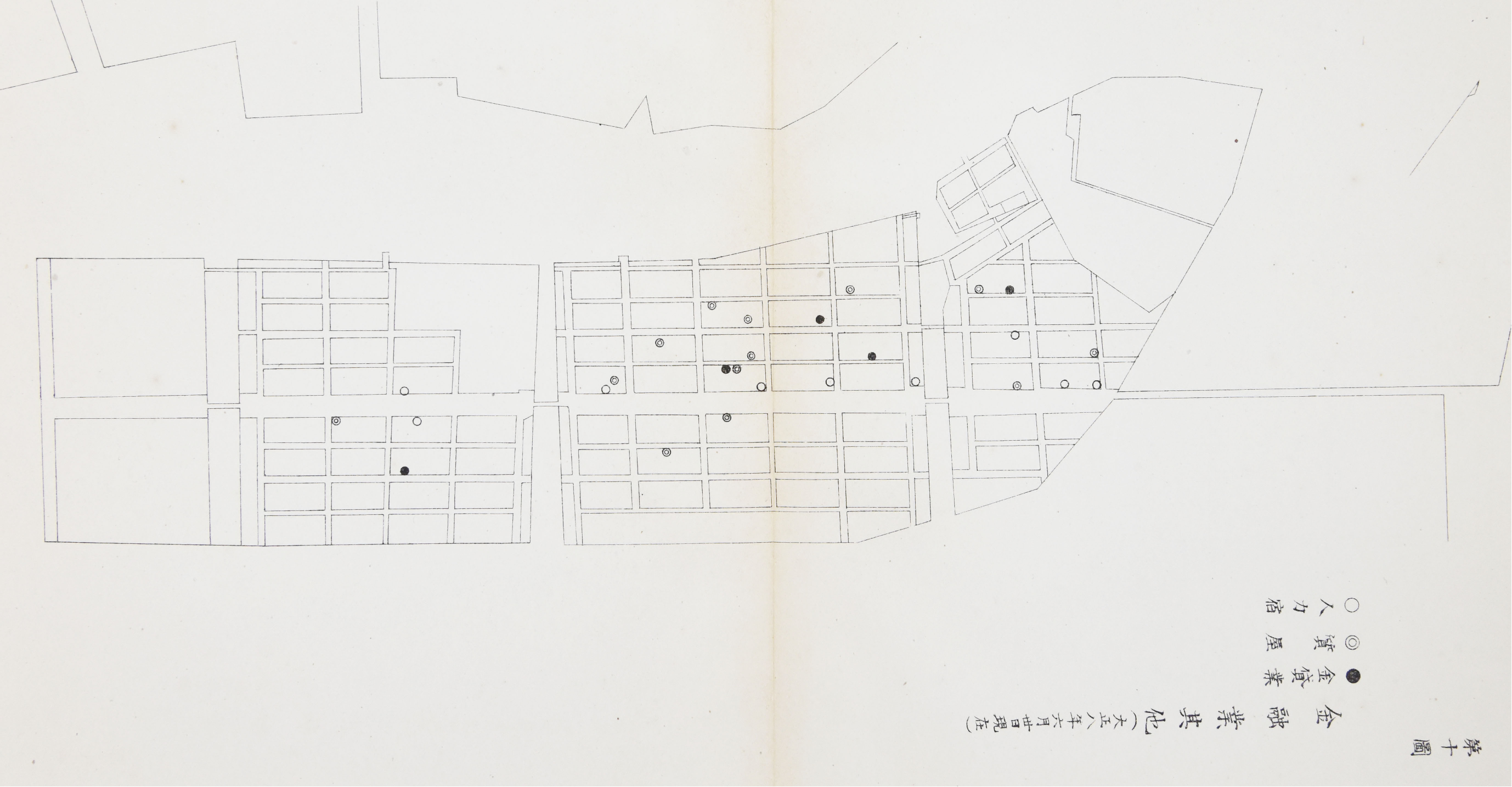}
\caption{Pawnshops and Money Lenders}
\label{fig:financial_institution}
\scriptsize{\begin{minipage}{450pt}
\setstretch{0.85}
Notes: 
This map shows the location of the pawnshops and moneylenders in Tsukishima.
The bullseye indicates the pawnshop (\textit{shichiya}).
The black circle shows the money lenders (\textit{kanekashi gy\=o}).
The white circle shows the rickshaw station (\textit{kuruma yado}).
The upper right corner indicates north.
Source: 
Department of Health, Ministry of the Interior 1923c, tenth map.
The tone was adjusted by the author using Adobe Photoshop 24.7.0.
\end{minipage}}
\end{figure}

According to the Tokyo City Office, pawnshops were primarily used by low-income workers (Tokyo City Social Affairs Bureau 1921, p.~9).
Lenders did not need to assess borrowers' creditworthiness, and borrowers faced limited risk of heavy debt because most pawned items were inexpensive clothes (Shibuya et al. 1982).
The average redemption rate was high, supported by relatively low interest rates regulated under the Pawnbroker Regulation Act of 1895.
Among the $88$ pawnshops in Kyobashi, the average redemption rate was approximately $94$\% ($401,222$/$427,265$ cases) in 1920 (Tokyo City Office 1922b, pp.~888--889).
In Tokyo, overall, the number of pawnshops was $1,334$ in 1918 and $1,261$ in 1919, with total transactions of $8,226,883$ and $7,573,406$, respectively.
These figures exceed three times the total population of Tokyo City, indicating that pawnshops were more accessible to workers than other financial institutions.
Figure~\ref{fig:financial_institution} shows that Tsukishima had $13$ pawnshops compared to five moneylenders.

However, because the items pawned were inexpensive, pawnshops primarily served short-term needs.
For example, in Kyobashi, the average loan per transaction was $7.8$ yen, approximately $10$\% of the monthly earnings of the THBS households.
Statistics on pawnshop users in Tokyo Prefecture in 1923 show that $1.7$\% worked in agriculture, $19.5$\% in commerce, and $15.6$\% in manufacturing.
The largest group of users consisted of day laborers and workers classified as ``miscellaneous,'' accounting for $63.4$\% (Tokyo Institute for Municipal Research 1926, p.~25).

Moneylenders were another lending institution(\textit{kinsen kashitsuke gy\=o}).
They were generally used by business owners, and their interest rates were significantly higher than those charged by pawnshops (Shibuya 2000, pp.~184; 248).
The Tokyo Institute for Municipal Research expressed concern over the growing number of unscrupulous moneylenders, many of whom were arrested by the National Police Agency (Tokyo Institute for Municipal Research 1925b, pp.~106--108).

Credit unions (\textit{shiny\=o kumiai}) were another option available to working-class households.
In Tokyo City, there were $215$ unions in 1923.
However, according to a survey of $184$ of these unions, the number of members from the manufacturing sector was only $5,662$ (Tokyo Institute for Municipal Research 1925c, p.~20).
This represented a small share of male workers in the manufacturing sector, as measured in the 1920 Population Census.

\subsection{Development of Tsukishima}\label{sec:seca_tsukishima}

\begin{figure}[htb]
\centering
\captionsetup{justification=centering,margin=1.5cm}
\includegraphics[width=0.39\textwidth]{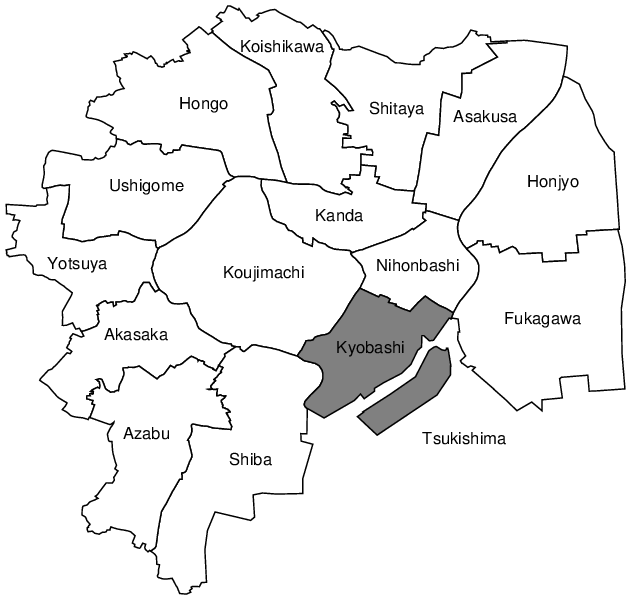}
\caption{Administrative Wards in Tokyo City}
\label{fig:map_tokyo}

\scriptsize{\begin{minipage}{450pt}
\setstretch{0.85}
Notes: The name in each lattice indicates the administrative ward in Tokyo.
The gray ward indicates the Kyobashi Ward and an island in the Kyobashi Ward indicates Tsukishima.
The border surrounded by Asakusa, Honjyo, Nihonbashi, and Fukagawa wards includes the Sumida River.
The borders of each ward are based on the administrative districts in 1920.
Geographical coordinate system is based on JGD2000/(B, L).
Source: Created by the author using the official shapefile (Ministry of Land, Infrastructure, Transport and Tourism, database).
\end{minipage}}
\end{figure}
\begin{figure}[htb]
\centering
\captionsetup{justification=centering,margin=1.5cm}
\includegraphics[width=0.52\textwidth]{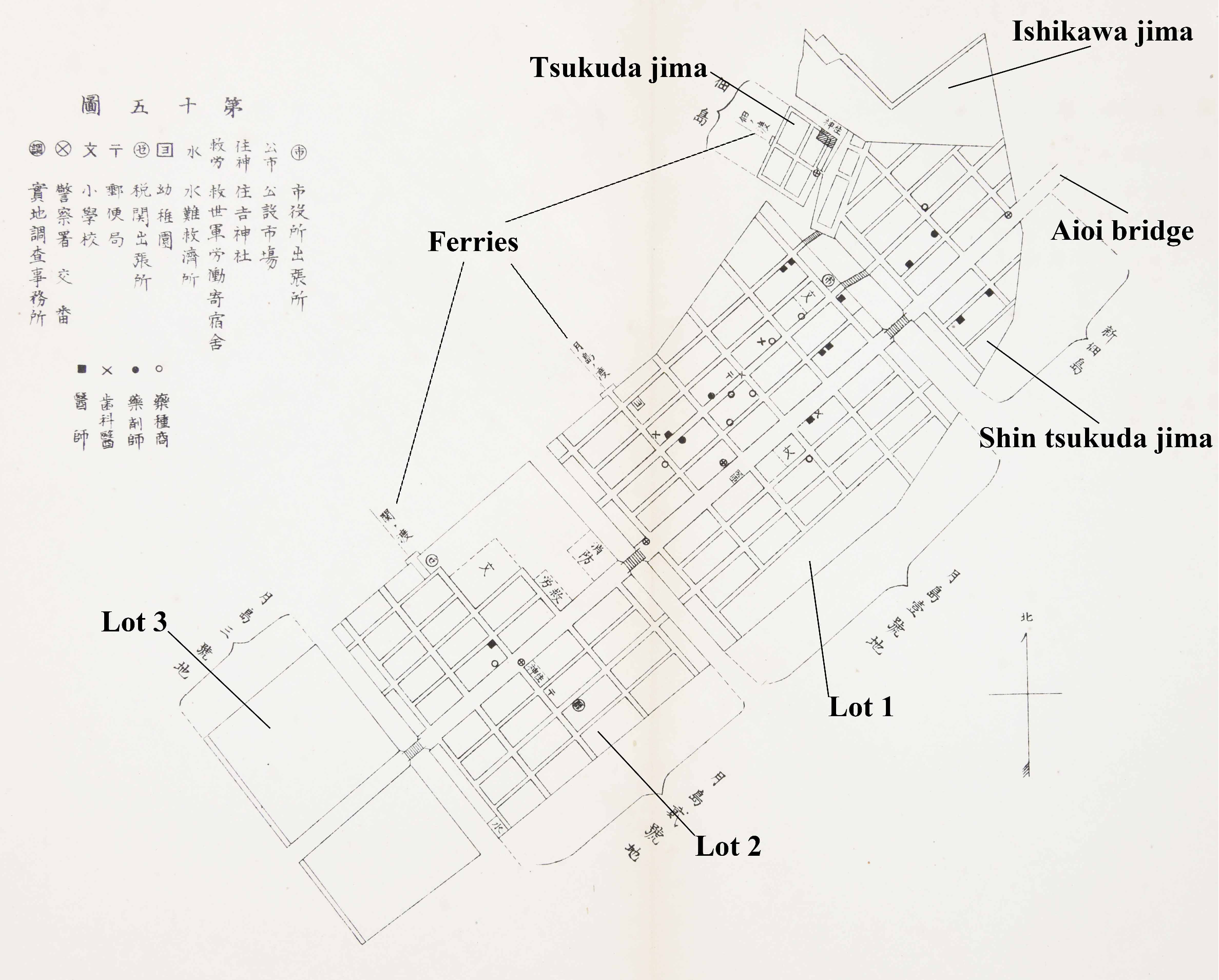}
\caption{Map of Tsukishima: Ishikawajima, Tsukudajima, Shin-tsukudajima, and Lots 1--3}
\label{fig:map_tsukishima}
\scriptsize{\begin{minipage}{450pt}
\setstretch{0.85}
Notes: Tsukishima contains Ishikawajima, Tsukudajima, Shin-tsukudajima, and Tsukishima Lots 1--3.
This map does not illustrate the northernmost part of Ishikawajima (see Figure~\ref{fig:tsukishima_factory} for this part).
``Ferries'' shows the Tsukishima to the mainland of Kyobashi Ward sea routes called Tsukuda, Tsukishima, and Kachidoki routes (right to left).
``Aioi bridge'' links Shin tsukudajima and Fukagawa ward.
Source: 
Department of Health, Ministry of the Interior 1923c, 15th map.
The author modified and adjusted the tone using Adobe Photoshop 24.7.0.
\end{minipage}}
\end{figure}

Tsukishima (translated as ``moon island'') is a reclaimed area on a sandbar of the Sumida River (Figure~\ref{fig:map_tokyo}).\footnote{The following description in this section is based on Ch\=u\=o Ward (1994, pp.~122--124).}
The islands of Ishikawajima and Tsukudajima at the northern end mark the starting point of Tsukishima's territory (Figure~\ref{fig:map_tsukishima}).
Ishikawajima was established in 1626 by Hachizaemon Ishikawa, a boatman who received a fief from the shogunate.
In 1790, Ishikawajima Prison (\textit{ninsoku yoseba}) was constructed to accommodate minor offenders and homeless individuals, providing a facility for reintegration into society.
After Matthew Perry's arrival in Japan in 1853, the shogunate instructed the Mito Clan to construct a shipyard on Ishikawajima.
A groundbreaking ceremony was held a month later, and a Western-style wooden sailing ship was delivered to the shogunate in 1856.
In 1866, the Ishikawajima Beacon was built on the west bank of the prison to signal the position of ships entering the Nihonbashi area.
In 1870, the prison was converted into an official place of imprisonment (\textit{toba}) and was renamed the Ishikawajima Jail Station of the Metropolitan Police Department in 1877.
The facility was eventually relocated to Sugamo in Kitatoshima County, Tokyo Prefecture, and the penitentiary on Ishikawajima was closed in 1895.

Adjacent to Ishikawajima lies Tsukudajima, where fishermen from Tsukuda in Osaka Prefecture settled in 1644 after receiving land from the shogunate.
In 1645, the Tsukuda Ferry began operations.
Throughout the Edo period, Tsukudajima became known as a fishing hub, supplying whitefish for the shogun's household.
In 1863, the Tsukudajima Battery was constructed on the southern side of the island as one of $13$ batteries built to defend Edo Bay.
This site later became the starting point for the reclamation of Lot 1 of Tsukishima.

In 1884, the Port of Tokyo Waterway Dredging Project
began to secure navigable passages by clearing sediment from the shipping channel, and the reclaimed land of Tsukishima was created using the dredged deposits.
Lot 1 was completed in 1891, and the Tsukishima Ferry began operations the same year.
Lot 2 was reclaimed in 1894, and Shin-tsukudajima (literally, new Tsukudajima) was established in 1896.
From the western shore of Shin-tsukudajima, the island offers views of the B\=os\=o Peninsula in Chiba Prefecture, making it a favored retreat for writers.
Cultural figures such as T\=oson Shimazaki and Kaoru Osanai stayed there from 1907 through the early Taish\=o period.
In 1903, the completion of the Aioi Bridge improved transportation access and brought water and electricity to Tsukishima, contributing to the growth of the local industry.
The Kachidoki Ferry began serving Lot 2 in 1905, and Lot 3 was finally completed in 1913.

\subsection{Housing in Tsukishima}\label{sec:seca_housing}

\begin{figure}[]
\centering
\subfloat[2-story tenement]{\label{fig:photo_hh1}\includegraphics[width=0.23\textwidth]{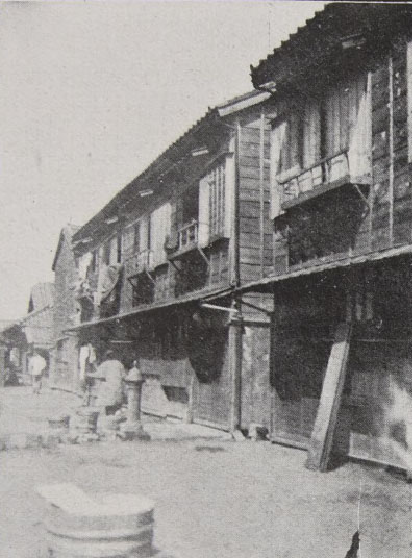}}
\hspace{1pc}
\subfloat[1-story tenement]{\label{fig:photo_hh2}\includegraphics[width=0.40\textwidth]{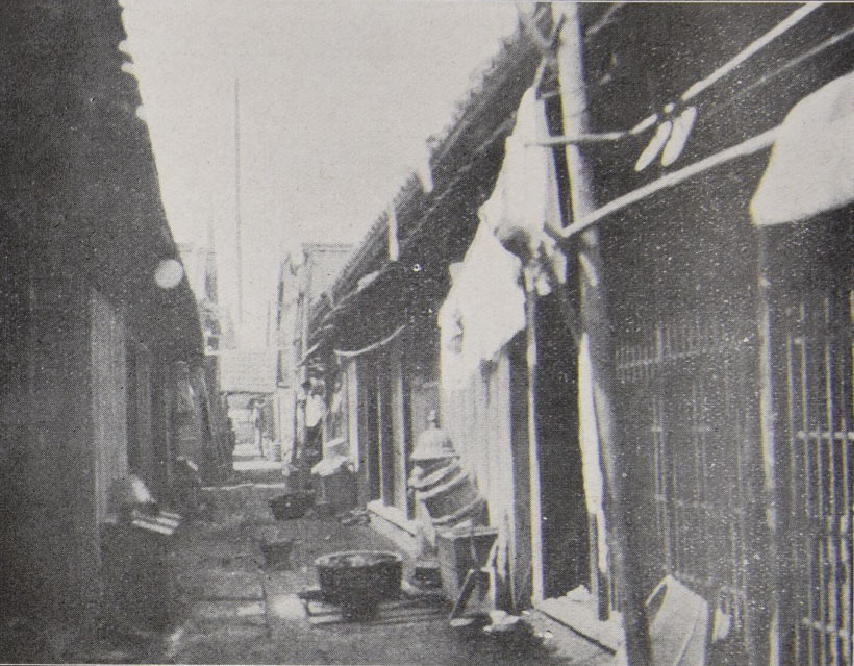}}
\caption{Photographs of the Tenements in Tsukishima}
\label{fig:housing}
\scriptsize{\begin{minipage}{450pt}
Note: Figures~\ref{fig:photo_hh1} and~\ref{fig:photo_hh2} show the two- and one-story tenements in Tsukishima, respectively.
Source: Department of Health, Ministry of the Interior 1923c, figures 62--63; 66.
\end{minipage}}
\end{figure}

Table~\ref{tab:house} summarizes the housing types recorded in the housing survey (Table~\ref{tab:tabb1}).
As shown, $88$ detached homes ($46$\%) were located in Tsukudajima.
Tsukudajima is reported to be ``superior to other areas in terms of housing quality, and the percentage of inferior housing is meager.''\footnote{The description in this section is based on the Department of Health, Ministry of the Interior (1923a, pp.~69--76).}
By contrast, Shin-tsukudajima had a low percentage of detached homes.
According to the report, it was ``inhabited by unskilled laborers'' and ``occupied the most inconvenient housing status of all the islands.''
Lot 1, considered the center of life in Tsukishima, consisted mainly of numerous tenement-style residences.
Lot 2, by comparison, had a relatively high percentage of single-family and two-family houses and appears to have been characterized more as a ``workers' residential area'' than Lot 1.
These findings suggest possible geographic differences in income distribution reflected in housing type.
Despite this, of the $989$ dwellings in Tsukishima, $799$ ($81\%$) were row houses, while only $190$ were single-family homes.
This indicates that most workers lived in similar tenements and experienced comparable living standards.
Figure~\ref{fig:housing} shows a photo of these tenement houses.

\begin{table}[htb]
\def\arraystretch{0.90}
\centering
\begin{center}
\caption{Housing Types in Tsukishima}
\label{tab:house}
\scriptsize
\scalebox{1.0}[1]{
\begin{tabular}{lrrrrrrrr}
\toprule[1pt]\midrule[0.3pt]
&&\multicolumn{4}{c}{Tenement (\textit{nagaya})}&&&\\
\cmidrule(rrrr){3-6}
&(A) Detached&\multicolumn{2}{c}{(B) 2 families}&\multicolumn{2}{c}{3+ families}&Total&(A)+(B)&\% \\
\cmidrule(rr){3-4}\cmidrule(rr){5-6}
&&2-story&1-story&2-story&1-story&&&\\\hline
Tsukudajima		&88	&32	&26	&6		&23	&175	&146	&83.4\\
Shin-tsukudajima	&16	&27	&10	&36	&59	&148	&53	&35.8\\
Lot 1				&58	&62	&74	&73	&200	&467	&194	&41.5\\
Lot 2				&28	&27	&43	&9		&92	&199	&98	&49.2\\
Total				&190	&148	&153	&124	&374	&989	&491	&49.6\\
\midrule[0.3pt]\bottomrule[1.0pt]
\end{tabular}
}
{\scriptsize
\begin{minipage}{390pt}
\setstretch{0.85}
Note:
This table summarizes the housing types in Tsukishima surveyed on June 30, 1919.\\
Source:
Department of Health, Ministry of the Interior 1923a, pp.~69--76.
\end{minipage}
}
\end{center}
\end{table}

\subsection{Machinery Factories in Tsukishima}\label{sec:seca_factory}

First, I use the results of the Primary School Children Survey, included in the Tsukishima Survey, to analyze the occupations of household heads.
This survey was conducted among 4th--6th graders at Tsukishima 1st and 2nd Elementary Schools in July 1919.
A total of $1,173$ children were in these grades, and the survey covered all $1,100$ children residing in Tsukishima.
Since no other elementary schools existed, this represents a complete survey of all 4th--6th graders in Tsukishima.
Although it excludes households without children in these grades, it remains a valuable source for understanding households with children.

Panel A of Table~\ref{tab:app_structure} presents the industrial structures.
The composition ratios are similar to the census-based statistics shown in Panel A of Table~\ref{tab:tab1}.
The report does not clarify whether siblings in the same household are counted more than once.
Nevertheless, the similarity between the results confirms that this survey provides reliable statistics for identifying overall trends in family households in Tsukishima.

Panel B of Table~\ref{tab:app_structure} classifies the ``social classes'' of household heads working in the manufacturing sector into six categories.
According to the report, large business owners are classified as the capitalist class; small business owners, free enterprise owners, and directors are grouped as the middle class; and workers are defined as the working class (Department of Health, Ministry of the Interior 1923a, p.~50).
Most household heads were workers, with only 1\% in the capitalist class and less than 20\% in the middle class.
Although the definition of ``executive'' is unclear, ``small business owners'' appear to refer to small factory owners whose earnings are similar to those of skilled worker households (Department of Health, Ministry of the Interior 1923a, p.~64).

\begin{table}[htb]
\def\arraystretch{0.78}
\centering
\captionsetup{justification=centering}
\begin{center}
\caption{Industrial Structure and Social Classes in Tsukishima: \\Evidence from Tsukishima Primary School Survey}
\label{tab:app_structure}
\footnotesize
\scalebox{0.95}[1]{
\begin{tabular}{lrrrrr}
\toprule[1pt]\midrule[0.3pt]
\multicolumn{6}{l}{\textbf{Panel A: Industrial Structure}}\\
Name of survey			&\multicolumn{2}{l}{Tsukishima Survey}&&&\\
Survey area				&\multicolumn{2}{l}{Tsukishima}&&&\\
Survey month and year	&\multicolumn{2}{l}{July 1919}&&&\\
Survey subject			&\multicolumn{5}{l}{Heads of all the students in grades 4--6}\\
						&\multicolumn{5}{l}{in Tsukishima 1st and 2nd Primary Schools}\\
				
\cmidrule(rr){2-3}
				&\# of heads	&\% share	&	&	&		\\\hline
Agriculture		&1				&0.1&&&\\
Fisheries		&13			&1.2&&&\\
Mining			&2				&0.2&&&\\
Manufacturing	&623			&58.6&&&\\
Commerce		&259			&24.4&&&\\
Transport		&72			&6.8&&&\\
Public service and professions	&55&5.2&&&\\
Other industry	&38			&3.6&&&\\
Observations	&1,100			&100&&&\\
&&&&&\\
\multicolumn{6}{l}{\textbf{Panel B: Worker Types of Heads in Manufacturing Sector}}\\
Name of survey			&\multicolumn{2}{l}{Tsukishima Survey}&&&\\
Survey area			&\multicolumn{2}{l}{Tsukishima}&&&\\
Survey month and year	&\multicolumn{2}{l}{July 1919}&&&\\
Survey subject			&\multicolumn{5}{l}{All students in grades 4--6}\\
					&\multicolumn{5}{l}{in Tsukishima 1st and 2nd Primary Schools}\\
\cmidrule(rrrrr){2-6}
					&Tsukudajima	&Shin-tsukudajima	&Lot 1		&Lot 2	&Total		\\\hline
Large business owner	&0		&0		&5			&3			&8		\\
Small business owner	&1		&10	&49		&13		&73	\\
Self-employed			&0		&0		&0			&0			&0		\\
Executive				&3		&13	&28		&7			&51	\\
Worker				&23	&94	&265		&108		&490	\\
Unemployed/Unknown	&0		&0		&0			&0			&0		\\
Total					&27	&117	&347		&131		&622	\\
\% share of workers		&85.2	&80.3		&76.4		&82.4		&78.8	\\\midrule[0.3pt]\bottomrule[1.0pt]
\end{tabular}
}
{\scriptsize 
\begin{minipage}{380pt}
\setstretch{0.85}
Notes:
This table summarizes the statistics from the complete survey on the 4th--6th graders in Tsukishima1st and 2nd Primary Schools in July 1919.
\textbf{Panel A} summarizes the heads' industries, and \textbf{Panel B} sorts the ``worker type'' of the heads in the manufacturing sector, respectively.
Source:
Department of Health, Ministry of the Interior 1923a, pp.~386--387.
\end{minipage}
}
\end{center}
\end{table}

In November 1920, $214$ factories existed in Tsukishima, of which $168$ (approximately $80$\%) were classified as machinery factories.
The official report notes that wooden pattern factories could be included in the machinery sector because they were part of the casting processes (Department of Health, Ministry of the Interior 1923a, p.~388).
Including the $24$ wooden pattern factories raises the share of machinery factories to approximately $90$\%.
Panel A of Table~\ref{tab:app_scale_type} uses data from the manufacturing survey in the Tsukishima Survey to classify the $168$ machinery factories by size and industrial sector.
First, more than $90$\% of the plants were concentrated in three sectors: machinery and equipment manufacturing, can manufacturing, and blacksmithing, with only $15$ plants in the casting sector.
Among the machinery factories, $113$ ($67$\%) employed fewer than $15$ workers.
Second, many smithing factories also had fewer than $15$ workers.
Hence, almost the entire smithing industry in Tsukishima was classified as small businesses.\footnote{The official report indicates that $43$ of the $47$ smithing factories were small factories and workplaces called (\textit{machi k\=oba}). The report does not clearly define \textit{machi k\=oba}. However, since their number consistently exceeds the count of factories with 15 or fewer workers, they likely include smaller factories operated as small businesses (Department of Health, Ministry of the Interior 1923a, pp.~392; 410--411).}

Panel B of Table~\ref{tab:app_scale_type} classifies $620$ workers based on the primary school survey data used in Table~\ref{tab:app_structure}.
Of these, $455$ are skilled workers and $165$ are unskilled.
Among the unskilled workers, 45 worked in factories and cottage industries, while 120 were day laborers considered low-income workers.
According to the same report, $490$ of the $620$ workers were industrial workers.
Day laborers are classified as non-industrial workers.
This is consistent with the calculation: $620$ workers minus $120$ day laborers equals $490$ industrial workers.
Therefore, approximately $90$\% of industrial workers ($455/490$) can be regarded as skilled workers: $65$\% worked in factories, $17$\% in cottage industries, and $18$\% were craftsmen.
Since workers in cottage and handicraft industries are classified as working in small factories, roughly $65$\% of these skilled workers likely worked in medium- to large-scale factories (with $15+$ workers).

Notably, this survey includes only household heads with primary school children aged 9--11 (i.e., 4th to 6th grades). This suggests that most household heads were in their thirties, an age range typically associated with skilled workers.
Therefore, the share of skilled workers among the entire working population may be lower than $90$\%.
Nevertheless, the data confirm that most Tsukishima workers in their 30s were skilled.

\begin{table}[htb]
\def\arraystretch{0.86}
\centering
\captionsetup{justification=centering}
\begin{center}
\caption{Scale of Machinery Factories and Type of Workers in Tsukishima}
\label{tab:app_scale_type}
\footnotesize
\scalebox{0.95}[1]{
\begin{tabular}{lrrrrr}
\toprule[1pt]\midrule[0.3pt]
\multicolumn{6}{l}{\textbf{Panel A: Scale of Machinery Factories}}\\
Name of survey			&\multicolumn{2}{l}{Tsukishima Survey}&&&\\
Survey area				&\multicolumn{2}{l}{Tsukishima}&&&\\
Survey month and year	&\multicolumn{5}{l}{November 1920}\\
Survey subject			&\multicolumn{5}{l}{All machinery factories}\\
\cmidrule(rrrrr){2-6}
							&Machinery and 	&Can	&Casting	&Smithing	&Total	\\
							&equipment		&&&&\\\hline
\# of workers $\geq 15$			&22				&15		&10		&8		&55	\\
\# of workers $< 15$				&44				&25		&5		&39		&113	\\
Total							&66				&40		&15		&47		&168	\\
\% shares						&39.3			&23.8	&8.9		&28.0	&100	\\
&&&&&\\
\multicolumn{6}{l}{\textbf{Panel B: Type of Workers (All Industries)}}\\
Name of survey			&\multicolumn{2}{l}{Tsukishima Survey}&&&\\
Survey area				&\multicolumn{2}{l}{Tsukishima}&&&\\
Survey month and year	&\multicolumn{2}{l}{July 1919}&&&\\
Survey subject			&\multicolumn{5}{l}{All students in grades 4--6}\\
						&\multicolumn{5}{l}{in Tsukishima 1st and 2nd Primary Schools}\\
\cmidrule(rrrrr){2-6}
						&Skilled workers&	&					&Unskilled workers&\\\hline
Factory workers			&295			&	&Factory workers	&42&	\\
Cottage industry		&76			&	&Cottage industry	&3&	\\
Handicraft industry		&84			&	&Day laborer 		&120&	\\
Total					&455			&	&					&165&	\\\midrule[0.3pt]\bottomrule[1.0pt]
\end{tabular}
}
{\scriptsize 
\begin{minipage}{406pt}
\setstretch{0.85}
Notes:
\textbf{Panel A} summarizes the number of machinery factories by the scale and type of the factory.
Statistics are from the manufacturing census conducted in the Tsukishima Survey (November 1, 1920).
The threshold of the scale (i.e., 15 workers) is based on the application of the Factory Act enacted in 1916.
\textbf{Panel B} sorted the heads by their skill level and job type.
Statistics are obtained from the Complete Survey on the 4th--6th graders in Tsukishima 1st and 2nd Primary Schools in July 1919.
Sources:
Panel A: Department of Health, Ministry of the Interior 1923a, pp.~386--387; 410--414.
Panel B: Department of Health, Ministry of the Interior 1923a, p.~149.
\end{minipage}
}
\end{center}
\end{table}

\subsection{Works in the Machinery Factories}\label{sec:seca_work}

Panel A of Table~\ref{tab:app_scale_type} summarizes the number of machinery factories by plant type.
Machinery and equipment manufacturing factories accounted for $39$\% of all machinery factories.
Typical jobs in these factories included finishers (or fitters) and turners.
A finisher is defined as ``a skilled worker responsible for finishing, assembling, and commissioning machine components'' and is considered ``one of the finest skilled workers because the role requires proficiency in all aspects to correct defects in machine making'' (Department of Health, Ministry of the Interior 1923a, p.~393).
The turner is described as the ``middle axis of machinists,'' a category that includes most semi-skilled and skilled workers.
Becoming a fully qualified finisher or lathe operator typically took at least five years.
Apprentices usually entered the workforce between ages $14$ and $18$, but did not remain in the same factory throughout their apprenticeship.
After acquiring basic skills for 1--2 years, they often moved to another factory to further develop their expertise (Department of Health, Ministry of the Interior 1923a, p.~394).
Wages were generally paid either on a piece-work or time basis.
In Tsukishima, time contracts were widely adopted and payment followed the ``residual time'' method, whereby ``if the work is completed within the contracted time, the remaining time is treated as a bonus.''
For example, if $12$ hours are contracted but the work is completed in $10$ hours, the remaining $2$ hours are paid as a bonus (Department of Health, Ministry of the Interior 1923a, pp.~396-397).
Wages were distributed twice a month, divided into two payment periods.
Average working hours were reported to be approximately $10$ hours per day, with occasional all-night shifts (Department of Health, Ministry of the Interior 1923a, p.~402).
Overtime averaged approximately two hours per day (Department of Health, Ministry of the Interior 1923a, p.~402).

Next, I provide a summary of the workforce in the canning factories, which accounted for $24$\% of all factories.
Canning factories were engaged in shipbuilding, constructing bridges, manufacturing steel frames for construction, and other work, including the fabrication and repair of gas reservoirs (Department of Health, Ministry of the Interior 1923a, p.~406).
Workers in these factories were called ``boiler makers'' and typically bent steel plates and joined them with rivets.
They were divided into rivetter, holder-up, plater, angle-ironsmith, and caulker.
Work was usually performed in groups, led by skilled angle-ironsmiths and caulkers.
The ``b\=osin'' of the rivetter served as the group leader and provided instructions.
Wages were paid on a piece-rate basis, with a fixed daily wage during the off-season.

Smithing was the second most common industry, accounting for $28$\% of the total.
Panel A of Table~\ref{tab:app_scale_type} shows that most smithing operations were relatively small.
Workers produced boulders, nuts, rivets, lathe stops, jigs, and spanners for can manufacturing and construction.
These small shops often acted as subcontractors supplying metal parts to larger factories (Department of Health, Ministry of the Interior 1923a, p.~412).
Three main jobs existed: \textit{yokoza}, \textit{sakite}, and \textit{tatara-fuki} (bellows-blower).
\textit{Yokoza} referred to self-employed skilled workers (\textit{oyakata}), while the other two referred to young, unskilled laborers.
In small operations, wages were usually determined as a percentage through voluntary agreement with the proprietor, and working hours were irregular (Department of Health, Ministry of the Interior 1923a, pp.~411--413).

\clearpage
\section{Data Appendix}\label{sec:secb}
\setcounter{figure}{0} \renewcommand{\thefigure}{B.\arabic{figure}}
\setcounter{table}{0} \renewcommand{\thetable}{B.\arabic{table}}

\subsection{Tsukishima Survey: The First Social Survey in Japan}\label{sec:secb1}

According to Kawai (1980), the methodology of social research in Japan was established in the 1930s.
During the 1910s and 1920s, ``numerous social surveys were conducted by various entities, including government agencies, research groups, local administrative bodies, private organizations, survey research institutes, and individual researchers.''
The Report on Field Survey in Tsukishima, Kyobashi Ward, Tokyo (called the Tsukishima Survey) represents a pioneering survey of urban communities (Sekiya 1970, p.~43).

First, I summarize the investigators, research questions, and subjects of the Tsukishima Survey.\footnote{Takano himself refers to this field survey as the ``Tsukishima Survey'' (Department of Health, Ministry of the Interior 1923a, p.~6). Considering this, I use the Tsukishima Survey as the name of this survey throughout this paper.}
The person responsible for the survey was Iwasabur\=o Takano (hereafter ``Takano''), a professor of social statistics and social policy at the Law School of Tokyo Imperial University.
Circa WWI, Takano observed widespread problems in labor, lifestyle, urban infrastructure, health and sanitation, public safety, and citizen welfare, driven by heavy and chemical industrialization, capitalist monopolization, and urban expansion.
In this context, Takano emphasized the importance of gathering objective statistics about workers through social surveys (Sekiya 1970, p.~33).
He conducted Japan's first household survey\footnote{This is called the ``Survey of Household Income and Expenditure of 20 Artisans and Workers in Tokyo,'' which was inspired by Charles Booth (Miyoshi 1980, p.~33). The original documents of this survey, preserved at Tokyo Imperial University, were destroyed in the Great Kant\=o Earthquake (Takano 1933, pp. 727--728). Before this household survey, Takano conducted a field survey in a poor area of Tokyo when he was a student at Tokyo Imperial University. The findings were reported in April 1894 as an exercise entitled ``East London in Tokyo'' (Kawai 1981b, p.~3).}
and had served on the seventh subcommittee (Rural Sanitation Subcommittee) of the Ministry of Home Affairs' Health and Sanitation Investigation Committee since 1916.\footnote{The Health and Sanitation Investigation Committee of the Ministry of Home Affairs was established on June 27, 1916, under the second Shigenobu Okuma Cabinet by issuing Imperial Ordinance No. 172. Initially, 34 members were appointed, and the first meeting was held on July 8. The reason for the establishment of this committee was the need for a policy response regarding the establishment and institutionalization of various investigation organizations and the penetration of the capitalist economy during the Meiji period. Another reason was the requirement for a public health response to the rising mortality rate, exemplified by the tuberculosis epidemic (Kawai 1981a, pp. 10--13).}
In 1917, the seventh subcommittee evolved into the Section on Urban and Rural Sanitary Conditions (Kawai 1981a, pp.~12--13).
On October 22, 1918, Takano submitted a proposal to the subcommittee for a field survey of urban sanitary conditions.
The survey aimed to determine the housing conditions, household budgets, and health status (including child health, live births, stillbirths, deaths, and diseases) in areas where skilled-worker families resided (Sekiya 1970, p.~6).
Initially, the survey area was planned as Yanagibashi Yokogawa-ch\=o in Honjo Ward.
However, at a committee meeting on November 23, the survey site was changed to Tsukishima in Kyobashi Ward, which had more skilled-worker households than Yokogawa-ch\=o (Kawai 1980, pp.~55--59).\footnote{Takeda (2015, pp.~75--76) provides more details on how the survey site was changed. Takano was looking for a commissioned researcher to conduct the survey and asked Sakuz\=o Yoshino, a colleague at Tokyo Imperial University, to recommend a suitable person. Yoshino approached his former student Kotora Tanahashi, who had become a member of the fraternity club, a mutual aid organization for workers, to accept the investigator position. The diary written by Kotora Tanahashi (``Tanahashi Kotora Diary'') is in the possession of the OISR, and Takeda (2015, pp.~76--77) used the diary to analyze the circumstances surrounding the change of survey location. Tanahashi eventually recommended Yoshitsuru Yamana, who was his classmate in high school and Tokyo Imperial University, as a researcher and considered Tsukishima to be the best research area because of the ease of selecting survey households using the fraternity's network. On October 26, Tanahashi and Yamana---together with the Kyobashi Branch of the fraternity club---conducted an on-site inspection of Tsukishima. Based on this inspection, Tanahashi and Yamana approached Takano about locating the survey station in Tsukishima. Finally, Takano and Yamana conducted an on-site inspection there on October 30 and decided on Tsukishima as the survey site.}

Although statistical sampling was not yet common at the time, Takano selected the survey site based on his judgment.
He stated: ``There can be no dispute that we have chosen Tokyo as the target of our metropolitan survey. However, it is impossible to survey the entire Tokyo area. We have no choice but to conduct a partial survey. For this purpose, it would be the best idea to select a representative area in Tokyo and survey that area narrowly but deeply'' (Department of Health, Ministry of the Interior 1923a, pp.~2--3).
Takano further emphasized the importance of focusing on the working class, which constitutes the majority of citizens, rather than the political and commercial center: ``Most of Tokyo citizens are wage earners who subsist on labor. As is usual in large cities, a large number of people are engaged in manual labor in Tokyo. However, the most genuine type of worker is the skilled worker. Therefore, the area where many families of skilled workers congregate can be regarded as a representative area of Tokyo. Further, it seemed to me that a study of the social and sanitary conditions in that area could explain a large part of the social and sanitary conditions in Tokyo'' (Department of Health, Ministry of the Interior 1923a, pp.~3).

Examining Takano's statement that ``the most genuine type of worker is the skilled worker'' and the fact that the main target of the Tsukishima Survey was skilled workers is important.\footnote{In other words, the Tsukishima Survey was not designed for poor households. Therefore, unlike the surveys conducted in the UK by Booth and Rowntree, investigating the standard of living was not the main objective of the survey (Sekiya 1970, p.~42).}
Kawai's description (1981a, pp.~13--18) is helpful in this regard.
As summarized in Section~\ref{sec:sec2}, Japan's mechanical and chemical industrialization during WWI led to a large concentration of city workers.
Consequently, the ratio of small-scale conventional factories based on ``master employees'' (as commonly seen in the textile industry) declined.
By contrast, the number of factories with a direct ``factory owner--employee'' relationship increased, particularly in the machinery and equipment, metal, and chemical industries.
Workers in the modern industrial sector acquired skills, formed households, and settled in cities, giving rise to a new urban worker lifestyle.
Although the city still had a class of workers engaged in traditional industries, Takano considered skilled workers in the modern industrial sector to be central in terms of labor force, labor movements, and urban life.

Sekiya also notes the rise of the labor movement.
As explained in Section~\ref{sec:sec2}, although wages rose during WWI, inflation caused relative wages to decline.
In 1916, Japan enacted its first labor protection legislation, the Factory Law.
However, the number of working days and hours remained unchanged, leading to a steady increase in labor disputes (Sekiya 1970, pp.~24--26).
Takano therefore believed that workers should organize labor unions to address problems not resolved through labor protection legislation.
Sekiya (1970, pp.~26--40) speculates that Takano's purpose in conducting the Tsukishima Survey was to obtain basic data for solving labor issues, specifically to identify problems that workers could address independently or through existing labor legislation.

Kawai (1981b, pp.~15; 17) assesses the $40$ households covered by the Tsukishima Survey as having ``ambiguous representativeness.''
No systematic description exists of how sample households were characterized among worker households in Tsukishima.
The official report noted: ``To find suitable persons to fill in the form, we must ask for the assistance of school principals, police officers, physicians, and laborers, and we have requested a meeting of these persons at the survey office to explain the purpose of the survey, and to solicit applicants for the distribution of the household account book" (Department of Health, Ministry of the Interior 1923a, pp. 8--9).
As noted, the target households were not randomly selected because statistical sampling was not yet prevalent.
However, as described in Section~\ref{sec:sec3}, the THBS households can be regarded as standard skilled factory-worker households in Tsukishima.
Therefore, Takano may have designed this household budget survey to capture representative households within the sampling framework of Tsukishima.

\begin{table}[htb]
\def\arraystretch{0.95}
\centering
\captionsetup{justification=centering}
\begin{center}
\caption{Survey Categories in the Field Study of Tsukishima Survey}
\label{tab:tabb1}
\footnotesize
\scalebox{1.0}[1]{
{\setlength\doublerulesep{2pt}
\begin{tabular}{lrr}
\toprule[1pt]\midrule[0.3pt]
									&Periods of the		&Chapter in the\\
									&field studies	&published report\\\hline
1.~Social mapping								&Nov. 1918--May 1919		&A2\\
2.~Physical examination of children				&Jan. 1919--July 1919		&3	\\
3.~Physical examination of workers				&June 1919					&3	\\
4.~Nutrition survey of the worker households		&May 1919					&3	\\
5.~Survey of tenement houses					&Autumn, 1919				&3	\\
6.~Survey of sanitation occupation				&June 1919					&3	\\
7.~Sanitary survey in elementary schools			&Unknown					&3	\\
8.~Factory labor survey						&Unknown					&4	\\
9.~Labor household survey					&Nov. 1918--Jan.1920		&2	\\
10.~Elementary school survey					&July 1919					&2	\\
11.~Restaurant survey						&Dec. 1919	\& April 1920	&2	\\
12.~Field survey of vaudeville					&Dec. 1919					&2	\\
13.~Survey of stalls and passersby				&Dec. 1919					&2	\\
14.~Photographing								&Oct. 1920--Nov.1920		&A2\\\midrule[0.3pt]\bottomrule[1pt]
\end{tabular}
}
}
{\scriptsize
\begin{minipage}{400pt}
\setstretch{0.85}
Notes:
The years/months listed in the first column (``Periods of the field studies'') are based on the description in the Department of Health, Ministry of the Interior (1923a, pp.~10--17).
Figures listed in the second column (``Chapter in the published report'') are based on the table of contents in the Department of Health, Ministry of the Interior (1923a, pp.~1--12; 1923b, pp.1--2).
``A2'' indicates the Appendix 2.
Sources: Department of Health, Ministry of the Interior (1923a, 1923b, 1923c).
\end{minipage}
}
\end{center}
\end{table}

Next, I provide an overview of the survey items and methodology.
The investigation was primarily divided into ``documentary'' and ``field'' surveys.
The former analyzed the characteristics of Tsukishima using published statistics, such as the Police Department Statistics Book (Department of Health, Ministry of the Interior 1923a, p.~10).
This was a pioneering approach that influenced subsequent urban studies (Kawai 1980, pp. 61--62).
The field survey included $14$ items (Table~\ref{tab:tabb1}).
According to the report, these can be grouped into four main categories: 1. workers' livelihood (chapter 2); 2. sanitary conditions (chapter 3); 3. labor conditions (chapter 4); and 4. social mapping and photography (appendix).
The survey was conducted over two years, from November 1918 to November 1920, with timing varying depending on the survey items (Table~\ref{tab:tabb1}).

A distinctive feature of the methodology was the use of young commissioned researchers.
The household budget and leisure life surveys were conducted by Yasunosuke Gonda, a faculty member of the Doitsugaku Ky\=okai Gakk\=o (Association for German Sciences) who had conducted research on cinematography and entertainment.
Tetsuo Hoshino, a medical doctor, oversaw the sanitary survey.
Yoshitsuru Yamana, a Tokyo Imperial University graduate who contributed to labor union organization, managed the labor survey.
Additionally, Toyotaro Miyoshi, a sociology student at Tokyo Imperial University, and others served as temporary employees (footnote~\ref{miyoshi}).
At the time, the Health Bureau of the Ministry of the Interior conducted several rural sanitary surveys, and Takano may have drawn on this experience in designing the urban sanitary survey (Kawai 1981a, pp.~8--9).
However, operating on the principle that ``health and sanitation surveys can at least encompass economic and social surveys,'' Takano included a broad range of survey items beyond sanitation (Department of Health, Ministry of the Interior 1923a, p.~56).
Kawai (1980, p. 55) credits the Tsukishima Survey as a notable pioneering effort: ``Going beyond a mere insurance and sanitation survey, it attempted to accurately depict the state of urban workers' lives as they became established in the city during modernization, with skilled workers as its core.''

\begin{figure}[]
\centering
\subfloat[Outside of the office]{\label{fig:office1}\includegraphics[width=0.38\textwidth]{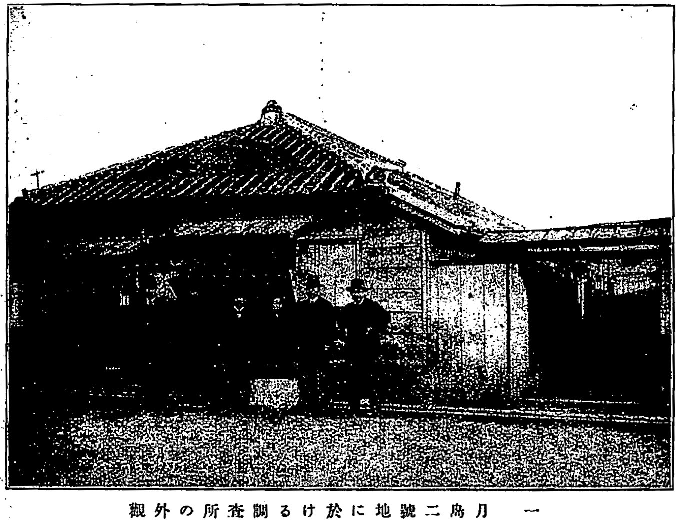}}
\subfloat[Inside of the office]{\label{fig:office2}\includegraphics[width=0.38\textwidth]{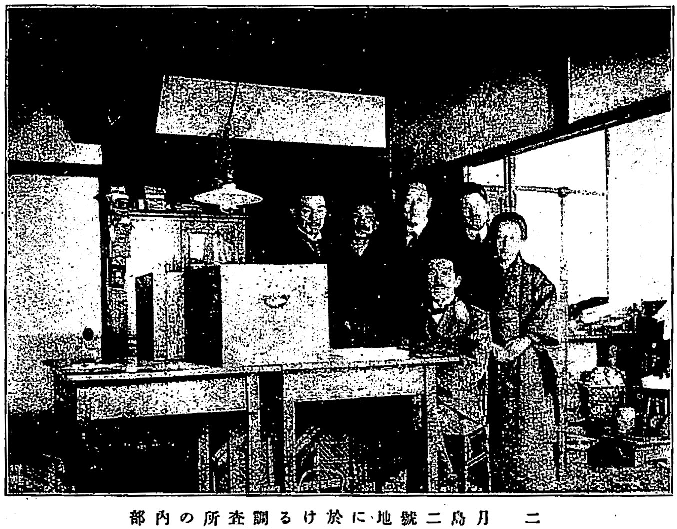}}
\caption{Photographs of the Survey Office in Tsukishima}
\label{fig:office}
\scriptsize{\begin{minipage}{450pt}
\setstretch{0.85}
Note: Figures~\ref{fig:office1} and~\ref{fig:office2} show the outside and inside of the survey office in Tsukishima, respectively.
Source: Department of Health, Ministry of the Interior 1923a, figures 1 and 2.
\end{minipage}}
\end{figure}

Takano stated, ``To be as complete as possible, it is best to continue the survey over as long a period as possible and to obtain accurate facts by recording the workers' household income and expenditures.''
For this reason, he believed that ``the best method of investigation would be to establish a survey station directly in the survey area and to have a full-time investigator reside there as much as possible, familiarize himself with the area and the people, and conduct the field survey'' (Department of Health, Ministry of the Interior 1923a, p.~4).
With the mediation of the Tsukishima police station, a survey office was rented at 9-3 Higashinakad\=ori, Tsukishima.
Yoshitsuru Yamana moved in from November 1918 and began contacting worker households (Department of Health, Ministry of the Interior 1923a, p.~7; Takeda 2015, p.~77).
The office remained open for the entire survey period, from November 1918 to December 1920. Figure~\ref{fig:office} shows photographs of the survey office.

Importantly, Takano and his team held \textit{weekly} meetings at the office throughout the survey period.
Toyotaro Miyoshi, who worked as an assistant on the survey, recalls:
\begin{quotation}
After the start of the survey office, all the members gathered once a week with Iwasaburo Takano present to discuss the implementation and progress of the survey and administrative matters.
Iwasaburo Takano would always attend each meeting, happily listening to the results of the week's research, inspecting the materials, providing detailed and thoughtful guidance, and usually ending the study session in a friendly atmosphere.
Even now, I can recall the warm face of sensei (i.e., Professor Takano) of those days.
It was always a friendly, laboratory-like atmosphere in which the Booth and Rowntree investigations were occasionally discussed.\\
 \hfill{Miyoshi (1980, p.~38)}
\end{quotation}
He also notes:
\begin{quotation}
When the Tsukishima Survey was launched, the Faculty of Economics (in the Tokyo Imperial University) was becoming independent, and he was swamped both inside and outside the university. However, his enthusiasm for the Tsukishima Survey was extreme, and he was always present on his business trips to Tsukishima. Moreover, he made every effort to carry out his duties with precision.\\
 \hfill{Miyoshi (1980, p.~40)}
\end{quotation}

Kawai (1981b, p.~28) observes no clear interconnections among the survey items, and that the survey ``ended up being several fragmentary surveys rather than a comprehensive and systematic survey of workers' lives.''
This may explain why previous academic studies have rarely used the Tsukishima Survey.
However, as explained earlier, Takano did not aim to conduct a large-scale social survey in Tsukishima.
Instead, he argued that the best method is to investigate a representative area ``narrowly but deeply,''
prioritizing the quality of the investigation over its scale.
In this light, the fact that researchers lived in the survey area and continuously organized and revised the results is significant.
Although the Tsukishima Survey is an early social survey in Japan, it is undoubtedly of high quality.

\begin{figure}[]
\centering
\captionsetup{justification=centering,margin=0.5cm}
\includegraphics[width=0.3\textwidth]{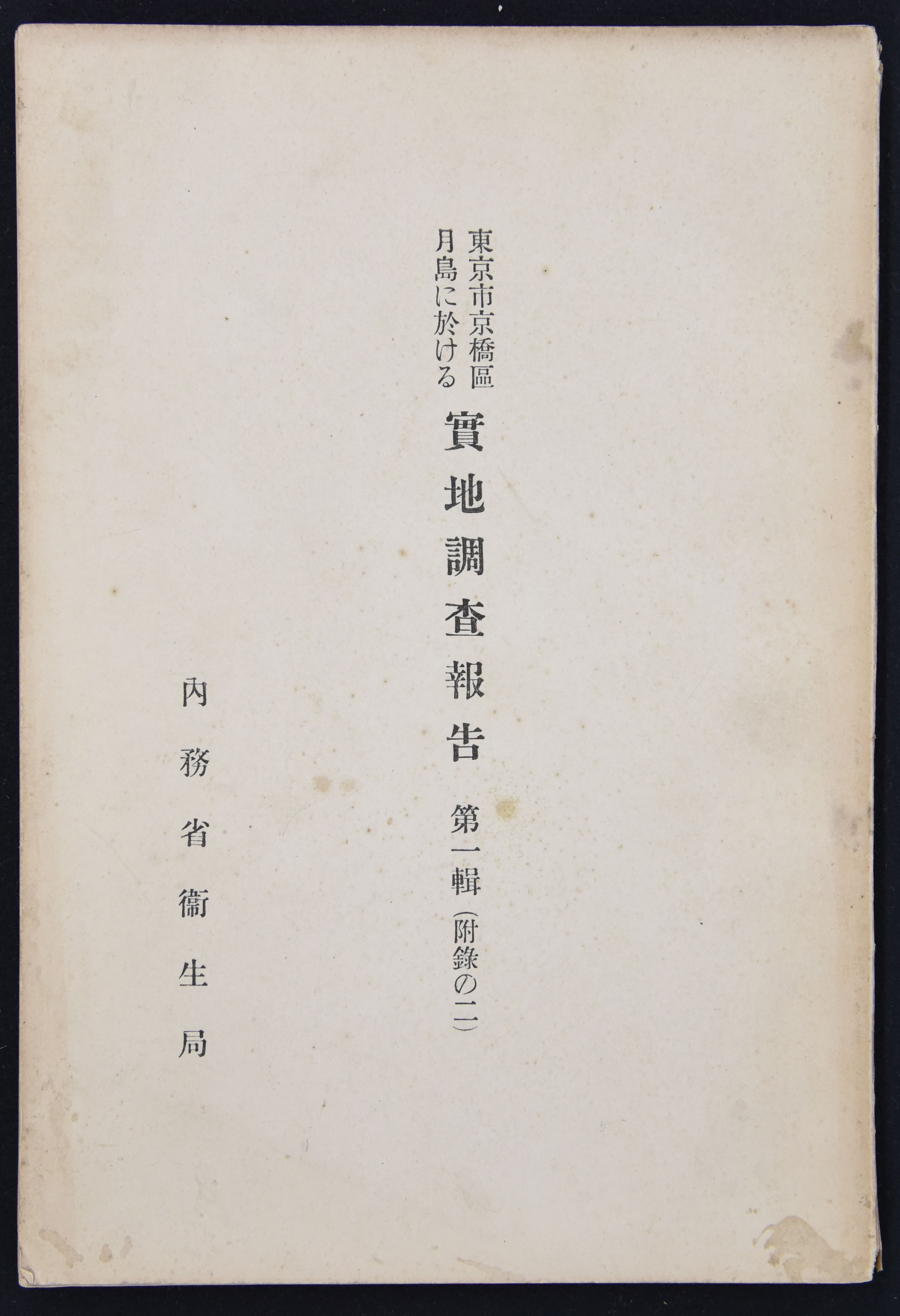}
\caption{Official Published Report of the Tsukishima Survey}
\label{fig:report}
\scriptsize{\begin{minipage}{450pt}
\setstretch{0.85}
Notes: 
This photograph shows an example of the official report of the Tsukishima Survey published by the Department of Health, Ministry of the Interior, in 1923.
Source: 
Department of Health, Ministry of the Interior 1923b, title page.
The tone was adjusted by the author using Adobe Photoshop 24.7.0.
\end{minipage}}
\end{figure}

Finally, I summarize the format of the published report and the unpublished materials.
The results of the Tsukishima Survey are officially documented in a report published in December 1921 by the Sanitary Bureau of the Ministry of the Interior, along with two appendices (Figure~\ref{fig:report} shows an example of the front cover).
The report, published under the title ``Report on Field Survey in Tsukishima, Kyobashi, Tokyo,'' consists of four sections.
The first section, a comprehensive review by Takano, outlines the survey.
The second section, written by Gonda (partly by Yamana), summarizes the living conditions of workers.
Hoshino authored the third section, which addresses workers' sanitary conditions.
In the fourth section, Yamana summarizes the labor situation in Tsukishima.
The two appendices complement the main report.
One contains statistical tables corresponding to the report, while the other organizes social maps and photographs.
The correspondence between survey items and the report is summarized in Table~\ref{tab:tabb1}.
These materials have been published by the Ministry of Home Affairs and are available at the Gonda Yasunosuke Library of the OISR and the University of Tokyo Library.
At the OISR, they are included in the Archives of Gonda Yasunosuke (7-2; 7-3; 7-4).

\subsection{Original Micro-spreadsheets from the Household Budget Survey in Tsukishima Survey}\label{sec:secb2}

The official report does not provide household-level information.
Fortunately, some of the original micro-spreadsheets are preserved in the OISR archives, as Takano was its first director.
This study uses the original micro-spreadsheets from the household budget survey (No. 9 in Table~\ref{tab:tabb1}), which Gonda used to write the second section of the official report.
Although these spreadsheets are not yet publicly available, I obtained permission from the OISR to use the original forms stored in the archives: the OISR, Archives of the Tsukishima Survey (THBS, unreleased).\footnote{I would like to express my sincere thanks to Professor Kazue Enoki at the OISR, for allowing me access to the documents. I wish to thank Ms. Mika Nakamura, an archivist at the institute, for her careful assistance during the research.}

\begin{figure}[h]
\centering
\captionsetup{justification=centering,margin=0.5cm}
\includegraphics[width=0.5\textwidth]{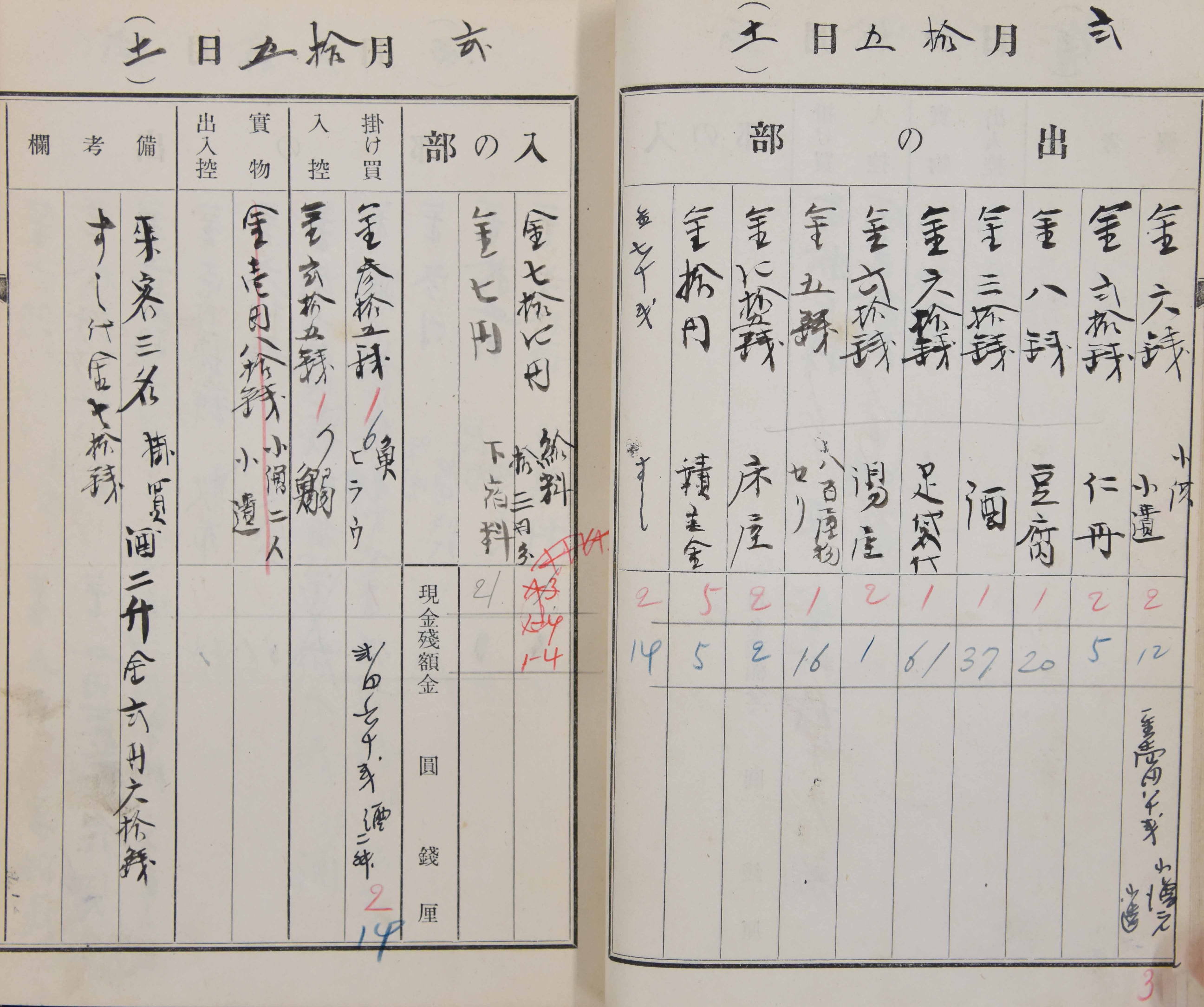}
\caption{An Example Page of the Budget Book\\ in the Tsukishima Household Budget Survey}
\label{fig:kakeibo}
\scriptsize{\begin{minipage}{450pt}
\setstretch{0.85}
Notes: 
This photograph depicts a sample page from the household expenses account of the THBS.
The right page documents the details of costs.
The left page includes the details of revenue and receiving and making payments on the goods (actual things, including gifts).
Source: 
The THBS, account \#6, 15th February, 1919.
The tone was adjusted by the author using Adobe Photoshop 24.7.0.
\end{minipage}}
\end{figure}

The budget book is titled the \textit{Kinsendeiri hikaech\=o} (money receipts and disbursements account book).
Its design is based on the form used in Takano's first household survey (Online Appendix~\ref{sec:secb1}).\footnote{Department of Health, Ministry of the Interior (1923a, p.~152).}
The book is completed daily and comprises four sections: income, expenses, receipts from credit purchases, and receipts and disbursements of actual goods.
Most entries appear in the income and expenses sections, indicating that daily life was largely sustained through wages and cash expenses.
Figure~\ref{fig:kakeibo} provides an example of the budget book.

As noted, the method for selecting households for the survey was not recorded in detail (Online Appendix~\ref{sec:secb1}).
According to the report, just over $90$ households were initially recruited through factories, the police station, and elementary schools in Tsukishima.
Of these, $50$ households completed the book for at least one month, but $10$ household's entries were considered incomplete and excluded from the report.
In addition to the $40$ budget books used for the report, the OISR holds another set marked as ``discarded.''
My survey indicates that about six incomplete books from worker households fall into this category.
Based on the condition of the remaining books, approximately $50$ households potentially participated in the survey.
Among the $40$ households included in the report, two maintained records for more than one year, $11$ households for 6--12 months, and the remaining $27$ households recorded data for less than six months (Department of Health, Ministry of the Interior 1923a, pp.~152--153).
Although the survey was initially planned for six months, the average bookkeeping period for these $40$ households was four months.

\subsection{Trimming}\label{sec:secb_trim}

\begin{figure}[htbp]
\centering
\subfloat[Daily panel dataset]{\label{fig:panels_daily}\includegraphics[width=0.32\textwidth]{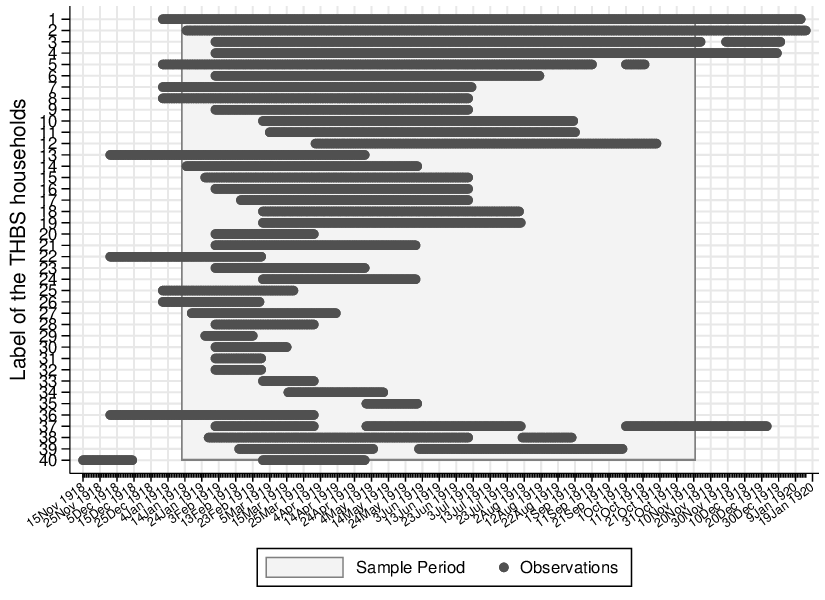}}
\subfloat[Semi-monthly panel dataset]{\label{fig:panels_semimonthly}\includegraphics[width=0.32\textwidth]{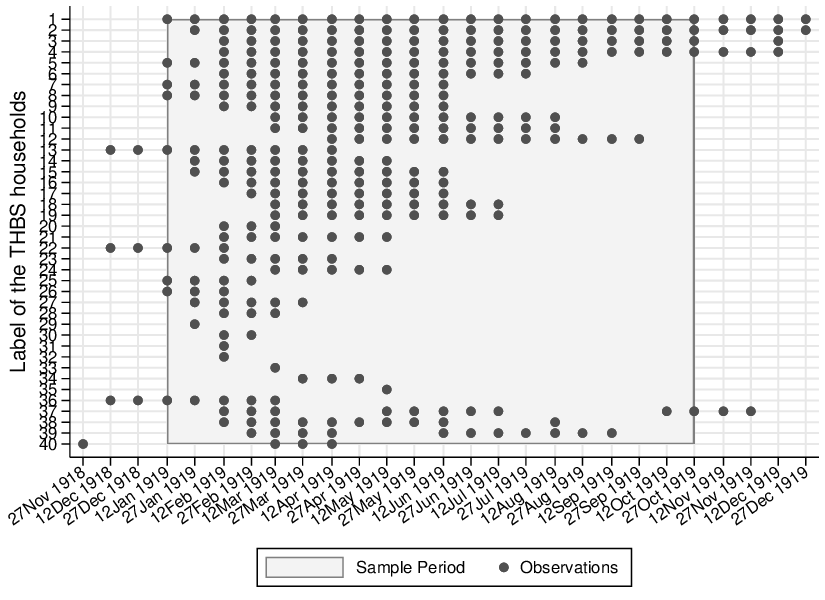}}
\subfloat[Monthly panel dataset]{\label{fig:panels_adjmonthly}\includegraphics[width=0.32\textwidth]{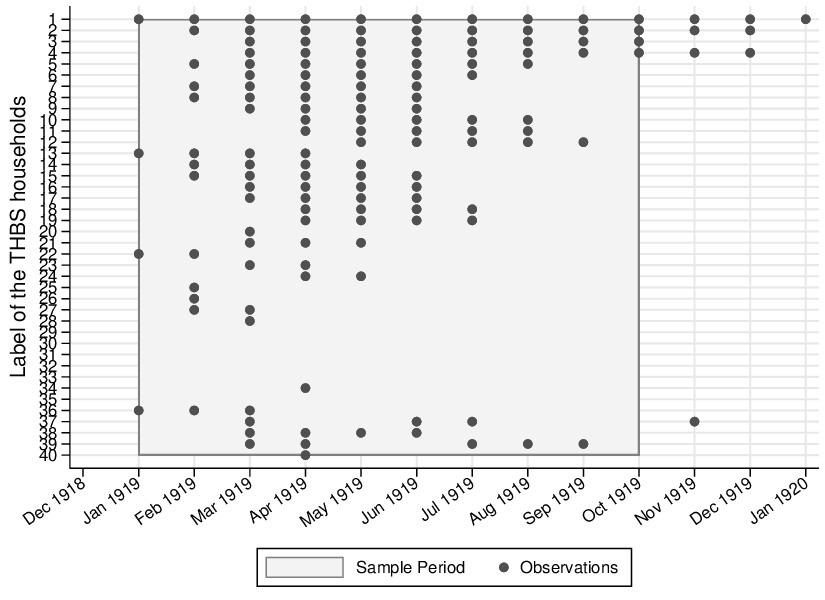}}
\caption{Structures of the THBS Datasets by Different Time Frequencies}
\label{fig:panel_structure}
\scriptsize{\begin{minipage}{450pt}
\setstretch{0.85}
Notes:
Figure~\ref{fig:panels_daily},~\ref{fig:panels_semimonthly}, and~\ref{fig:panels_adjmonthly} show the observations in the daily, adjusted semi-monthly, and adjusted monthly panel datasets, respectively.
For example, a marker in Figure~\ref{fig:panels_daily} indicates an observation in a specific year-month-date cell.
Strictly, semi-monthly, and monthly datasets include both panel and cross-sectional units because aggregation shall trim fractions in the edges of the survey period.
Figure~\ref{fig:panels_semimonthly} shows 35-panel units and five cross-sectional units (\#29; \#31-33; \#35).
Figure~\ref{fig:panels_adjmonthly} shows 28-panel units and six cross-sectional units (\#20; \#25; \#26; \#28; \#34; \#40).
Five cross-sectional units in the semi-monthly panel do not appear in this figure (\#29; \#31-33; \#35).
Note that unit \#30 having two semi-monthly cells is not included because those cells are separated into different adjusted months (February and March, respectively).
Source: Created by the author using the THBS sample.
\end{minipage}}
\end{figure}

I digitized all $40$ household budget books.
Throughout this digitization process, I found that there are a few misreporting issues in the budget books.

First, there is a household that does not include income information (household \#37).
Although the official report by Gonda reported the average monthly income for this household (Department of Health, Ministry of the Interior 1923a, p.~183), no documentation enabled me to determine how this average income was calculated.
Hence, I excluded this household's budget book.
Similarly, a household does not include information on the payments to the credit purchases (household \#39).
I excluded this household's budget book because the expenditure series became systematically smaller due to the omission of the payment information.

Second, the final pages of the budget books are sometimes omitted.
This may be because the vendor who bound the budget books accidentally truncated the last page.
All these incomplete sheets could not be included in the analysis.

Third, several households did not complete their budget books in the first and/or last month of the survey period.
For example, the budget book of household \#1 has statistics from January 1, 1919 to January 12, 1920.
Similarly, the book of household \#29 contains statistics from January 26 to February 23, 1919.
In semi-monthly and monthly datasets, these censored budget books may not be included in the aggregation.
In the latter case, for instance, the budget sheets of January 26 and after February 11, 1919, are not included in a semi-month cell because the semi-month cell is defined from the 27th to the 11th of the next month, in this case.
Thus, household \#29 only has a cross-sectional observation and could not be included in my analytical sample.

Figure~\ref{fig:panel_structure} summarizes the structure of the $40$ THBS households.
Figures~\ref{fig:panels_daily} and~\ref{fig:panels_semimonthly} show the observations in the daily and semi-monthly datasets, respectively.
As explained, there are five cross-sectional units (\#29; \#31-33; \#35) in the semi-monthly dataset.
Therefore, there are 35 units with a panel structure.
After excluding two units without complete information on income and credit purchases (\#37; \#39), I obtained the 33 THBS households as the analytical sample for my panel data analysis.

Regarding the time dimension, I trimmed all the observations before January 12, 1919, and after November 11, 1919, because the number of surveyed households is substantially small in the edge periods.
For example, Figure~\ref{fig:panels_semimonthly} shows that the number of cross-sectional units is three or less in both periods, causing substantial fluctuations in the net income.
Consequently, my analytical sample includes $33$ households measured between January 12 and November 11, 1919.
The average number of semi-months per unit is $8.8$ (Std. Dev. $= 5.0$).
The characteristics of the $33$ THBS households are summarized in detail in Section~\ref{sec:sec3}.

Regarding the adjusted monthly series, I further exclude several units that have insufficient observations for constructing the monthly panel: 26 THBS households are included in my adjusted monthly panel dataset.
Figure~\ref{fig:panels_adjmonthly} shows the observations in the adjusted monthly dataset.

\subsection{Testing the Potential Influence of the Lack of Balance}\label{sec:secb_balancing_test}

\def\arraystretch{0.95}
\begin{table}[htb]
\begin{center}
\captionsetup{justification=centering}
\caption{Results for the Balancing Tests
\label{tab:balancing}
}
\scriptsize
\scalebox{1.0}[1]{
\begin{tabular}{lrrr}
\toprule[1pt]\midrule[0.3pt]
&\multicolumn{3}{c}{DV: Indicator variable for the households}\\
&\multicolumn{3}{c}{with shorter time-series observations}\\
\cmidrule(rrr){2-4}
&\multicolumn{3}{c}{Threshold:}\\
\cmidrule(rrr){2-4}
&(1) Median		&(2) 75 percentile 	&(3) 25percentile \\
&($150$ days)	&($184$ days) 		&($79$ days) \\\hline
Size										&-0.243	&0.238		&-0.135\\
										&[0.244]	&[0.254]	&[0.294]\\
Children aged 6--12 (\%)						&0.017		&-0.011	&0.026\\
										&[0.020]	&[0.019]	&[0.022]\\
Children aged 13--16 (\%)						&0.029		&-0.050	&0.057\\
										&[0.038]	&[0.045]	&[0.041]\\
Men aged 17+ (\%)							&-0.008	&0.008		&-0.010\\
										&[0.019]	&[0.018]	&[0.026]\\
Intercept									&1.149		&-0.576	&-0.164\\
										&[2.107]	&[2.048]	&[2.805]\\\hline
Zero slope ($p$-value)						&0.710		&0.753		&0.160\\
Maximized Log-likelihood						&-21.8		&-18.5		&-15.2\\
Pseudo $R$-squared						&0.046		&0.046		&0.171\\
Number of households						&33		&33		&33\\\midrule[0.3pt]\bottomrule[1pt]
\end{tabular}
}
{\scriptsize
\begin{minipage}{300pt}
\setstretch{0.85}
The results from Probit models are reported.
Robust standard errors are in brackets.\\
Notes: 
The dependent variable is an indicator variable that takes one if the household has shorter time-series observations.
Column 1 uses the median of the number of days per unit as the threshold.
The result from the regression using the average number of days per unit ($144$ days) as the threshold is identical to that listed in column 1.
Column 2 uses the 75 percentile of the number of days per unit as the threshold.
Column 3 uses the 25 percentile of the number of days per unit as the threshold.
All family size variables are the time-series average in each unit over the sample period.
The proportion of children aged 0--5 years (\%) is used as the reference group.
Wald $\chi^{2}$ statistics $p$-values for the null of the zero-slope hypothesis are reported in the sixth row.
The results from the expanded regressions including the head's age and earnings are materially similar.
The null of the zero-slope hypothesis is not rejected in all the specifications at the conventional level (not reported).
\end{minipage}
}
\end{center}
\end{table}

If the difference in the preference for consumption predicted large attritions, there must be statistically significant correlations between the family size variables and attrition.
In other words, the units with shorter (longer) time-series observations have different preferences from those with longer (shorter) time-series observations.
Table~\ref{tab:balancing} summarizes the results of the balancing tests.
The dependent variable is an indicator variable that takes one if the household has time-series observations less than the median (column 1), 75 percentile (column 2), and 25 percentile (column 3).
In Column 1, all the estimated coefficients on the family size variables are close to zero and statistically insignificant, and the Wald statistics $p$-value suggests the null results.
This is unchanged if I use the third and first quantiles in columns (2) and (3), respectively.
Therefore, the family characteristics are similar between the households with short-term observations and those with long-term observations.
This result supports the evidence that the lack of balance in the THBS dataset is unlikely to lead to selection bias.

\subsection{Comparing Distributions of the Household Size}\label{sec:secb_size}

\begin{figure}[]
\centering
\captionsetup{justification=centering,margin=1.5cm}
\includegraphics[width=0.45\textwidth]{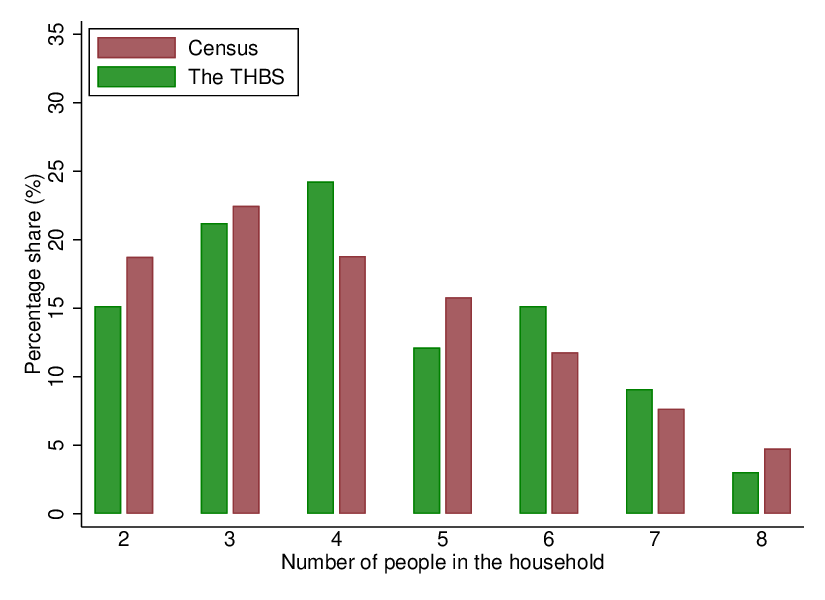}
\caption{Distribution of the Household Size:\\ Population Census v. the THBS sample}
\label{fig:hist_size}
\scriptsize{\begin{minipage}{450pt}
\setstretch{0.85}
Notes:
This figure shows the distribution of household size measured in the 1920 Population Census (red) and the THBS sample (green).
The Pearson chi-squared test does not reject the null hypothesis of the equality of distributions with $p$-value of $0.948$.
The share in each household size measured in the census is used as the theoretical probability in calculating $\chi^{2}$ test statistic (Pearson 1900).
Sources: Created by the author using the Tokyo City Office (1922a, pp.~262--283) and the THBS dataset.
\end{minipage}}
\end{figure}

Figure~\ref{fig:hist_size} illustrates the household size distributions measured in the census and THBS sample.
The Pearson chi-squared test does not reject the null hypothesis of the equality of distributions with $p$-value of $0.948$.
This supports the evidence that the THBS sample has a similar household size distribution to that of the entire Tsukishima population (Section~\ref{sec:sec31}).
As described in Section~\ref{sec:sec32}, while the THBS sample does not cover households with one person and with a very large number of people, it covers most of the classes in the household sizes with reasonable approximation to the population distribution.\footnote{The number of households with one people is $362$, accounting for only $6$\% of the total number of households. Similarly, the number of households with nine or more people is $294$, accounting for only $5$\% (Section~\ref{sec:sec32}).}

\subsection{Estimating Average Head's Monthly Income}\label{sec:secb_est_wage}

The time-series figures of average wages in prewar Japan were provided by the Long-Term Economic Statistics (hereafter LTES).
Specifically, the LTES estimates the average national salary series of ``general occupations,'' ``manufacturing sector,'' and ``manufacturing occupations (by medium classification).''
For the manufacturing occupations category, the LTES reports the average wage per day of factory workers in Tokyo and Osaka cities from 1917 to 1922 (Ohkawa et al. 1967, p.~255).
These values were estimated using the ``Annual Statistics of the City of Tokyo'' (hereafter, the ASCT) and ``Osaka City Statistical Table.''
However, the household survey includes the household head's monthly income but no information on the daily wage.
Hence, the ``average daily wage per factory worker'' estimated in the LTES is not useful for comparing the heads' average monthly income from the THBS.
In this section, I use available wage and labor statistics to estimate the average monthly income of male factory worker households in Tokyo and compare it to the THBS heads' average monthly earnings.

To clarify the problem in estimating the monthly earnings, I first estimate average monthly earnings in the simplest way.
To do so, I use the ASCT (Volume 17), which provides wage statistics as of the end of 1918, as a case study.
For each industrial sector, the ASCT (Volume 17) provides the ``average daily wage of male factory workers aged 15 and over,'' ``the number of male factory workers aged 15 and over,'' and ``average number of working days per year'' for four factory categories: factories with ten or fewer workers, and factories using or not using motive power.
Using the ``number of workers over 15 years old'' in each factory category cell ($c$) as weight, I can obtain the average wage per month for a worker in each industrial sector.
For example, in the manufacturing sector, this can be calculated as:
\begin{equation}\label{eqn:wage1}
\textit{Wage}^{15+}_{Machine} = \frac{\sum_{c}\{(\textit{Daily Wage}^{15+}_{Machine, c} \times \textit{Working Days}_{Machine, c}) \times \textit{Workers}^{15+}_{Machine, c}\}}{12\sum_{c}\textit{Workers}^{15+}_{Machine, c}},
\end{equation}
which is estimated to be $27.8$ yen.
This is the most straightforward estimate using daily wages, the number of workers, and the number of working days.
Thus, it has a few critical issues that cause a downward bias in the estimate.

\begin{figure}[]
\centering
\captionsetup{justification=centering,margin=1.5cm}
\includegraphics[width=0.45\textwidth]{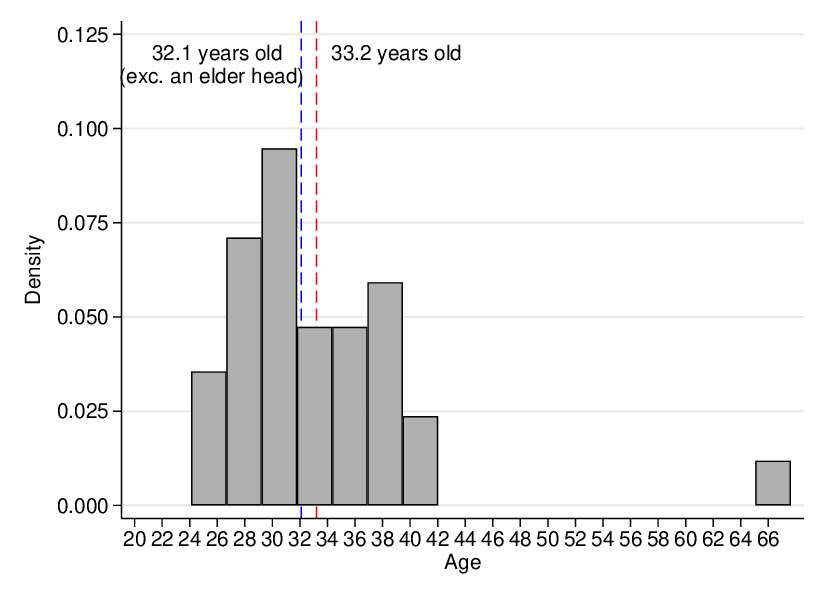}
\caption{Distribution of the THBS Heads' Age}
\label{fig:hist_hage}
\scriptsize{\begin{minipage}{450pt}
\setstretch{0.85}
Notes:
This figure shows the density of the THBS heads' age calculated from the THBS daily panel dataset.
The dashed line (red) indicates the average age of all the heads (33.2 years old).
The dashed line (blue) shows the average age of the heads, excluding an elder head aged 67.6 (32.1 years old).
The difference between the two average figures is statistically insignificant with $p=0.4087$.
Source: Created by the author using the THBS dataset.
\end{minipage}}
\end{figure}

The first issue is that the wages of factory workers are aggregated in a ``15 years old or older'' category in the ASCT (Volume 17).
Figure~\ref{fig:hist_hage} shows the THBS heads' age distribution.
The sample mean is $33.2$ years old, which is unchanged if I exclude a head aged more than $60$ (i.e., $32.1$ years old).
This means that I should refer to the average daily wage for the 30s to 40s for comparison.
However, since many workers are in their twenties in the distribution of entire workers, the average wages reported in the ASCT are considerably lower than the average wage for factory workers at approximately 30 years old.
This means that I need the wage profiles by age to correct this downward bias.\footnote{The ASCT (Volume 18), which contains statistics for 1919, no longer lists such wages by age group (average wages only). As is clear, the same problem arises in this case.}

Second, there is no consideration of ancillary pay such as bonus and allowance.
As explained later, these wages share a non-negligible proportion of the head's earnings.
Thus, estimation excluding ancillary wages would have a strong downward bias.

In the following subsections, I summarize the procedure for correcting these factors leading to the downward bias in the monthly income estimation.

\subsubsection{Daily Wage Distribution by Ages}\label{sec:secb_est_wage1}

As the early wage surveys focused only on the average wage for entire workers, official statistics in the 1910s never included average wage statistics by age (bin).
In the Statistical Data Field Survey Law (Law No.~52) in April 1922, average wages by age group were surveyed for the first time in the Labor Statistics Field Survey in October 1924 conducted by the Social Affairs Bureau of the Ministry of Home Affairs (Committee for the History of the Labor Movement 1964, p.~14).
The result of this survey in Tokyo was published as the Labor Statistics Table for Tokyo City and Suburbs (hereafter, LST) by the Tokyo City Statistics Division in 1926.

\begin{figure}[]
\centering
\captionsetup{justification=centering,margin=1.5cm}
\subfloat[1918]{\label{fig:wage1918}\includegraphics[width=0.44\textwidth]{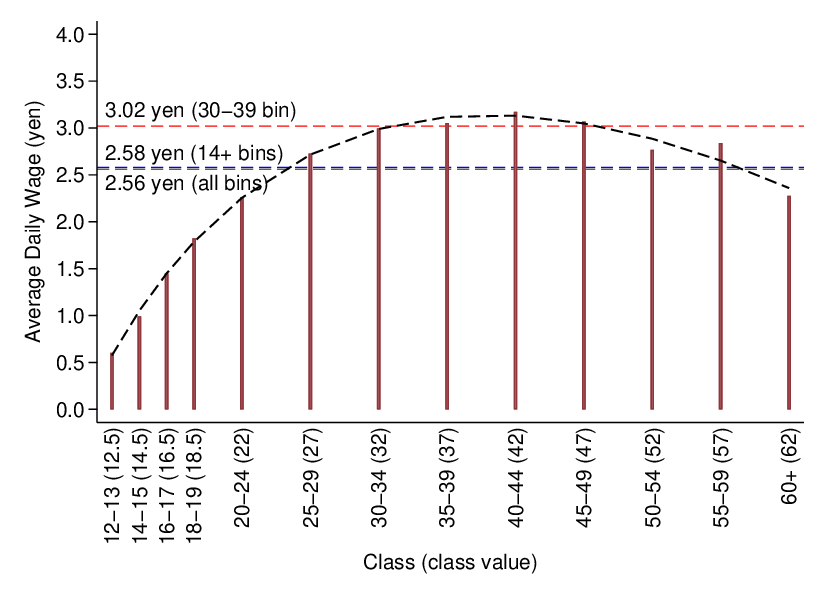}}
\subfloat[1919]{\label{fig:wage1919}\includegraphics[width=0.44\textwidth]{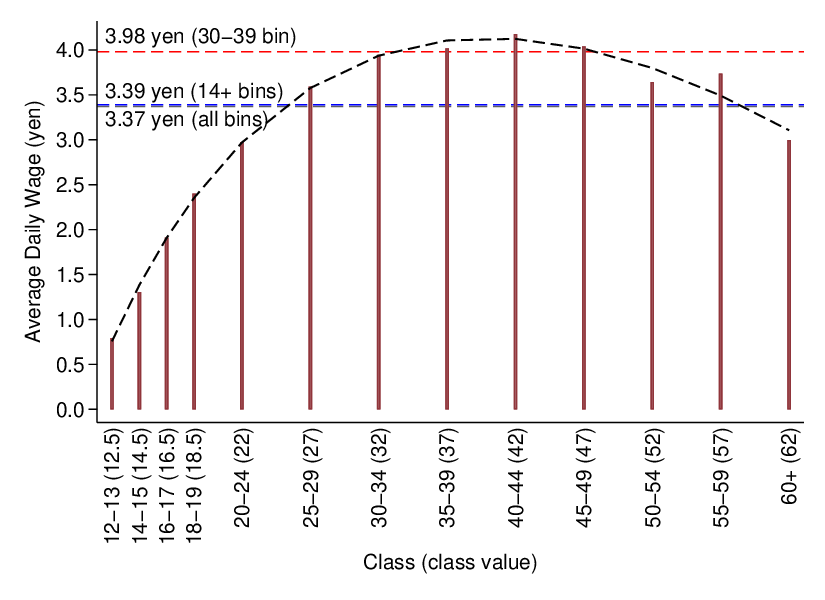}}
\caption{Average Daily Wage by Age Bins}
\label{fig:hist_daily_wage}
\scriptsize{\begin{minipage}{450pt}
\setstretch{0.85}
Notes:
This figure shows the average daily wage distribution by age bins based on the manufacturing census in 1924.
Figures~\ref{fig:wage1918} and~\ref{fig:wage1919} show the distributions under 1918 and 1919 prices, respectively.
The weighted average daily wage for a 30--39-year-old bin is shown in red.
The weighted averages for all and 14+ bins are shown in gray and blue lines, respectively.
The fitted fractional polynomial models are shown in the dashed lines.
For the 1918 price data, the fitted model is $\hat{y}^{\textit{Daily Wage}} = -10.01 + 4.15 a^{1/2}-0.33 a^{2}$ with $R^{2}=0.99$.
For the 1919 price data, the fitted model is $\hat{y}^{\textit{Daily Wage}} = -13.18 + 5.47 a^{1/2}-0.43 a^{2}$ with $R^{2}=0.99$.
Sources:
Created by the author using the Tokyo City Statistics Division (1926a, pp.~16; 244--245).
Data on the consumer price index for the cities are obtained from Ohkawa et al. (1967, p.~135).
\end{minipage}}
\end{figure}


Based on the LST, Figure~\ref{fig:hist_daily_wage} shows the average daily wages of male factory workers in Tokyo City's machinery and equipment manufacturing industry by age group.
In the LST, there are $13$ age bins, and the number of workers and average wages are listed for each bin.
The average daily wages have an inverted-U-shape, which is a similar life-cycle pattern to the factory workers observed in late 19th century England (Horrell and Oxley 2000, p.~42).
Figure~\ref{fig:wage1918} shows the distribution of wages adjusted to 1918 prices using the consumer price index (CPI) for urban areas (Ohkawa et al. 1967, p.~135).
The weighted average for ages $14$ and older is $2.58$ yen, which is about the same as the weighted average for daily wages of $2.56$ yen.
The $2.58$ yen is between the average values for the 20--24 age bin and the 25--29 age bin.
Thus, it is likely to represent the daily wage of workers at approximately $25$.
This is consistent with the fact that workers are concentrated in the twenties.
However, the weighted average for the 30--39 age bin is $3.02$ yen, which deviates significantly from $2.58$ yen.
The distribution of wages at 1919 prices shows a similar result (Figure~\ref{fig:wage1919}).
This means that the unweighted average leads to the erroneous conclusion that the sample households are upwardly biased.

To address this issue, I use the following steps to estimate the distribution of wages of the factory workers in Tokyo City in 1918--1919, the observation period of the THBS sample.
I use 1918 as an example below.

First, the distribution of daily wages by age group was predicted from statistics obtained from the LST.
Following Royston and Altman (1994), I consider a fractional polynomial of degree $m$ as follows:
\begin{equation}\label{eqn:fpr}
\phi_{m} (a; \bm{\varphi}, \mathbf{p}) = \varphi_{0} + \sum_{j=1}^{m} \varphi_{j}a^{(p_{j})},
\end{equation}
where $a$ ($>0$) denotes class value of age bin, $\bm{\varphi}$ is a vector of coefficients and $\mathbf{p}$ is a vector of powers, satisfying $p_{1}<p_{2}<...<p_{m}$.
$a^{(p_{j})}$ indicates the Box-Tidwell transformation, taking $\text{ln}(a)$ if $p_{j} = 0$ and $a^{p_{j}}$ if $p_{j} \neq 0$.
As suggested in Royston and Altman (1994, p.~433), the models requiring a degree higher than two ($m>2$) are rare in practice.
The wage by age bin distribution illustrated in Figure~\ref{fig:hist_daily_wage} clearly shows a simple quadratic curve.
Thus, I consider a fractional polynomial function with degree two as follows:
\begin{equation}\label{eqn:fpr}
\phi_{2} (a; \varphi_{0}, \varphi_{1}, \varphi_{2}, p_{1}, p_{2}) = \varphi_{0} + \varphi_{1}a^{(p_{1})} + \varphi_{2}a^{(p_{2})}.
\end{equation}
To model the function, I consider a set of powers $\mathscr{P} = \{-2, -1, -0.5, 0, 0.5, 1, 2, 3\}$ following Royston and Altman (1994, p.~434).

The maximum likelihood estimation is used to search for the most suitable candidate of the coefficients for a given $m$.
Since $m=2$, there are ${}_8 \mathrm{H}_2$ ($=36$) repeated combinations for the candidates of the parameters.
By adding the $8$ simple candidates under the nested models (i.e., the models with degree-1 fractional polynomial) to be conservative, I run $44$ regressions in total to find out the model with the highest maximized log-likelihood.
The best power vector suggested is $(p_{1}, p_{2})=(0.5, 1)$.\footnote{The deviance, defined as the twice negative (maximized) log-likelihood, is calculated to be $-31.5$. Royston and Altman (1994) proposed the deviance difference, $D(m, \mathbf{p})-D(m, \mathbf{\tilde{p}})$, where $\mathbf{\tilde{p}}$ is the best power vector to compare the different models. This statistic has an asymptotic chi-squared distribution with $m$ degree of freedom. The estimated deviance difference with the simplest linear model (i.e., $m=1$ and $p_{1}=1$) is $54.6$ with $p$-value = $0.000$. Similarly, the difference with an another non-linear model with $m=1$ and $p_{1}=-2$ is $34.2$ with  $p$-value = $0.000$.}
The projection model fitted using the best power vector is as follows:
\begin{equation}\label{eqn:fpr_fit}
\hat{y}^{\textit{Daily Wage}} = -10.01 + 4.15 a^{\frac{1}{2}} -0.33 a^{1}, \quad R^{2}=0.99.
\end{equation}
The dashed curve in Figure~\ref{fig:hist_daily_wage} is the predicted distribution in equation~\ref{eqn:fpr_fit}.
The fitting suggests that the predicted distribution works properly, especially below the 50--54 bin.

\begin{figure}[]
\centering
\captionsetup{justification=centering,margin=1.5cm}
\subfloat[1918]{\label{fig:wage1918pred}\includegraphics[width=0.44\textwidth]{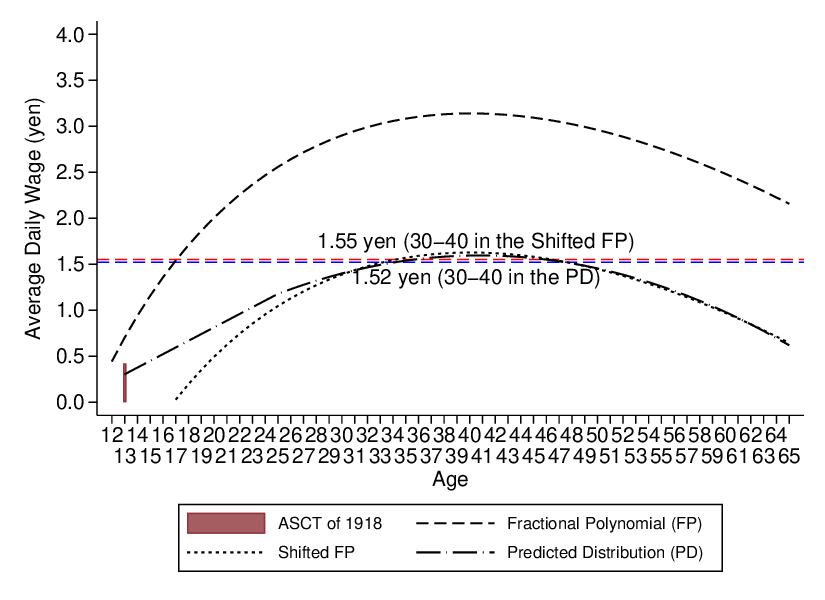}}
\subfloat[1919]{\label{fig:wage1919pred}\includegraphics[width=0.44\textwidth]{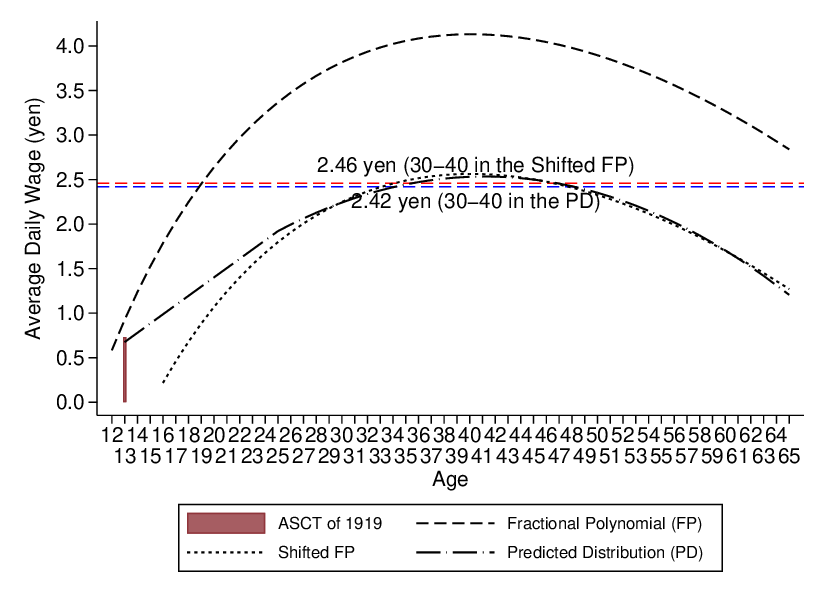}}
\caption{Daily Wage by Ages Distribution:\\ Shifted Fractional Polynomial}
\label{fig:hist_daily_wage_def_pred}
\scriptsize{\begin{minipage}{450pt}
\setstretch{0.85}
Notes:
The short dashed curve illustrates the fractional polynomial (FP) in Figure~\ref{fig:hist_daily_wage}.
The dotted curve indicates the FP shifted by the difference in the average daily wage between the ASCT and LST.
The ASCT documents the wage statistics for four different categories for each manufacturing sector: the scale of the factory (10+ or below ten workers) and engines (use or not).
I calculated the weighted average of the daily wage using the number of factory workers in each age bin and the category as weights.
The long-dashed curve shows the FP refitted using the average daily wage for 12--14-year-old bin ($0.42$ yen) measured in the ASCT and the daily wages for 25--65-year-old suggested in the shifted FP.
The average daily wage for a 12--14-year-old bin is only available for the ASCT of 1918.
Therefore, this figure for the 1919 price is calculated using the ratio between the average daily wage ($1.05$ yen) and that for 12--14 years old bin ($0.42$) in 1918 as $1.80 \times (0.42/1.05)$, where $1.80$ is the average daily wage measured in the ASCT of 1919.
The red dashed line is the average daily wage for 30--40-year-old in the shifted FP.
The blue dashed line is the average daily wage for 30--40-year-old in the refitted FP.
Sources: Created by the author using the Tokyo City Office (1921, pp.~726--757); Tokyo City Statistics Division (1926a, pp.~16; 244--245).
\end{minipage}}
\end{figure}

Next, I shift the predicted distribution based on the difference in the average daily wages in the ASCT and LST.
The weighted average of the daily wage is $2.56$ yen for the LST, whereas that for the ASCT is $1.05$ yen, meaning that there is approximately $1.5$ yen difference.\footnote{For the LST, I use the number of factory workers in each age bin as weight. The ASCT reports the average daily wage for four different categories of factories in each manufacturing sector: the factory scale (ten+ or below ten workers) and engines (use or not). I calculate the weighted average of the daily wage using the number of factory workers in each age bin (12--14 and 15+) and in each category as weights.}
Thus, I shift the functional polynomial function by this amount of difference to yield the proxy distribution for 1918.\footnote{This operation, therefore, assumes that the wage-by-age distributions in 1918 and 1924 are reasonably similar. In the LST (i.e., the statistic on which the predicted distribution is based), the weighted average for those aged $14$ and older was $3.25$ yen. In comparison, the overall weighted average is $3.23$ yen, a difference of only $0.02$ yen. A similar trend can be observed in the statistics obtained from the ASCT: the weighted average for those aged $15$ and older is $1.14$, while the overall average is $1.13$, a difference of only $0.01$ yen. This suggests that the wage distribution in the ASCT is similar to that in the LST.}
Figure~\ref{fig:wage1918pred} illustrates this operation: dotted curve is a fractional polynomial shifted from the original fractional polynomial shown as a short dashed curve.

The average daily wage for 12--14-year-old bin is reported to be $0.42$ yen in the ASCT, whereas that for 12--13-year-old bin in the LST is $0.60$ yen.
This difference (i.e., $0.18$ yen) is smaller than the difference in the weighted average for all age bins (i.e., $1.5$ yen).
This implies that the daily wage by age distribution in the ASCT could be smoother (i.e., flatter) than that of the LST.
Thus, the shifted fractional polynomial would provide stronger shrinkage for those bins of younger age.
In other words, the shifted curve shall get closer to the actual (unobservable) wage distribution as age increases.

The long dashed curve in Figure~\ref{fig:wage1918pred} examines this relationship by fitting another fractional polynomial model using the 12--14-year old bin figure of the ASCT and the predicted wages for 25--65 from the shifted fractional polynomial.\footnote{The fitted model is as follows: $\hat{y}^{\textit{Daily Wage}} = -1.19 + 0.14 a^{1} -0.002 a^{2}, \quad R^{2}=0.98.$ The deviance is reported to be $-149$. I used the 25--65 age range for the prediction based on the age bins used in the LST.}
Let this curve predict wage distribution for the true distribution.
The average daily wage for the 30--40-year-old bin in the shifted fractional polynomial is $1.55$ yen, which is similar to that for the same-age bin in the predicted wage distribution (i.e., $1.52$ yen).
However, the wage gaps for the younger ages show more significant differences.
If our interest is in the late 20s to the early 40s, it is plausible to employ the predicted wages for the wage comparison in Section~\ref{sec:sec3}.

Figure~\ref{fig:wage1919pred} illustrates the result of the same analysis from the wage data of 1919 price, reflecting the increasing trend in price between 1918 and 1919.
The average daily wage for a 30--40-year-old bin in the shifted fractional polynomial is $2.46$ yen.
Again, this is similar to that for the predicted wage distribution (i.e., $2.42$ yen).

\subsubsection{Ancillary Wage}\label{sec:secb_est_wage2}

\begin{figure}[h]
\centering
\captionsetup{justification=centering,margin=1.5cm}
\includegraphics[width=0.45\textwidth]{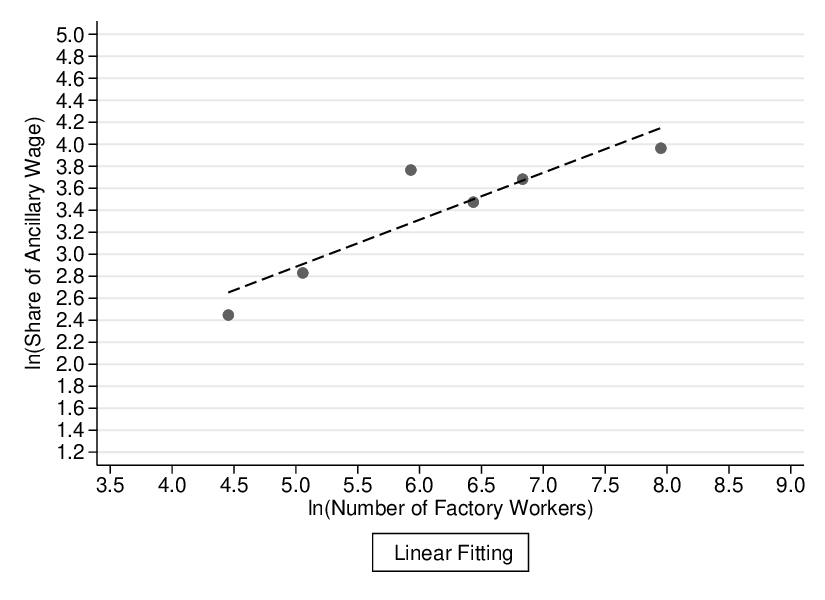}
\caption{Log-Relationship between the Number of Workers and Share of Ancillary Wage: Original Data from the Factory Survey\\ by Kitazawa (1924)}
\label{fig:scatter_fs}
\scriptsize{\begin{minipage}{450pt}
\setstretch{0.85}
Notes:
This scatter plot shows the relationship between the log-transformed number of factory workers and the log-transformed share of ancillary wage.
The raw data are listed in Table~\ref{tab:tabb2}.
The dashed line indicates the linear fit using the log-transformed variables: the fitted regression is $\hat{y}=0.747+0.428x$ with $R^{2}=0.82$.
The predictions are materially similar if I use the fractional polynomial using the raw data on the number of factory workers and share of ancillary wage in Table~\ref{tab:tabb2}.
In the case of fitting based on the raw data, the fitted fractional polynomial model is $\hat{y}^{\textit{Daily Wage}} = 53.6 -399.3 a^{-1/2}+(2.84\times10^{-10}) a^{3}$ with $R^{2}=0.87$.
I prefer to use linear regression herein because it can provide more precise fitted values for small-and medium-scale factories.
Source: Kitazawa (1924, pp.~20; 58--62).
\end{minipage}}
\end{figure}

The daily wage estimated in Section~\ref{sec:secb_est_wage1} is a fixed salary and thus, does not include additional salaries such as bonuses and allowances (hereafter, wages other than regular salaries are collectively referred to as ancillary wages).
However, there are no systematic statistics on the ancillary wages throughout the prewar period (Committee for the History of the Labor Movement 1964, pp.~14--15).
Moreover, since the Tsukishima Survey targets skilled worker households, information on the ancillary wages paid to skilled workers is necessary.
Fortunately, I found a survey report named Wage Survey Report edited by Kitazawa (1924).
This report includes the result of a survey on the wages for machine finishers (fitters) and turners at six machine factories in Tokyo from May--October 1921.
Thus, this is particularly useful material to know the wage structure of skilled factory workers.

According to Kitazawa (1924), there are two methods of paying wages to workers in the machine industry: one is to use hourly wages to calculate fixed salaries, and the other is to combine hourly wages with piece-rate wages, with the former being the most common (p.~33).
Despite this, wages paid in both methods are usually divided into fixed salaries and other ancillary wages, and there is not much difference in essence (p.~39).
Fixed wages are salaries added from a fixed hourly wage rate, while ancillary wages include bonuses and overtime pay, night shift pay, and labor premiums (profit sharing).
In the case of a combination of hourly and piece-rate pay systems, the ancillary wage includes a share of the piece-rate profit.

Panel A in Table~\ref{tab:tabb2} summarizes the survey results.
The surveyed factories included $86$, $157$, $376$, $623$, $928$, and $2,837$ workers, ranging from small to large plants.
The percentage of skilled workers (finishers and turners) in the surveyed factories ranged from $27$\% to $77$\%, and the portion of monthly ancillary wages was 12--53\%.
Figure~\ref{fig:scatter_fs} is a linear regression fit using log-transformed data on the percentage of ancillary wages and the number of workers.
There is a positive correlation between the factory size and ancillary wages.
This is consistent with the historical fact that larger factories have more extensive benefits than smaller ones (Odaka 1999).

Next, I estimate the percentage of ancillary wages at the Tsukishima factories.
One of the most frequently used statistics on factories in Tokyo is the Handbook of Factories (\textit{k\=ojy\=o ts\=uran}), which covers all the factories with ten or more workers as of January 1, 1920.
This handbook is the most comprehensive source for ascertaining the number of factories and their sizes \textit{circa} 1919.
This document does not cover the number of tiny working places with less than nine workers.
Importantly, however, the purpose of Panel C in Table~\ref{tab:tab2} is to compare the average monthly wage of the skilled factory workers in Tsukishima and the THBS heads' average monthly earnings.
Since the THBS heads did not work in the small workplaces, the censoring at the left tail of the factory size distribution is unlikely to influence this purpose.

Panel B of Table~\ref{tab:tabb2} summarizes the distributions of the number of factories, the total number of workers, and the estimated average share of ancillary wage in Tokyo City (Column 1) and Tsukishima (Column 2).
The average ancillary wage at the Tsukishima factories is $12$\%, which is slightly larger than the value for the entire city of Tokyo ($9$\%).
The percentage shares of the number of factory workers by each bin are listed in Columns (1) and (2) of Panel C in Table~\ref{tab:tabb2}.
As shown, Tsukishima factories tended to have larger factories than those of the entire city, which is reflected in the slightly larger estimate in the average ancillary wage for the Tsukishima factories.
In Column (3) of Panel C in Table~\ref{tab:tabb2}, the number of THBS heads whose occupation, factory name, and factory size are available are listed.
While the percentage shares of the heads in each bin show a similar distribution to that of the Tsukishima factories, the share of middle (large) sized factories is larger (smaller) than the population.
If interpreted literally, this tendency implies that the (unobservable) true average ancillary wage for the THBS heads may be slightly lower than the population mean.
However, note that the purpose of Panel C in Table~\ref{tab:tab2} is to compare the average monthly wage of the skilled factory workers employed in the Tsukishima factories to the THBS heads' average monthly earnings.
Therefore, the purpose herein is to estimate the average ancillary wage for the target population, for example, the Tsukishima factories, but not to calculate the average ancillary wage for the THBS sample itself.

\begin{table}[hbtp]
\def\arraystretch{0.95}
\centering
\captionsetup{justification=centering}
\begin{center}
\caption{Estimating the Average Share of Monthly Ancillary Wage}
\label{tab:tabb2}
\scriptsize
\scalebox{1.0}[1]{
{\setlength\doublerulesep{2pt}
\begin{tabular}{lrrrrrr}
\toprule[1pt]\midrule[0.3pt]
\multicolumn{7}{l}{\textbf{Panel A: Original Data on the Share of Ancillary Wage}}\\
\multicolumn{7}{l}{Document: Report of Factory Survey by Kitazawa (1924)}\\
\multicolumn{7}{l}{Survey subject: Six Machinery Factories in Tokyo City}\\
\multicolumn{7}{l}{Survey month and year: May to October 1921}\\
			&Number of 	&Share of		&Share of 	&Share of &&\\
			&factory		&skilled			&fixed		&ancillary&&\\
Factory		&workers		&workers (\%)	&wage (\%)		&wage (\%)&&\\
\cmidrule(rrrrr){1-5}
A			&157			&44			&83		&17&&\\
B			&86			&53			&88		&12&&\\
C			&376			&27			&57		&43&&\\
D			&623			&77			&68		&32&&\\
E			&2,837			&NA			&47		&53&&\\
F			&928			&41			&60		&40&&\\
&&&&&&\\
&&&&&&\\
\multicolumn{7}{l}{\textbf{Panel B: Estimating the Average Share of Ancillary Wage}}\\
\multicolumn{7}{l}{Document: Handbook of Factories (1921 edition)}\\
\multicolumn{7}{l}{Survey subject: Machinery Factories with 10+ workers in Tokyo City}\\
\multicolumn{7}{l}{Survey month and year: January 1920}\\
&\multicolumn{6}{c}{Machinery Factories}\\
\cmidrule(rrrrrr){2-7}
&\multicolumn{3}{c}{(1) Tokyo City}&\multicolumn{3}{c}{(2) Tsukishima}\\
\cmidrule(rrr){2-4}\cmidrule(rrr){5-7}
							&Number		&Number 		&Average share 		&Number		&Number		&Average share	\\
							& of factories	&of factory		&of ancillary 		&of factories	&of factory		&of ancillary		\\
Factory size					&				&workers		&wage (\%)			&				&workers		&wage (\%)		\\\hline

10--99 workers				&760			&18,760		&8					&39			&1,252		&9\\
100--199 workers			&52			&7,155			&17				&5				&713		&18\\
200+ workers				&34			&21,764		&31				&3				&3,851		&38\\
Total						&846			&47,679		&9					&47			&5,816		&12\\
&&&&&&\\
&&&&&&\\
\multicolumn{7}{l}{\textbf{Panel C: Comparisons of the Distributions of Factory Size}}\\
&\multicolumn{6}{c}{Machinery Factories}\\
\cmidrule(rrrrrr){2-7}
&\multicolumn{2}{c}{(1) Tokyo City}&\multicolumn{2}{c}{(2) Tsukishima}&\multicolumn{2}{c}{(3) THBS}\\
\cmidrule(rr){2-3}\cmidrule(rr){4-5}\cmidrule(rr){6-7}
							&Number		&\% share 	&Number	&\% share	&Number	&\% share	\\
							& of factory		&			&of factory 	&			&of THBS	&			\\
Factory size					&workers		&			&workers	&			&heads		&			\\\hline
10--99 workers				&18,760		&39		&1,252		&22		&6			&21		\\
100--199 workers			&7,155			&15		&713		&12		&7			&25		\\
200+ workers				&21,764		&46		&3,851		&66		&15		&54		\\
Total						&47,679		&100		&5,816		&100		&28		&100		\\\midrule[0.3pt]\bottomrule[1pt]
\end{tabular}
}
}
{\scriptsize
\begin{minipage}{410pt}
\setstretch{0.85}
Notes:
\textbf{Panel A}:
This panel shows the size of factories and the share of fixed and ancillary monthly wages in six factories surveyed in Kitazawa (1924).
The number of factory workers surveyed in factories A--F are 14, 23, 9, 11, 11, and 16, respectively.
The share of these surveyed skilled workers to all skilled workers is 20.3 in Factory A, 50.0 in Factory B, 5.3 in Factory C, 2.3 in Factory D, and 4.2 in Factory F, respectively.
This figure for factory E is unavailable because the number of skilled workers is unknown (Kitazawa 1924, p.~21).
\textbf{Panel B}: 
Subcolumns 1--2 and 4--5 summarize the number of factories and total number of workers in Tokyo City and Tsukishima, respectively (Handbook of Factories 1921).
Subcolumns 3 and 6 show the average share of ancillary monthly wage predicted in Figure~\ref{fig:scatter_fs} for Tokyo City and Tsukishima, respectively.
\textbf{Panel C}:
Columns 1--3 summarize the number of factory workers and its share in percentage points in Tokyo City, Tsukishima, THBS sample, respectively.
The statistics in Columns 1 and 2 (same figures used in Panel B) are from the Handbook of Factories (1921).
In Column 3, the THBS heads who worked in the machinery factories are considered.
There are 28 heads whose factory names and size are available.
For a blacksmith included in these heads, I assume the number of workers to be ten, based on the report (Department of Health, Ministry of the Interior 1923a, p.~412).
Sources: 
Panel A: Kitazawa (1924, pp.~20; 58--62).
Panel B: The number of factory workers is taken from the Ministry of Agriculture and Commerce (1921, pp.~613--629; 674).
The average shares of ancillary wage are calculated using Kitazawa (1924, pp.~20; 58--62). See also Figure~\ref{fig:scatter_fs}.
Panel C: Data on the number of factory workers and factories are from the Ministry of Agriculture and Commerce (1921, pp.~613--629; 674).
Data on the number of heads are from the THBS sample.
\end{minipage}
}
\end{center}
\end{table}

\subsubsection{Estimating Monthly Income of Adult Male Factory Workers}\label{sec:secb_est_wage3}

\begin{table}[hbtp]
\def\arraystretch{0.95}
\captionsetup{justification=centering}
\begin{center}
\caption{Estimating the Monthly Income of Adult Male \\ Factory Workers in Machinery Sector}
\label{tab:wage}
\footnotesize
\scalebox{1.0}[1]{
\setlength\doublerulesep{2pt}
\begin{tabular}{lrrr}
\toprule[1pt]\midrule[0.3pt]
&\multicolumn{3}{c}{Weight for the Ancillary Wage}\\
\cmidrule(rrr){2-4}
&(1) Tsukishima		&(2) Tokyo City	&	\\
&$\vartheta=0.12$	&$\vartheta=0.09$		&	\\\hline
\multicolumn{4}{l}{\textbf{Panel A:  Components in Equation~\ref{eqn:wage_est_full}: $w^{Daily}_{j}$, $l_{j}$, and $n_{m}$}}\\
Average daily wage ($w^{Daily}_{j}$)	&&&		\\
$w^{Daily}_{1918}$					&1.52&1.52&\\
$w^{Daily}_{1919}$					&2.42&2.42&\\
Average annual working days ($l_{j}$)	&&&		\\
$l_{1918}$						&313	&313	&\\
$l_{1919}$						&322	&322	&\\
Frequency weights ($n_{m}$)			&&&\\
$n_{1}$							&5	&5	&\\
$n_{2}$							&11	&11	&\\
$n_{3}$							&19	&19	&\\
$n_{4}$							&22	&22	&\\
$n_{5}$							&21	&21	&\\
$n_{6}$							&18	&18	&\\
$n_{7}$							&11	&11	&\\
$n_{8}$							&8	&8	&\\
$n_{9}$							&5	&5	&\\
$n_{10}$							&4	&4	&\\
&&&\\
\multicolumn{4}{l}{\textbf{Panel B:  Estimated Monthly Earning: $w^{Total}$}}\\
$w^{\textit{Total}}$					&56.0	&54.5	&		\\\midrule[0.3pt]\bottomrule[1pt]
\end{tabular}
}
{\scriptsize
\begin{minipage}{320pt}
\setstretch{0.85}
Notes:
1. Average daily wage ($w^{Daily}_{j}$): The average daily wage of male factory workers in the skilled age range (aged 30--40) in the machinery sector is calculated using two official reports of the manufacturing censuses documented in the ASCT and LST.
Details in the calculation steps are summarized in Online Appendix~\ref{sec:secb_est_wage1}.\\
2. Average annual working days ($l_{j}$): 
The ASCT documents the number of factories and average annual working days for each machinery sub-sector in two factory-size bins.
I use the number of factories in each machinery sub-sector and factory size bin as the weights to calculate the weighted average of the annual working days in the entire machinery sector.\\
3. Weight for the ancillary wage ($\vartheta$): The average share of monthly ancillary wage in the machinery sector is calculated based on the factory survey report in Kitazawa (1924).
The calculation steps are described in Online Appendix~\ref{sec:secb_est_wage2}.
Column 1 uses the share based on the factory size distribution in Tsukishima, whereas column 2 uses that in the entire Tokyo City.\\
4. Frequency weights ($n_{m}$) are based on the number of cross-sectional observations between January and October 1919 in the adjusted monthly panel dataset.\\
5. $w^{\textit{Total}}$ is calculated using equation~\ref{eqn:wage_est_full}.\\
Sources:
Data used to calculate the average daily wage are from the Tokyo City Office (1921, pp.~726-757); Tokyo City Office (1922, pp.~732-733); Tokyo City Statistics Division (1926a, pp.~16; 244--245).
Data used to calculate the average annual working days are from the Tokyo City Office (1921, pp.~726--741) and Tokyo City Office (1922, pp.~716--725).
Data used to calculate the average share of monthly ancillary wage are from Kitazawa (1924, pp.~20; 58--62) and the Ministry of Agriculture and Commerce (1921, pp.~613--629; 674).
\end{minipage}
}
\end{center}
\end{table}

Following the institution described in Online Appendix~\ref{sec:secb_est_wage2}, the monthly wage of the male workers in the skilled age range ($w^{\textit{Total}}$) is defined as follows:
\begin{equation}\label{eqn:wage_est1}
w^{\textit{Total}} = w^{\textit{Regular}}+w^{\textit{Ancillary}},
\end{equation}
where $w^{\textit{Regular}}>0$ is the monthly regular wage and $w^{\textit{Ancillary}} \geq 0$ indicates the monthly ancillary wage.
Online Appendix~\ref{sec:secb_est_wage2} shows that $w^{\textit{Ancillary}}$ can be calculated as the product of the monthly regular wage and weight for the ancillary wage, $\vartheta \in (0, 1)$.\footnote{Strictly, this weight may not be required to be in this range. I use this range because all the cases reported in Kitazawa (1924) take the weights between zero and one.}
The monthly wage equation is then specified as follows:
\begin{equation}\label{eqn:wage_est_fin}
w^{\textit{Total}} = (1 + \vartheta) w^{\textit{Regular}}.
\end{equation}

To calculate equation~\ref{eqn:wage_est_fin}, I have to conduct a few final tunings in my wage data.
First, I fix the deviations in the timing of the manufacturing censuses and the THBS observations. 
Let $w^{\textit{Regular}}_{j}$ be the monthly regular wage calculated using wage statistics measured in year $j \in \{1918, 1919\}$ as follows:
\begin{equation}\label{eqn:wage_regular1}
w^{\textit{Regular}}_{j} = w^{\textit{Daily}}_{j} l_{j}/12,
\end{equation}
where $w^{\textit{Daily}}_{j}$ is the average daily wage of male factory workers in the skilled age range and $l_{j}$ is the average annual working days (Online Appendix~\ref{sec:secb_est_wage1}).
The THBS were conducted from November 1918 to January 1920, and my THBS sample was observed from December 1918 to December 1919 (Section~\ref{sec:sec3}).
However, $w^{\textit{Regular}}_{j}$ is calculated using the wage statistics measured in the manufacturing census conducted in \textit{December} of 1918 and 1919.
This means that if I used $w^{\textit{Regular}}_{1919}$ alone for equation~\ref{eqn:wage_est_fin}, it shall provide upward-biased wage due to the deviation between the timing of the THBS observations and of the manufacturing census.
To manage this issue, I use the weighted average between $w^{\textit{Regular}}_{1918}$ and $w^{\textit{Regular}}_{1919}$ to yield monthly wage series as follows:
\begin{equation}\label{eqn:wage_regular2}
w^{\textit{Regular}}_{m} = \{(12 - m) w^{\textit{Regular}}_{1918} + m w^{\textit{Regular}}_{1919}\}/12,
\end{equation}
where $m \in \{0, 1, 2,...,12\}$ corresponds to the year-month cells from December 1918 to December 1919.
This way of weighting assumes that the average daily wage of machinery workers had increased during this period.
This assumption is plausible given the historical fact that the average wage in the manufacturing industry indeed showed a secular increasing trend \textit{circa} 1920 in Tokyo City (Ohkawa et al. 1967, p.~255).

Second, I use the number of observations in each year-month cell as the frequency weight.
Let $n_{m}$ be the number of households measured in year-month cell $m$.
Then, the average monthly wage can be defined as:
\begin{equation}\label{eqn:wage_regular3}
w^{\textit{Regular}} = \frac{\sum_{m} n_{m} w^{\textit{Regular}}_{m}}{\sum_{m}n_{m}},
\end{equation}
where $m \in \{\text{December}~1918, \text{January}~1919,..., \text{December}~1919\}$ is the year-month cell and $n_{m}$ is the frequency weight for year-month cell $m$.

Substitute equation~\ref{eqn:wage_regular1},~\ref{eqn:wage_regular2}, and~\ref{eqn:wage_regular3} into equation~\ref{eqn:wage_est_fin} to yield the final specification:
\begin{equation}\label{eqn:wage_est_full}
w^{\textit{Total}} = (1 + \vartheta) \frac{\sum_{m} n_{m} \{(12 - m) w^{\textit{Daily}}_{1918} l_{1918} + m w^{\textit{Daily}}_{1919} l_{1919}\}}{12^{2}\sum_{m}n_{m}}.
\end{equation}
Panel A in Table~\ref{tab:wage} summarizes all the components in equation~\ref{eqn:wage_est_full}.
In Column 1 of Panel B in Table~\ref{tab:wage}, I calculate the average monthly wage under the share of the ancillary wage for the factories in Tsukishima ($\vartheta$).
The estimate is $56$ yen, which is similar to the average and median of monthly earnings in the THBS heads ($59$ and $56$ yen, respectively; see Panel C of Table~\ref{tab:tab2}).
Therefore, I can conclude that the deviation from the population mean ($56$ yen) is not too large to infer the mean household strategies of the skilled workers in the machinery factories in Tsukishima.
Given that the average daily wage of the male factory workers in the skilled age range is $2.5$ yen (Figure~\ref{fig:hist_daily_wage_def_pred}), $3$ yen difference from the average monthly income is approximately one day's earnings.
This is consistent with the findings in my descriptive analyses that show the similarities in the distribution of family structure, average household size, and share of ancillary wage between the THBS households and Tsukishima households (Section~\ref{sec:sec3}; Online Appendix~\ref{sec:secb_est_wage2}).

Column 2 uses the share of the ancillary wage for the entire Tokyo City ($\vartheta=0.09$) to calculate the average monthly wage.
The estimate is $54.5$ yen, which is slightly smaller than that in Column 1, reflecting the difference in the weights for the ancillary wages.
If I view this figure as an unbiased estimate for the entire Tokyo City, the difference in the average factory size between the entire Tokyo City and Tsukishima may be reflected in this deviation (Panel B in Table~\ref{tab:tabb2}).

\subsection{Measurement Errors in the Unadjusted Panel Dataset}\label{sec:secb_me}

\begin{figure}[]
\centering
\captionsetup{justification=centering,margin=0.25cm}
\subfloat[Unadjusted Semi-monthly Panel Dataset: Income and expenditure]{\label{fig:tssm_minc_mexp_raw_app}\includegraphics[width=0.4\textwidth]{tssm_minc_mexp_raw.eps}}
\subfloat[Unadjusted Semi-monthly Panel Dataset: Net income]{\label{fig:tssm_mexpdef_raw_app}\includegraphics[width=0.4\textwidth]{tssm_mexpdef_raw.eps}}\\

\subfloat[Shifted Semi-monthly Panel Dataset: Income and expenditure]{\label{fig:tssm_minc_mexp_shift}\includegraphics[width=0.4\textwidth]{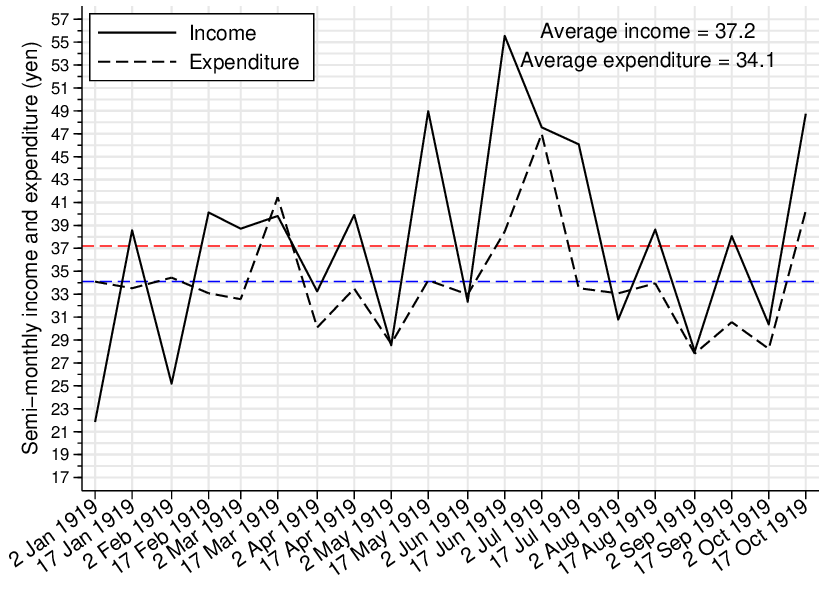}}
\subfloat[Shifted Semi-monthly Panel Dataset: Net income]{\label{fig:tssm_mexpdef_shift}\includegraphics[width=0.4\textwidth]{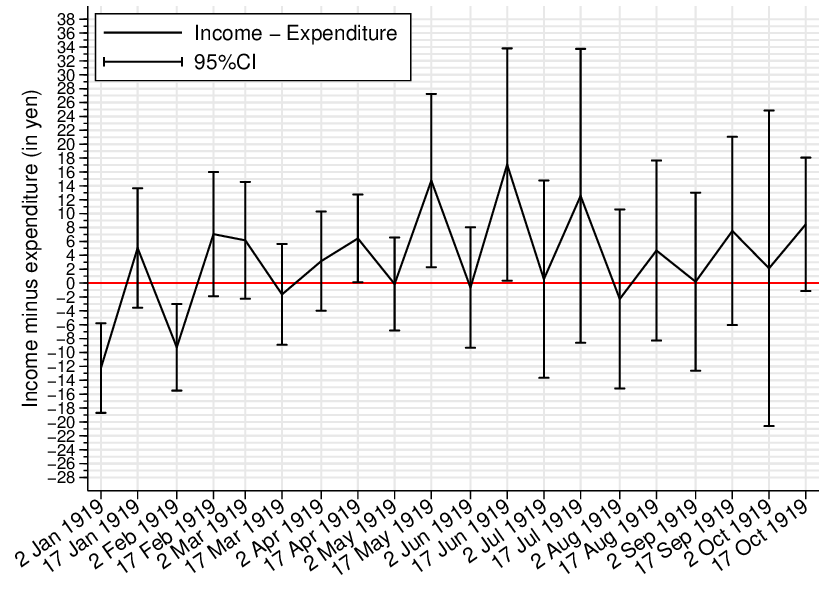}}\\

\subfloat[Adjusted Semi-monthly Panel Dataset: Income and expenditure]{\label{fig:tssm_minc_mexp_app}\includegraphics[width=0.4\textwidth]{tssm_minc_mexp.eps}}
\subfloat[Adjusted Semi-monthly Panel Dataset: Net income]{\label{fig:tssm_mexpdef_app}\includegraphics[width=0.4\textwidth]{tssm_mexpdef.eps}}\\

\caption{Comparisons of Average Semi-monthly Income and Expenditure\\ from Different Aggregations}
\label{fig:ts_overview_com}
\scriptsize{\begin{minipage}{450pt}
\setstretch{0.85}
Note:
Figures~\ref{fig:tssm_minc_mexp_raw_app},~\ref{fig:tssm_minc_mexp_shift},~\ref{fig:tssm_minc_mexp_app} show the average semi-monthly income and expenditure calculated from unadjusted, shifted, and adjusted panel datasets, respectively.
Unadjusted series defines the days from the 1st to the 15th of each month as the first half, whereas the days from the 16th to the end of each month as the second half.
The shifted series defines the days from the 2nd to the 16th of each month as the first half, whereas the days from the 17th to the 1st of the following month are the second half.
Adjusted series is the baseline definition in the main text, which uses the days from 12th to 26th in each month as the ``first-half'' and those from 27th to 11th in the next month as the ``second-half.''
As explained in Section~\ref{sec:sec3}, the raw series contains considerable measurement errors due to the timing of payday.
The shifted series can manage the measurement errors from the timing of payday but cannot adjust the difference in the payday timing and expenses.
In other words, the shifted series ignores the households' decision on the consumption schedule in each semi-month.
The adjusted series reflecting the exact timing of the household expenses can manage measurement errors.
Figures~\ref{fig:tssm_mexpdef_raw_app},~\ref{fig:tssm_mexpdef_shift},~\ref{fig:tssm_mexpdef_app} illustrate the average semi-monthly income minus expenditure calculated from unadjusted, shifted, and adjusted panel datasets, respectively.
Source: Created by the author using the THBS sample.
\end{minipage}}
\end{figure}

Figure~\ref{fig:ts_overview_com} explains how different methods of aggregation influence the time-series in income and expenditure.
Figures~\ref{fig:tssm_minc_mexp_raw_app},~\ref{fig:tssm_minc_mexp_shift},~\ref{fig:tssm_minc_mexp_app} show the average semi-monthly income and expenditure calculated from unadjasted, shifted, and adjusted panel datasets, respectively.

The unadjusted (calender-month) series defines the days from the 1st to the 15th of each month as the first half and the days from the 16th to the end of each month as the second half.
This series shall contain substantial measurement errors due to the mismatches in the income and expenditure data explained in Section~\ref{sec:sec33}.
The time-series plots of average income in Figure~\ref{fig:tssm_minc_mexp_raw_app} have a substantially large variance and fluctuate over time.
This is because some semi-monthly income is assigned to the wrong semi-month cells.
In these wrong cells, the measured incomes take unnaturally greater values than usual.
Moreover, the time-series plots of average expenditure do not chase those of average income.
For example, the average expenditure in the second half of June greatly differs from the average income in the same cell.
Consequently, the average net income (i.e., income minus expenditure) in each semi-month takes unnaturally large positive values in a few cells (Figure~\ref{fig:tssm_mexpdef_raw_app}).

The shifted series defines the days from the 2nd to the 16th of each month as the first half and the days from the 17th to the 1st of the following month as the second half.
Although the shifted series can partly manage the measurement errors in income, it does not adjust the difference in the timing of paydays and expenses.
Figure~\ref{fig:tssm_minc_mexp_shift} obtains a much smaller variance in the average incomes over time because all the paydays are now fixed to assign suitable semi-month cells.
However, the expenditure series still does not chase the income series well because the payday is placed on the \textit{latter} edge in each semi-month cell.
In other words, the average expenditure in each semi-month reflects a large part of the consumption based on the income of the previous semi-month cell.
Since the average incomes take a better series than Figure~\ref{fig:tssm_minc_mexp_raw_app}, the average net income offers better time-series plots (Figure~\ref{fig:tssm_mexpdef_shift}).
However, it still has a few cells with large negative and large values due to the deviations of expenditure from income.

The adjusted series uses the days from the 12th to 26th of each month as the ``first half'' and those from the 27th to 11th of the following month as the ``second half.''
By shifting the entire semi-month cells, this adjusted series can manage the measurement errors from the payday shifts and can fix the miss-matching in the timing of income and expenditure.
In other words, each payday is assigned to a suitable semi-month cell, and each semi-month cell reflects the household's decision on consumption in response to the latest income.
Figure~\ref{fig:tssm_minc_mexp_app} shows similar time-series plots of average income in Figure~\ref{fig:tssm_minc_mexp_shift}, in which has a corrected semi-monthly average income series only.
Notably, the expenditure series are now chasing the income series in Figure~\ref{fig:tssm_minc_mexp_app} because each semi-month cell's expenditure is based on the income in the same cell.
Accordingly, Figure~\ref{fig:tssm_mexpdef_app} now shows a smoother series with a smaller variance in the net income.

These illustrations support the evidence that the adjusted semi-monthly datasets can offer cleaner within variations in income and expenditure because they can fix the measurement errors and reflect the households' consumption strategies based on their latest incomes measured in the same semi-month cell.

\subsection{Testing Attenuation Bias in the Unadjusted Panel Dataset}\label{sec:secb_ab}

I assess the validity of the aggregation method used in this study.
Adjusting the aggregate data to account for the timing of paydays reveals more transparent relationships between income and expenditure than the unadjusted data (Section~\ref{sec:sec33}).
To demonstrate this mechanism, I listed the estimates from the unadjusted monthly panel dataset in Column (2) of Table~\ref{tab:cs_baseline_ab}.
The estimate for total consumption is $0.26$, which is much lower than the estimate for the adjusted series ($0.43$) in Column (1).
In addition, most subcategories show systematically smaller estimates than those from the adjusted data listed in Column (2).
For example, the estimate for the food category is now $0.10$ and statistically insignificant.
This means that measurement errors in the unadjusted aggregation lead to strong attenuation bias, which results in false negatives (i.e., false rejections of the null of full risk-sharing hypothesis) in most of the consumption categories.
I, therefore, specified a generating process of systematic measurement errors in the aggregate survey data, which has attracted attention in the macroeconomics literature (Cochrane 1991; Dynarski et al. 1997; Gervais and Klein 2010; Nelson 1994).
Moreover, several subcategories select values close to one or large negative values, which are no longer economically interpretable.\footnote{For example, the estimate for housing was negative. The households paid a fixed rent amount monthly. Given that the rent payday was fixed, the shifts in wage paydays might create cells with less income with a fixed amount of rent. In the case of seasonal consumption categories, the shifts may coincide with the timing of seasonal spending. The THBS households tended to purchase work clothes in March to prepare for the start of the new fiscal year in April. They also spent more on clothing in June, likely to buy summer clothes. The shift of paydays to these months may have created a strong but false positive correlation.}
This means that the miss-assignment of paydays in the unadjusted data may cause critical statistical inference issues in a few subcategories.

\def\arraystretch{0.95}
\begin{table}[htb]
\begin{center}
\caption{Testing Attenuation Bias in the Unadjusted Panel Dataset}
\label{tab:cs_baseline_ab}
\scriptsize
\scalebox{1.0}[1]{
\begin{tabular}{lrlcrlc}
\toprule[1pt]\midrule[0.3pt]
&\multicolumn{3}{c}{(1) Adjusted Monthly Panels}&\multicolumn{3}{c}{(2) Unadjusted Monthly Panels}\\
\cmidrule(rll){2-4}\cmidrule(rll){5-7}
&\multicolumn{2}{c}{Disposable income}&\multirow{2}{*}{Obs.}&\multicolumn{2}{c}{Disposable income}&\multirow{2}{*}{Obs.}\\
\cmidrule(rl){2-3}\cmidrule(rl){5-6}
&Coef.&Std. error&&Coef.&Std. error\\\hline
Total consumption 						&0.433&[0.059]***	&123	&0.261	&[0.054]***	&141\\
\hspace{10pt}Food						&0.293&[0.066]***	&114	&0.100	&[0.072]		&132\\
\hspace{10pt}Housing					&0.013&[0.084]		&105	&-0.263	&[0.167]		&124\\
\hspace{10pt}Utilities						&0.343&[0.236]		&110	&0.197	&[0.334]		&130\\
\hspace{10pt}Furniture					&-0.114&[0.706]	&98	&0.250	&[0.339]		&112\\
\hspace{10pt}Clothes					&0.347&[0.424]		&112	&0.806	&[0.262]***	&132\\
\hspace{10pt}Education					&0.162&[0.128]		&110	&0.069	&[0.146]		&128\\
\hspace{10pt}Medical expenses			&0.194&[0.120]		&114	&-0.128	&[0.094]		&132\\
\hspace{10pt}Entertainment expenses		&0.716&[0.181]***	&113	&0.194	&[0.153]		&132\\
\hspace{10pt}Transportation				&0.605&[0.264]**	&97	&0.362	&[0.291]		&115\\
\hspace{10pt}Gifts						&0.469&[0.197]**	&113	&0.622	&[0.306]*		&130\\
\hspace{10pt}Miscellaneous				&0.522&[0.252]**	&110	&0.246	&[0.267]		&126\\\midrule[0.3pt]\bottomrule[1pt]
\end{tabular}
}
{\scriptsize
\begin{minipage}{355pt}
\setstretch{0.85}
***, **, and * denote statistical significance at the 1\%, 5\%, and 10\% levels, respectively.
Standard errors in brackets are clustered at the household level.\\
Notes: 
This table shows the results of equation~\ref{eqn:eq_cs}.
The regressions of the 11 measures of log-transformed consumption on log-transformed disposable income, the family size control, household-fixed effect, and time-fixed effect are listed.
Columns 1 and 2 show the results for the adjusted monthly and unadjusted monthly panel datasets, respectively.
The estimates in column (1) are the same as those reported in Table~\ref{tab:cs_baseline}.
The estimated coefficients on log-transformed disposable income are listed in the columns named ``Coef.''.
\end{minipage}
}
\end{center}
\end{table}

\subsection{Considering Child Share as a Preference Shock Variable}\label{sec:secb_child}

As explained in the main text, the child share variable exhibits more within-variations than that of the family size.
Table~\ref{tab:cs_baseline_child} presents the results from specifications that include the share of children under the primary school age range, specifically those aged 0--5, as an alternative preference shock variable.
Panels A and B show the results for the consumption smoothing and risk-coping strategy regressions, respectively.
The results are materially similar to the main results (Tables~\ref{tab:cs_baseline} and~\ref{tab:risk}).
Note that the results remain robust to different definitions of age bins (not reported).

\def\arraystretch{0.90}
\begin{table}[htbp]
\begin{center}
\centering
\captionsetup{justification=centering}
\caption{Results for the Alternative Specification: \\Conditioning on the Share of Children Aged 0--5}
\label{tab:cs_baseline_child}
\scriptsize
\scalebox{1.0}[1]{
\begin{tabular}{lrlcrlcrlc}
\toprule[1pt]\midrule[0.3pt]
&\multicolumn{3}{c}{(1) Semi-monthly Panels}&\multicolumn{3}{c}{(2) Monthly Panels}\\
\cmidrule(rll){2-4}\cmidrule(rll){5-7}
&\multicolumn{2}{c}{Disposable income}&\multirow{2}{*}{Obs.}&\multicolumn{2}{c}{Disposable income}&\multirow{2}{*}{Obs.}\\
\cmidrule(rl){2-3}\cmidrule(rl){5-6}
\textbf{Panel A: Consumption Smoothing}&Coef.&Std. error&&Coef.&Std. error&\\\hline
Total consumption 						&0.361&[0.034]***	&278	&0.440&[0.058]***	&123	\\
\hspace{10pt}Food						&0.130&[0.028]***	&259	&0.277&[0.062]***	&114	\\
\hspace{10pt}Housing					&0.097&[0.413]		&134	&0.015&[0.079]		&105	\\
\hspace{10pt}Utilities						&-0.005&[0.140]	&223	&0.339&[0.232]		&110	\\
\hspace{10pt}Furniture					&0.431&[0.234]*	&169	&0.172&[0.665]		&98	\\
\hspace{10pt}Clothes					&0.381&[0.189]*	&238	&0.319&[0.412]		&112	\\
\hspace{10pt}Education					&0.103&[0.050]**	&236	&0.166&[0.127]		&110	\\
\hspace{10pt}Medical expenses			&0.037&[0.066]		&258	&0.182&[0.122]		&114	\\
\hspace{10pt}Entertainment expenses		&0.304&[0.085]***	&250	&0.695&[0.176]***	&113	\\
\hspace{10pt}Transportation				&0.281&[0.110]**	&187	&0.613&[0.267]**	&97	\\
\hspace{10pt}Gifts						&0.524&[0.107]***	&234	&0.476&[0.192]**	&113	\\
\hspace{10pt}Miscellaneous				&-0.079&[0.148]	&214	&0.506&[0.253]*	&110	\\\hline

&&&&&&\\
&\multicolumn{3}{c}{(1) Semi-monthly Panels}&\multicolumn{3}{c}{(2) Monthly Panels}\\
\cmidrule(rll){2-4}\cmidrule(rll){5-7}
&\multicolumn{2}{c}{Head's earning}&\multirow{2}{*}{Obs.}&\multicolumn{2}{c}{Head's earning}&\multirow{2}{*}{Obs.}\\
\cmidrule(rl){2-3}\cmidrule(rl){5-6}
\textbf{Panel B: Risk-coping Strategy} &Coef.&Std. error&&Coef.&Std. error&\\ \hline
\multicolumn{7}{l}{Panel B-1: Savings, insurance, borrowing, and gifts (yen)}\\
Net savings					&-0.043	&[0.018]**		&289  	&-0.035	&[0.028]	&124	\\
Net insurance					&0.001	&[0.013]		&289  	&0.002	&[0.032]	&124	\\
Net borrowing					&-0.021	&[0.027]		&289  	&-0.026	&[0.017]	&124	\\
Net credit purchase				&-0.129	&[0.039]***	&289 	 &-0.021	&[0.042]	&124	\\
Net gifts						&-0.052	&[0.020]**		&289 	 &-0.006	&[0.029]	&124	\\
&&&&&&\\
\multicolumn{7}{l}{Panel B-2: Labor supply adjustments \& sales of miscellaneous assets (yen)}\\
Other members' earnings			&0.001	&[0.010]		&289		&0.036	&[0.033]	&124	\\
Sales of miscellaneous assets		&-0.000	&[0.000]		&289		&0.000	&[0.001]	&124	\\

\midrule[0.3pt]\bottomrule[1pt]
\end{tabular}
}
{\scriptsize
\begin{minipage}{380pt}
\setstretch{0.85}
***, **, and * denote statistical significance at the 1\%, 5\%, and 10\% levels, respectively.
Standard errors in brackets are clustered at the household level.\\
Notes: 
\textbf{Panel A}: This table presents the results of equation~\ref{eqn:eq_cs}.
It reports regressions of 11 measures of log-transformed consumption on log-transformed disposable income, the family size control, the household fixed effect, and the time fixed effect.
Columns 1 and 2 show the results for the adjusted semi-monthly and adjusted monthly panel datasets, respectively.
The estimated coefficients on log-transformed disposable income are shown in the columns labeled ``Coef.''
\textbf{Panel B}: This table shows the results of equation~\ref{eqn:eq_risk}.
The proportion of children aged 0--5, household fixed effects, and time-fixed effects are included in all the regressions.
Columns 1 and 2 show the estimates for the adjusted semi-monthly and adjusted monthly panel datasets, respectively.
The estimated coefficients on the head's earnings are listed in the ``Coef.'' columns.
Panel B-1: 
Each net income variable is defined as income minus expenditure.
``Net savings'': withdrawal amount minus deposits to savings.
``Net insurance'': insurance payments received minus the expenses paid for insurance.
``Net borrowing'': the amount borrowed minus debt payments.
This category includes net income from both pawnshop and other lending institutions.
``Net credit purchase'' refers to the amount of credit purchases minus credit redeemed.
``Net gifts'': total gifts received, including both pecuniary and non-pecuniary, minus any payments for gifts.
Panel B-2: ``Other family members' earnings'': total income earned by all family members except the household head.
``Sales of miscellaneous assets'': include the sale of goods such as old newspapers and empty bottles.
All dependent variables are in yen.
\end{minipage}
}
\end{center}
\end{table}

\subsection{Distribution of the Idiosyncratic Shocks}\label{sec:secb_is}

\begin{figure}[htbp]
\centering
\captionsetup{justification=centering,margin=1.5cm}
\includegraphics[width=0.45\textwidth]{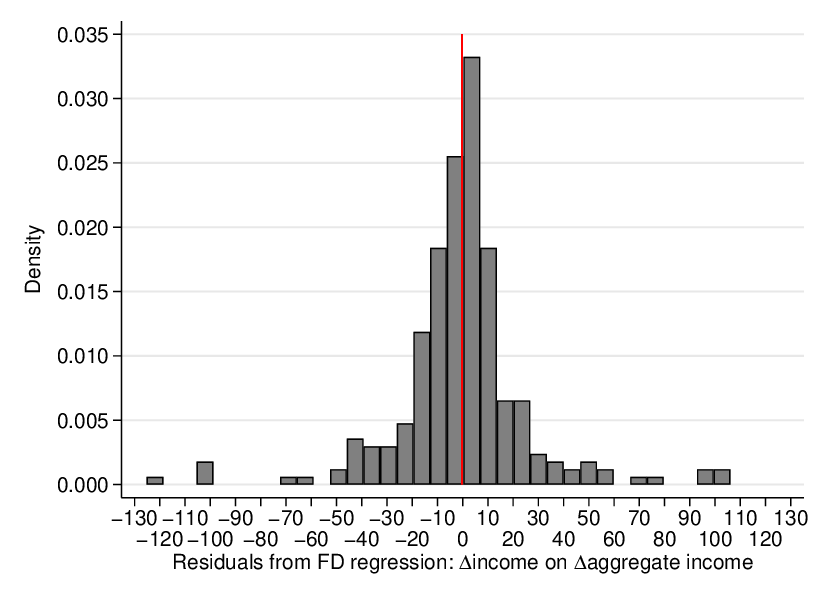}
\caption{Distribution of the Idiosyncratic Shocks}
\label{fig:hist_is}
\scriptsize{\begin{minipage}{450pt}
\setstretch{0.85}
Notes: 
This figure shows the distribution of idiosyncratic shocks in the adjusted semi-monthly THBS panel dataset.
The idiosyncratic shocks are estimated as the residuals from the regression of the first difference in income on the first difference in aggregate income.
The red line indicates the mean of the idiosyncratic shocks ($-0.235$).
Source: 
Created by the author using the THBS sample.
\end{minipage}}
\end{figure}

Figure~\ref{fig:hist_is} illustrates the distribution of the idiosyncratic shocks predicted using the semi-monthly panel dataset.\footnote{I use the semi-monthly panel dataset to overview the entire distribution of the idiosyncratic shocks rather than the daily panel datasets. This is because the daily panels include a set of consecutive household-date cells with no income. These cells cannot be used to predict the idiosyncratic shocks as the residuals are economically meaningfully, and thus it is practically impossible to illustrate the idiosyncratic shocks based on the daily dataset.}
The idiosyncratic shocks are calculated as residuals from the regression in the first difference in semi-monthly income on the first difference in aggregate semi-monthly income.
Let $n_{t}$ be the number of households in semi-month $t$ in the year 1919.
For household $i$ in semi-month $t$ cell, the residual is then defined as follows:
\begin{equation}\label{eqn:is}
\Delta \hat{\upsilon}_{i,t} = \Delta \textit{Income}_{i,t} - \hat{\alpha} \Delta \overline{\textit{Income}}_{.,t},
\end{equation}
where $\overline{\textit{Income}}_{.,t} = \sum_{i=1}^{n_{t}}\frac{\textit{Income}_{i,t}}{n_{t}}$ and $\hat{\alpha}$ is the estimated coefficient.
Figure~\ref{fig:hist_is} shows that the residuals satisfy the zero-average property.
However, this figure also indicates that the residuals have rich variations, supporting that the THBS households had experienced both favorable and adverse idiosyncratic shocks during the survey periods.

\subsection{Unit Root and Autocorrelations}\label{sec:secb_st_sc}

Table~\ref{tab:ts_test} presents the results of the Fisher-type panel unit root tests for the total expenditure in the semi-monthly (Column 1) and monthly (Column 2) panel datasets.
Both columns show that the null hypothesis of unit roots in all the panels is rejected at the conventional level.
The results of the unit root tests for consumption subcategories and variables used to test the risk-coping mechanisms are not reported in this table because those are materially similar results.
In Section~\ref{sec:sec4}, I used the cluster-robust variance-covariance estimator in the regressions to allow arbitrary autocorrelations (Arellano 1987).
Despite this, I confirmed that the null hypothesis of no serial correlation is not rejected at the conventional level in Wooldridge's autocorrelation test for almost of the consumption and net income variables (Wooldridge 2002).

\def\arraystretch{0.95}
\begin{table}[htb]
\begin{center}
\captionsetup{justification=centering}
\caption{Results of the Unit Root Tests
\label{tab:ts_test}
}
\footnotesize
\scalebox{1.0}[1]{
\begin{tabular}{lrr}
\toprule[1pt]\midrule[0.3pt]
Test statistics					&(1) Semi-Monthly Panels 	&(2) Monthly Panels	\\\hline
$P$-statistic $p$-value			&0.0000			&0.0001		\\
$Z$-statistic $p$-value			&0.0000			&0.0000		\\
$L^{*}$-statistic $p$-value			&0.0000			&0.0000		\\
$P_{m}$-statistic $p$-value		&0.0000			&0.0000		\\\midrule[0.3pt]\bottomrule[1pt]
\end{tabular}
}
{\scriptsize
\begin{minipage}{450pt}
\setstretch{0.85}
Notes: The results of the Fisher-type panel unit root tests based on augmented Dickey-Fuller (ADF) tests are reported in this table.
The null hypothesis is that all the panels contain unit roots, whereas the alternative hypothesis is that at least one panel is stationary.
In all the specifications, the process under the null hypothesis is assumed to be a random walk with drift. 
The demeaned data are used to address the effect of cross-sectional dependence. 
The number of lagged differences in the ADF regression equation reported is set as one because including two or more two lags reduces a large part of the panel units available for computing the test statistics due to the lack of balance of the panels.
I confirm that including two lags in the adjusted semi-monthly panels does not change the results.
Including two lags in the adjusted monthly panels and three lags in the adjusted semi-monthly panels is impossible.
See Choi (2001) for the details of the tests.
These tests need both the first-differenced dependent variable and the lagged dependent variable.
Therefore, the units with less than four time-series observations are excluded in the regressions.\\
Source: Calculated by the author using the THBS sample.
\end{minipage}
}
\end{center}
\end{table}

\subsection{Risk Preferences}\label{sec:secb_preference}

The potential influences of heterogeneous risk preference on standard risk-sharing regressions are discussed in a few works (Schulhofer-Wohl 2011; Mozzoco and Saini 2012).
In my empirical setting, the heterogeneity in the risk preference is assumed to be removed by using the household fixed-effect.
When the risk preferences are differenced over households like $\sigma = \sigma_{i}$, equation~\ref{eqn:cs_level} becomes:
\begin{equation}\label{eqn:cs_level_hetero}
\log c_{i,t}=\log c_{t}^{a} + \frac{1}{\sigma_{i}}(\log \omega_{i}-\omega^{a})+(\theta_{i,t} - \theta_{t}^{a}).
\end{equation}
The first differencing removes the second term with the coefficient of relative risk aversion in the right-hand side of this equation, even though it varies over the households.

Schulhofer-Wohl (2011) shows a specific case (utility function) that the heterogeneous risk preference may lead to bias in the estimation of the income elasticity under the standard risk-sharing regressions and proposes a parametric test for that heterogeneity.
This test is valid under a particular utility function.
Specifically, the bias could be raised in the case that the coefficient of relative risk aversion interacts with the aggregate shock in the first-order condition for consumption because the interaction means that the aggregate shock cannot be removed using the time-fixed effect in the risk-sharing regressions.\footnote{Specifically, Schulhofer-Wohl (2011, pp.~928--929) considers the utility maximization problem under the following weighted sum of discounted expected utilities:
\begin{equation}\label{eqn:sw_ump}
\sum_{i} \dot{\alpha}_{i} E \left[\sum_{t}\dot{\beta}^{t} \frac{[c_{i,t}(s_{t})]^{1-\dot{\gamma}_{i}}}{1-\dot{\gamma}_{i}}\right],
\end{equation}
where $\dot{\alpha}_{i}$ is Parate weight and $\dot{\gamma}_{i}$ is the coefficient of relative risk aversion for household $i$. The log-transformed first-order condition becomes:
\begin{equation}\label{eqn:sw_ump_foc}
\text{log} c_{i,t} = \frac{\text{log} \dot{\alpha}_{i}}{\dot{\gamma}_{i}} - \frac{\text{log} \dot{\lambda}_{t}}{\dot{\gamma}_{i}},
\end{equation}
where $\dot{\lambda}_{t}$ indicates the Lagrangian multiplier. This result implies that the effects of the aggregate shocks ($\dot{\lambda}_{t}$) depend on the household's attitudes toward risk and that the first-differencing does not remove the second term in equation~\ref{eqn:sw_ump_foc}.}
As explained, this setting differs from the one used in this current study.
Mozzoco and Saini (2012) consider a more general case for testing the homogeneity of risk preferences.
Their test requires a panel dataset with a very long time dimension that is not applicable to my THBS dataset.\footnote{The data used in their study covers the household-level monthly panel dataset surveyed in rural India between 1975 and 1985. The households with fewer than 80 data points are not included in the homogeneity tests (Mozzoco and Saini 2012, p.~453). See also Shrinivas and Fafchamps (2018) and Mozzoco and Saini (2018) for the subsequent discussions.}

An alternative way to test potential heterogeneity in the risk preference is to use direct measures on the risk preferences of workers.
Schulhofer-Wohl (2011) uses the results of experimental risk-tolerance questions on the workers' risk preferences in the Health and Retirement Study to test whether the correlations between earnings and aggregate shocks differ across high- and low-risk-tolerance workers.\footnote{This question is asking if the respondents would be willing to take a risk to get a new job with a higher income, even if there was a 50-50 chance of a reduction in income by a third (Schulhofer-Wohl 2011, P.~931).}
They found that the high-risk-tolerance workers have more procyclical incomes than the low-risk-tolerance workers.

Since the THBS has no experimental questions, it is impossible to apply the same test.
Given this limitation, I compare the correlations between the heads' income and aggregate consumption over sub-groups to obtain insights into the potential difference in the heads' attitudes toward risks.
For household $i$ and semi-month $t$, the specification is characterized as follows:
\begin{equation}\label{eqn:hetero_pref}
\log \tilde{y}_{i,t}= \xi_{1} \log c_{t}^{a} +  \xi_{2} d_{i} \log c_{t}^{a} + o x_{i,t} + \iota_{i} + \varepsilon_{i,t},
\end{equation}
where $\tilde{y}_{i,t}$ is head's earnings, $c_{t}^{a}$ is aggregate consumption, $x_{i,t}$ is the household size, $\iota_{i}$ is the household fixed-effect, and $\varepsilon_{i,t}$ is a random error term.
The aggregate consumption is calculated as the cross-sectional average of the total expenditure among the THBS sample, given that the population mean of consumption expenditures among all the skilled factory workers' households in Tsukishima is unavailable.

$d_{i}$ is an indicator variable that takes one if a household $i$ belongs to a specific group that might have had different preferences on their jobs than the other heads in the sample.
Firstly, I test the hypothesis that the heads who work in the large-scale factories have different risk preferences.
Although there are no specific anecdotes behind this, the workers in the large-scale factories might have had a lower risk tolerance than those in the small- and medium-scale enterprises.
Thirteen heads who worked in the Ishikawajima Shipyard and governmental factory are classified into this group.
Secondly, I test the hypothesis that the heads who work in the smithing and the non-machinery factories have different risk preferences.
As described in Online Appendix~\ref{sec:seca_factory}, skilled workers in the smithing sector could be small business owners.\footnote{Workers in the canning factories might have had a different preference on the risk preferences (Online Appendix~\ref{sec:seca_work}). However, my THBS sample does not include the workers in canning factories.}
The workers in non-machinery factories might also have different job preferences than the machinery factory workers.
This group has only three heads: one worked in a smithing factory, and two worked in food and paper-making factories.

\def\arraystretch{0.95}
\begin{table}[h!]
\begin{center}
\captionsetup{justification=centering}
\caption{Results for Testing Heterogeneous Risk Preferences
\label{tab:hetero_pref}
}
\scriptsize
\scalebox{1.0}[1]{
\begin{tabular}{lrrrr}
\toprule[1pt]\midrule[0.3pt]
&\multicolumn{4}{c}{DV: ln(head's earnings)}\\
\cmidrule(rrrr){2-5}
&\multicolumn{4}{c}{Analytical Period}\\
\cmidrule(rrrr){2-5}
&(1) Full		&(2) Full 	&(3) Jan. to Sept.&(4) Jan. to Sept.\\\hline
Aggregate Consumption			&0.277*	&0.273	&0.361**	&0.362*	\\
							&[0.158]	&[0.198]	&[0.164]	&[0.198]	\\
Aggregate Consumption 			&-0.041	&		&-0.061	&		\\
$\times$ Large Scale Factory		&[0.453]	&		&[0.415]	&		\\
Aggregate Consumption 			&		&-0.089	&		&-0.178	\\
$\times$ Smithing/Non-machinery	&		&[0.215]	&		&[0.215]	\\\hline
Observations					&261		&261		&239		&239		\\\midrule[0.3pt]\bottomrule[1pt]
\end{tabular}
}
{\scriptsize
\begin{minipage}{340pt}
\setstretch{0.85}
***, **, and * denote statistical significance at the 1\%, 5\%, and 10\% levels, respectively.
Standard errors in brackets are clustered at the household level.\\
Notes:
This table shows the results of equation~\ref{eqn:hetero_pref} for the adjusted semi-monthly panel dataset.
`Aggregate Consumption' indicates the cross-sectional sample average of the total expenditures.
`Large Scale Factory' indicates an indicator variable that takes one if the head worked in the Ishikawajima Shipyard or governmental factory.
`Smithing/Non-machinery' indicates an indicator variable that takes one if the head worked in the smithing, food, or paper-making factories.
All the regressions include the family size variable and household fixed effect.
Columns 1--2 show the results for the entire sample period.
Columns 3--4 show the results for the period between January and September 1919 to remove year-month cells with the smaller (cross-sectional) number of households (Online Appendix Figure~\ref{fig:panel_structure}).
\end{minipage}
}
\end{center}
\end{table}

Table~\ref{tab:hetero_pref} presents the results.
In Column (1), the estimated coefficient on the interaction term is close to zero and statistically insignificant.
Thus, the sensitivities to aggregate consumption are similar across different scales of factories.
Column (2) further shows that the estimated coefficient on the interaction term is not statistically significant under an alternative proxy of risk tolerance.
It suggests that the heads who work in the smithing and non-machinery factories show similar responses to the aggregate shock.
Since the aggregate consumption is the cross-sectional sample average, the aggregate variable may contain noise in the period with the small number of cross-sectional observations.
To manage this potential issue, I exclude October and November 1919 in Columns (3) and (4).
As shown, the estimated coefficients on the interaction terms are largely unchanged in both columns.

This finding aligns with the similarity of these household heads, as the THBS targeted skilled factory workers within a geographically narrow area.
Therefore, it is plausible that the THBS households had similar attitudes toward the risks.

\clearpage
\section{Empirical Analysis Appendix}\label{sec:secc}
\setcounter{figure}{0} \renewcommand{\thefigure}{C.\arabic{figure}}
\setcounter{table}{0} \renewcommand{\thetable}{C.\arabic{table}}

\subsection{Conceptual Framework}\label{sec:secc0}

Consider the economy where household $i = 1,...,N$ receives an uncertain income at time $t$ depending on the state of the world $s_{t} \in S_{t}$ as $y_{i,t}(s_{t})$.
For infinite time horizon, the state $s_{t}$ realizes with probability $\pi(s_{t})\in [0, 1]$.
The weighted sum of the expected lifetime utility of $N$ households in the economy is expressed as:
\begin{equation}\label{eqn:of}
\sum_{i=1}^N\omega_{i}\sum_{t=0}^\infty \rho^t\sum_{s_{t}\in S_{t}} \pi(s_{t})u(c_{i,t}(s_{t}), \theta_{i,t}(s_{t})), 
\end{equation}
where $\omega_{i}$ is the social planner's weight representing the reciprocal of the marginal utility of each household ($0<\omega_{i}<1$ and $\sum_{i=1}^{N} \omega_{i}=1$), $0<\rho^t<1$ is the discount factor, $c_{i,t}(s_{t})$ is consumption, and $\theta_{i,t}(s_{t})$ is a preference shock.
For all states $s_{t}$, the aggregate resource constraint is given by:
\begin{equation}\label{eqn:rc}
\sum_{i=1}^N c_{i,t}(s_{t})=\sum_{i=1}^N e_{i,t}(s_{t}),
\end{equation}
where $e_{i,t}(s_{t})$ is household $i$'s exogenous endowment at time $t$.

Following Mace (1991), I assume that the aggregate idiosyncratic shocks approach zero as $N$ gets large.
Consider the situation in which the household's endowment can be decomposed into the permanent component ($e_{i,t}^{P}$), shock linked to macroeconomic trends ($\varrho_{i,t}^{a} (s_{t})$), and idiosyncratic shock ($\varsigma_{i,t} (s_{t})$) as:
\begin{equation}\label{eqn:edm}
e_{i,t}(s_{t}) = e_{i,t}^{P} + \varrho_{i,t}^{a} (s_{t}) + \varsigma_{i,t} (s_{t}).
\end{equation}
Given the random nature of the idiosyncratic shock ($\forall t, E[\varsigma_{i,t} (s_{t})|s_{t}]=0$), the sample average of the idiosyncratic shock satisfies $N^{-1} \sum_{i=1}^{N} \varsigma_{i,t}(s_{t}) \overset{p}{\to} 0$ as $N$ gets large ($\because$ WLLN).
This means that the aggregate endowment is written without the idiosyncratic shock as: $e_{t}(s_{t})^{a} = e_{t}^{P, a} + \varrho_{t}^{a} (s_{t})$.
Intuitively, this implies that the idiosyncratic shocks are shared among an economy with numerous households.

Subject to equation(\ref{eqn:rc}), the social planner maximizes equation(\ref{eqn:of}) by choosing an allocation of consumption across households.
The first-order condition at time $t$ is derived as follows:
\begin{equation}\label{eqn:foc}
\omega_{i}\rho^{t}\pi(s_{t})u_{c}(c_{i,t}(s_{t}), \theta_{i,t}(s_{t}))=\lambda(s_{t}),
\end{equation}
where $\lambda$ is the Lagrange multiplier for the single lifetime resource constraint.
To derive my empirical specification, I assume a constant relative risk aversion preference as:
\begin{equation}\label{eqn:uf}
u(c_{i,t}(s_{t}), \theta_{i,t}(s_{t})) = \frac{c_{i,t}^{1-\sigma}(s_{t})}{1-\sigma}\exp(\sigma \theta_{i,t}(s_{t})),
\end{equation}
where $\sigma > 0$ is the coefficient of relative risk aversion.
Below, I consider a particular history of the state of the world so that the subscript $t$ can represent dependence on the state (Hayashi et al. 1996).
This simplifies the notation in equation~\ref{eqn:foc} as:
\begin{equation}\label{eqn:foc_crra}
\omega_{i}\rho^{t}\pi_{t}c_{i,t}^{-\sigma}\text{exp}(\sigma \theta_{i,t})=\lambda_{t}.
\end{equation}
Taking the log of equation (\ref{eqn:foc_crra}) and aggregating over households, consumption for household $i$ is expressed as: 
\begin{equation}\label{eqn:cs_level}
\log c_{i,t}=\log c_{t}^{a} + \frac{1}{\sigma}(\log \omega_{i}-\omega^{a})+(\theta_{i,t} - \theta_{t}^{a}), 
\end{equation}
where
\begin{equation}\label{eqn:agg}
c_{t}^{a}=\exp \left( \frac{1}{N} \sum_{i=1}^N \log c_{i,t} \right),~~~~~\omega^a=\frac{1}{N} \sum_{i=1}^N \log \omega_{i},~~~~~\theta_{t}^{a}=\frac{1}{N} \sum_{i=1}^N \theta_{i,t}.
\end{equation}
The first-difference in equation (\ref{eqn:cs_level}) eliminates the household time-constant effect to yield:
\begin{equation}\label{eqn:cs_fd}
\log c_{i,t} - \log c_{i,t-1} = \log c_{t}^{a} - \log c_{t-1}^{a} + (\theta_{i,t} - \theta_{i,t-1}) + (\theta_{t}^{a}-\theta_{t-1}^{a}). 
\end{equation}
Individual income is a conventional proxy variable for the preference shock (Mace 1991).
Using the growth rate of household income $\log y_{i,t} - \log y_{i,t-1}$ as a proxy for idiosyncratic shock and adding a measurement error in consumption $\epsilon_{i,t}$, the empirical specification can be characterized as follows:
\begin{equation}\label{eqn:reg1}
\log c_{i,t} - \log c_{i,t-1}=\alpha_1 (\log c_{t}^{a} - \log c_{t-1}^{a})+\alpha_2 (\log y_{i,t} - \log y_{i,t-1})+\epsilon_{i,t}.
\end{equation}

The aggregate consumption growth in equation(\ref{eqn:reg1}) is assumed to reflect macroeconomic shocks in Tsukishima in the case of my empirical setting.
As discussed in Section~\ref{sec:sec34}, the consumption of the THBS households captures the overall macroeconomic trend.
Despite this, interpreting $\alpha_1$ is difficult because the THBS households include the factory workers' households in Tsukishima alone.
To manage this potential issue, I use a two-way fixed-effects model instead of the first-difference model, following Cochrane (1991).
Another potential issue is that $\epsilon_{i,t}$ may include household-specific time-varying preference shock, which might be correlated with idiosyncratic income shock.
Family size influences the household's preference because the marginal utility of consumption increases with family size (Jappelli and Pistaferri 2017).
As explained, family size was stable during the sample period in the THBS households, meaning that the preference shifts rarely occurred (Section~\ref{sec:sec34}).
To be conservative, however, I include the number of family members as a control variable for representing potential preference shock.

Considering these modifications, the empirical specification can finally be characterized as follows:
\begin{equation}\label{eqn:reg2}
\log c_{i,t} = \gamma \log y_{i,t} + \delta x_{i,t} + \mu_{i} + \phi_{t}+ u_{i,t},
\end{equation}
where $c_{i,t}$ is consumption, $y_{i,t}$ is disposable income, $x_{i,t}$ is family size, $\mu_{i}$ is the household fixed effect, $\phi_{t}$ is the time fixed effect, and $u_{i,t}$ is a random error term.
The estimate of $\gamma$ shall range from zero (i.e., idiosyncratic shocks are perfectly insured) to one (for the absence of insurance).\footnote{I use the conventional within estimator. The bias-corrected estimator for the two-way fixed-effects model in a difference-in-differences (TWFE-DID) design with continuous treatment over multiple periods remains underdeveloped (de Chaisemartin and D'Haultf\oe uille 2023). Therefore, previous studies have not yet incorporated recent theoretical developments in the TWFE-DID framework.}

To test the risk-coping mechanisms, I consider the following representation of individual consumption following Fufchamps and Lund (2003).
\begin{equation}\label{eqn:rs_set}
c_{i,t} = y_{i}^{P} + y_{i,t}^{T} + r_{i,t},
\end{equation}
where $ y_{i}^{P}$ is permanent income and $y_{i,t}^{T}$ is transitory income.
$r_{i,t}=\bm{z}'_{i,t} \bm{1}$ is a linear combination of the net temporary incomes expressed using a $k \times 1$ net income vector ($\bm{z}$) and a vector of ones.
Substituting equation~\ref{eqn:rs_set} into equation~\ref{eqn:cs_level} and taking the first-difference removes the time-constant components, including permanent income.
It follows that my empirical specification can be characterized as follows:
\begin{equation}\label{eqn:eq_risk_app}
r_{i,t} = \kappa \tilde{y}_{i,t} + \eta x_{i, t} + \nu_{i} + \zeta_{t} + \epsilon_{i,t},
\end{equation}
where $\tilde{y}_{i,t}$ is the head's earning, $x_{i, t}$ is the family size control, $\nu_{i}$ indicates the household-fixed effect, $\zeta_{t}$ indicates the time-fixed effect, and $\epsilon_{i,t}$ is a random error term.
I herein use the head's earnings ($\tilde{y}_{i,t}$) as the idiosyncratic shock variable instead of the disposable income.\footnote{I consider the labor-supply adjustment variable and sales of miscellaneous assets as the additional dependent variables in addition to the net income variables $r_{i,t}$ (Section~\ref{sec:sec42}). The former variable is the total earnings of the family members except for the head, meaning that the simultaneous issue arises if I use the disposable income in this equation.}
The regression is separately run for each component in the net income vector ($\bm{z}$).

\subsection{Raw Relationship between Income and Expenditures}\label{sec:secc1}

\begin{figure}[h!]
\centering
\captionsetup{justification=centering}
  \subfloat[Total]{\label{fig:hist_dexp_sm}\includegraphics[width=0.17\textwidth]{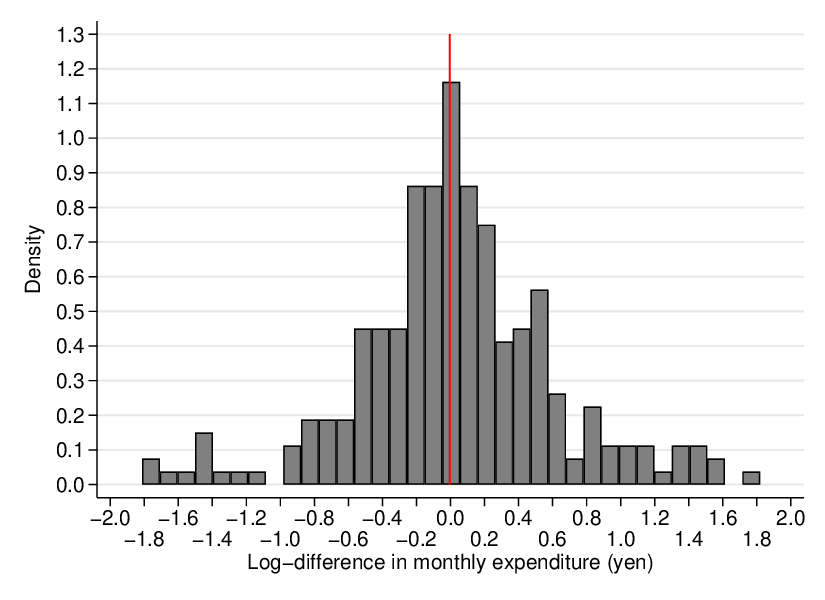}}
  \subfloat[Food]{\label{fig:hist_dfood_sm}\includegraphics[width=0.17\textwidth]{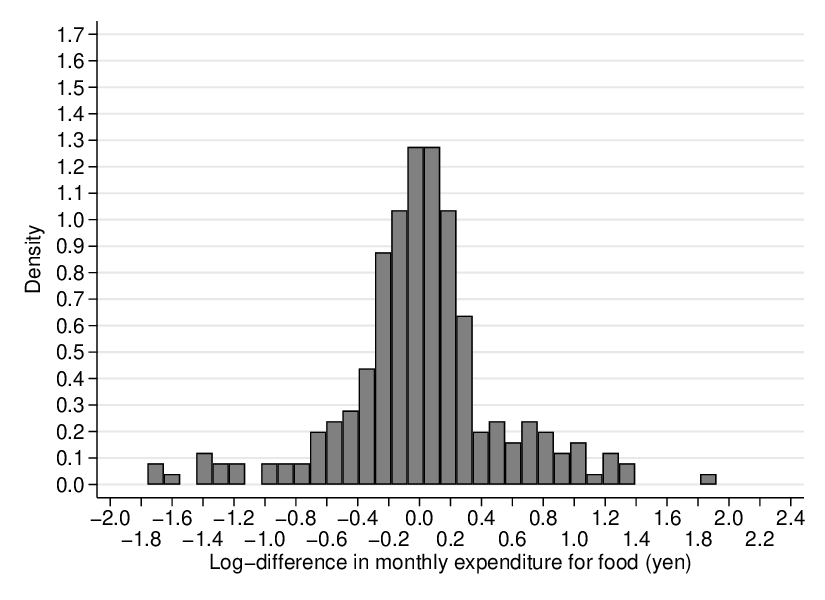}}
  \subfloat[Housing]{\label{fig:hist_dhousing_sm}\includegraphics[width=0.17\textwidth]{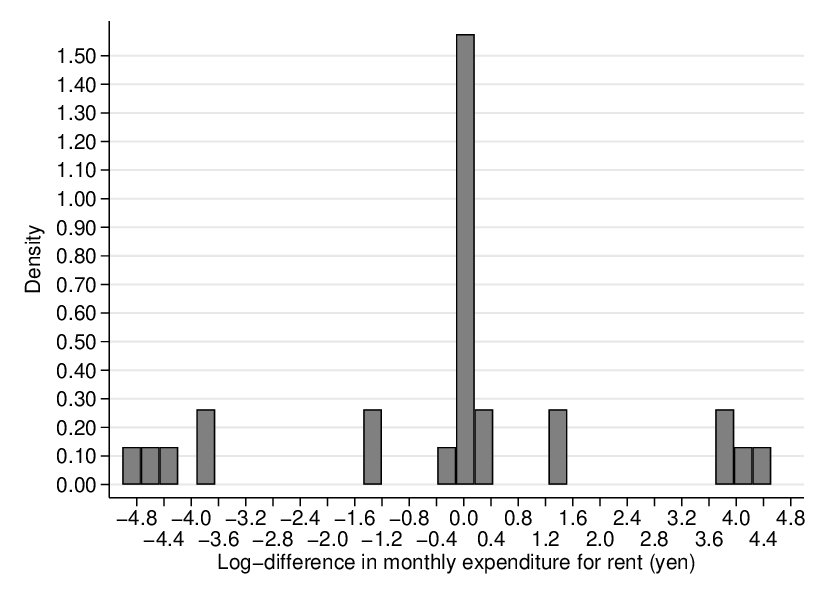}}
  \subfloat[Utilities]{\label{fig:hist_dutilities_sm}\includegraphics[width=0.17\textwidth]{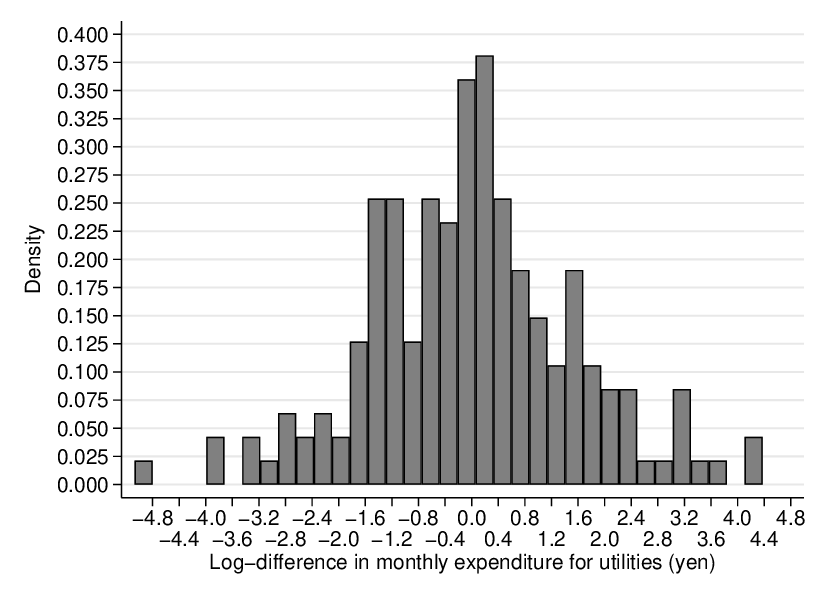}}
  \subfloat[Furniture]{\label{fig:hist_dfurniture_sm}\includegraphics[width=0.17\textwidth]{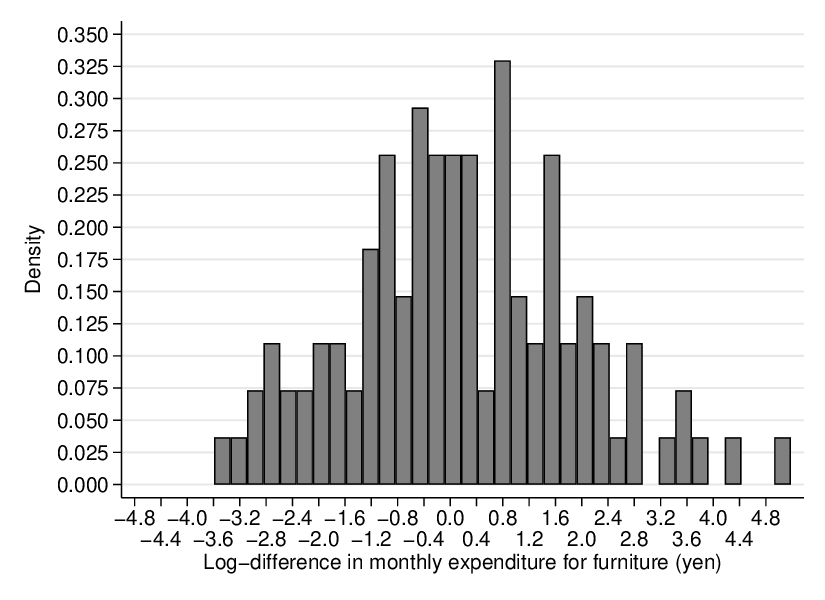}}
  \subfloat[Clothing]{\label{fig:hist_dclothing_sm}\includegraphics[width=0.17\textwidth]{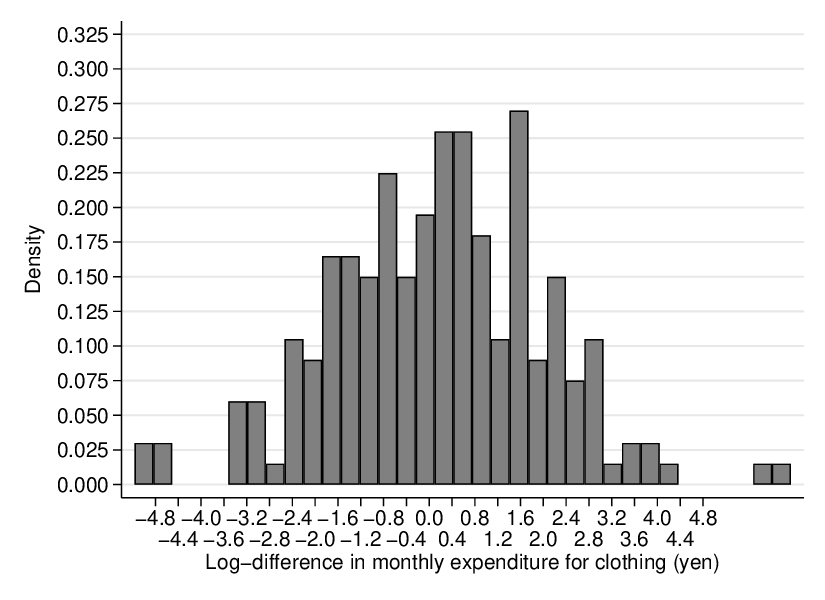}}\\
  \subfloat[Education]{\label{fig:hist_deducation_sm}\includegraphics[width=0.17\textwidth]{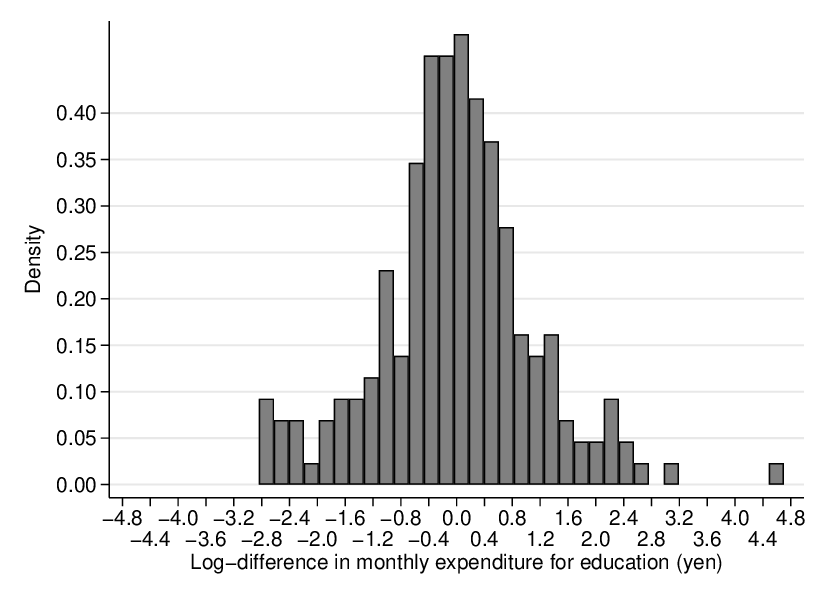}}
  \subfloat[Medical]{\label{fig:hist_dmedical_sm}\includegraphics[width=0.17\textwidth]{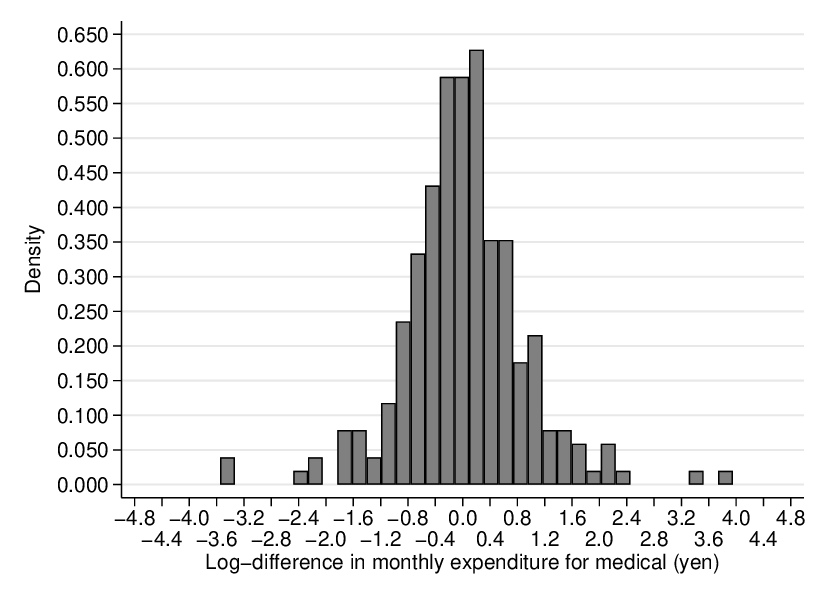}}
  \subfloat[Entertainment]{\label{fig:hist_dentertainment_sm}\includegraphics[width=0.17\textwidth]{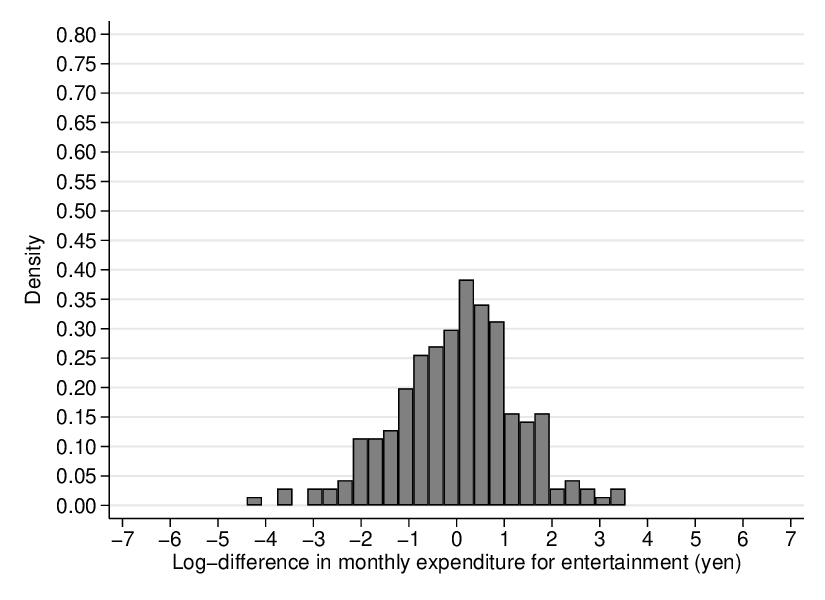}}
  \subfloat[Transportation]{\label{fig:hist_dtransportation_sm}\includegraphics[width=0.17\textwidth]{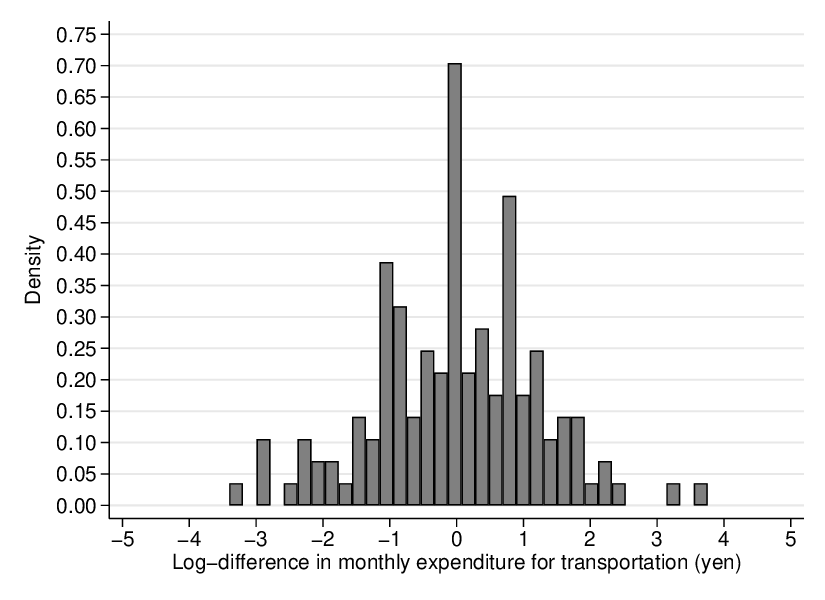}}
  \subfloat[Gifts]{\label{fig:hist_dgift_sm}\includegraphics[width=0.17\textwidth]{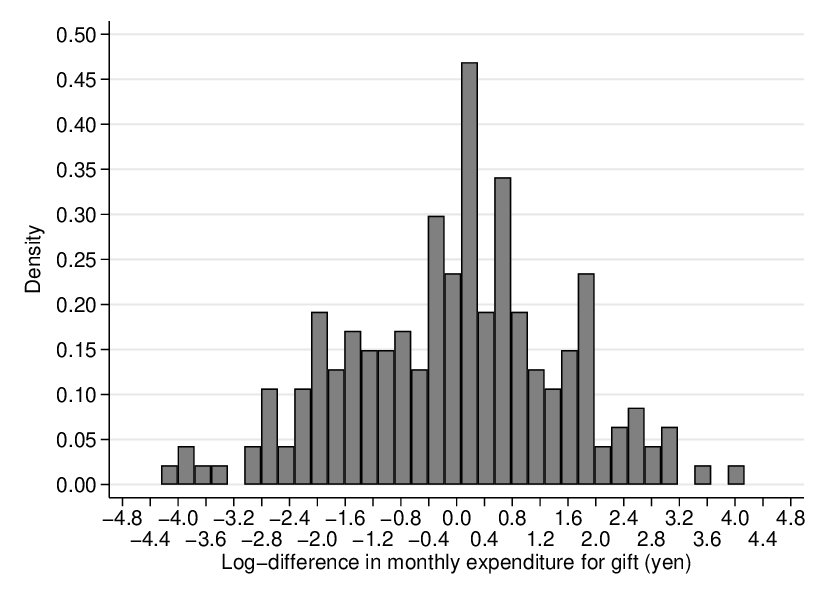}}
  \subfloat[Miscellaneous]{\label{fig:hist_dmiscellaneous_sm}\includegraphics[width=0.17\textwidth]{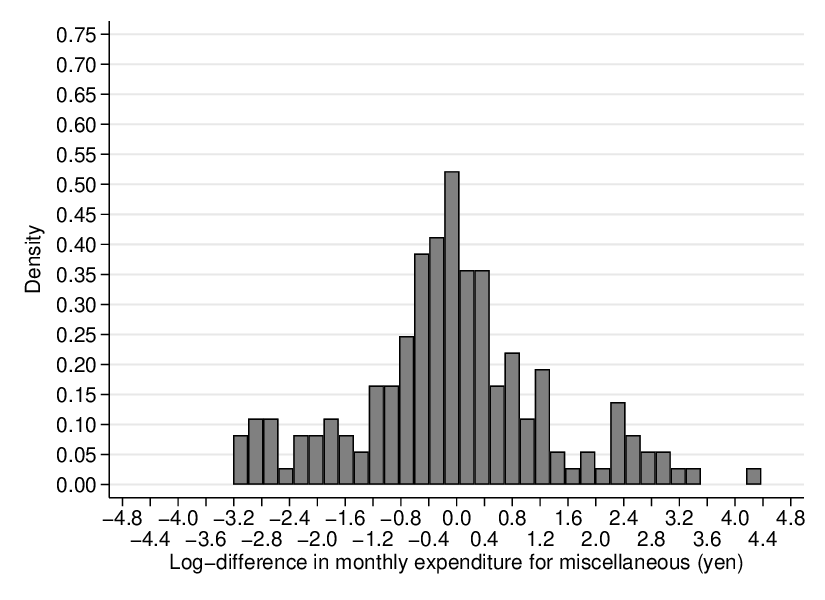}}
\caption{Distributions of the Log-differences of the Semi-monthly Expenditures}
\label{fig:hist_exp}
\scriptsize{\begin{minipage}{450pt}
\setstretch{0.85}
Notes:
The distribution of the log-difference in adjusted semi-monthly total expenditure is shown in Figure~\ref{fig:hist_dexp_sm}.
Figures~\ref{fig:hist_dfood_sm}--\ref{fig:hist_dmiscellaneous_sm} show the distributions of the log-differences in adjusted semi-monthly expenditure for the 11 subcategories listed in panel A of Table~\ref{tab:sum}.
Source: Created by the author using the THBS sample.
\end{minipage}}
\end{figure}
\begin{figure}[h!]
\centering
\captionsetup{justification=centering}
  \subfloat[Total]{\label{fig:scat_dexp_sm}\includegraphics[width=0.17\textwidth]{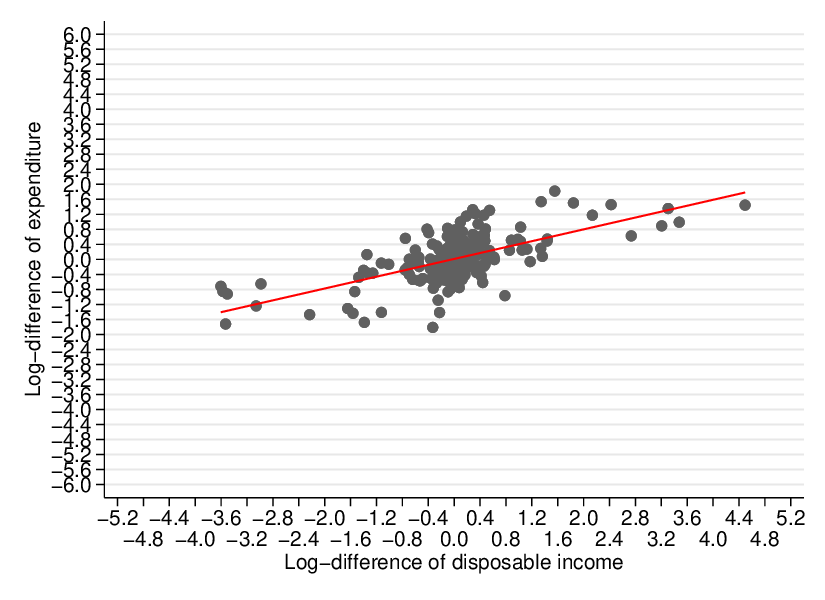}}
  \subfloat[Food]{\label{fig:scat_dfood_sm}\includegraphics[width=0.17\textwidth]{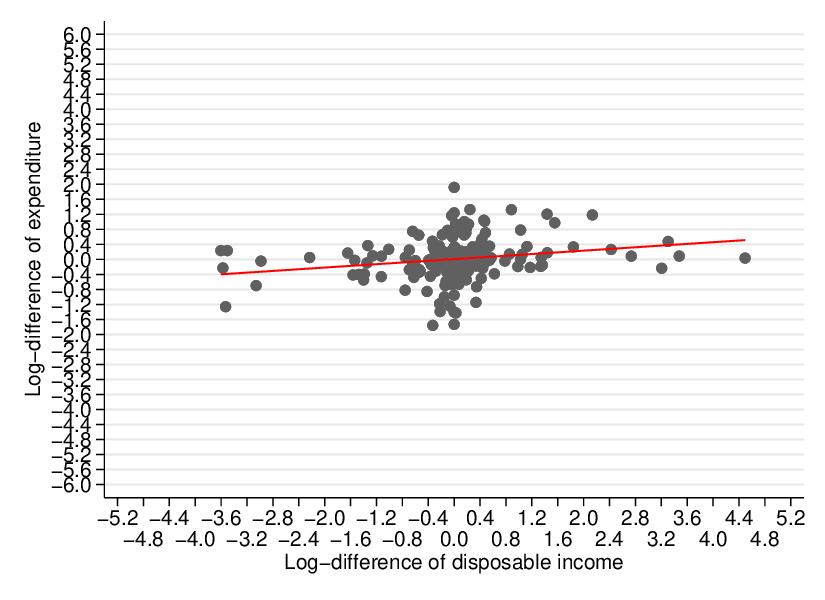}}
  \subfloat[Housing]{\label{fig:scat_dhousing_sm}\includegraphics[width=0.17\textwidth]{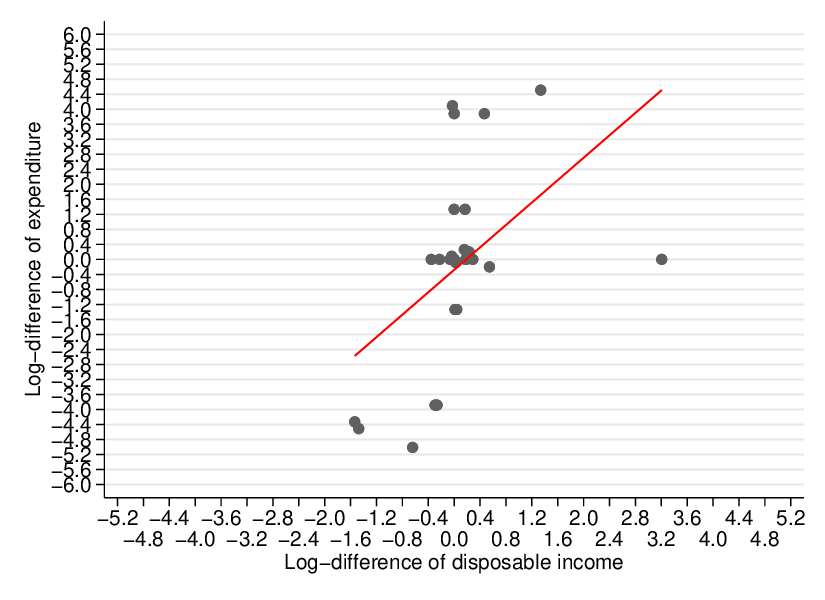}}
  \subfloat[Utilities]{\label{fig:scat_dutilities_sm}\includegraphics[width=0.17\textwidth]{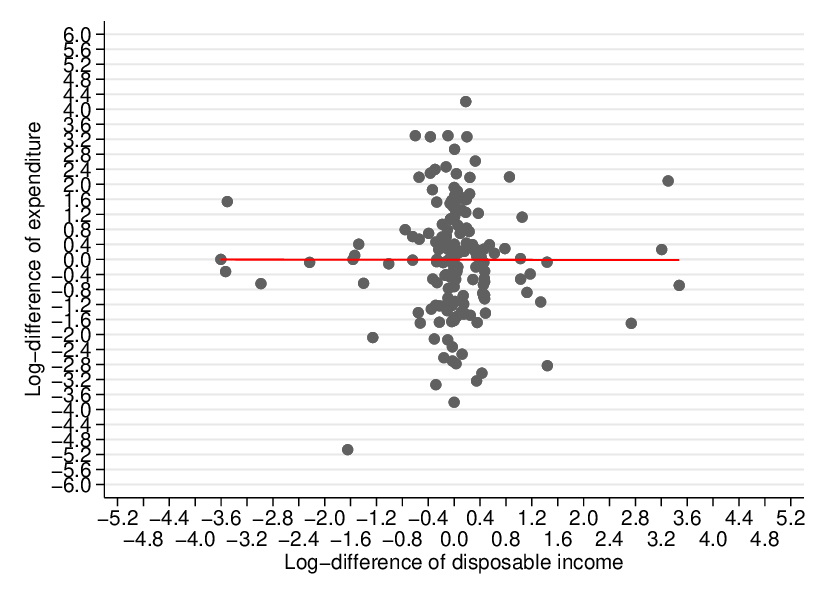}}
  \subfloat[Furniture]{\label{fig:scat_dfurniture_sm}\includegraphics[width=0.17\textwidth]{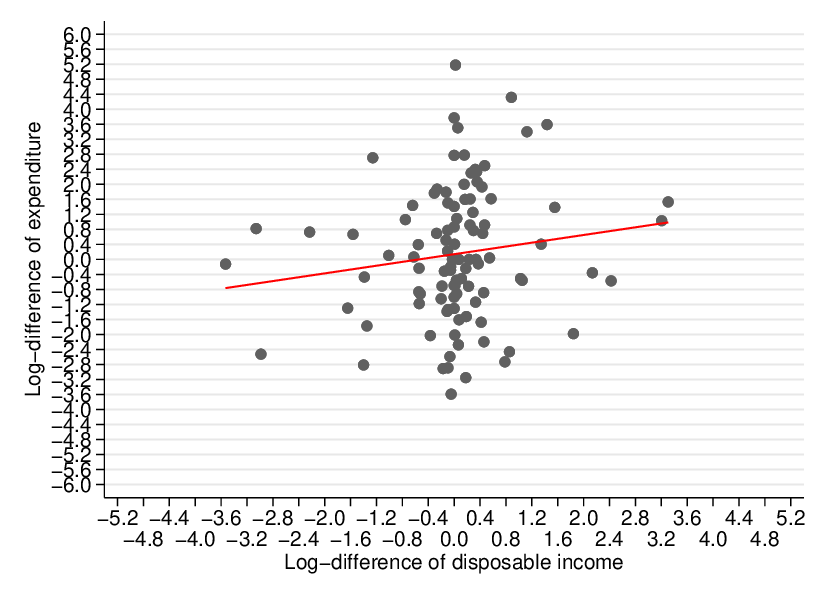}}
  \subfloat[Clothing]{\label{fig:scat_dclothing_sm}\includegraphics[width=0.17\textwidth]{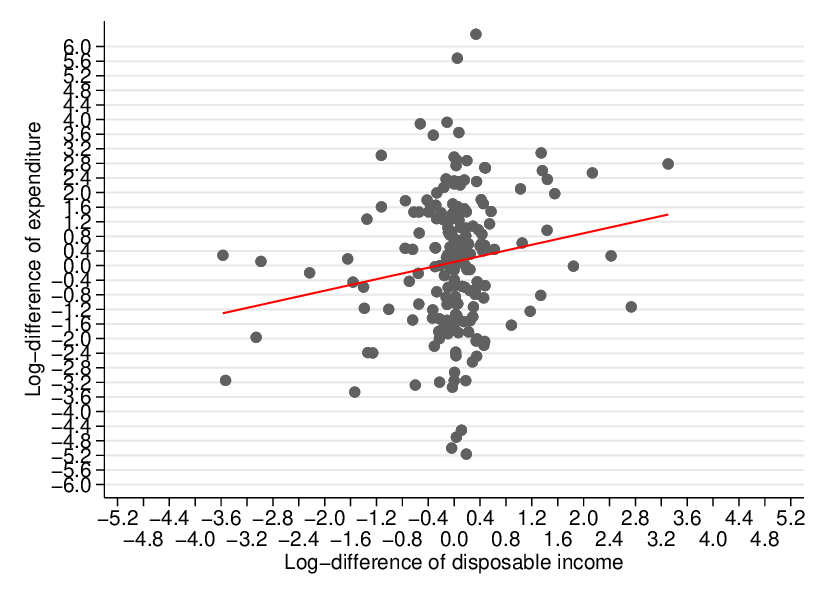}}\\
  \subfloat[Education]{\label{fig:scat_deducation_sm}\includegraphics[width=0.17\textwidth]{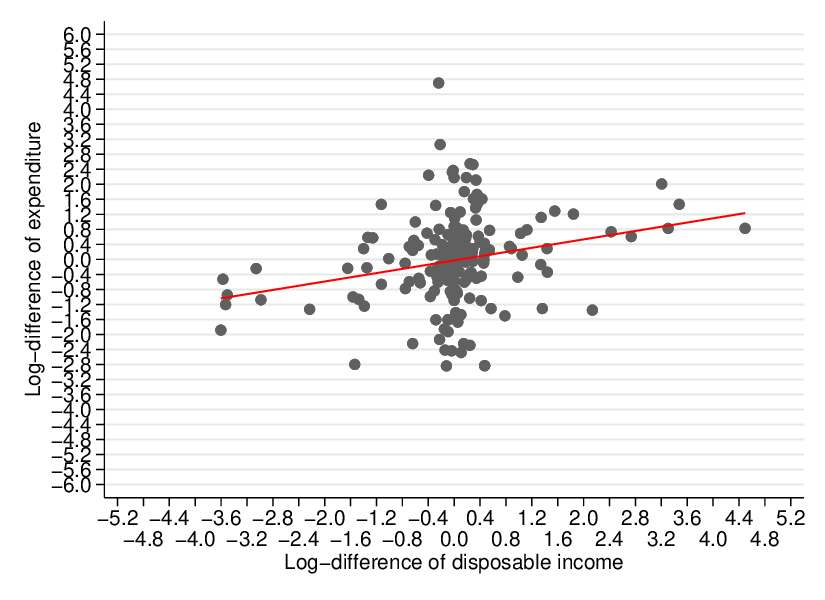}}
  \subfloat[Medical]{\label{fig:scat_dmedical_sm}\includegraphics[width=0.17\textwidth]{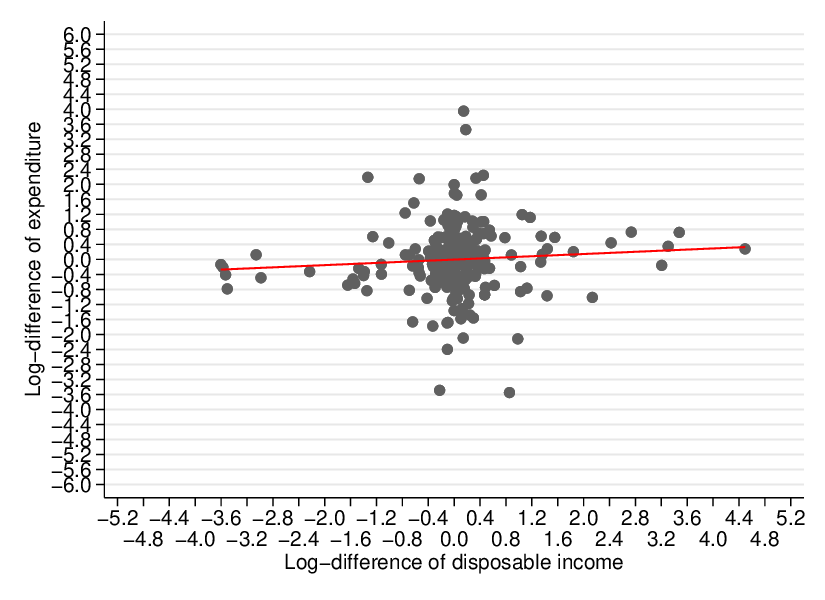}}
  \subfloat[Entertainment]{\label{fig:scat_dentertainment_sm}\includegraphics[width=0.17\textwidth]{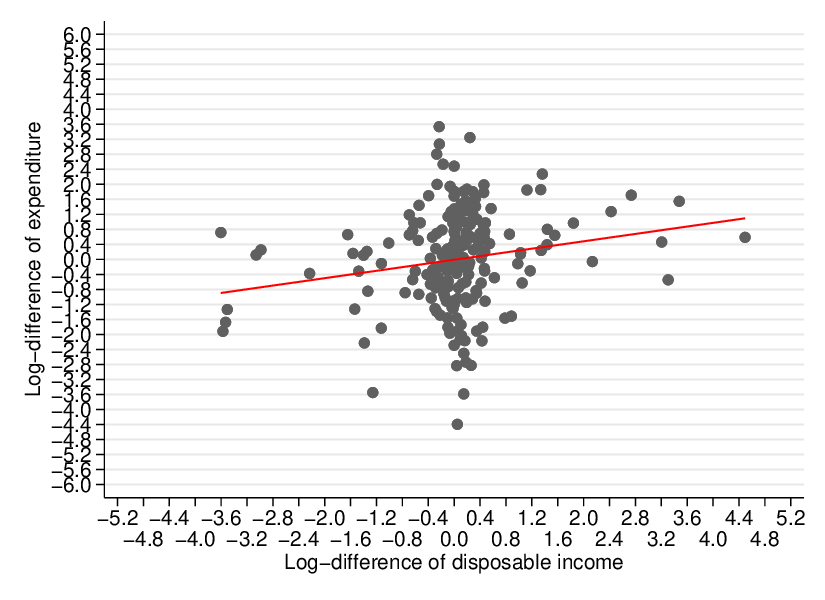}}
  \subfloat[Transportation]{\label{fig:scat_dtransportation_sm}\includegraphics[width=0.17\textwidth]{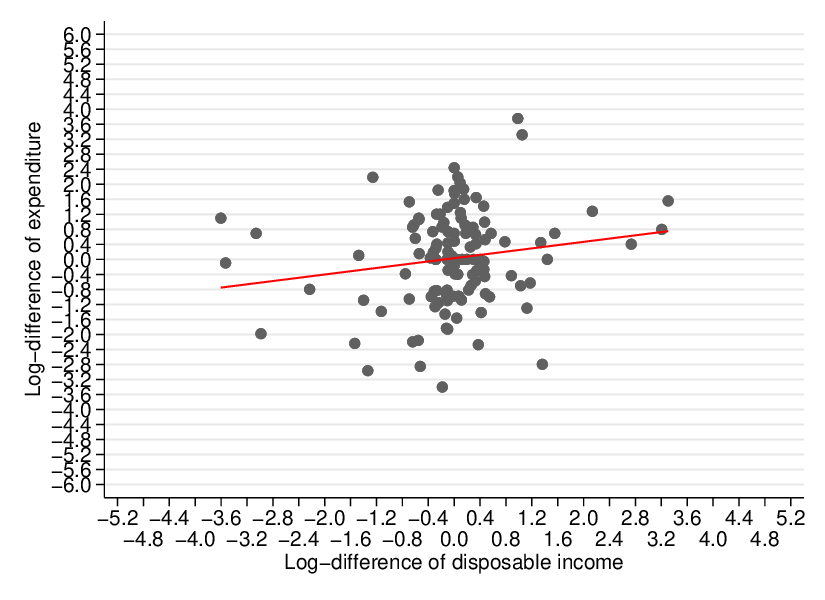}}
  \subfloat[Gifts]{\label{fig:scat_dgift_sm}\includegraphics[width=0.17\textwidth]{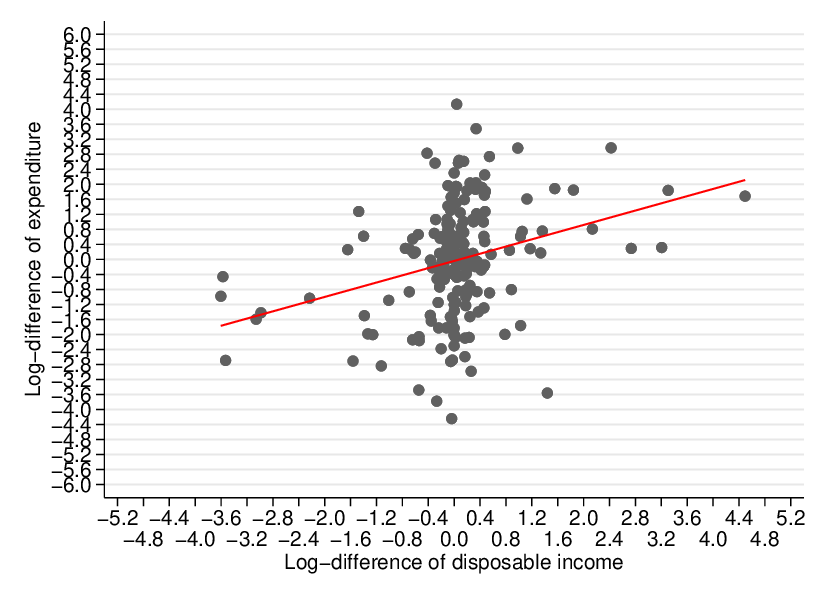}}
  \subfloat[Miscellaneous]{\label{fig:scat_dmiscellaneous_sm}\includegraphics[width=0.17\textwidth]{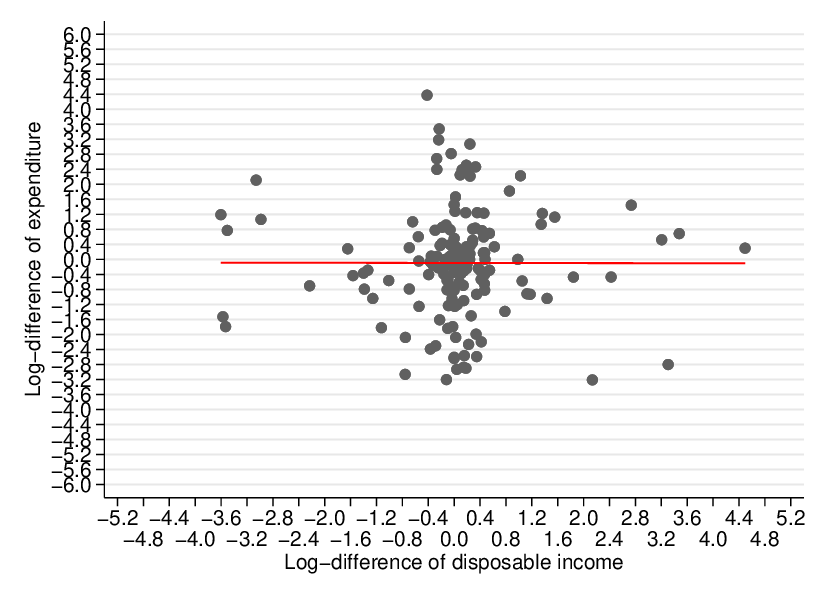}}
\caption{Raw Relationship between the Log-differences\\ in Semi-monthly Income and Expenditure}
\label{fig:scat_exp}
\scriptsize{\begin{minipage}{450pt}
\setstretch{0.85}
Notes:
Figure~\ref{fig:scat_dexp_sm} illustrates the relationship between the log-difference in adjusted semi-monthly disposable income and expenditure.
Figures~\ref{fig:scat_dfood_sm}--\ref{fig:scat_dmiscellaneous_sm} illustrate the relationships between the log-differences in adjusted semi-monthly disposable income and the 11 expenditure subcategories listed in Panel A in Table~\ref{tab:sum}. 
The range of the y-axis is fixed across all the figures for comparability.
Source: Created by the author using the THBS sample.
\end{minipage}}
\end{figure}
\begin{figure}[h!]
\centering
\captionsetup{justification=centering}
  \subfloat[Total]{\label{fig:hist_dexp_am}\includegraphics[width=0.17\textwidth]{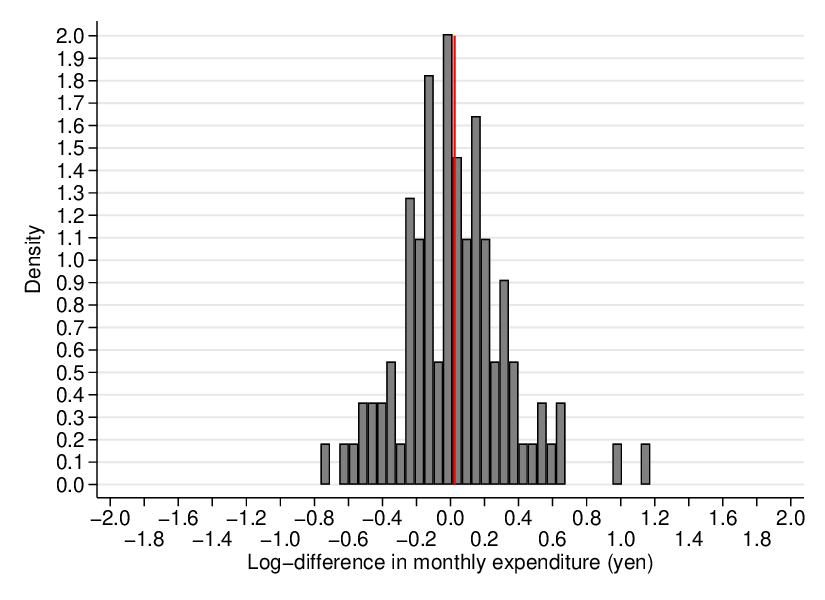}}
  \subfloat[Food]{\label{fig:hist_dfood_am}\includegraphics[width=0.17\textwidth]{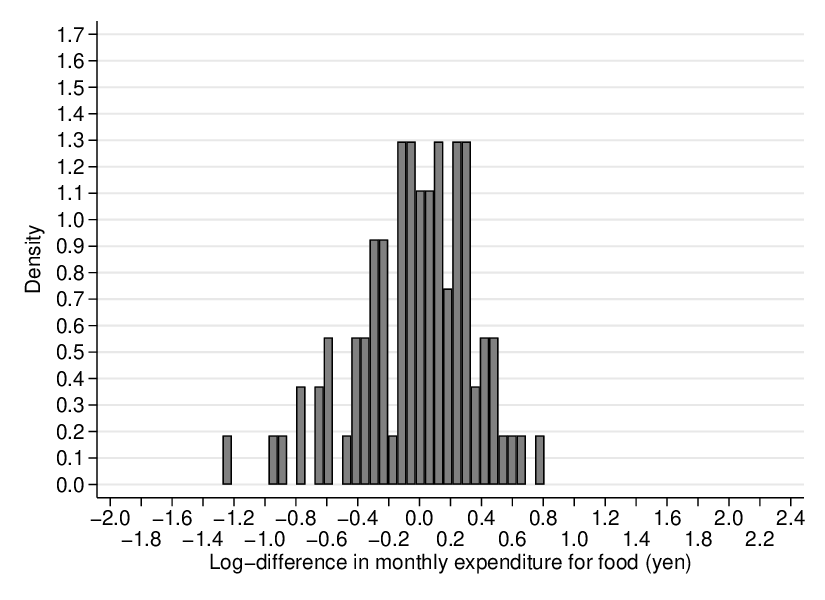}}
  \subfloat[Housing]{\label{fig:hist_dhousing_am}\includegraphics[width=0.17\textwidth]{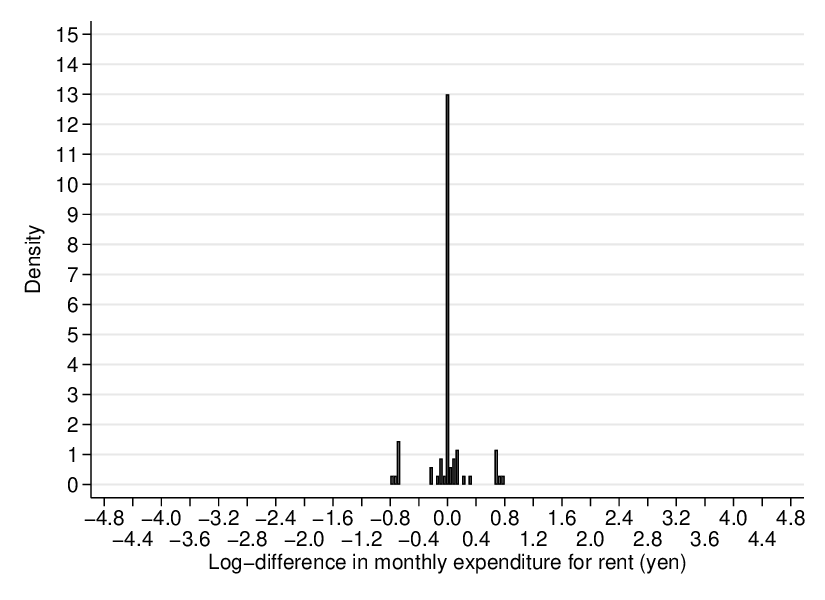}}
  \subfloat[Utilities]{\label{fig:hist_dutilities_am}\includegraphics[width=0.17\textwidth]{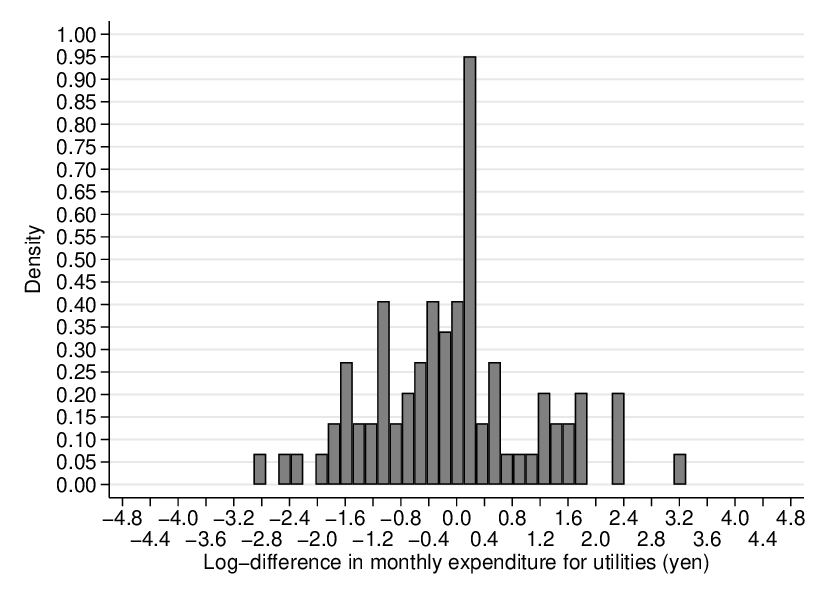}}
  \subfloat[Furniture]{\label{fig:hist_dfurniture_am}\includegraphics[width=0.17\textwidth]{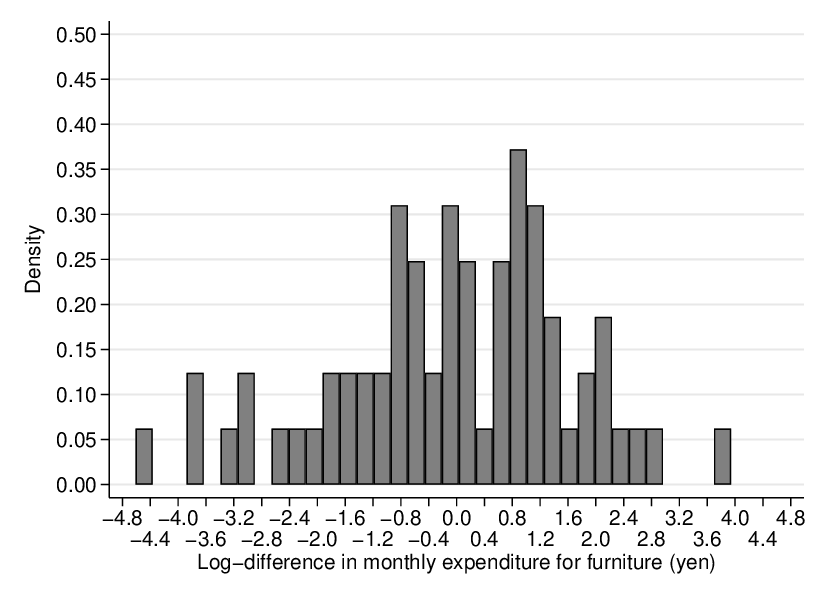}}
  \subfloat[Clothing]{\label{fig:hist_dclothing_am}\includegraphics[width=0.17\textwidth]{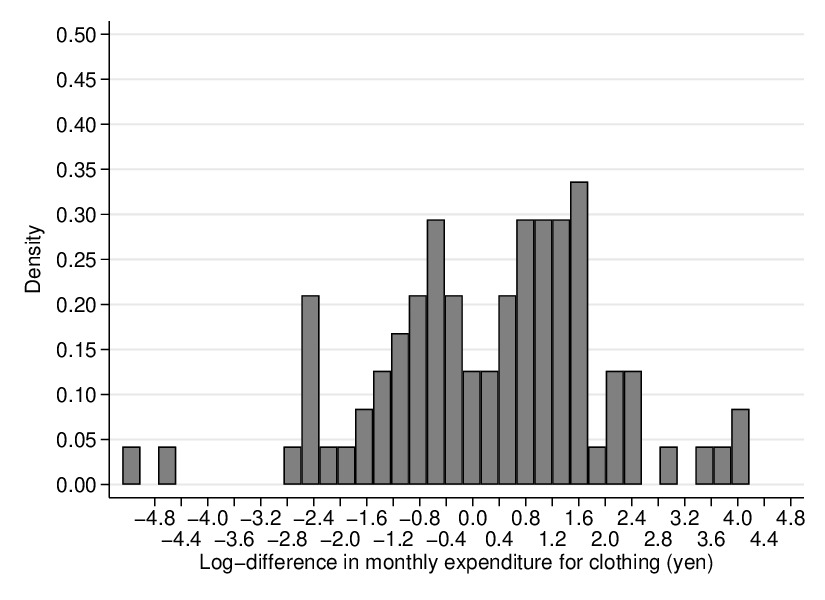}}\\
  \subfloat[Education]{\label{fig:hist_deducation_am}\includegraphics[width=0.17\textwidth]{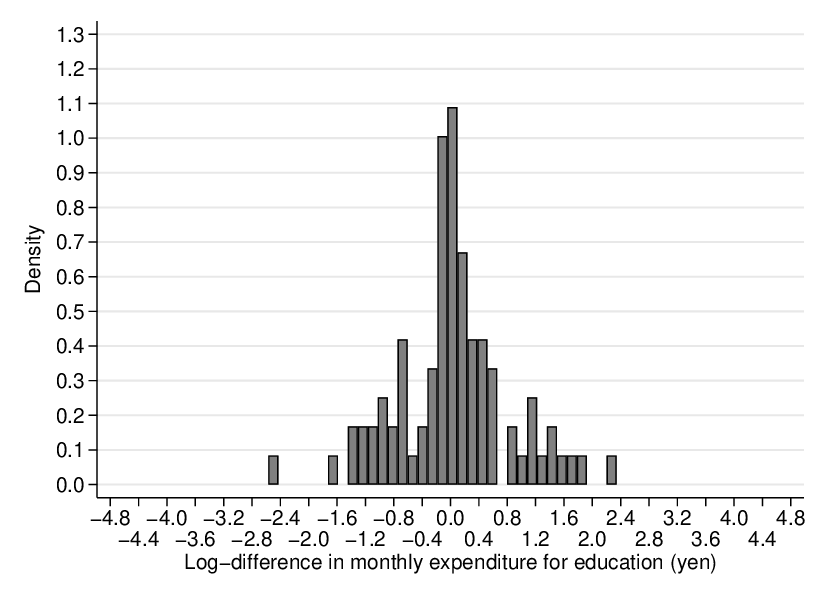}}
  \subfloat[Medical]{\label{fig:hist_dmedical_am}\includegraphics[width=0.17\textwidth]{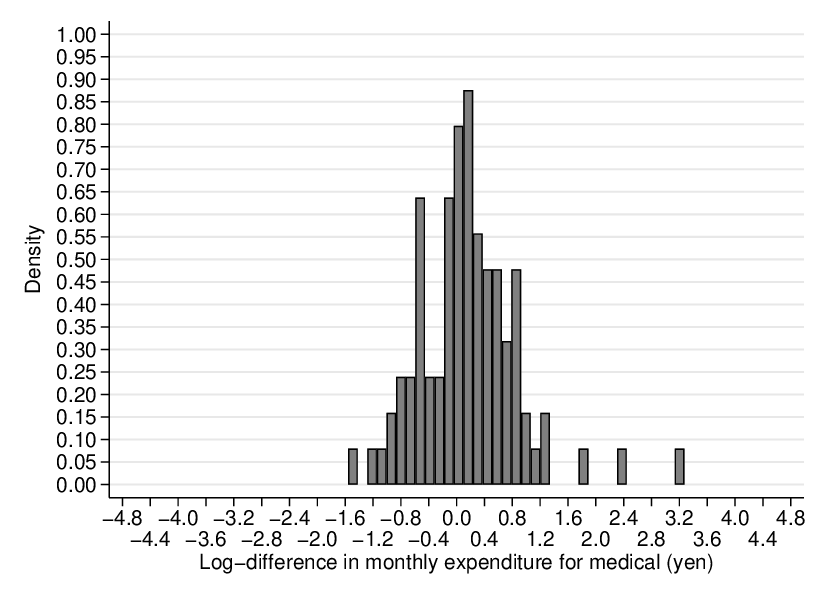}}
  \subfloat[Entertainment]{\label{fig:hist_dentertainment_am}\includegraphics[width=0.17\textwidth]{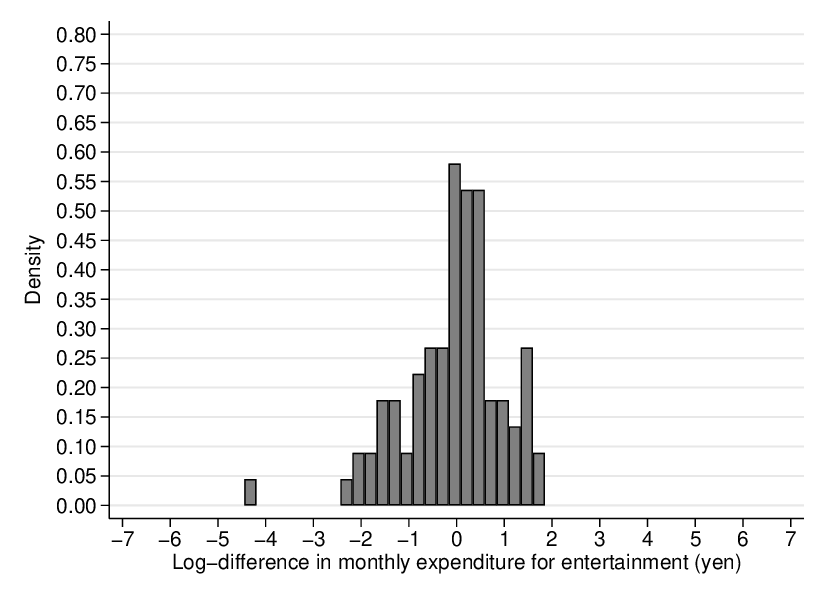}}
  \subfloat[Transportation]{\label{fig:hist_dtransportation_am}\includegraphics[width=0.17\textwidth]{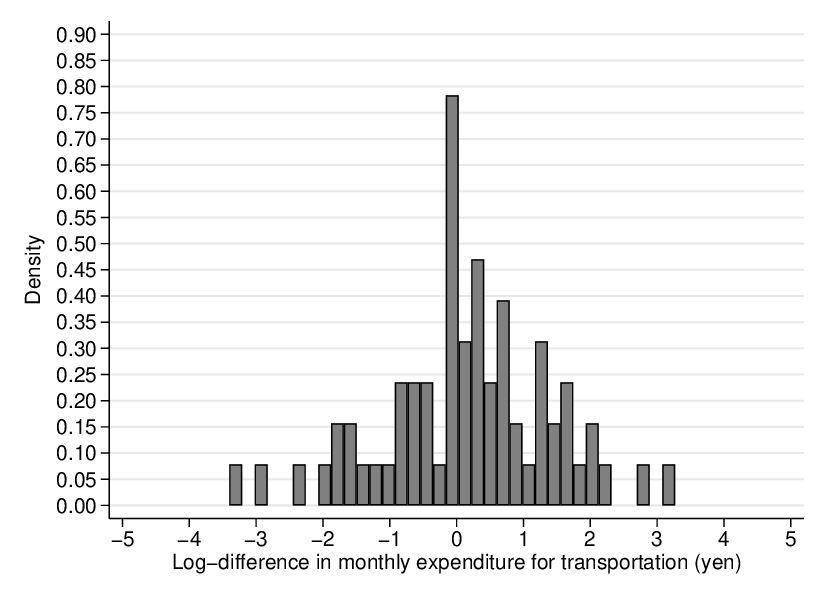}}
  \subfloat[Gifts]{\label{fig:hist_dgift_am}\includegraphics[width=0.17\textwidth]{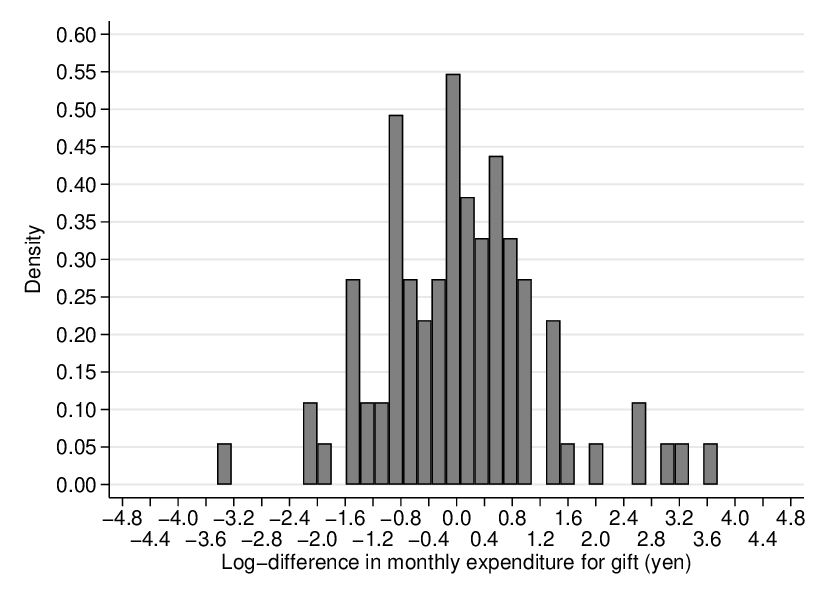}}
  \subfloat[Miscellaneous]{\label{fig:hist_dmiscellaneous_am}\includegraphics[width=0.17\textwidth]{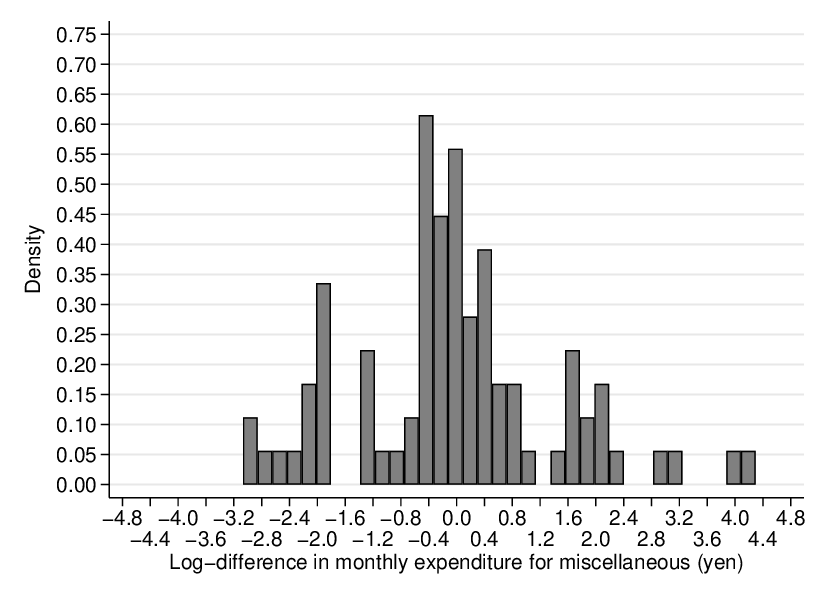}}
\caption{Distributions of the Log-differences of the Monthly Expenditures}
\label{fig:hist_exp_am}
\scriptsize{\begin{minipage}{450pt}
\setstretch{0.85}
Notes:
The distribution of the log-difference in adjusted monthly total expenditure is shown in Figure~\ref{fig:hist_dexp_am}.
Figures~\ref{fig:hist_dfood_am}--\ref{fig:hist_dmiscellaneous_am} show the distributions of the log-differences in adjusted monthly expenditure for the 11 subcategories listed in panel A of Table~\ref{tab:sum}.
Source: Created by the author using the THBS sample.
\end{minipage}}
\end{figure}
\begin{figure}[h!]
\centering
\captionsetup{justification=centering}
  \subfloat[Total]{\label{fig:scat_dexp_am}\includegraphics[width=0.17\textwidth]{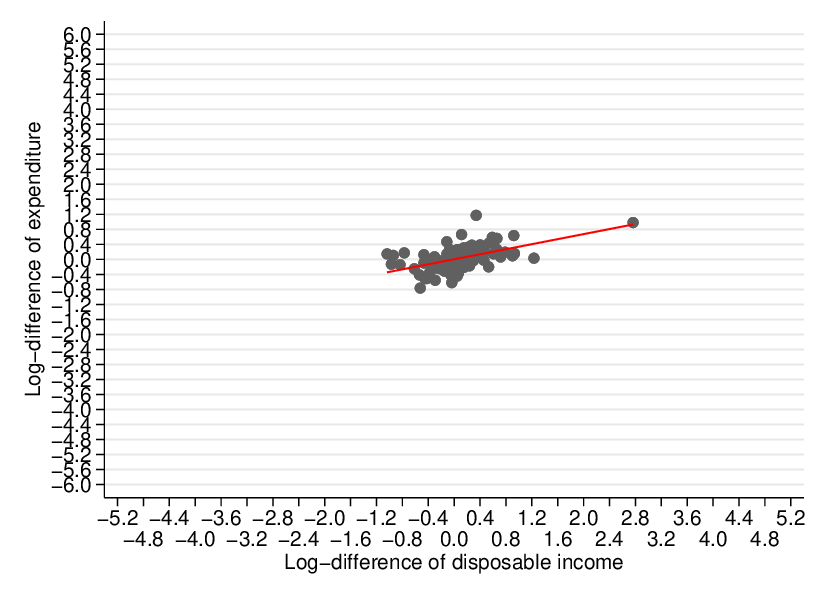}}
  \subfloat[Food]{\label{fig:scat_dfood_am}\includegraphics[width=0.17\textwidth]{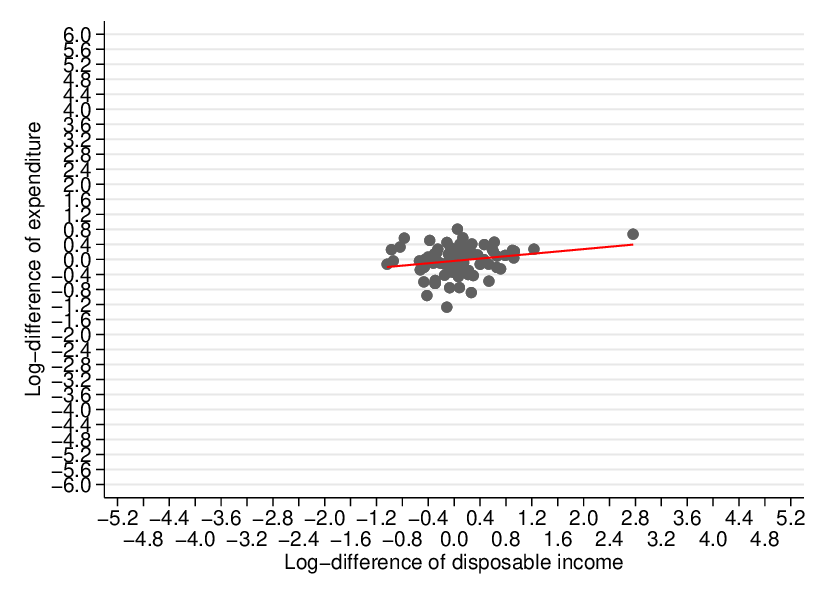}}
  \subfloat[Housing]{\label{fig:scat_dhousing_am}\includegraphics[width=0.17\textwidth]{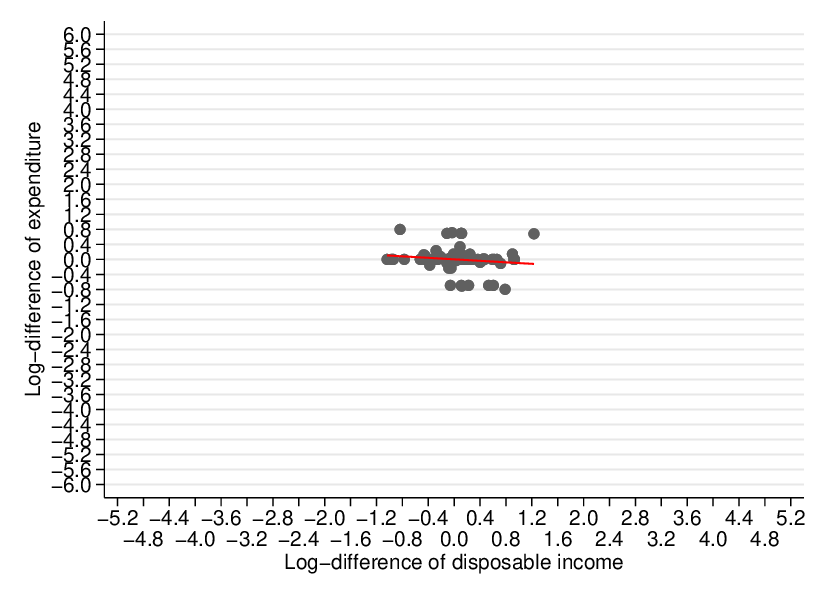}}
  \subfloat[Utilities]{\label{fig:scat_dutilities_am}\includegraphics[width=0.17\textwidth]{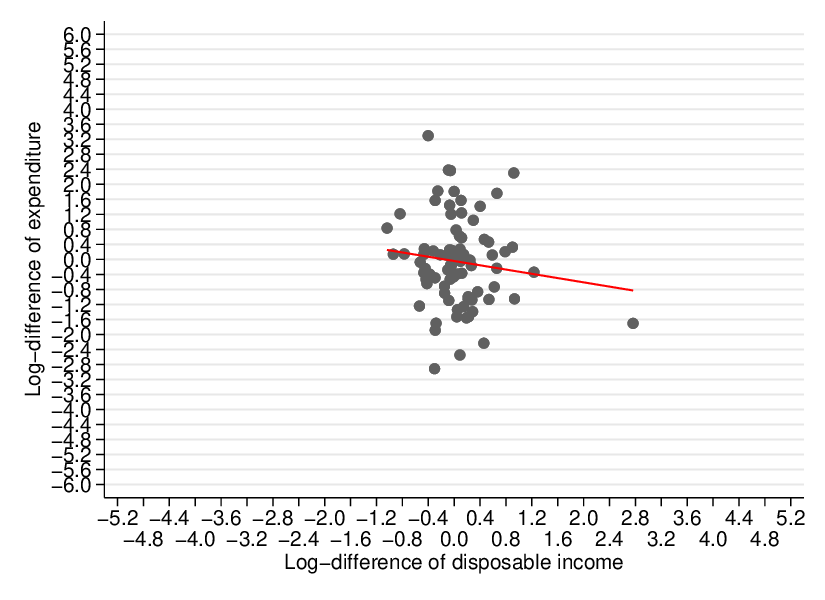}}
  \subfloat[Furniture]{\label{fig:scat_dfurniture_am}\includegraphics[width=0.17\textwidth]{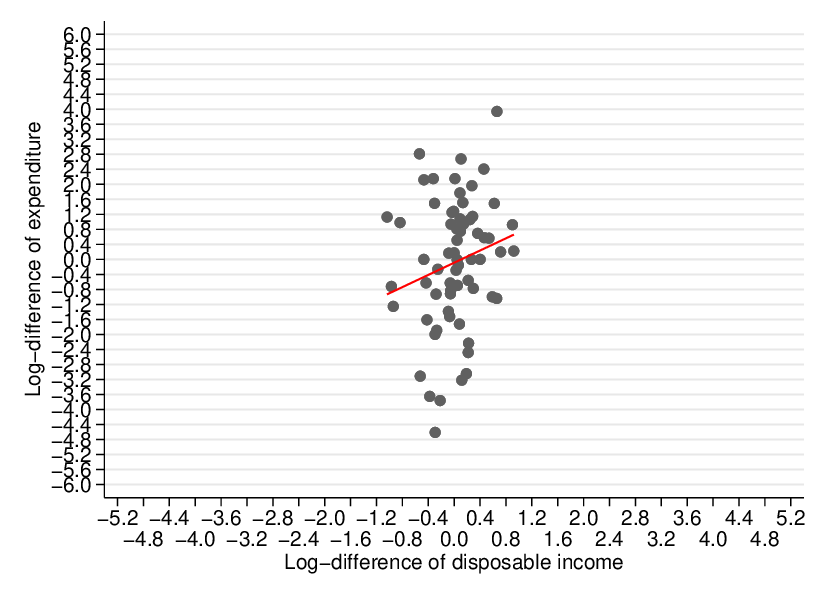}}
  \subfloat[Clothing]{\label{fig:scat_dclothing_am}\includegraphics[width=0.17\textwidth]{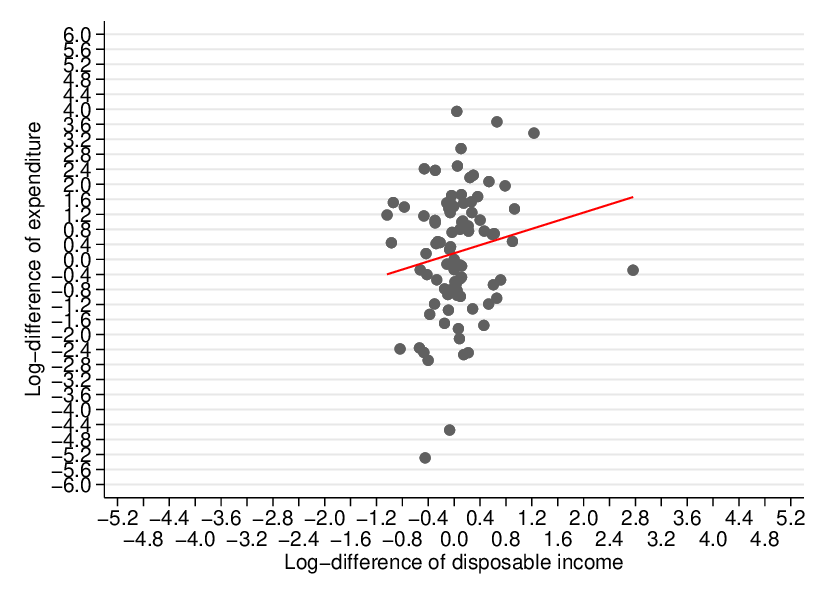}}\\
  \subfloat[Education]{\label{fig:scat_deducation_am}\includegraphics[width=0.17\textwidth]{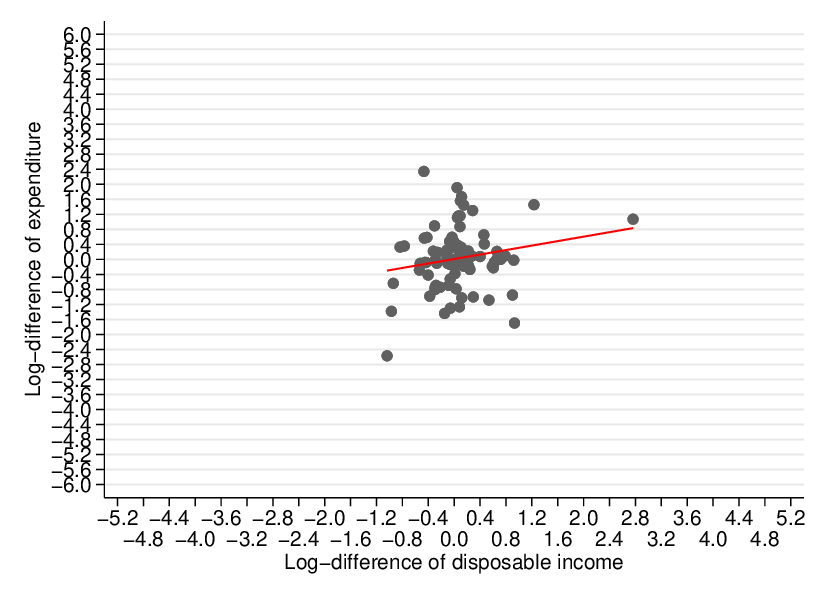}}
  \subfloat[Medical]{\label{fig:scat_dmedical_am}\includegraphics[width=0.17\textwidth]{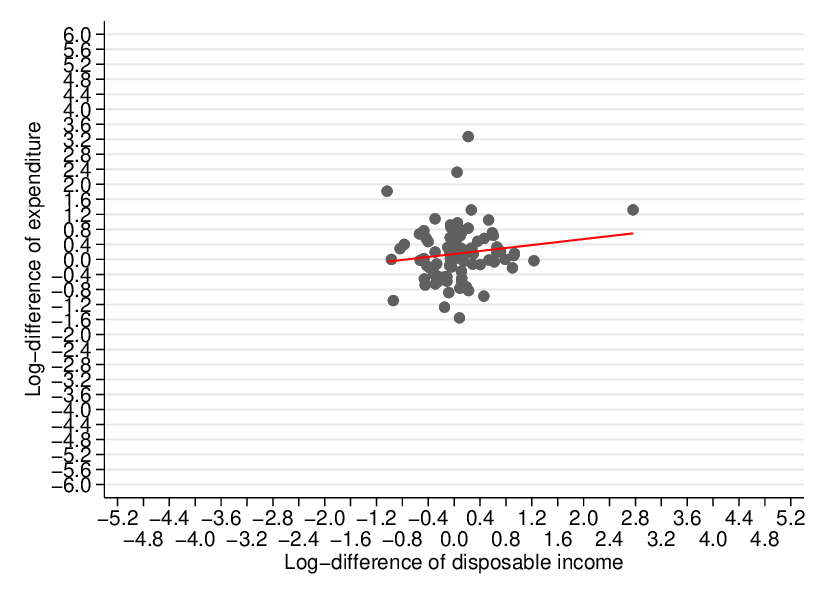}}
  \subfloat[Entertainment]{\label{fig:scat_dentertainment_am}\includegraphics[width=0.17\textwidth]{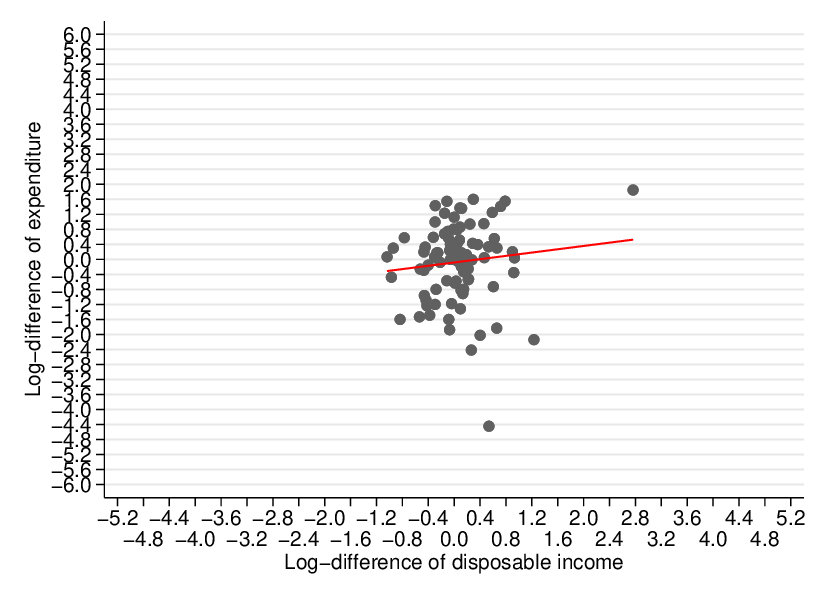}}
  \subfloat[Transportation]{\label{fig:scat_dtransportation_am}\includegraphics[width=0.17\textwidth]{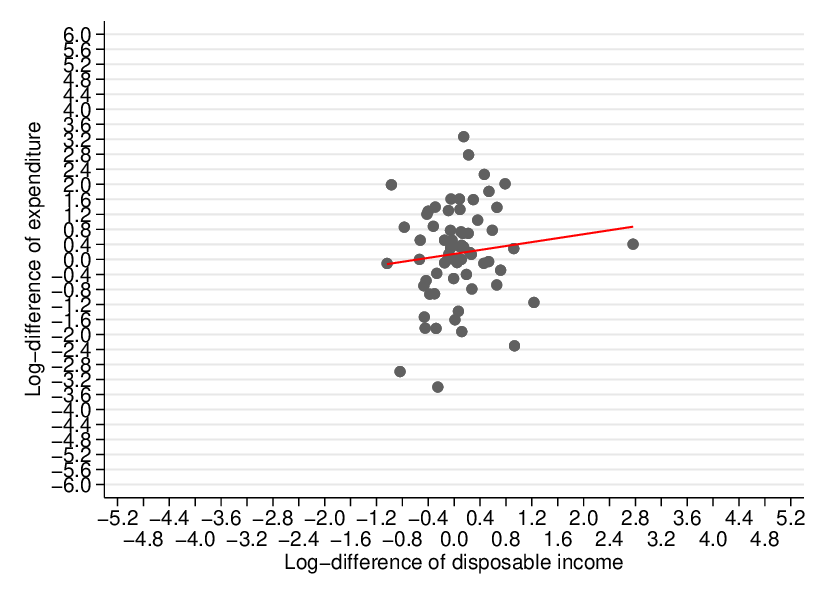}}
  \subfloat[Gifts]{\label{fig:scat_dgift_am}\includegraphics[width=0.17\textwidth]{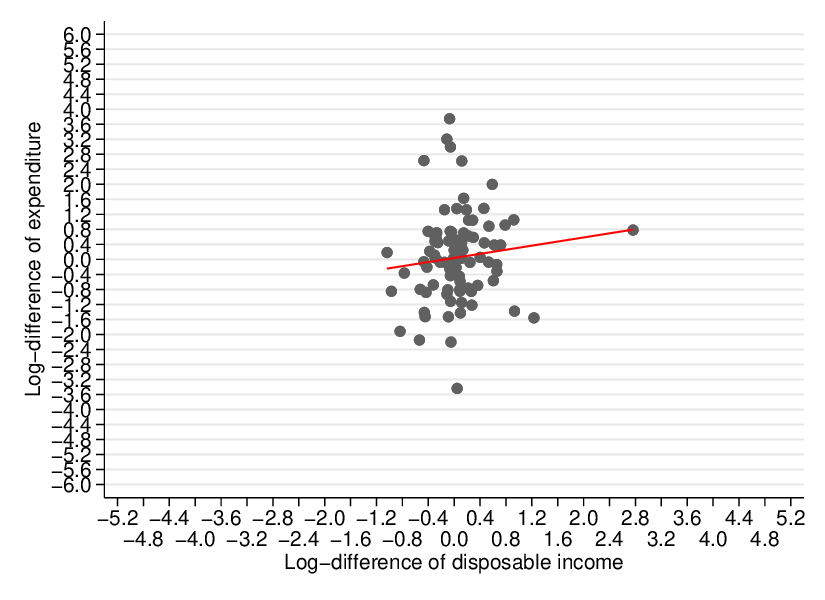}}
  \subfloat[Miscellaneous]{\label{fig:scat_dmiscellaneous_am}\includegraphics[width=0.17\textwidth]{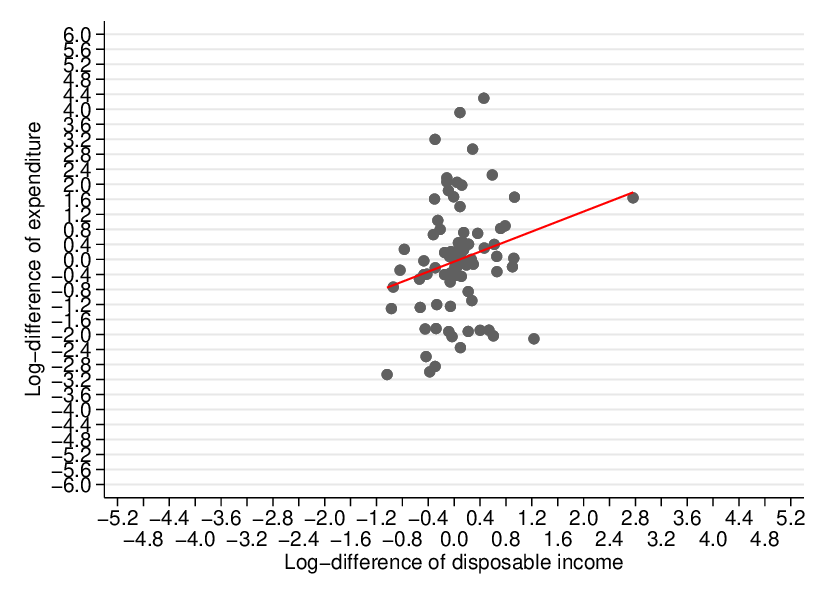}}
\caption{Raw Relationship between the Log-differences\\ in Monthly Income and Expenditure}
\label{fig:scat_exp_am}
\scriptsize{\begin{minipage}{450pt}
\setstretch{0.85}
Notes:
Figure~\ref{fig:scat_dexp_am} illustrates the relationship between the log-difference in monthly disposable income and expenditure.
Figures~\ref{fig:scat_dfood_am}--\ref{fig:scat_dmiscellaneous_am} illustrate the relationships between the log-differences in monthly disposable income and the 11 expenditure subcategories listed in Panel A in Table~\ref{tab:sum}. 
The range of the y-axis is fixed across all the figures for comparability.
Sources: Created by the author using the THBS sample.
\end{minipage}}
\end{figure}

Figure~\ref{fig:hist_exp} summarizes the densities of the log-differences (LD) of the semi-monthly expenditures.
Figure~\ref{fig:hist_dexp_sm} illustrates the LD distribution of the total expenditure, whereas Figures~\ref{fig:hist_dfood_sm}--\ref{fig:hist_dmiscellaneous_sm} summarize the LD distributions of the 11 subcategories of the expenditure.
These figures confirm no specific outliers in all these expenditure categories.

Figure~\ref{fig:scat_exp} summarizes the scatter plots between the LD of the adjusted semi-monthly disposable income and expenditures.
Figure~\ref{fig:scat_dexp_sm} illustrates the relationship between the LD of the adjusted semi-monthly disposable income and total expenditure.
Figures~\ref{fig:scat_dfood_sm}--\ref{fig:scat_dmiscellaneous_sm} illustrate the correlations between the LD of the adjusted semi-monthly disposable income and 11 expenditure subcategories.

Generally, rent was paid once per month.
This means that the semi-monthly dataset fails to capture the variations in rent payment because it includes some semi-monthly cells with zero values.
In addition, the water bills are included in the rent subcategory but not in the utility subcategory.
This means that a semi-month cell takes a minimal value when the payment of rent is not included in that cell.
Figure~\ref{fig:hist_dhousing_sm} shows that the LD of rent generates some larger values in the semi-monthly panel dataset due to those water bill payments.
These cells taking large LD values substantially increase the slope, as shown in Figure~\ref{fig:scat_dhousing_sm}.

Given this technical issue, the monthly panel dataset is suitable for estimating the income elasticity for the rent category.
In fact, Figure~\ref{fig:scat_dhousing_am} confirms that the scatter plot from the monthly variation shows a flatter relationship.

\subsection{Testing Heterogeneity: Accessibility to Retailers}\label{sec:secc_heterogeneity}

\begin{figure}[htb]
\centering
\captionsetup{justification=centering,margin=1.5cm}
\includegraphics[angle =0, width=0.8\textwidth]{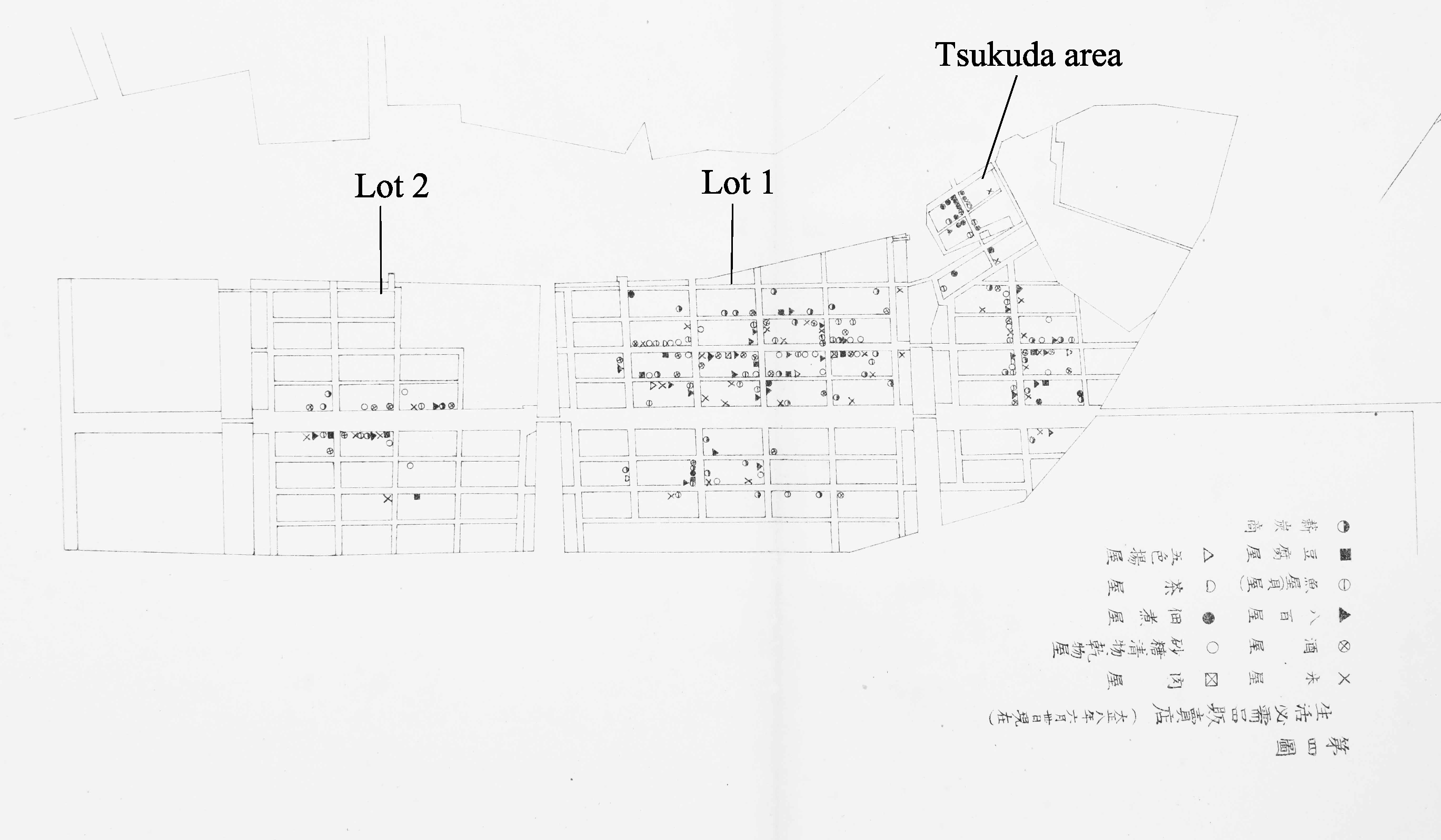}
\caption{Retail Shops for Daily Necessities: Food and Fuel Retailers}
\label{fig:map_retailers}
\scriptsize{\begin{minipage}{450pt}
\setstretch{0.85}
Notes: 
This map shows the locations of retail shops for daily necessities (\textit{seikatsu hitsujyuhin hanbaiten}) in Tsukishima.
Cross mark: rice shop (\textit{kome ya}).
Circle with a cross: liquor shop (\textit{saka ya}).
Black triangle: greengrocer (\textit{yao ya})
Circle with a minus: fish store (\textit{sakana ya}).
Black square: \textit{t\=ofu} store.
Half-black circle: firewood and charcoal store (\textit{shintan ya}).
Square with a cross: butcher store (\textit{niku ya}).
White circle: sugar, pickles, and dried food store (\textit{sat\=o, tsukemono, kanbutsu ya}).
Black circle: \textit{tsukudani} shop.
\textit{tsukudani} is a food simmered in sweet soy sauce.
Half circle: tea shop (\textit{cha ya}).
White triangle: \textit{goshikiage} shop.
Note: Lot 3, the southernmost lot, had no households in 1919.
\textit{Goshikiage ya} is a fried fish paste containing vegetables.
Source: 
Department of Health, Ministry of the Interior 1923c, fourth map.
The tone was adjusted by the author using Adobe Photoshop 26.6.1.
\end{minipage}}
\end{figure}

The approximate locations of retail shops for daily necessities are reported in the official document and shown in Figure~\ref{fig:map_retailers}.
Density varies across lots: the number of retail shops per $1,000$ population is $8.8$ in the Tsukuda area, $7.8$ in Lot 1, and $4.6$ in Lot 2.
Lot 2 has a lower density because it is relatively newer than the others (Online Appendix~\ref{sec:seca_tsukishima}).

First, I include the density of retail shops in each lot as a control variable in Equation~\ref{eqn:eq_risk}.
Because this variable is time-constant, I interact it with the time dummies.
Column 1 of Table~\ref{tab:r_cp_hetero} shows an estimate of $-0.132$, which is close to the baseline estimate of $-0.128$ (Table~\ref{tab:r_cp}).

Second, I expand Equation~\ref{eqn:eq_risk} by adding an interaction term between the head's earnings and retail shop density.
In this expanded model, a statistically significantly negative coefficient on the interaction term would indicate that households in denser areas rely more on credit purchases.
However, Column 2 shows that the estimated coefficient on the interaction term is $-0.004$ and not statistically significant.
In Column 3, I use an alternative definition of density, adopting the number of households as the denominator to account for variation in family size.
The estimated coefficient on the interaction term remains near zero ($-0.002$).
In both columns, the marginal effect of the head's earnings is estimated at $-0.137$, similar to the estimate in Column 1.

The total length of the residential area in Tsukishima is roughly $1.4$ kilometers with a width of $500$ meters (Ch\=u\=o Ward 1994, p.~38).
Thus, although the living areas are divided into three lots, residents can easily access shops in different lots on foot (Department of Health, Ministry of the Interior 1923a, pp.~40--46).

\def\arraystretch{1.0}
\begin{table}[htb]
\begin{center}
\captionsetup{justification=centering,margin=1.5cm}
\caption{Testing Heterogeneity: Density of the Retailers in Each Lot}
\label{tab:r_cp_hetero}
\scriptsize
\scalebox{0.93}[1]{
\begin{tabular}{lD{.}{.}{-2}D{.}{.}{-2}D{.}{.}{-2}D{.}{.}{-2}D{.}{.}{-2}D{.}{.}{-2}}
\toprule[1pt]\midrule[0.3pt]
&\multicolumn{6}{c}{Dependent Variable: Net Credit Purchase (income minus expenditure, yen)}\\
\cmidrule(rrrrrr){2-7}
&\multicolumn{3}{c}{Semi-monthly Panel}&\multicolumn{3}{c}{Adjusted Monthly Panel}\\
\cmidrule(rrr){2-4}\cmidrule(rrr){5-7}
&\multicolumn{1}{c}{(1)}	&\multicolumn{1}{c}{(2)}	&\multicolumn{1}{c}{(3)}
&\multicolumn{1}{c}{(4)}	&\multicolumn{1}{c}{(5)}	&\multicolumn{1}{c}{(6)}	\\\hline
Head's Earnings				&-0.132$***$	&-0.107		&-0.099		&-0.026	&0.010	&0.013	\\
							&[0.044]		&[0.117]		&[0.135]		&[0.061]	&[0.160]	&[0.190]	\\
Head's Earnings $\times$	Density	&			&-0.004		&-0.002		&		&-0.003	&-0.001	\\
							&			&[0.020]		&[0.007]		&		&[0.026]	&[0.009]	\\\hdashline
Marginal Effect ($\partial r_{it}/\partial \tilde{y}_{it}$)	&			&-0.137$***$	&-0.137$***$	&		&-0.008	&-0.008	\\
							&			&[0.045]		&[0.044]		&		&[0.051]	&[0.051]	\\\hline
Density $\times$ Time FEs	&\multicolumn{1}{c}{Yes}	&\multicolumn{1}{c}{No}	&\multicolumn{1}{c}{No}	&\multicolumn{1}{c}{Yes}	&\multicolumn{1}{c}{No}		&\multicolumn{1}{c}{No}	\\
Family Size 				&\multicolumn{1}{c}{Yes}	&\multicolumn{1}{c}{Yes}	&\multicolumn{1}{c}{Yes}	&\multicolumn{1}{c}{Yes}	&\multicolumn{1}{c}{Yes}		&\multicolumn{1}{c}{Yes}	\\
Denominator of the Density	&\multicolumn{1}{c}{Population}	&\multicolumn{1}{c}{Population}	&\multicolumn{1}{c}{Households}	&\multicolumn{1}{c}{Population}	&\multicolumn{1}{c}{Population}	&\multicolumn{1}{c}{Households}	\\
Observations				&\multicolumn{1}{c}{279}	&\multicolumn{1}{c}{279}	&\multicolumn{1}{c}{279}	&\multicolumn{1}{c}{120}	&\multicolumn{1}{c}{120}	&\multicolumn{1}{c}{120}	\\
\midrule[0.3pt]\bottomrule[1pt]
\end{tabular}
}
{\scriptsize
\begin{minipage}{445pt}
\setstretch{0.85}
***, **, and * denote statistical significance at the 1\%, 5\%, and 10\% levels, respectively.
Robust standard errors are reported in brackets.\\
Notes: 
Columns 1 and 4 report the estimated coefficient on the head's earnings in the expanded regression of Equation~\ref{eqn:eq_risk}, which includes interaction terms between retailer density and time dummies as additional controls.
Columns 2, 3, 5, and 6 present results from the expanded regression of Equation~\ref{eqn:eq_risk}, which includes the interaction between the head's earnings and retailer density.
The density variable in Columns 2 and 5 is measured per population, whereas it is measured per household in Columns 3 and 6.
Columns 1--3 show results for the semi-monthly panel, while Columns 4--6 show results for the adjusted monthly panel.
``Marginal Effect ($\partial r_{it}/\partial \tilde{y}_{it}$)'' indicates the marginal effect of the head's earnings at the mean of the density variable.
All regressions control for family size.
Households with missing residence information are excluded from the analytical sample.
Results remain unchanged if these households are assumed to reside in any of the three lots.
A few households living on the mainland in Kyobashi Ward are assigned to the nearest lot of Tsukishima.
Sources: See the main text for the THBS sample.
Data on the number of people and households in the three lots are from the Department of Health, Ministry of the Interior (1923b, p.~2).
The number of retail shops in each lot is from the Department of Health, Ministry of the Interior (1923c, fourth map).
\end{minipage}
}
\end{center}
\end{table}

\subsection{Testing the Roles of Cooperatives and Pawnshops}\label{sec:secc_coop}

\def\arraystretch{0.95}
\begin{table}[h!]
\begin{center}
\captionsetup{justification=centering,margin=1.5cm}
\caption{Testing Roles of the Cooperatives and Pawnshops}
\label{tab:r_coop}
\scriptsize
\scalebox{0.98}[1]{
\begin{tabular}{lD{.}{.}{-2}D{.}{.}{-2}D{.}{.}{-2}D{.}{.}{-2}D{.}{.}{-2}D{.}{.}{-2}}
\toprule[1pt]\midrule[0.3pt]
\multicolumn{7}{l}{\textbf{Panel A: Estimates from the Baseline Specification}}\\
&\multicolumn{6}{c}{Dependent Variable:}\\
\cmidrule(rrrrrr){2-7}
&\multicolumn{3}{c}{Prices of Purchased Items}&\multicolumn{3}{c}{}\\
&\multicolumn{3}{c}{from Cooperatives (yen)}&\multicolumn{3}{c}{Income from Pawnshops (yen)}\\
\cmidrule(rrr){2-4}\cmidrule(rrr){5-7}
&\multicolumn{1}{c}{}	&\multicolumn{1}{c}{(1)}	&\multicolumn{1}{c}{(2)}
&\multicolumn{1}{c}{}	&\multicolumn{1}{c}{(3)}	&\multicolumn{1}{c}{(4)}	\\\hline
Head's Earnings				&&-0.001		&-0.006		&&0.000	&0.005	\\
							&&[0.002]		&[0.004]		&&[0.002]	&[0.006]	\\\hline
Type of Panels 				&\multicolumn{1}{c}{}	&\multicolumn{1}{c}{Semi-monthly}	&\multicolumn{1}{c}{Monthly}	&\multicolumn{1}{c}{}	&\multicolumn{1}{c}{Semi-monthly}		&\multicolumn{1}{c}{Monthly}	\\
Family Size 				&\multicolumn{1}{c}{}	&\multicolumn{1}{c}{Yes}	&\multicolumn{1}{c}{Yes}	&\multicolumn{1}{c}{}	&\multicolumn{1}{c}{Yes}		&\multicolumn{1}{c}{Yes}	\\
Observations				&\multicolumn{1}{c}{}	&\multicolumn{1}{c}{289}	&\multicolumn{1}{c}{124}	&\multicolumn{1}{c}{}	&\multicolumn{1}{c}{289}	&\multicolumn{1}{c}{120}	\\
&&&&&&\\
\multicolumn{7}{l}{\textbf{Panel B: Testing Heterogeneity using Density of the Pawnshop in Each Lot}}\\
&\multicolumn{6}{c}{Dependent Variable: Income from Pawnshops (yen)}\\
\cmidrule(rrrrrr){2-7}
&\multicolumn{3}{c}{Semi-monthly Panel}&\multicolumn{3}{c}{Adjusted Monthly Panel}\\
\cmidrule(rrr){2-4}\cmidrule(rrr){5-7}
&\multicolumn{1}{c}{(1)}	&\multicolumn{1}{c}{(2)}	&\multicolumn{1}{c}{(3)}
&\multicolumn{1}{c}{(4)}	&\multicolumn{1}{c}{(5)}	&\multicolumn{1}{c}{(6)}	\\\hline
Head's Earnings				&-0.003		&0.013		&0.018		&0.001	&0.113	&0.101	\\
							&[0.003]		&[0.022]		&[0.025]		&[0.004]	&[0.126]	&[0.115]	\\
Head's Earnings $\times$	Density	&			&-0.079		&-0.032		&		&-0.686	&-0.182	\\
							&			&[0.129]		&[0.044]		&		&[0.753]	&[0.203]	\\\hdashline
Marginal Effect ($\partial r_{it}/\partial \tilde{y}_{it}$)	&&0.001		&0.001		&		&0.006	&0.007	\\
							&			&[0.002]		&[0.003]		&		&[0.009]	&[0.010]	\\\hline
Density $\times$ Time FEs	&\multicolumn{1}{c}{Yes}	&\multicolumn{1}{c}{No}	&\multicolumn{1}{c}{No}	&\multicolumn{1}{c}{Yes}	&\multicolumn{1}{c}{No}		&\multicolumn{1}{c}{No}	\\
Family Size 				&\multicolumn{1}{c}{Yes}	&\multicolumn{1}{c}{Yes}	&\multicolumn{1}{c}{Yes}	&\multicolumn{1}{c}{Yes}	&\multicolumn{1}{c}{Yes}		&\multicolumn{1}{c}{Yes}	\\
Denominator of the Density	&\multicolumn{1}{c}{Population}	&\multicolumn{1}{c}{Population}	&\multicolumn{1}{c}{Households}	&\multicolumn{1}{c}{Population}	&\multicolumn{1}{c}{Population}	&\multicolumn{1}{c}{Households}	\\
Observations				&\multicolumn{1}{c}{279}	&\multicolumn{1}{c}{279}	&\multicolumn{1}{c}{279}	&\multicolumn{1}{c}{120}	&\multicolumn{1}{c}{120}	&\multicolumn{1}{c}{120}	\\\midrule[0.3pt]\bottomrule[1pt]
\end{tabular}
}
{\scriptsize
\begin{minipage}{445pt}
\setstretch{0.86}
***, **, and * denote statistical significance at the 1\%, 5\%, and 10\% levels, respectively.
Robust standard errors are reported in brackets.\\
Notes:
\textbf{Panel A}: 
This panel reports results from Equation~\ref{eqn:eq_risk}, using prices of purchased items from cooperatives (Columns 1--2) or temporary income from pawnshops (Columns 3--4) as the dependent variable.
Columns 1 and 3 use the semi-monthly panel, while Columns 2 and 4 use the adjusted monthly panel.
All regressions control for family size.
\textbf{Panel B}: 
Columns 1 and 4 present the estimated coefficient on the head's earnings in the expanded regression of Equation~\ref{eqn:eq_risk}, which includes interaction terms between pawnshop density and time dummies as additional controls.
Columns 2, 3, 5, and 6 present results from the expanded regression, accounting for the interaction between the head's earnings and pawnshop density.
The density variable in Columns 2 and 5 is measured per population, while in Columns 3 and 6, it is measured per household.
Columns 1--3 show results for the semi-monthly panel, and Columns 4--6 show results for the adjusted monthly panel.
``Marginal Effect ($\partial r_{it}/\partial \tilde{y}_{it}$)'' indicates the marginal effect of the head's earnings at the mean of the density variable.
Households with no address are excluded from the analytical sample.
Results are unchanged if these households are assumed to reside in any of the three lots.
A few households living on the mainland in Kyobashi Ward are assigned to the nearest lot of Tsukishima.
Sources: See the main text for the THBS sample.
Data on the number of people and households in the three lots are from the Department of Health, Ministry of the Interior (1923b, p.~2).
The number of pawnshops in each lot is from the Department of Health, Ministry of the Interior (1923c, 15th map).
\end{minipage}
}
\end{center}
\end{table}

In the THBS sample, five households used cooperatives to buy daily commodities in a few semi-monthly cells.
Although this is a small portion, cooperatives may have offered some compensation to their members.
To examine this possibility, I use the total prices of items purchased from cooperatives as the dependent variable in Equation~\ref{eqn:eq_risk}.
If such purchases increase following an idiosyncratic shock, the estimated coefficient on the head's income would turn negative and become statistically significant.
Columns 1 and 2 in Panel A of Table~\ref{tab:r_coop} show that the estimates are negative but close to zero and statistically insignificant in both the semi-monthly and adjusted monthly panels.

Similarly, a few THBS households used pawnshops in a couple of semi-monthly cells.
I examine the role of pawnshops using the nominal income from pawnshops in Equation~\ref{eqn:eq_risk}.\footnote{Nominal income is used instead of net income because the THBS budget book summarizes payments in a column labeled payment (\textit{bensai}), which includes amounts paid not only to pawnshops but also to other lending institutions.}
If workers rely on pawnshops to mitigate hardships, the estimated coefficient on the head's income would be negative.
Columns 3 and 4 in Panel B show that the estimates are near zero and statistically insignificant in both the semi-monthly and adjusted monthly panels.
This result aligns with the main finding for the borrowing category (Table~\ref{tab:risk}).

In Panel B of the same table, I examine potential heterogeneity in pawnshop usage following the approach in Table~\ref{tab:r_cp_hetero}.
The official report provides the locations of pawnshops, allowing me to calculate their density, measured as the number of pawnshops per $1,000$ people, in each lot: $0.13$, $0.18$, and $0.16$ in the Tsukuda area, Lot 1, and Lot 2, respectively.
Column 1 includes density as a control variable in Equation~\ref{eqn:eq_risk}.
Column 2 adds an interaction term between the head's income and the density variable.
Column 3 tests sensitivity by using the number of households as an alternative denominator for density.
In all columns, the estimated coefficients are near zero and not statistically significant.
Columns 4--6, using the adjusted monthly panel, produce the same results.
This pattern aligns with the findings in Table~\ref{tab:r_cp_hetero}, consistent with Tsukishima's geographically narrow layout.

\subsection{Retail Sales in Tokyo City: A Retail Store Survey of 1935}\label{sec:secc_1935survey}

Table~\ref{tab:cp_tokyo_full} summarizes the contribution of regular customers to their total sales, social classes of the customers, and sales methods in the retailers selling white rice, fish, fruit and vegetable, and fuels (firewood and charcoal) by scale of business.
In the original document, the figures for ``Total'' in the firewood and charcoal category are missing in ``Sales Method (\%).''
For these three cells, I filled the averages over the three business-size categories in each sales method: 20.4\% for ``Cash,'' 78.4\% for ``Credit purchase,'' and 1.3\% for ``Monthly installment.''

\def\arraystretch{0.94}
\begin{table}[htb]
\begin{center}
\captionsetup{justification=centering,margin=1.5cm}
\caption{Customers, Sales Methods, and Competitiveness in the Retailers in Tokyo City}
\label{tab:cp_tokyo_full}
\scriptsize
\scalebox{0.82}[1]{
\begin{tabular}{llrrrrrrrrr}
\toprule[1pt]\midrule[0.3pt]

Selling Item		&Scale of	&Number of 	&Share of sales		&\multicolumn{3}{l}{Social Class of}			&\multicolumn{3}{l}{Sales Method (\%)}	&Number of \\
\cmidrule(rrr){8-10}
				&Business	&stores			&for regular		 	&\multicolumn{3}{l}{the Regular Customer (\%)}	&Cash		&Credit		&Monthly	&competitors\\
\cmidrule(rrr){5-7}
				&			&reported		&customer (\%)		&Upper		&Middle		&Other				&			&purchase	&installment	&reported		\\\hline
White Rice		&Small		&19			&69.7				&12.1		&49.5		&38.4				&29.5		&70.2		&0.3			&7.3		\\
				&Medium	&55			&69.5				&17.5		&41.6		&40.9				&26.9		&72.6		&0.5			&8.7		\\
				&Large		&27			&81.7				&15.9		&51.5		&32.6				&26.7		&71.0		&2.3			&8.8		\\
				&Total		&101			&73.1				&16		&45.7		&38.2				&26.9		&71.5		&1.7			&8.5		\\
				&&&&&&&&&&\\
Fish				&Small		&34			&64.5				&5.9		&53.2		&40.9				&42.7		&57.3		&0				&5.1		\\
				&Medium	&20			&62.6				&10		&58.0		&32.0				&43.8		&56.2		&0				&5.3		\\
				&Large		&5				&72.0				&0			&86.0		&14.0				&44.8		&55.2		&0				&3.4		\\
				&Total		&59			&64.6				&6.8		&57.6		&35.6				&43.9		&56.1		&0				&5.0		\\
				&&&&&&&&&&\\
Greengrocer		&Small	&46			&42.2				&8			&39.2		&52.7				&59.7		&40.3		&0				&6.5		\\
				&Medium	&17			&57.1				&8.8		&43.5		&47.6				&65.0		&35.0		&0				&6.1		\\
				&Large		&3				&36.7				&33.3		&40.0		&26.7				&79.7		&20.3		&0				&3.7		\\
				&Total		&66			&45.4				&9.4		&40.4		&50.2				&65.8		&34.2		&0				&6.2		\\
				&&&&&&&&&&\\
Firewood and charcoal&Small		&78		&68.5				&13.5		&50.7		&35.8				&25.0		&74.2		&0.8			&8.4		\\
				&Medium	&17			&77.7				&23.5		&49.1		&27.4				&19.3		&79.1		&1.6			&8.9		\\
				&Large		&5				&33.3				&23.0		&60.0		&17.0				&16.8		&81.8		&1.4			&10.6		\\
				&Total		&100			&51.1				&15.7		&50.9		&33.4				&20.4		&78.4		&1.3			&8.6		\\\midrule[0.3pt]\bottomrule[1pt]
\end{tabular}
}
{\scriptsize
\begin{minipage}{445pt}
\setstretch{0.85}
Notes: This table summarizes the statistics on the sales in the retailers selling white rice, fish, fruit and vegetables, and firewood and charcoal in Tokyo surveyed in 1935.\\
1.~Small, middle, and large-scale businesses include retailers with annual sales of less than 10,000, 10,000--30,000, and 30,000+ yen, respectively.\\
2.~The number of stores reporting the share of regular customers (\%) is 90, 56, 61, and 93 in the white rice, fish, greengrocers, and firewood and charcoal categories, respectively.\\
3.~The share of regular customers is defined as the percentage of sales for regular customers in the annual total sales in 1935.\\
4.~The definition of the social classes (``Upper''; ``Middle''; ``Other'') is not clearly defined in the report. Therefore, this must be based on the subjective judgment of the retailers.\\
5.~The share of sales method is based on the percentage of sales using each method in the annual total sales in 1935.\\
6.~The number of competitors indicates the number of peers within a three-cho (approximately 327 meters) radius.\\
Source: Tokyo Chamber of Commerce and Industry 1937, pp.~legend; 20--21; questionnaire.
\end{minipage}
}
\end{center}
\end{table}

In Table~\ref{tab:r_comp}, I analyze the relationship between credit purchases and competitiveness among local retailers using the selling item by scale of business cells reported in Table~\ref{tab:cp_tokyo_full}.
In Column (1), I regressed the sales in credit purchases (in percentage points) on the number of local competitors.
I then conditioned the business scales in Column (2).
The $F$-statistic $p$-value for joint zero-slope hypotheses for these scale dummies is estimated to be $0.990$.
In Column (3), I included the selling item dummies instead of the scale dummies.
The $F$-statistic $p$-value for joint zero-slope hypotheses is rejected at the conventional level, and the measure of fit is relatively high compared with the other two models reported in Columns (2) and (3).
The estimate is $3.420$, suggesting that a one unit increase in the local competitors increases the use of credit purchase by roughly $3.4$ percentage points.
Given that the standard deviation of the local competitors is approximately $2.3$, the estimated magnitude corresponding to the one standard deviation increase would be $7.9$ ($3.420 \times 2.3$).
This is roughly $41$\% of the standard deviation of the dependent variable ($19.2$) and thus, is an economically meaningful magnitude.

\def\arraystretch{1.0}
\begin{table}[htb]
\begin{center}
\captionsetup{justification=centering,margin=1.5cm}
\caption{Use of Credit Purchases and Competitiveness among Local Retailers}
\label{tab:r_comp}
\scriptsize
\scalebox{1.0}[1]{
\begin{tabular}{lD{.}{.}{-2}D{.}{.}{-2}D{.}{.}{-2}}
\toprule[1pt]\midrule[0.3pt]
&\multicolumn{3}{c}{Dependent Variable: Sales in Credit Purchases (\%)}\\
\cmidrule(rrr){2-4}
&\multicolumn{1}{c}{(1)}	&\multicolumn{1}{c}{(2)}		&\multicolumn{1}{c}{(3)}\\\hline
Number of Competitors			&7.190$***$		&7.254$***$	&3.420$**$	\\
							&[2.041]		&[2.145]		&[1.310]	\\\hline
Sample mean of the DV			&59.4		&59.4		&59.4		\\
Model						&\multicolumn{1}{c}{OLSE}	&\multicolumn{1}{c}{OLSE}	&\multicolumn{1}{c}{OLSE}	\\
Number of observations			&\multicolumn{1}{c}{12}		&\multicolumn{1}{c}{12}		&\multicolumn{1}{c}{12}		\\
Number of retail shops (weight)			&\multicolumn{1}{c}{326}		&\multicolumn{1}{c}{326}		&\multicolumn{1}{c}{326}		\\
Scale FEs			 			&\multicolumn{1}{c}{No}		&\multicolumn{1}{c}{Yes}		&\multicolumn{1}{c}{No}		\\
Selling item FEs		 		&\multicolumn{1}{c}{No}		&\multicolumn{1}{c}{No}		&\multicolumn{1}{c}{Yes}		\\
$F$-statistic $p$-value for the FEs	&\multicolumn{1}{c}{}			&\multicolumn{1}{c}{0.990}	&\multicolumn{1}{c}{0.000}	\\
R-squared						&\multicolumn{1}{c}{0.577}	&\multicolumn{1}{c}{0.578}	&\multicolumn{1}{c}{0.986}	\\
\midrule[0.3pt]\bottomrule[1pt]
\end{tabular}
}
{\scriptsize
\begin{minipage}{330pt}
\setstretch{0.85}
***, **, and * denote statistical significance at the 1\%, 5\%, and 10\% levels, respectively.
Robust standard errors are reported in brackets.\\
Notes: 
The observations used in this table are from the selling item by scale of business cells reported in Table~\ref{tab:cp_tokyo_full}.
The dependent variable in Columns (1)--(3) is the percentage of sales under credit purchases.
``Number of competitors'' indicates the number of peers within a three-cho (approximately 327 meters) radius.
The sample mean and standard deviation of the dependent variable are $59.4$ and $19.2$, respectively.
The sample mean of the number of competitors is $6.89$ (Std. Dev. = $2.26$).
All the regressions are weighted by the number of stores in each selling item by scale of business cell (see third column of Table~\ref{tab:cp_tokyo_full}).
The scale of business dummies are added in Column 2, whereas the selling item dummies are included in Column 3.
$F$-statistics $p$-values for these dummies are reported in Columns 2 and 3.
The standard errors are based on the HC2 estimator proposed by Horn et al. (1975).
\end{minipage}
}
\end{center}
\end{table}

\subsection{Retail Sales in Tokyo City: A Small and Medium Scale Enterprises Survey of 1930}\label{sec:secc_1930survey}

Table~\ref{tab:cp_tokyo_loss} summarizes the use of credit purchases and the associated losses documented in the 1930 Survey Report.
Following the report's classification, items are divided into six categories: white rice, other grains, vegetables, fruits, fish, and fuel.
The total number of retailers in each category is $157$, $7$, $67$, $16$, $46$, and $64$, respectively.
These retailers used credit purchases as their primary selling method.
Notably, almost all retailers using credit purchases experienced sales losses.
For example, $99$\% of white rice retailers reported losses, while the minimum share was $97$\% for fuel retailers.
This indicates that credit purchase was generally a loss-making sales method.

Table~\ref{tab:cp_tokyo_loss} also shows that many cases have unknown loss amounts.
Overall, most losses appear to be concentrated in bins of less than $5$\% of sales.
For instance, $87$\% ($97$ retailers) of the $111$ rice retailers with available data suffered losses amounting to $5$\% of sales.
Similarly, $65$\% ($32$ retailers) of the $49$ fuel retailers fell into the same loss category.
Systematic statistics on the percentage of total profits corresponding to these losses are unavailable.
However, a large portion of the losses is capped at $5$\% of total sales across all retailer types.
This suggests that, although retailers could not entirely avoid losses, they may have set prices to account for anticipated losses.

\def\arraystretch{0.90}
\begin{table}[htb]
\begin{center}
\captionsetup{justification=centering,margin=1.5cm}
\caption{Sales Methods and Losses on Sales under Credit Purchases in Tokyo City}
\label{tab:cp_tokyo_loss}
\scriptsize
\scalebox{1.0}[1]{
\begin{tabular}{lrrrrrr}
\toprule[1pt]\midrule[0.3pt]

							&\multicolumn{6}{c}{Selling Item}\\
\cmidrule(rrrrrr){2-7}
										&White rice	&Other grains	&Vegetables	&Fruit	&Fish	&Fuel\\\hline
Total number of retailers						&157			&7			&67			&16		&46		&64\\
&&&&&&\\
Cash payment only							&2			&0			&10			&8		&8		&3\\							
(never used the credit purchase)	&&&&&&\\
&&&&&&\\
Retailers using credit purchases				&155			&7			&57			&8		&58		&61\\	
\hspace{10pt}Losses among sales amount (\%)&&&&&&\\
\hspace{30pt}$\leq$ 1\%						&59			&1			&14			&4		&10		&18\\
\hspace{30pt}2--5\%							&38			&1			&19			&2		&8		&14\\
\hspace{30pt}6--10\%						&8			&1			&4			&0		&3		&12\\
\hspace{30pt}11--20\%						&3			&0			&0			&0		&1		&3\\
\hspace{30pt}21--30\%						&2			&0			&1			&0		&0		&0\\
\hspace{30pt}30+\%							&0			&0			&0			&2		&0		&0\\
\hspace{30pt}Unknown						&44			&4			&19			&0		&16		&12\\
\hspace{10pt}No loss						&1			&0			&0			&0		&0		&2\\
&&&&&&\\
Retailers using credit purchases (\%)				&99			&100			&85		&50		&79		&95\\
Share of retailers experienced losses (\%)			&99			&100			&100		&100		&100		&97\\\midrule[0.3pt]\bottomrule[1pt]
\end{tabular}
}
{\scriptsize
\begin{minipage}{400pt}
\setstretch{0.85}
Note: This table summarizes the statistics on the sales methods and losses on sales under credit purchases among the retailers measured in Tokyo City surveyed in 1930. Source: Tokyo City Office 1932, pp.~454--469.
\end{minipage}
}
\end{center}
\end{table}

Lastly, Table~\ref{tab:tabc_1930} summarizes the industrial and family structures in 1930 Tokyo using the official report of the Population Census.
Column 1 summarizes the share of male workers in each industry, showing a similar distribution as that in 1920 (Column 1 of Panel A in Table~\ref{tab:tab1}).
While the development of the city throughout the 1920s slightly increased the share of commercial industries, manufacturing and commercial industries covered approximately 75\% (77\% in 1920).
Column 2 confirms that the average household size, share of married males, sex ratio, and average age of males are in a similar range to those measured in the 1920 Population Census (Column 1 of Panel B in Table~\ref{tab:tab1}).
These show that not only the industrial structure but also the average household characteristics had not changed essentially between 1920 and 1930.
While Tokyo experienced the Great Kanto Earthquake in 1923, the population structure did not change dramatically, except for a temporary evacuation soon after the earthquake.

\begin{table}[htb]
\def\arraystretch{1.0}
\centering
\begin{center}
\caption{Industrial and Family Structures in Tokyo: 1930 Population Census}
\label{tab:tabc_1930}
\scriptsize
\scalebox{0.94}[1]{
\begin{tabular}{lrrrr}
\toprule[1pt]\midrule[0.3pt]
\multicolumn{2}{l}{}						&(1) Industrial structure		&&(2) Family Structure		\\
\multicolumn{2}{l}{Name of survey}			&1930 Population census		&&1930 Population census	\\
\multicolumn{2}{l}{Survey area}				&Tokyo City				&&Tokyo City				\\
\multicolumn{2}{l}{Survey subject}			&Complete survey			&&Complete survey			\\
\multicolumn{2}{l}{Survey month and year}	&October 1930				&&October 1930			\\\hline

\multicolumn{2}{l}{Agriculture}						&0.5		&Average household size (in person)	&5.1		\\ 
\multicolumn{2}{l}{Fisheries}						&0.1		&Share of married males (\%)			&34.2	\\
\multicolumn{2}{l}{Mining}							&0.0		&Sex ratio (males/females)			&1.2		\\
\multicolumn{2}{l}{Manufacturing}					&37.2	&Average age of males				&25.2	\\
\multicolumn{2}{l}{Commerce}						&37.3	&&\\
\multicolumn{2}{l}{Transport}						&7.2		&&\\
\multicolumn{2}{l}{Public service and professions}		&13.5	&&\\ 
\multicolumn{2}{l}{Housework}						&0.6		&&\\
\multicolumn{2}{l}{Other industry}					&3.6		&&\\\midrule[0.3pt]\bottomrule[1pt]
\end{tabular}
}
{\scriptsize
\begin{minipage}{430pt}
\setstretch{0.85}
Notes:
\textbf{Panel A}: This panel summarizes industrial structures based on the occupations of male workers recorded in the 1930 Population Census.
Each share is calculated as the number of males working as regular employees in each sector divided by the total number of male workers (percentage).
Unemployed males (\textit{mushoku}) are excluded from these figures.
\textbf{Panel B}: The average household size is calculated as the total number of people divided by the number of households.
The share of married males is calculated as the number of married males living with/without spouses divided by the total number of males (percentage).
The average age of males is calculated from the population tables by age group reported in the census.
For the open-ended age category over 65, the value is rounded to $67$, following the range of the previous category (i.e., 60--64).
All figures in this panel exclude a small number of quasi-households (\textit{jyun setai}) such as students in dormitories and patients in hospitals.
Sources:
Column 1: Statistics Bureau of the Cabinet (1933, pp.~52--53).
Column 2: Statistics Bureau of the Cabinet (1933, pp.~14; 6--7; 357).
\end{minipage}
}
\end{center}
\end{table}

\subsection{Testing Relationship between Volatility of Earnings and Use of Credit Purchases}\label{sec:secc_volatility}

\def\arraystretch{0.95}
\begin{table}[htb]
\begin{center}
\captionsetup{justification=centering,margin=1.5cm}
\caption{Volatility of Earnings and Use of Credit Purchases}
\label{tab:r_cp}
\scriptsize
\scalebox{0.95}[1]{
\begin{tabular}{lD{.}{.}{-2}D{.}{.}{-2}D{.}{.}{-2}D{.}{.}{-2}D{.}{.}{-2}D{.}{.}{-2}}
\toprule[1pt]\midrule[0.3pt]
&\multicolumn{6}{c}{Dependent Variable: Indicator Variable for}\\
\cmidrule(rrrrrr){2-7}
&\multicolumn{3}{c}{Non-CP Users}&\multicolumn{3}{c}{Non-CP + Low-CP Users}\\
\cmidrule(rrr){2-4}\cmidrule(rrr){5-7}
&\multicolumn{1}{c}{(1)}	&\multicolumn{1}{c}{(2)}		&\multicolumn{1}{c}{(3)}		&\multicolumn{1}{c}{(4)}&\multicolumn{1}{c}{(5)}		&\multicolumn{1}{c}{(6)}		\\\hline
CV of Head's Earnings	&0.003		&-0.002	&-0.027	&0.019		&0.008		&-0.005	\\
						&[0.012]	&[0.014]	&[0.050]	&[0.013]	&[0.017]	&[0.049]	\\\hline
Sample mean of the DV	&0.12		&0.12		&0.12		&0.24		&0.24		&0.24	\\
Model					&\multicolumn{1}{c}{LPM}	&\multicolumn{1}{c}{LPM}	&\multicolumn{1}{c}{Probit}	&\multicolumn{1}{c}{LPM}	&\multicolumn{1}{c}{LPM}	&\multicolumn{1}{c}{Probit}	\\
Number of THBS households	&\multicolumn{1}{c}{33}	&\multicolumn{1}{c}{33}	&\multicolumn{1}{c}{33}	&\multicolumn{1}{c}{33}	&\multicolumn{1}{c}{33}	&\multicolumn{1}{c}{33}	\\\hline
Family Size 					&\multicolumn{1}{c}{Yes}	&\multicolumn{1}{c}{Yes}	&\multicolumn{1}{c}{Yes}	&\multicolumn{1}{c}{Yes}	&\multicolumn{1}{c}{Yes}	&\multicolumn{1}{c}{Yes}	\\
Children aged 6--12 (\%) 		&\multicolumn{1}{c}{Yes}	&\multicolumn{1}{c}{Yes}	&\multicolumn{1}{c}{Yes}	&\multicolumn{1}{c}{Yes}	&\multicolumn{1}{c}{Yes}	&\multicolumn{1}{c}{Yes}	\\
Children aged 13--16 (\%) 	&\multicolumn{1}{c}{Yes}	&\multicolumn{1}{c}{Yes}	&\multicolumn{1}{c}{Yes}	&\multicolumn{1}{c}{Yes}	&\multicolumn{1}{c}{Yes}	&\multicolumn{1}{c}{Yes}	\\
Adults aged 17+ (\%) 		&\multicolumn{1}{c}{Yes}	&\multicolumn{1}{c}{Yes}	&\multicolumn{1}{c}{Yes}	&\multicolumn{1}{c}{Yes}	&\multicolumn{1}{c}{Yes}	&\multicolumn{1}{c}{Yes}	\\
Head's age			 		&\multicolumn{1}{c}{No}	&\multicolumn{1}{c}{Yes}	&\multicolumn{1}{c}{Yes}	&\multicolumn{1}{c}{No}	&\multicolumn{1}{c}{Yes}	&\multicolumn{1}{c}{Yes}	\\
Head's earnings			 	&\multicolumn{1}{c}{No}	&\multicolumn{1}{c}{Yes}	&\multicolumn{1}{c}{Yes}	&\multicolumn{1}{c}{No}	&\multicolumn{1}{c}{Yes}	&\multicolumn{1}{c}{Yes}	\\
Zero slope ($p$-value)		&\multicolumn{1}{c}{0.467}	&\multicolumn{1}{c}{0.718}	&\multicolumn{1}{c}{0.336}	&\multicolumn{1}{c}{0.515}	&\multicolumn{1}{c}{0.519}	&\multicolumn{1}{c}{0.451}	\\
\midrule[0.3pt]\bottomrule[1pt]
\end{tabular}
}
{\scriptsize
\begin{minipage}{380pt}
\setstretch{0.85}
***, **, and * denote statistical significance at the 1\%, 5\%, and 10\% levels, respectively.
Robust standard errors are reported in brackets.\\
Notes: 
The dependent variable in Columns (1)--(3) is the indicator variable for the households who did not use any credit purchases during the sample period (non-CP users).
The dependent variable in Columns (4)--(6) is the indicator variable for the non-CP users and households less than the first quartile in the sample average of credit purchases.
Columns (1)--(2) and (4)--(5) use the linear probability model (LPM), whereas Columns (3) and (6) use the Probit model.
``CV of Head's Earnings'' indicates the coefficient of variations in the head's semi-monthly earnings.
The sample mean of the CV is $5.76$ (Std. Dev. = $5.72$).
All the regressions include the family composition variables: average family size, share of children aged 6--12, share of children aged 13--16, and share of adults aged 17+ (share of children aged 0--5 is used as a reference group).
The average head's age and average head's earnings are added to the control variables in Columns 3 and 6.
$F$-statistics $p$-values for the null of the zero-slope hypothesis are reported in Columns (1)--(2) and (4)--(5).
$\chi^{2}$-statistics $p$-values for the null of the zero-slope hypothesis are listed in Columns (3) and (6).
The standard errors for the LPM are based on the HC2 estimator proposed by Horn et al. (1975).
\end{minipage}
}
\end{center}
\end{table}

Column (1) of Table~\ref{tab:r_cp} presents the result for Equation~\ref{eqn:eq_cc}.
The estimated coefficient is close to zero and statistically insignificant.
A potential concern may be the correlations between the volatility of earnings and the head's age and earnings level.
Column (2) shows the result from the extended model, including the head's age and earnings in the controls, which shows the robustness of the baseline estimate in Column (1).
Another potential issue is the functional form assumption in the linear probability model.
Column (3) then lists an alternative non-linear model, including the full set of control variables used in Column (2).
It confirms that the estimate is not statistically significantly different from zero.
Next, I changed the threshold for the dependent indicator variable to consider a wider range of households including the households with less than the first quartile of the average credit purchases.
Columns (4)--(6) summarize the results in the same column layouts as Columns (1)--(3), confirming that the estimates are still close to zero in all the specifications.
Finally, the last row in this table shows that the null of the joint zero slope hypothesis is strongly rejected in all the regressions.
Overall, there is no apparent trend in the characteristics of households that use credit for their purchases.
This suggests that retailers may not have considered the volatility of earnings when providing credit to factory workers.
This is consistent with the retailers' motivation to secure their local customers discussed earlier, and with the fact that factory workers commonly used credit for purchases.\footnote{Despite this, low-income workers might have had more difficulty using credit for purchases. A famous essay drawn by a primary school student in the 1930s depicts an example of a rice retailer who refused to buy rice under credit purchase (Toyoda 1969, pp.~346--355).}

\renewcommand{\refname}{References used in the Appendices}

\renewcommand{\refname}{Documents and Database used in the Appendices}

\renewcommand{\refname}{Archival Sources}

\end{document}